\documentclass[fleqn,10pt]{wlscirep}
\usepackage[utf8]{inputenc}
\usepackage[T1]{fontenc}
\usepackage{lineno}
\usepackage{bm}
\usepackage{adjustbox}
\usepackage{listings}         
\newcommand{\Ki}{ ~  \\ ~ \\}

\newcommand{\RmonthlyOne}{{\ttfamily Rmonthly}}
\newcommand{\RmonthlyTwo}{{\ttfamily Rmonthly2000}}
\newcommand{\CmonthlyOne}{{\ttfamily Cmonthly}}
\newcommand{\CmonthlyTwo}{{\ttfamily Cmonthly2000}}
\newcommand{\RyearlyOne}{{\ttfamily Ryearly}}
\newcommand{\RyearlyTwo}{{\ttfamily Ryearly2000}}
\newcommand{\CyearlyOne}{{\ttfamily Cyearly}}
\newcommand{\CyearlyTwo}{{\ttfamily Cyearly2000}}
\newcommand{\RmonthlyNew}{{\ttfamily Rmonthly2019}}
\newcommand{\CmonthlyNew}{{\ttfamily Cmonthly2019}}

\newcommand{\Rr}[1]{{\color{red} \bf #1}}

\newcommand{\RModel}[1]{{\M$_{#1, R}$}}
\newcommand{\CModel}[1]{{\M$_{#1,\chi^2}$}}
\newcommand{\M}{$\mathcal{M}$}
\newcommand{\UM}{{\color{black} {\bf \mathrm{UM}}}}         
\newcommand{\IF}{{\color{black} {\bf \mathrm{IF}}}}         
\newcommand{\AD}{{\color{black} {\bf \mathrm{AD}}}}         
\newcommand{\LP}{{\color{black} {\bf \mathrm{LP}}}}         
\newcommand{\Webcolour}{\color{blue}}
\newcommand{\Link}[2]{\href{#1}{\Webcolour #2}}
\newcommand{\PRtext}[1]{{{\ttfamily #1}}}
\newcommand{\REJ}{$<\!10^{-16}$}
\newcommand{\PR}[1]{{\color{black} {\ttfamily \tiny #1}}}
\newcommand{\PM}{\!\!\pm\!\!}

\newcommand{\TM}{\!\!\times\!\!}

\newcommand{\DW}{{\color{black}$^{2\times}$}}
\newcommand{\SignalOne}{S$_{11^{\mathrm{y}}}$} 
\newcommand{\SignalTwo}{S$_{10^{\mathrm{y}}}$} 
\newcommand{\SignalThree}{S$_{11.^{\mathrm{y}}86}$}
\newcommand{\SignalFour}{S$_{110^{\mathrm{y}}}$}
\newcommand{\SignalFive}{S$_{10.^{\mathrm{y}}6}$}
\newcommand{\SignalSix}{S$_{8.^{\mathrm{y}}4}$}
\newcommand{\SignalSeven}{S$_{53^{\mathrm{y}}}$}
\newcommand{\SignalEight}{S$_{8^{\mathrm{y}}}$}
\newcommand{\SignalNine}{S$_{66^{\mathrm{y}}}$}
\newcommand{\Rar}[1]{$\overrightarrow{#1}$}
\newcommand{\Lar}[1]{$\overleftarrow{#1}$}
\newcommand{\RevisedText}[1]{{\color{black} #1}}
\newcommand{\SM}{{\color{black} Non-stationary Stochastic Model}}
\newcommand{\SMul}{{\color{black} Stationary Multi-periodic Model}}
\newcommand{\Detect}{{\color{black} Consistency Criterion}}
\newcommand{\Dhypothesis}{H$_{\mathrm{Dynamo}}$}
\newcommand{\Phypothesis}{H$_{\mathrm{Planet}}$}
\newcommand{\TIDRel}{{R_{\mathrm{Tidal}}}}
\newcommand{\Pmer}{P_{\mathrm{M}}}
\newcommand{\Pven}{P_{\mathrm{V}}}
\newcommand{\Pear}{P_{\mathrm{E}}}

\newcommand{\Pjup}{P_{\mathrm{J}}}
\newcommand{\PMV}{P_{\mathrm{M,V}}}
\newcommand{\PME}{P_{\mathrm{M,E}}}
\newcommand{\PMJ}{P_{\mathrm{M,J}}}
\newcommand{\PVE}{P_{\mathrm{V,E}}}
\newcommand{\PVJ}{P_{\mathrm{V,J}}}
\newcommand{\PEJ}{P_{\mathrm{E,J}}}
\newcommand{\PMVE}{P_{\mathrm{M,V,E}}}
\newcommand{\PMVJ}{P_{\mathrm{M,V,J}}}
\newcommand{\PVEJ}{P_{\mathrm{V,M,J}}}
\newcommand{\AmpOne}{A_{\mathrm{11^y}}}
\newcommand{\AmpTwo}{A_{\mathrm{10^y}}}
\newcommand{\AmpThree}{A_{\mathrm{11^y.86}}}
\newcommand{\AmpFour}{A_{\mathrm{110^y}}}
\newcommand{\AmpFive}{A_{\mathrm{10^y.6}}}

\newcommand{\Supplementary}{{\color{black}Supplementary }}

\newcommand{\aap}{{Astron. Astrophys.}}
\newcommand{\aaps}{{Astron. Astrophys. Suppl.}}

\newcommand{\apj}{Astrophys. J.}

\newcommand{\apss}{{Astrophys. Space Sci.}}

\newcommand{\grl}{{Geophys. Res. Lett.}}

\newcommand{\mnras}{{ Mon. Not. R. Astron. Soc.}}
\newcommand{\nat}{{Nature}}

\newcommand{\solphys}{{Sol. Phys.}}
 
\newcommand{\ssr}{{Space Sci. Rev.}}

\chardef\us=`\_
\title{Do the planets cause the sunspot cycle?}
\author[1,*]{Lauri Jetsu}
\affil[1]{Department of Physics, P.O. Box 64,
FI-00014 University of Helsinki, Finland}
\affil[*]{lauri.jetsu@helsinki.fi}

\renewcommand{\Ki}{}
\begin{abstract}
The mainstream dynamo models predict that
the sunspot cycle is non-stationary and stochastic. \Ki
    The official Solar Cycle Prediction Panel 
    forecasts only the ongoing sunspot cycle
    because any
    forecast beyond one cycle is
    considered impossible. \Ki
    We analyse the sunspot data using
    our Discrete Chi-square Method (DCM). \Ki
    This method can
    detect many periodic signals
    superimposed on an arbitrary trend. \Ki
    We detect the extremely significant strong
    10, 11, and 11.86 years signals. \Ki
    The Discrete Fourier Transform (DFT)
    cross-check confirms that these signals
    are certainly real. \Ki
    The interference of the
      10 and 11 years signals can
      cause the detected
      long 110 years signal in real data.
      \Ki
      Long periods
      or interference can
      never appear in simulated non-stationary stochastic
      sunspot data.
      \Ki
     \Ki
Our deterministic
DCM model predictions are more accurate and longer
    than the official Solar Cycle Prediction Panel
    forecast. \Ki
    The DCM models can predict the past prolonged activity minima,
    like the Maunder minimum era,
    although we do not have data from these periods.
    \Ki
    We claim
    that the sunspot data are stationary and multi-periodic,
    and therefore deterministic.
    \Ki
    \Ki 
    We  connect the detected signals
    to the orbital periods of Mercury, Venus,
    the Earth
    and Jupiter. 
    \Ki 
    If the planets cause the sunspot cycle,
    the exoplanets may cause the starspot cycles
    observed in other chromospherically active stars. \\ ~ \\
    {\bf Keywords} Sunspots, Time series analysis, Solar cycle,
    Solar-planetary interaction 
  \end{abstract}
 \begin{document}

\flushbottom
\maketitle
\thispagestyle{empty}

\section*{Introduction
     \label{SectIntro}} 

In the year 1844,
Schwabe discovered the 10 years cycle
in the number of sunspots\cite{Sch44}.
Only eight years later,
Wolf revised
the cycle period to 11.1 years\cite{Wol52}.
This period is not constant
but varies between
8 and 17 years\cite{Las95}.
The amplitude of the sunspot number modulation
follows a cycle of about 80 years\cite{Gle45}.
All solar cycles have not been discovered
  from the sunspots, like
  the $P_{\mathrm{Rieger}}=154$ days
  Rieger cycle discovered from the solar
  flares \cite{Rie84} and the 205 years Suess-de Vries
  cycle discovered from
  the terrestrial radiocarbon $^{14}{\mathrm{C}}$
  record\cite{Wag01}. 
  These observations indicate
  that the sunspot data are multi-periodic.
  Multiple periods have been detected in the sunspot
  data with different time series analysis
  methods\cite{Oht94,Zhu18,Jay22}.
There is no ``phase-lock'' in
the observed sunspot cycle minimum and the maximum
epochs\cite{Wei23,Bis23}.
Such a ``phase-lock'' would not exist, if the sunspot data
were multi-periodic.


The solar surface magnetic field
is strongest in the sunspots.
The polarity of this field is reversed
at the beginning of each solar cycle. Therefore,
the field geometry returns to its
original state during the Hale cycle of
about 22 years\cite{Hal19}.
One persistent solar cycle regularity has been
the Gnevyshev–Ohl rule: 
every odd numbered sunspot cycle has had a higher
intensity (integral)
than the preceding even numbered sunspot
cycle\cite{Gne48}.

There have been prolonged periods when very few sunspots,
or even none at all, were observed.
The four most recent prolonged grand sunspot minima 
have been
classified\cite{Uso07}. 
The Dalton minimum\cite{Kom04}
was not classified as a grand sunspot minimum.
We give the epochs of these five most recent prolonged activity
minima in our Table \ref{TableEvents}.
The Earth's climate temperature rises during
strong solar activity\cite{Van99}.
Thus, the current lower level of solar activity
is not contributing to
the human-induced climate change.

Parker presented the dynamo model,
where the solar magnetic field arises from interaction
between internal differential rotation and convection\cite{Par55}.
Babcock\ and Leighton presented the first
models of how this dynamo may generate the sunspot
{cycle\cite{Bab61,Lei69}.
One recent review of the current dynamo models
of the solar cycles concludes that
all physical mechanisms causing the
solar cycles are not yet fully understood\cite{Cha10}.
For example, the physical processes   
regulating the magnetic amplitude
of the solar cycle are still unknown.
It is also not known how the sunspot-forming
magnetic flux ropes are produced by
the dynamo-generated mean magnetic field.
Very recently, it was
proposed that
  the solar magnetic cycles are caused
  by a dynamo resulting from near-surface
  magneto-rotational instability \cite{Vas24}. 

The official sunspot number
forecast for the ongoing solar cycle 25
was made in the year 2019 by
the Solar Cycle Prediction Panel,
representing NOAA, NASA and the International
Space Environmental Services (ISES).
These forecasts are made only for one cycle at the time,
because it is widely accepted that
this stochastic variability is unpredictable
beyond one solar cycle\cite{Pet20}.
Nevertheless,
countless sunspot cycle predictions are constantly
published\cite{Asi23,Jav23,Kra23}.
The physical models for solar cycle prediction
were reviewed recently\cite{Bho23}. 

In the year 1859,
Wolf already argued that the variations in the sunspot
cycle period may depend on Jupiter, Saturn,
the Earth and
Venus\cite{Wol59}. 
There have been numerous
attempts to find the connection
between the planetary motions and the sunspot
numbers\cite{Sch1911,Han78}.
In one recently formulated solar-dynamo model,
the planetary tidal forcing causes the sunspot cycle\cite{Ste19}.
The planetary tidal forces may alter the solar
structure and cause the sunspot cycle\cite{Sca22}.
It has been argued that the weak periodic tidal interaction
signal of planets requires constant amplification
because internal
solar dynamo mechanisms can otherwise mask this
signal\cite{Cha22}.

The problem with the
deterministic ``planetary-influence-theory''
is that long-term sunspot cycle predictions fail.
At the moment, in the year 2025,
the ``solar-dynamo-theory'' of a stochastic sunspot cycle
is mainstream, while the 
``planetary-influence-theory'' of
a deterministic sunspot cycle is not.
Here, we present successful 
deterministic sunspot number predictions,
which are more accurate
and extend over longer time intervals
than the official 
Solar Cycle Prediction Panel forecast.

\begin{table}[h]
  \caption{Recent prolonged periods of
    weak sunspot activity.}
  \label{TableEvents} 
  \begin{center}
\begin{tabular}{lcl}
  \hline
Minimum               &
Duration               \\
\hline
  Dalton\cite{Kom04}     & 1790 - 1830 \\
  Maunder\cite{Uso07}  & 1640 - 1720  \\ 
  Spörer\cite{Uso07}     & 1390 - 1550 \\ 
  Wolf\cite{Uso07}        & 1270 - 1340 \\ 
  Oort\cite{Uso07}        & 1010 - 1070  \\ 
  \hline
\end{tabular}
\end{center}
\end{table}

{

\section*{Results}

\begin{table}[h]
  \caption{Signals in monthly and yearly sunspot numbers.
    (1) Signal abbreviation.
    (2) \RmonthlyTwo ~sample results for pure sines
    \Supplementary Table \ref{TableRmonthly2000K410R14}).
    All six detected signals
    are given in order of decreasing
    peak to peak amplitude $A$.
    (3-13) Results for other samples with chosen $K_2$ values.
    Double sinusoid signals are denoted with ``\DW''.
    We give one example of how the contents of this table
    are connected to \Supplementary  Tables
    \ref{TableRmonthly2000K410R14}-\ref{TableCyearlyK410R14}.
    Signal \SignalOne ~is detected
    in sample \RmonthlyTwo ~(Column 2).
    Rank=''(1)'' means that this \SignalOne ~signal
    is strongest
    of all six signals in \Supplementary 
    Table  \ref{TableRmonthly2000K410R14}.
    Signal \SignalOne ~has period
    $P=11.0324$ $\pm0.0048$ and 
    peak to peak amplitude $A=98.8\pm2.6$.} %
    \label{TableCompare} 
  \begin{tiny}
     \begin{center}
       \begin{adjustbox}{angle=90}
               \begin{tabular}{ccccccccccccc}
        \hline
(1)          &
(2)          &
(3)          &
(4)          &
(5)          &
(6)          &
(7)          &
(8)          &
(9)          &
(10)         &
(11)         &
(12)         &
(13)         \\
              &
\RmonthlyTwo  &
\RmonthlyOne  &
\RmonthlyTwo  &
\RmonthlyOne  &
\CmonthlyTwo  &
\CmonthlyOne  &
\CmonthlyTwo  &
\CmonthlyOne  &
\RyearlyTwo   &
\RyearlyOne   &
\CyearlyTwo   &
\CyearlyOne   \\
              &
$K_2=1$       &
$K_2=1$       &
$K_2=2$       &
$K_2=2$       &
$K_2=1$       &
$K_2=1$       &
$K_2=2$       &
$K_2=2$       &
$K_2=1$       &
$K_2=1$       &
$K_2=1$       &
$K_2=1$       \\
Signal   &
Rank (-) &
Rank (-) &
Rank (-) &
Rank (-) &
Rank (-) &
Rank (-) &
Rank (-) &
Rank (-) &
Rank (-) &
Rank (-) &
Rank (-) &
Rank (-) \\
         &
$P$ (y)  &
$P$ (y)  &
$P$ (y)  &
$P$ (y)  &
$P$ (y)  &
$P$ (y)  &
$P$ (y)  &
$P$ (y)  &
$P$ (y)  &
$P$ (y)  &
$P$ (y)  &
$P$ (y)  \\
         &
$A$ (-)  &
$A$ (-)  &
$A$ (-)  &
$A$ (-)  &
$A$ (-)  &
$A$ (-)  &
$A$ (-)  &
$A$ (-)  &
$A$ (-)  &
$A$ (-)  &
$A$ (-)  &
$A$ (-)  \\
\hline
             & (1)               & (1)              & (1)              & (1)                  & (1)             & (1)              &  (1)               & (1)                & (1)              & (1)            & (1)            & (1)            \\
\SignalOne   &$11.0324\PM0.0048$ &$11.0033\PM0.0064$&$11.0234\PM0.0068$&$10.9878\PM0.0051    $&$10.805\PM0.013$ &$10.8585\PM0.0048$&$10.807\PM0.012    $&$21.667\PM0.014$\DW & $11.006\PM0.017$ &$10.981\PM0.020$&$10.813\PM0.029$&$10.863\PM0.022$\\
             & $98.8\PM2.6      $ &$101.5\PM2.5     $&$102.9\PM2.7     $&$104.6\PM2.3         $&$118.2\PM3.7    $&$117.5\PM2.9$     &$123.5\PM3.0       $&$132.6\PM3.4       $& $90.4\PM4.7$     &$98.3\PM7.8    $&$119.2\PM8.5$   &$119.3\PM7.5   $\\
\hline
             & (2)               & (2)              &(2)               & (2)                  & (2)             & (2)              & (2)                &                    & (2)              & (2)            & (2)            & (2)            \\
\SignalTwo   &$9.9842\PM0.0075 $ &$10.0001\PM0.0081$&$9.9882\PM0.0061 $&$20.0062\PM0.0088$\DW &$10.196\PM0.027$ &$10.0658\PM0.0077$&$10.196\PM0.017    $&                    &$9.980\PM0.016$   &$9.975\PM0.017 $&$10.184\PM0.064$&$10.058\PM0.026$\\
             &$77.4\PM2.2      $ &$65.5\PM2.1      $&$78.6\PM2.6      $&$75.2\PM2.7          $&$61.9\PM3.9    $ &$59.3\PM2.8      $&$62.1\PM3.7        $&                    &$64.6\PM7.6$      &$62.8\PM6.1    $&$62.3\PM6.8$    &$61.0\PM9.6    $\\
\hline
             &(3)                & (3)               & (4)               & (4)                 &                 & (4)              &                   &                    & (5)              & (5)            &                &   (4)          \\
\SignalThree & $11.846\PM0.012   $ &$11.807\PM0.012 $&$23.686\PM0.018$\DW&$11.770\PM0.011     $&                 &$11.863\PM0.021  $&                   &                    &$11.862\PM0.034$  &$11.820\PM0.027$&                &$11.856\PM0.068$\\
             & $60.0\PM2.2       $ &$56.4\PM2.4     $&$69.4\PM2.7      $&$62.7\PM2.6           $&                &$43.3\PM2.0      $&                   &                    &$42.6\PM4.1$      &$40.5\PM3.3    $&                &$43.8\PM7.9$    \\ 
\hline
             & (4)               &(4)               & (3)              & (3)                  &                 & (3)              &                    &                   & (4)              & (4)            &                &   (3)           \\
\SignalFour  &$96.2\PM1.0      $ &$99.92\PM0.57   $&$104.16\PM0.59   $&$104.23\PM0.68        $&                 &$116.7\PM1.3     $&                    &                   &$100.3\PM2.1$     &$101.4\PM2.4   $&                &$115.6\PM3.6$    \\
             &$50.1\PM2.8       $ &$51.8\PM2.3     $&$70.4\PM3.6      $&$68.1\PM2.7          $&                 &$52.9\PM2.5      $&                    &                   &$52.4\PM6.3$      &$51.7\PM4.9    $&                &$53.8\PM5.2$     \\
\hline
             & (5)                & (5)               & (5)             & (5)                 &                 &                  &                    &                    & (3)              & (3)            &                &                \\
\SignalFive  & $10.541\PM0.014   $&$10.569\PM0.020 $&$10.5372\PM0.0075$&$10.5407\PM0.0060    $&                 &                  &                    &                    &$10.620\PM0.033$  &$10.659\PM0.031$&                &                \\
             & $45.0\PM 1.8      $&$51.0\PM1.8     $&$51.2\PM1.8      $&$55.6\PM2.3          $&                 &                  &                    &                    &$52.4\PM8.5$      &$59.9\PM8.6    $&                &                \\
\hline
             &                   &                  & (6)              &  (6)             &                     &                  &                    &                    & (7)              & (7)            &                &                \\
\SignalSix   &                   &                  & $16.781\PM0.034 $\DW&$16.753\PM0.027$\DW&                 &                  &                    &                    &$8.460\PM0.027$   &$8.466\PM0.016 $&                &                \\
             &                   &                  &$44.9\PM4.0     $&$41.2\PM3.1       $&                     &                  &                    &                    &$33.0\PM4.4$      &$31.1\PM4.8    $&                &                \\
\hline
             &                   &  (6)             &                  &                      &                 &                  &                    &                    & (6)              & (6)            &                &                \\
\SignalSeven &                   &$52.66\PM0.27    $&                  &                      &                 &                  &                    &                    &$53.24\PM0.78$    &$53.83\PM0.76  $&                &                \\
             &                   & $33.9\PM2.0     $&                  &                      &                 &                  &                    &                    &$35.3\PM6.3      $&$34.3\PM3.6    $&                &                \\
\hline
             & (6)               & (7)              & (8)              & (8)                  &                 & (5)              &                    &                    &                  &                &                &     (5)        \\
\SignalEight  &$8.1054\PM0.0084$  &$8.1087\PM0.0076$&$8.173\PM0.011    $&$8.169\PM0.014       $&                 &$8.009\PM0.017   $&                   &                    &                  &                &                &$8.005\PM0.058$ \\
             & $36.5\PM1.7     $  &$33.6\PM1.6     $&$37.6\PM4.8       $&$35.5\PM3.3          $&                 &$25.6\PM2.5      $&                    &                    &                 &                &                &$24.0\PM7.0$     \\
\hline
             &                    &                  & (7)              & (7)                 &                 &                  &                    &                    &  (8)             & (8)            &                &                 \\
\SignalNine  &                    &                  & $136.43\PM0.99$\DW&$143.39\PM0.99$\DW  &                 &                  &                    &                    &$65.4\PM1.3$      &$66.7\PM1.4   $ &                &                 \\
             &                    &                  & $42.3\PM2.2   $  &$37.3\PM2.7         $&                 &                  &                    &                    &$29.0\PM4.7$      &$26.7\PM4.4   $ &                &                 \\
\hline 
\end{tabular}
\end{adjustbox}
\end{center}
\end{tiny}
\addtolength{\tabcolsep}{+0.05cm}
\end{table}

\subsection*{DCM period detections}
\label{SectDCMAnalysis}

\subsubsection*{Periods between 5 and 200 years}
\label{SectionLong}

We analyse eight data samples
and detect 
only nine signals in 
the period range between 5 and 200 years
(\Supplementary  Section  \ref{SectDCMone}:
  Tables
  \ref{TableRmonthly2000K410R14}
  -
  \ref{TableCyearlyK410R14}).
  Eight pure sine DCM models and four double wave DCM models
    are tested.
    The small number of detected signals
  leaves very little room for inconsistency. 
The periods and the amplitudes
of the detected nine signals
are summarised in Table \ref{TableCompare},
where we introduce
the following signal abbreviations
\SignalOne,
\SignalTwo,
\SignalThree,
\SignalFour,
\SignalFive,
\SignalSix,
\SignalSeven,
\SignalEight ~and
\SignalNine.
The numerical values in the subscripts of our
signal abbreviations refer to the signal
period in Earth years.
These signals are arranged into
the order of decreasing strength,
or equivalently, of decreasing amplitude.
We use the signal ranking based on the
signal amplitudes in the first analysed sample
(Table \ref{TableCompare}: \RmonthlyTwo),
because this order for signal amplitudes 
varies slightly in different samples.
For each particular sample, 
we give the exact signal
ranking order
in Table \ref{TableCompare}.
Our DCM detects the same signals for

\begin{itemize}
  \item[] Monthly and yearly data
  \item[] Non-weighted and weighted data
  \item[] Pure sine and double wave models
  \item[] All data samples and their subsets:
    predictive data samples
  \end{itemize}

  \noindent
The \SignalOne ~signal is detected first in every sample. 
The Fisher-test critical levels for the other signal
detections are $Q_{\mathrm{F}}$ \REJ 
~(Equation \ref{EqFisher},  \Supplementary 
Tables \ref{TableRmonthly2000K410R14}
-
\ref{TableCyearlyK410R14}).
This $10^{-16}$ limit is the highest achievable
    accuracy for the computational Fisher-test subroutine
    f.cdf in scipy.optimize
   python library.
Therefore, all DCM signal detections are extremely
significant.
The two strongest detected signals in all samples
are the \SignalOne ~and \SignalTwo ~signals,
and always in this particular order.

\subsubsection*{Periods between 0.1 and 5 years}

The orbital and synodic
  periods of inner planets are shorter than 5 years.
We can not search for these short periods in the
yearly sunspot data, because their numerical values
are close to, or below,
the one-year window period.
However, the monthly sunspot data can be used to
search for these shorter planetary periods.
Our DCM analysis in the period range between
0.1 and 5 years reveals no signs of these inner
planet orbital or synodic periods 
(\Supplementary Section  \ref{SectDCMtwo}).

\subsection*{DFT period detections}
\label{SectDFTAnalysis}

DFT pre-whitening technique\cite{Zhu18} 
is used to cross-check DCM analysis results
(\Supplementary Section \ref{SectDFT}).
We simulate data similar to the sunspot data
(\Supplementary Section \ref{SectDFTdata}).
When DCM and DFT are applied to these simulated data,
DCM analysis succeeds better than DFT analysis
(\Supplementary Sections 
\ref{SectDFTEqual} and
\ref{SectDFTUnequal}).
Regardless of its limitations,
DFT detects the same periods as DCM
(\Supplementary Section \Ref{SectDFTResults}).

\begin{figure}[h] 
\vspace{0.02\textwidth}
\centerline{\hspace*{0.005\textwidth}
 \includegraphics[width=0.25\textwidth,clip=]{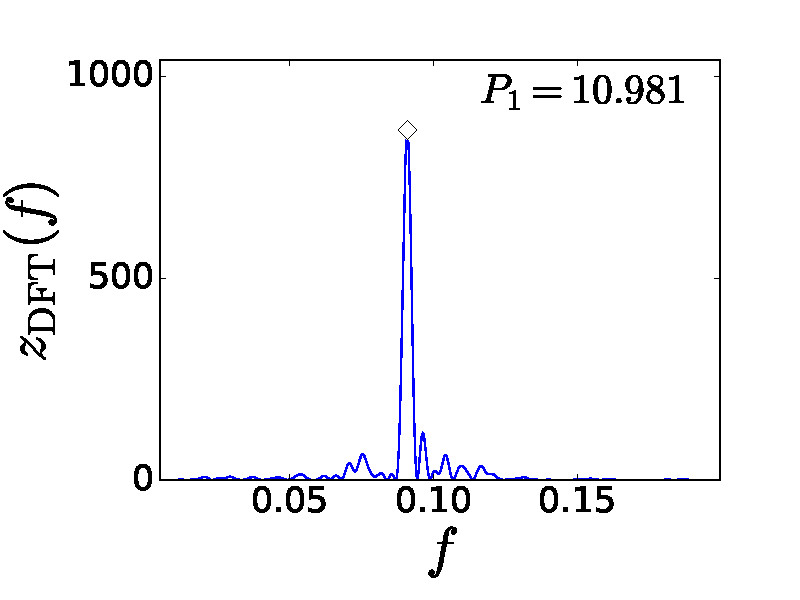} 
 \hspace*{-0.01\textwidth}
 \includegraphics[width=0.25\textwidth,clip=]{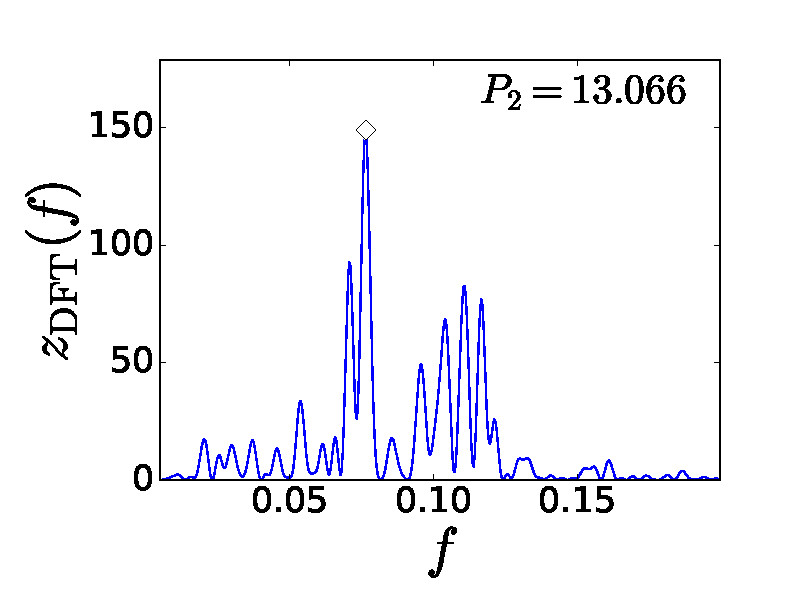}  
 \hspace*{-0.01\textwidth}
 \includegraphics[width=0.25\textwidth,clip=]{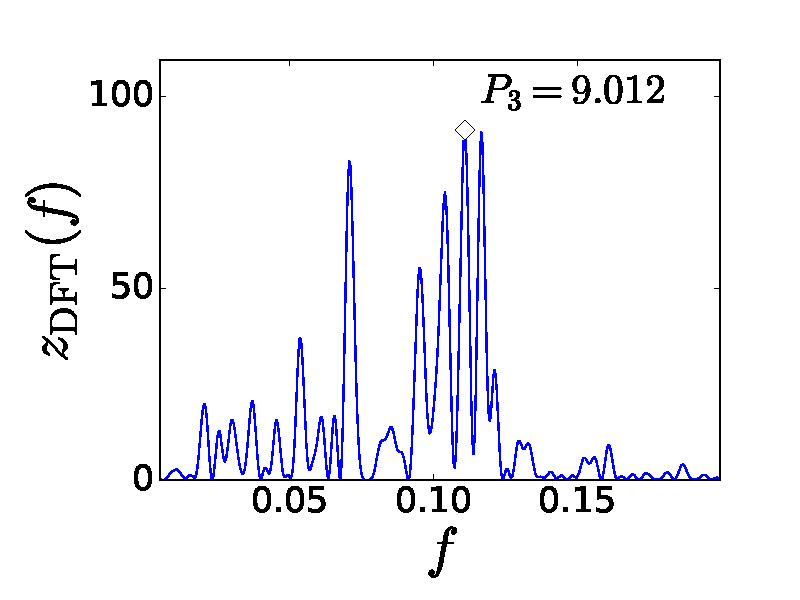}  
 \hspace*{-0.01\textwidth}
 \includegraphics[width=0.25\textwidth,clip=]{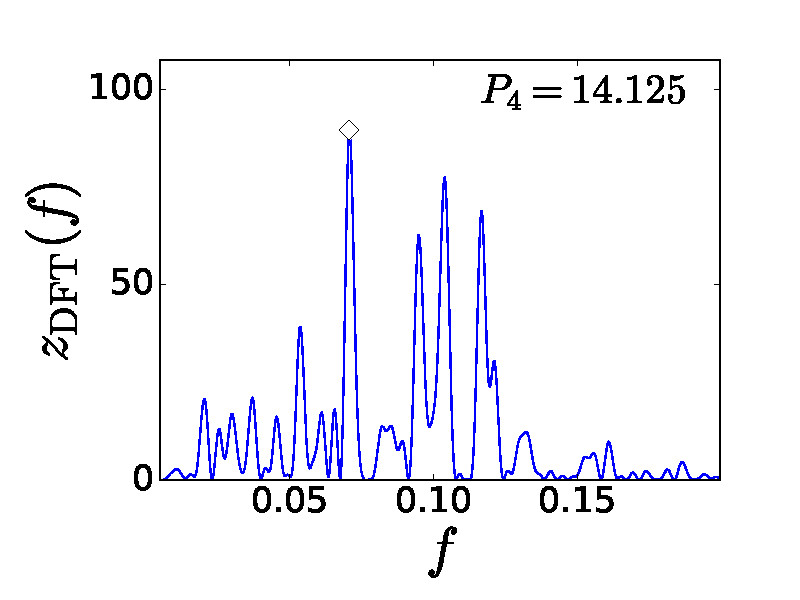}  
 }
\vspace{-0.22\textwidth}
\centerline{\normalsize \bf 
\hspace{0.11\textwidth}   \color{black}{(a)}
\hspace{0.23\textwidth}  \color{black}{(b)}
\hspace{0.22\textwidth}  \color{black}{(c)}
\hspace{0.22\textwidth}  \color{black}{(d)}
\hfill}
\vspace{0.21\textwidth}
\centerline{\hspace*{0.005\textwidth}
 \includegraphics[width=0.25\textwidth,clip=]{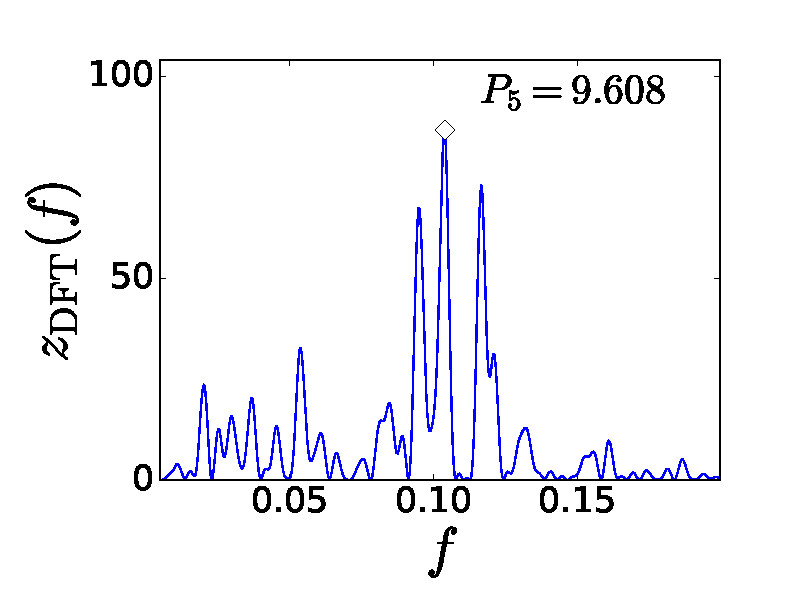} 
 \hspace*{-0.01\textwidth}
 \includegraphics[width=0.25\textwidth,clip=]{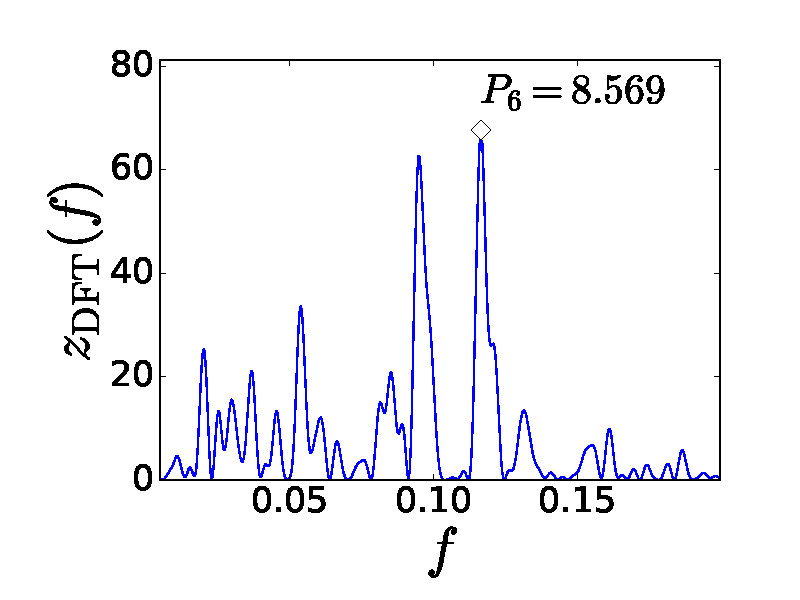} 
 \hspace*{-0.01\textwidth}
 \includegraphics[width=0.25\textwidth,clip=]{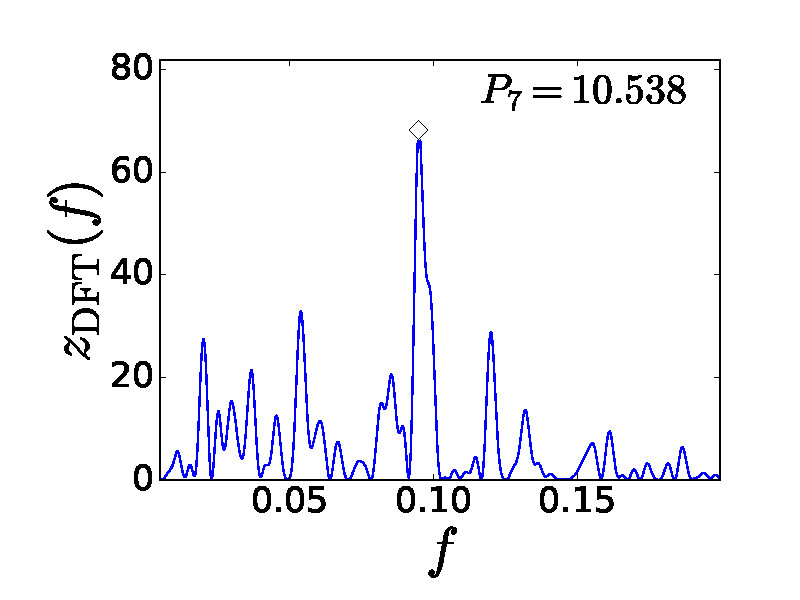} 
 \hspace*{-0.01\textwidth}
 \includegraphics[width=0.25\textwidth,clip=]{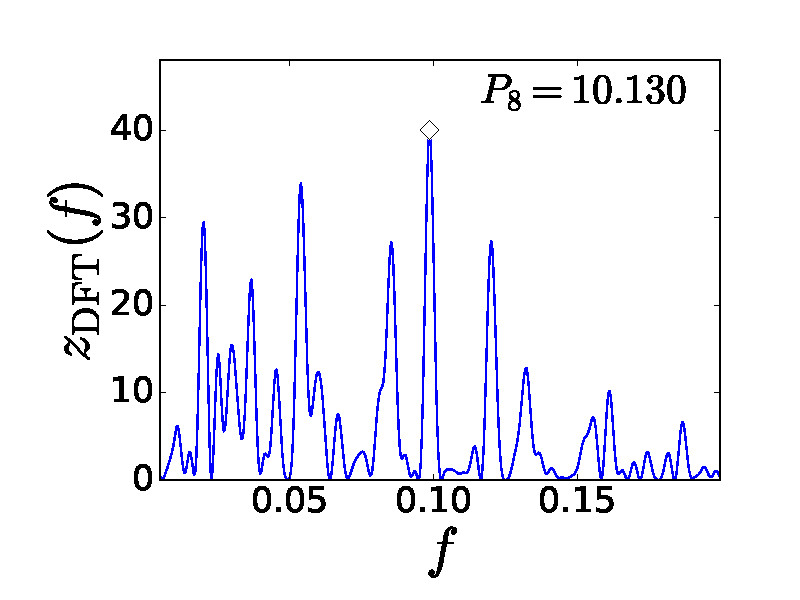} 
  }
\vspace{-0.22\textwidth}
\centerline{\normalsize \bf 
\hspace{0.11\textwidth}   \color{black}{(e)}
\hspace{0.23\textwidth}  \color{black}{(f)}
\hspace{0.22\textwidth}  \color{black}{(g)}
\hspace{0.22\textwidth}  \color{black}{(h)}
  \hfill}
\vspace{0.18\textwidth}
\caption{DFT analysis periodograms for an arbitrary
  simulated non-stationary
  stochastic dataset containing 24 sunspot cycles.
  (a) DFT periodogram for original data.
  Diamond denotes peak of periodogram at frequency
  of best period $P_1=10.981$ years.
  Units are $[f]=$ 1/y and $[z_{\mathrm{DFT}}]=$
  dimensionless.
  (b) Second $P_1=13.066$ years signal detected
  from residuals of pure sine model for original data. 
  (c-h) Signals detected  from residuals of next pure sine models. 
} 
\label{FigDFTReviewC24}
\end{figure}

\begin{figure}  
\vspace{0.02\textwidth}
\centerline{\hspace*{0.005\textwidth}
 \includegraphics[width=0.25\textwidth,clip=]{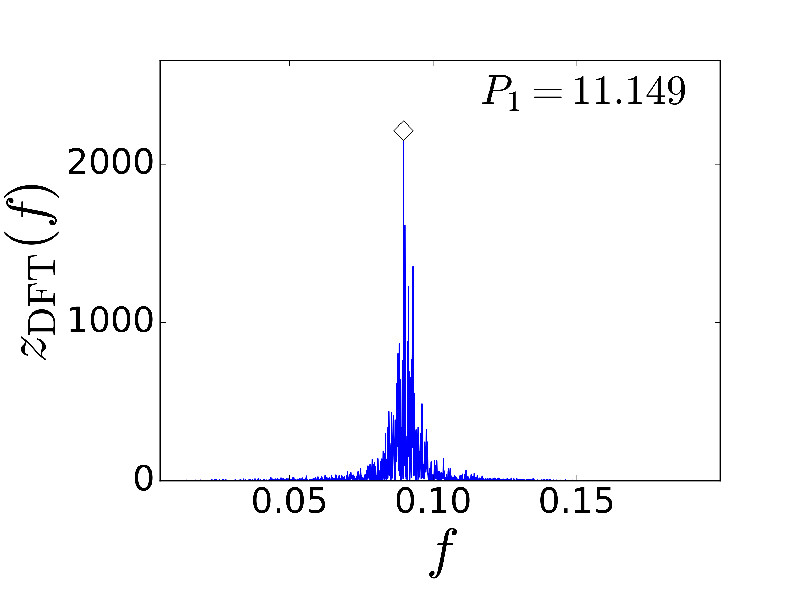} 
 \hspace*{-0.01\textwidth}
 \includegraphics[width=0.25\textwidth,clip=]{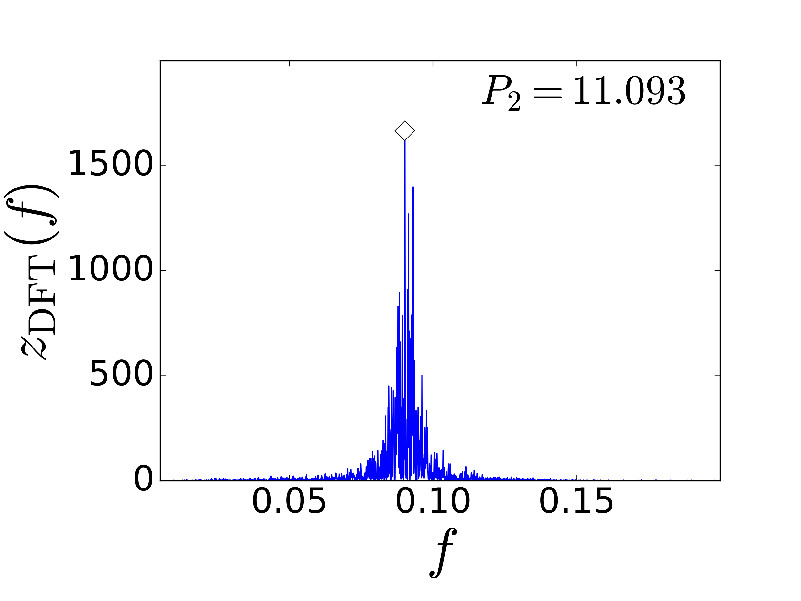} 
 \hspace*{-0.01\textwidth}
 \includegraphics[width=0.25\textwidth,clip=]{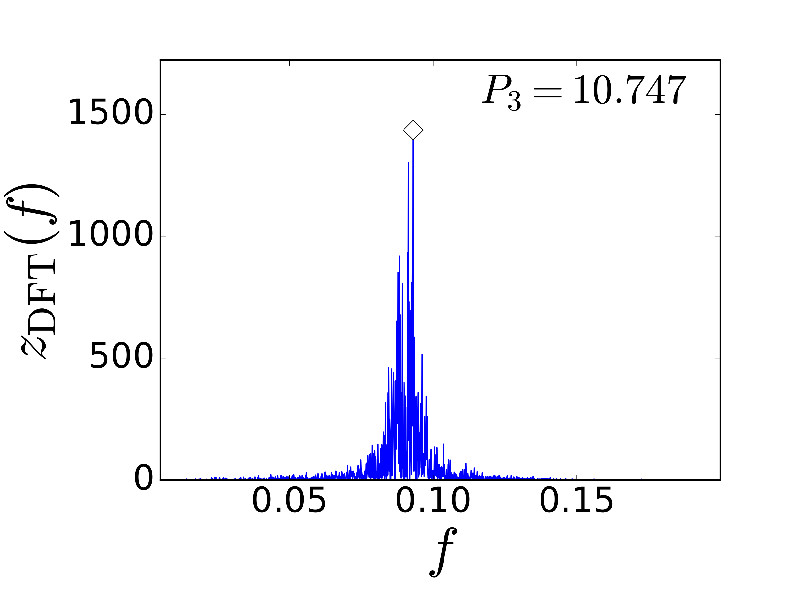} 
 \hspace*{-0.01\textwidth}
 \includegraphics[width=0.25\textwidth,clip=]{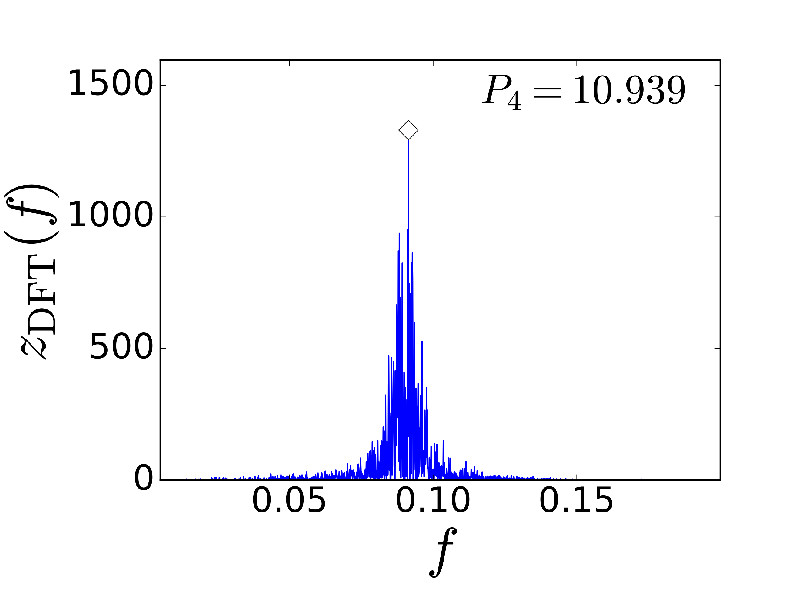} 
 }
\vspace{-0.22\textwidth}
\centerline{\normalsize \bf 
\hspace{0.11\textwidth}   \color{black}{(a)}
\hspace{0.23\textwidth}  \color{black}{(b)}
\hspace{0.22\textwidth}  \color{black}{(c)}
\hspace{0.22\textwidth}  \color{black}{(d)}
\hfill}
\vspace{0.21\textwidth}
\centerline{\hspace*{0.005\textwidth}
 \includegraphics[width=0.25\textwidth,clip=]{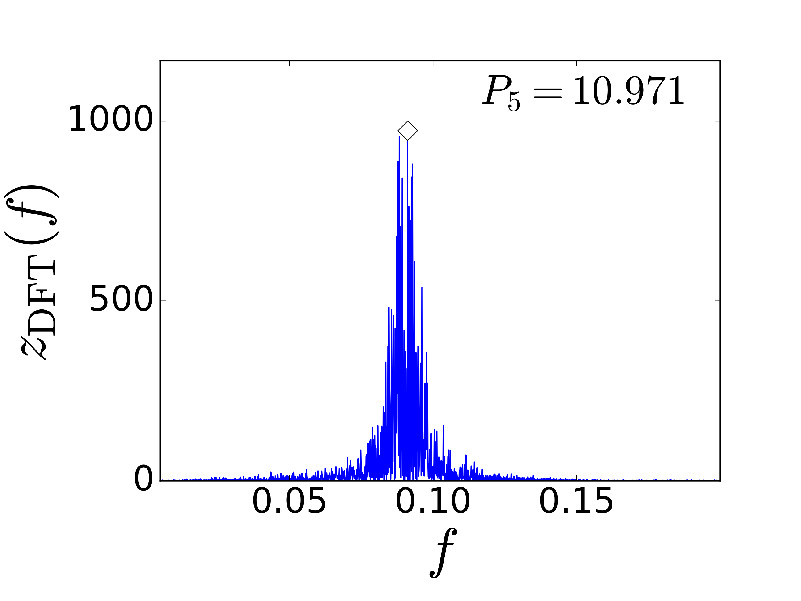} 
 \hspace*{-0.01\textwidth}
 \includegraphics[width=0.25\textwidth,clip=]{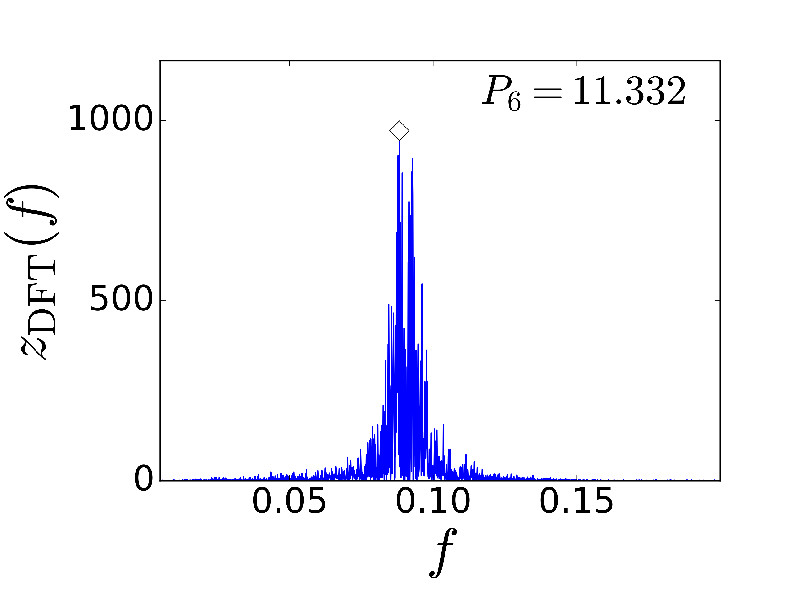} 
 \hspace*{-0.01\textwidth}
 \includegraphics[width=0.25\textwidth,clip=]{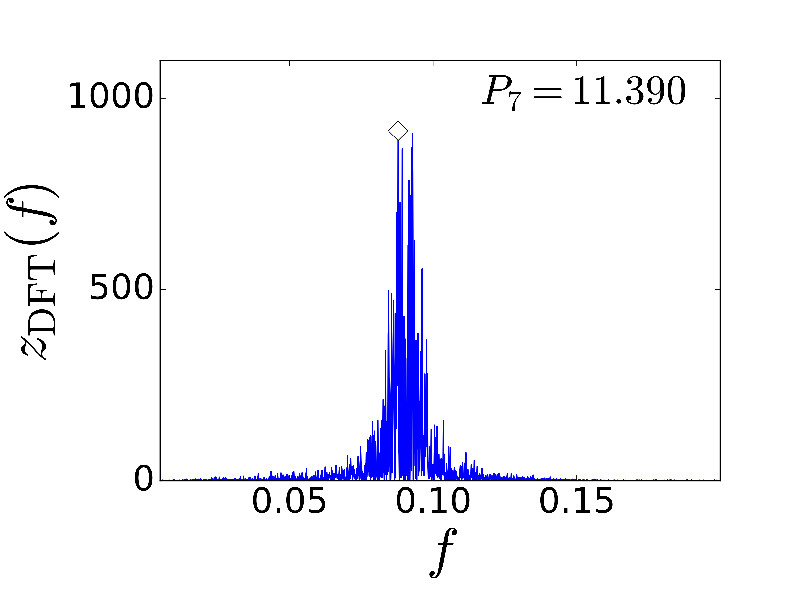} 
 \hspace*{-0.01\textwidth}
 \includegraphics[width=0.25\textwidth,clip=]{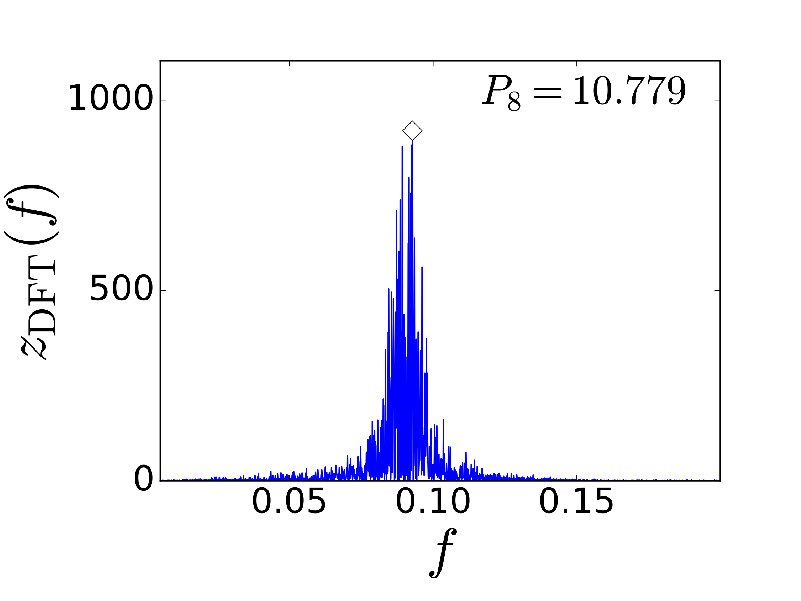} 
  }
\vspace{-0.22\textwidth}
\centerline{\normalsize \bf 
\hspace{0.11\textwidth}   \color{black}{(e)}
\hspace{0.23\textwidth}  \color{black}{(f)}
\hspace{0.22\textwidth}  \color{black}{(g)}
\hspace{0.22\textwidth}  \color{black}{(h)}
  \hfill}
\vspace{0.18\textwidth}
\caption{DFT pre-whitening analysis periodograms for an arbitrary
  simulated non-stationary
  stochastic dataset containing 1000 sunspot cycles.
  Notations are as in Fig. \ref{FigDFTReviewC24}.
} 
\label{FigDFTReviewC1000}
\end{figure}

\begin{figure}  
\vspace{0.02\textwidth}
\centerline{\hspace*{0.005\textwidth}
 \includegraphics[width=0.80\textwidth,clip=]{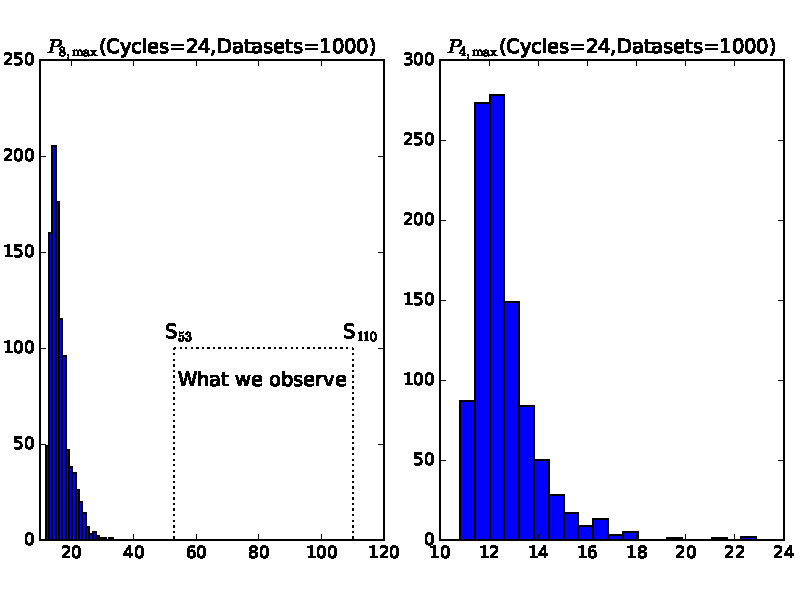} 
 }
\vspace{-0.61\textwidth}
\centerline{\normalsize \bf 
\hspace{0.30\textwidth}   \color{black}{(a)}
\hspace{0.35\textwidth}  \color{black}{(b)}
\hfill}
\vspace{0.55\textwidth}
\centerline{\hspace*{0.005\textwidth}
 \includegraphics[width=0.80\textwidth,clip=]{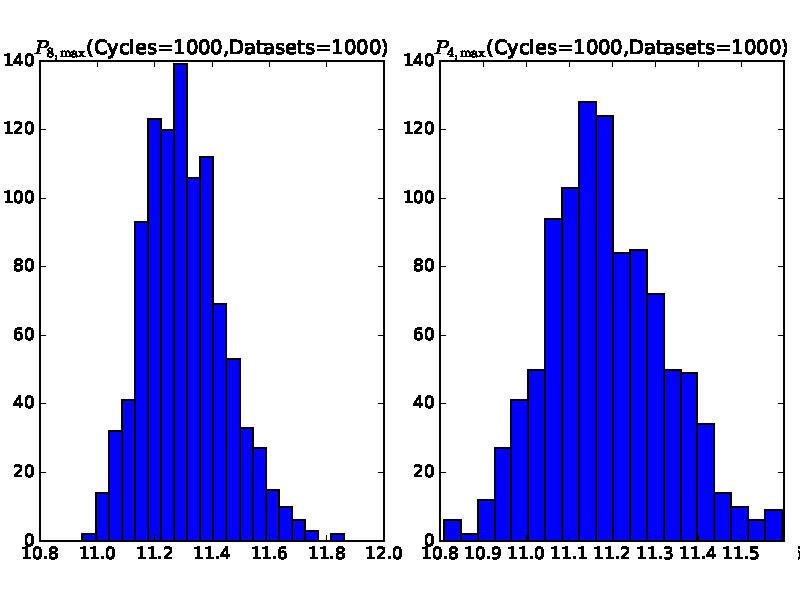} 
 }
\vspace{-0.61\textwidth}
\centerline{\normalsize \bf 
\hspace{0.30\textwidth}   \color{black}{(c)}
\hspace{0.35\textwidth}  \color{black}{(d)}
\hfill}
\vspace{0.55\textwidth}
\caption{Longest 
  $P_{\mathrm{8,max}}$ and $P_{\mathrm{4,max}}$ periods
  that \SM ~can reproduce (20 bins).
  (a-b) One thousand simulated
  stochastic datasets containing 24 sunspot cycles.
  Dotted black lines in ``(a)'' denote observed
    \SignalSeven ~and \SignalFour ~signals.
  (c-d) One thousand simulated
  stochastic datasets containing 1000 sunspot cycles.  
} 
\label{FigDFTDistributions}
\end{figure}

\subsection*{Signals are not artefacts detected
from non-stationary stochastic data}
      \label{SectArtefact}

  \renewcommand{\Ki}{}
    \RevisedText{In this section, we  simulate artificial sunspot data 
     using  the \SM,  which mimics the observed solar cycle.
     \Ki
        We will show that this model can not reproduce all DCM and DFT period
        detections.
      \Ki}

    \renewcommand{\Ki}{}
      Our whole DCM and DFT time series analysis
      is based on the ``basic assumption''
      that the sunspot data are stationary and multi-periodic.
      \Ki
       We call our model the \SMul. 
    \Ki
      The results of our analysis are wrong,
      if the sunspot data are non-stationary and stochastic,
      in which case there can not be any multi-periodicity.
      \Ki
      In this section, we simulate stochastic non-stationary datasets.
      \Ki
      The simulated cycle shape is a pure sine.
      \Ki
      We assume that the period and the amplitude of
      every sunspot cycle are stochastic.
      \Ki
      The simulated random sunspot cycle lengths
      are uniformly distributed between 9 and 14 years.
      \Ki
      The uniformly distributed simulated random sine curve amplitudes,
      the half of the peak to peak amplitudes,
      are between 50 and 150.
      \Ki
      We compute these simulated datasets for the monthly time points
      and add noise having a standard deviation of 42,
      like
      in our other simulations
      (\Supplementary Equation \ref{EqSimData}).
      \Ki 
      Our simulated datasets are
      non-stationary, weakly stochastic and
      contain no multi-periodicity.

    \renewcommand{\Ki}{}
      First, we simulate non-stationary stochastic
      sunspot datasets containing 24 sunspot cycles.
      \Ki
      For every simulated
      dataset, we
      perform an exactly the same kind of
      DFT pre-whitening analysis as for
      the real data
      (\Supplementary Fig. \ref{FigDFTSunSpots}).
      \Ki 
      The eight DFT periodograms for one such arbitrary simulated
      non-stationary stochastic
      dataset are shown in Fig. \ref{FigDFTReviewC24}.
      \Ki
      The DFT pre-whitening
      detects many extremely significant periods
      at both sides of 11 years.
      \Ki
      Except for the first
      original simulated data
      periodogram 
      (Fig. \ref{FigDFTReviewC24}a),
      all other periodograms show more than
      one dominant peak
      (Fig.s \ref{FigDFTReviewC24}b-h).
      \Ki
      This supports the conclusion
      that the many DFT
      periodogram peaks for the real data 
      (\Supplementary Fig. \ref{FigDFTSunSpots})
      may be artefacts arising
      from the analysis of a too short non-stationary
      stochastic dataset.
      \RevisedText{ Soon we shall show that this conclusion is wrong.}
      
     \renewcommand{\Ki}{}
      \Ki
      Then, we simulate non-stationary stochastic
      datasets containing 1000 sunspot cycles.
      \Ki
      The eight DFT periodograms for one such arbitrary
      simulated dataset are shown in Fig. \ref{FigDFTReviewC1000}.
      \Ki
      All eight
      periodograms for this longer simulated dataset
      display only one broad dominant peak.
      \Ki
      This supports the conclusion
      that our many real data 
      periodogram peaks 
      (\Supplementary Fig. \ref{FigDFTSunSpots})
      may be artefacts
      that will vanish when the non-stationary
      stochastic dataset becomes longer.
      \RevisedText{Soon we shall show that this conclusion is also wrong.}
      
    \renewcommand{\Ki}{}
      \Ki
       The above non-stationary stochastic dataset
      simulations were \RevisedText{presented to us}.
      \Ki
      These simulations were claimed to prove
      that all our DCM and DFT analysis
      results \RevisedText{can not be trusted}.
      \Ki
      These simulations had the apparent potential to make or
      break our efforts.
      \Ki
      We can not, however, wait for some
      $1000 \times 11$ years
      to find out whether our period
      detections are artefacts or not.
    \renewcommand{\Ki}{}
      \Ki
      \RevisedText{
        Fortunately,
        it is easy to show that
        the \SM ~is not the best model for the sunspot data.
        \Ki
        We test the following \Detect
        \begin{itemize}
        \item[] ``The best model  {\it for} the data is the model that
          consistently
          reproduces
       {\it all} periods that the time series analysis methods detect {\it from}
       the data.''
       \end{itemize}
       \Ki
       No one can question this criterion.
       \Ki
        Our next simulations will
        show that the \SM ~fails the \Detect.
        \Ki
        \Ki}

    \renewcommand{\Ki}{}
      \RevisedText{We} simulate one thousand 
      non-stationary stochastic
      datasets containing 24 sunspot cycles.
      \Ki
      We repeat the same DFT search for the eight
      best periods in every dataset.
      \Ki
      Our notation for the
      largest period detected
      in each dataset is $P_{\mathrm{8,max}}$.
      \Ki
      We also take the four first strongest
      periods detected in each dataset,
      and select their largest period value  $P_{\mathrm{4,max}}$.
      \Ki
      The $P_{\mathrm{8,max}}$ and $P_{\mathrm{4,max}}$
      distributions in all
      one thousand simulated non-stationary
      stochastic datasets containing 24 sunspot cycles
      are shown on Fig.s \ref{FigDFTDistributions}a-b.
      \Ki
      The $P_{\mathrm{8,max}}$ distribution concentrates
      close to 15 years, and all values are below 34 years.
      \Ki
      The peak of the $P_{\mathrm{4,max}}$ distribution is
      at about 12 years, and not a single value exceeds 24 years.
    \renewcommand{\Ki}{}
       \RevisedText{\Ki
         In these \SM ~simulations, all detected periods
         are shorter than the periods of the
         extremely significant
         fourth strongest
         \SignalFour ~signal and
         seventh strongest
         \SignalSeven ~signal detected from real sunspot data.
         \Ki
         These two long period signals can not be artefacts because
         the \SM ~simulations fail to reproduce them.
         \Ki
         The black dotted lines in
           Fig. \ref{FigDFTDistributions}a highlight how much
           the longest simulated \SM ~
           period  $P_{\mathrm{8,max}}$ distribution  contradicts the
           longest \SignalFour 
           ~and \SignalSeven ~signal
           periods detected from the observed real sunspot data.
           \Ki
           The emphasis is on the word ``observed''.
         \Ki
       \Ki}

    \renewcommand{\Ki}{}
      Finally, we simulate one thousand non-stationary stochastic
      datasets containing 1000 sunspots cycles
      (Fig.s \ref{FigDFTDistributions}c-d).
      \Ki
      The  $P_{\mathrm{8,max}}$ and  $P_{\mathrm{4,max}}$
      distributions concentrate close to 11 years.
      \Ki
      The largest possible values for $P_{\mathrm{8,max}}$
      are close to Jupiter's period. The
       $P_{\mathrm{4,max}}$ values are shorter.
       \RevisedText{\Ki
         All detected periods
         in these \SM ~simulations of 1000 sunspot cycles
         are much shorter than the periods of
         the extremely significant real data
         \SignalFour ~and \SignalSeven ~signals.
         \Ki
         There is no way that
         these two long period signals can be artefacts
         caused by the \SM.
         \Ki
       \Ki}

           For three reasons,
           our DCM and DFT analysis results for the real 
           data rule out the \SM ~alternative.

      \begin{itemize}

    \renewcommand{\Ki}{}
      \item[1.] Long periods:
        The period of our seventh strongest \SignalSeven ~signal
      detected in the {\it real data} exceeds
      all  $P_{\mathrm{8,max}}$ values in 
      {\it all simulated} non-stationary stochastic datasets
      (Figs. \ref{FigDFTDistributions}a and \ref{FigDFTDistributions}c).
      \Ki
      Our fourth strongest \SignalFour ~signal period detected
      in the {\it real data} is certainly
      outside the $P_{\mathrm{4,max}}$ distribution limits
      of {\it all simulated} non-stationary
      stochastic datasets
      (Figs. \ref{FigDFTDistributions}b
      and \ref{FigDFTDistributions}d).
      \Ki
      Neither one of these longer period
      \SignalFour ~and \SignalSeven ~signals
      {\it never} appears alone,
      let alone together, in any simulated
      non-stationary stochastic dataset.
      \Ki
      Nevertheless, both of these two
      longer periods are detected
      in the {\it real data}.
      
  \renewcommand{\Ki}{}
     \RevisedText{ \Ki
       These results confirm that the  \SM ~fails the \Detect.
       \Ki
       The \SMul ~passes the \Detect ~because it
       can reproduce {\it all} periods
       that the DCM and DFT time series analysis methods detect
       {\it from} the real data.
      \Ki
      Hence, the  \SMul ~is a better model
      for the data than the \SM.
      \Ki
       In general, a model can not be correct if it
       reproduces only some detected signals (e.g., short periods),
       but not all detected signals (e.g., long periods).
      In other words, one can not arbitrarily pick and choose
      which signals are real and which ones are artefacts.
      \Ki
       }
      
    \renewcommand{\Ki}{}
      \item[2.] Interference:
      The reason for not detecting
      the \SignalFour ~signal in the simulated
      non-stationary stochastic datasets is simple.
      \Ki
      The interference of the \SignalOne ~and \SignalTwo ~signals
      in the {\it real data} can cause this \SignalFour ~signal
      having a period close to the Gleissberg cycle period.
      \Ki
      Furthermore, interference connects all our
      five strongest DCM and DFT signal detections.
      \Ki
      We give their $\pm1$ round synodic period connections
\begin{equation}
P_{\pm}= [P_1^{-1} \pm P_2^{-}]^{-1}
\label{EqConnections}
\end{equation}
in Table \ref{TableConnections},
which also shows that
the interference of
  \SignalThree ~and  \SignalFour ~signals
  can cause the \SignalFive ~signal.
  \Ki
  Hence, the \SignalOne, \SignalTwo ~and \SignalThree
  ~signals may be the only real ones.
  \Ki
  We refer to these five strongest signals
  as the`''interference quintet''.
   \Ki
   This kind of interference is {\it impossible}
   for the \SM.
   \RevisedText{There is a zero-probability that {\it all}
     five strongest signals detected from some random
     non-stationary stochastic dataset are artefacts, which
     can be connected
     by a single simple formula (Equation \ref{EqConnections}).
   Yet, this is what we detect from the real sunspot data.}

    \renewcommand{\Ki}{}

\item[3.] Predictability:
     \Ki
     We will present many successful predictions.
     \Ki
      Here, we mention only two of them:
      our Maunder minimum era predictions
      (Figs. \ref{FigCmonthlySine}b and \ref{FigCyearlySine}b),
      and
      our current sunspot cycle 25 predictions
      (Fig. \ref{FigCycleComparison})
      that are
      better than the official Solar Cycle Panel prediction.
      \Ki
      These predictions would be {\it impossible}
      for non-stationary stochastic sunspot data.
      \Ki
      
      \end{itemize}

      \noindent
      Ironically,
      the \SM ~simulations 
      confirm that the DCM and DFT period detections can not be artefacts.
      \Ki
      Our ``basic assumption'' of stationary
      and multi-periodic sunspot data is
      correct.

       \begin{table}[h]
       \caption{Interference quintet.
         (1) $P_1$ period. (2-5) $P_2$ period.
         Next lines: $P_{\pm}$ periods
         for $P_1$ and $P_2$ combinations  (Equation \ref{EqConnections}).
    Signals \SignalTwo ~and \SignalOne ~connect to signal \SignalFour.
    Signals \SignalThree ~and \SignalFour ~connect to signal \SignalFive.
    Signal \SignalFive ~has no direct connection to signals
    \SignalTwo ~or \SignalOne.}
  \label{TableConnections} 
\begin{center}
  \begin{tabular}{ccccc}
\hline
(1) &
(2) &
(3) &
(4) &
(5) \\
\hline
 &$P_2=$ \SignalFive&$P_2=$ \SignalOne &$P_2=$ \SignalThree &$P_2=$ \SignalFour \\
\hline
$P_1=$ \SignalTwo  &$P_{-}=177$&$P_{-}=110 $&$P_{-}=63.8$&$P_{-}=11.0$\\
                   &$P_{+}=5.15$&$P_{+}=5.24$&$P_{+}=5.42$&$P_{+}=9.17$ \\
\hline
$P_1=$ \SignalFive &           &$P_{-}=292 $&$P_{-}=99.8$&$P_{-}=11.7$ \\
                   &           &$P_{+}=5.40$&$P_{+}=5.60$&$P_{+}=9.67$ \\
\hline
$P_1=$ \SignalOne  &           &            &$P_{-}=152$&$P_{-}=12.2$ \\
                   &           &            &$P_{+}=5.71$&$P_{+}=10.0$ \\
\hline 
$P_1=$ \SignalThree&           &            &           &$P_{-}=13.3$ \\
                   &           &            &           &$P_{+}=10.7$ \\ 
           \hline
  \end{tabular}
  \end{center}
\end{table}

\subsection*{Predictions}
  \label{SectPredictions}

   The
   two alternative hypotheses for the solar cycle
  are

\begin{itemize}

\item[] \Dhypothesis: {\it ``The \SM ~is correct.
    Therefore, deterministic
    sunspot number predictions
    longer than one solar cycle fail.''}

\item[] \Phypothesis: {\it ``The \SMul ~is correct.
    Therefore, deterministic sunspot number predictions 
    longer than one solar cycle can succeed.''} 

\end{itemize}

\begin{figure}[h] 
  \centerline{\includegraphics[width=0.5\textwidth,clip=]{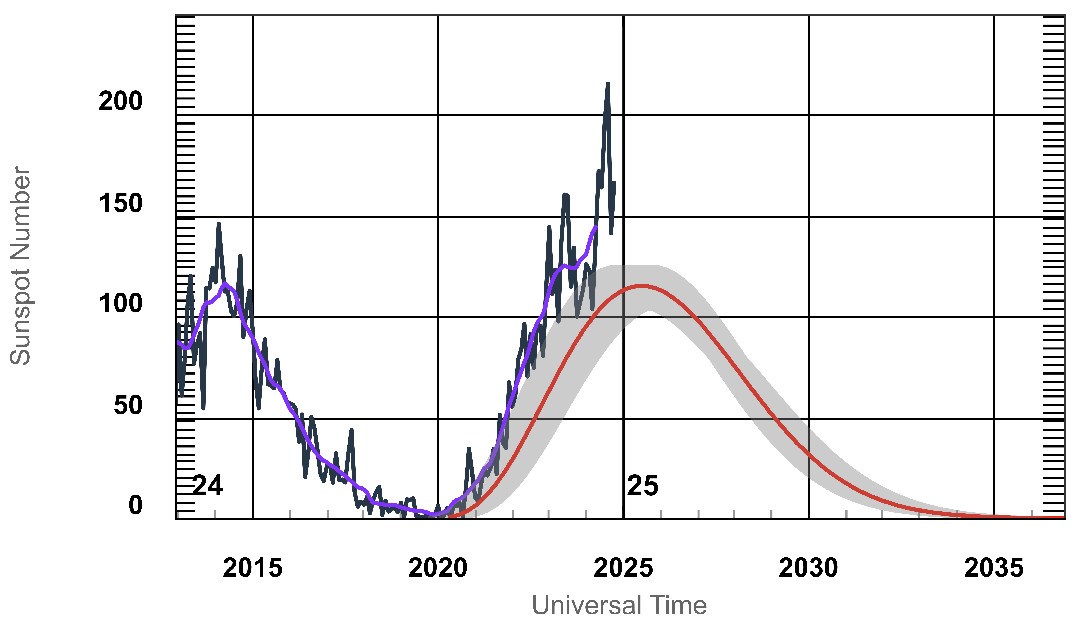}} 
  \caption{Solar Cycle Prediction Panel forecast for solar cycle 25.
    Red curve shows predicted monthly sunspot numbers
    after December 2019.
    Grey shaded area shows prediction error.
    Black curve 
    denotes observed monthly sunspot
    number up to October 2024.
    Blue line denotes observed smoothed 13 months mean
    $y_{13}$ up to April 2024.
    (Source:
    \Link{https://www.swpc.noaa.gov/products/solar-cycle-progression}
    {Space Weather Prediction Center on November 21, 2024,
      https://www.swpc.noaa.gov/products/solar-cycle-progression})
    }  
\label{FigISES}
\end{figure} 

\subsubsection*{Solar Cycle Prediction Panel forecast
  \label{SectPanel}}

It is crucial to check
what level of predictability has already been reached.
In the year 2019,
the Solar Cycle Prediction Panel
representing NOAA, NASA and the International
Space Environmental Services (ISES)
made the official forecast
for solar cycle 25.
The red curve in
Fig. \ref{FigISES} shows the synthesis of about
50 different forecasts received from
the scientific community.
During the current sunspot cycle 25,
  the maximum difference between the observed $y_{13}$
  and the predicted $g_{13}$ 
  smoothed thirteen month mean 
  has been 
\begin{eqnarray}
  s_{\mathrm{ISES}}= |y_{13}-g_{13}|_{\mathrm{max}}=|\epsilon|_{\mathrm{max}}= 52.0
  \label{EqISES}
\end{eqnarray}
on March 2024.
This gives an
estimate for the accuracy
that could be achieved
in sunspot number predictions for the past five years
after 2019.
We do not argue that this is the best achievable accuracy,
but this is the current best official published forecast.
However, this accuracy would certainly be worse
  if this forecast covered more than one solar cycle.
The $s_{\mathrm{ISES}}$ parameter
of Equation \ref{EqISES}
is hereafter referred to as
the ``five year ISES limit''.
If our model prediction deviates less
than $s_{\mathrm{ISES}}$ from the predicted observations,
we state that the prediction is
``within the ISES limit''.
In such cases, our prediction $g(t)$ deviates
less than $s_{\mathrm{ISES}}$ from $y_{13}$.

 We evaluate predictivity $z_{\mathrm{pred}}$
 for the six pairs of
 predictive and predicted data
 (Equation \ref{EqPredz}).
  The six all data samples are used to
  predict $m_{\mathrm{pred}}$ for
  the past prolonged activity minima
(Equation \ref{EqPredMean}).
We will show many predictions
extending
to time intervals before and
after the analysed samples
(Figs.
\ref{FigRmonthly2000Sine},
\ref{FigRmonthly2000Double},
\ref{FigCmonthlySine},
\ref{FigRyearlySine} and
\ref{FigCyearlySine}).
The red continuous and dotted lines
to the denote the model and the model errors,
respectively.
The sliding thirty years mean of this model and the sliding
mean error
are denoted with the green continuous and dotted curves.
The thirty years sliding mean
predictions are always more
accurate than the model predictions
    (Figs.
    \ref{FigRmonthly2000Sine}-\ref{FigCyearlySine}:
  red curves).
  Therefore, the green curve errors
  (dotted green lines) are always
  smaller than the red curve errors
  (dotted red lines).
  The red model curve predictions are shown only for the first
  few cycles after the newest, most accurate sunspot data.
  Since the oldest sunspot data are sparse and inaccurate, 
  we show no red curve predictions before these data.
  The thirty years sliding mean green curve predictions
  are shown for the time intervals before and after the sunspot data.
To ensure easy readability of all
Figs. \ref{FigRmonthly2000Sine}-\ref{FigCyearlySine},
we surround the realm of our predictions
with  black dash-dotted rectangles.
We also use long black horizontal arrows to
highlight the direction
of these predictions in time.
The dash-dotted cyan
lines in all model residuals $\epsilon_i$ plots
denote the five year ISES limits 
$\pm s_{\mathrm{ISES}}$  (Equation \ref{EqISES}).

     \subsubsection*{Non-weighted monthly sunspot data
\label{SectNonWeighted}}


{\bf Pure sines.} Sample \RmonthlyTwo ~contains the predictive data.
The predicted data are
the \RmonthlyOne ~sample observations
after the year 2000.
The whole \RmonthlyOne ~sample represents all data.

The smallest
predicted test statistic $z_{\mathrm{pred}}$
(Equation \ref{EqPredz})
value
is achieved for the five pure sines \M=5 model
(Fig. \ref{FigRmonthly2000Sine}a).
Predictability is better for one signal than for two signals.
Then, this $z_{\mathrm{pred}}$
predictability improves for three,
four and five signals.
This $z_{\mathrm{pred}}$ improvement trend
  indicates that
  these five first signals are deterministic, not stochastic.
Six signals give nearly the same predictability.
For more than six signals, predictability becomes worse.
Furthermore, the seven and eight signal models
are unstable (\Supplementary Table  \ref{TableRmonthly2000K410R14}:
``$\UM$'' models \M=7 and 8).
We compute  $z_{\mathrm{pred}}$ values also for
these unstable models, just to verify,
if predictability stops improving due to instability.

For the predictive \RmonthlyTwo ~data sample,
the five signal prediction after 2000 
succeeds quite well until a clear deviation
from the data occurs in
the year 2013 (Fig. \ref{FigRmonthly2000Sine}b).
Most of the blue dots denoting the
  the predictive model residuals are
  within the horizontal cyan dash dotted
  lines of ISES limit (Equation \ref{EqISES}).

The five pure sines model can reproduce the Dalton minimum
because it is inside the predictive
data  (Fig. \ref{FigRmonthly2000Sine}c).
There is a dip at the end of the Maunder minimum,
but the prediction
for the duration of the Maunder minimum fails.

The DCM pure sine model analysis results for \RmonthlyOne ~are
given in \Supplementary Table \ref{TableRmonthlyK410R14}. 
This all data sample gives
the predicted mean level 
$m_{\mathrm{pred}}$
(Equation \ref{EqPredMean})
values for the past
activity minima.
These predictions are not convincing
(Fig. \ref{FigRmonthly2000Sine}d).
Only the blue and yellow circles denoting Sporer and
Oort minima are below the black horizontal dotted
line denoting the mean level $m=81.6$ 
of all sunspot numbers.
\RmonthlyOne ~sample pure sines prediction
indicates that the current prolonged
low solar activity level has just ended
 (Fig. \ref{FigRmonthly2000Sine}e: green curve).
The long-term past and future predictions are shown
in  Fig. \ref{FigRmonthly2000Sine}f.

\begin{figure}  
\vspace*{-0.08\textwidth}
  \centerline{\hspace*{0.015\textwidth}
         \includegraphics[width=0.515\textwidth,clip=]{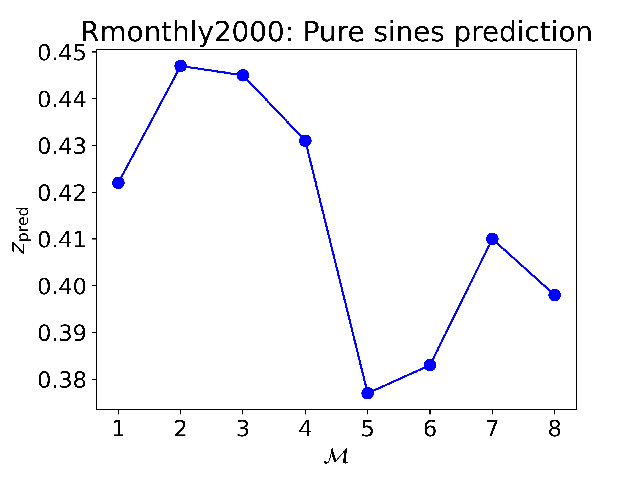} 
         \hspace*{-0.03\textwidth}
         \includegraphics[width=0.515\textwidth,clip=]{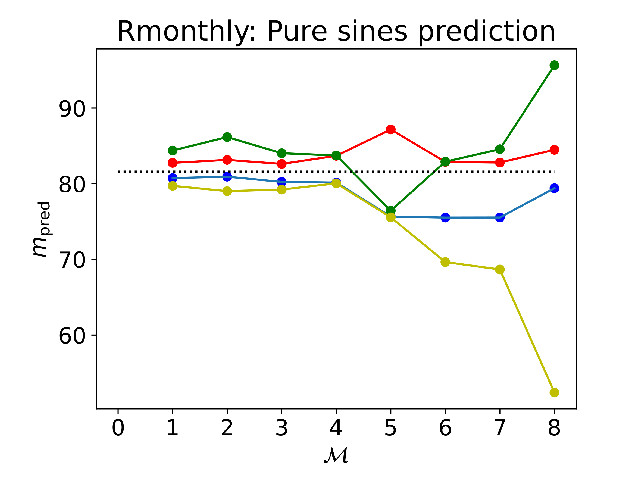}       
        }
\vspace{-0.33\textwidth}
\centerline{\Large \bf 
\hspace{0.38\textwidth}  \color{black}{(a)}
\hspace{0.43\textwidth}  \color{black}{(d)}
\hfill}
\vspace{0.29\textwidth}    
\centerline{\hspace*{0.015\textwidth}
         \includegraphics[width=0.515\textwidth,clip=]{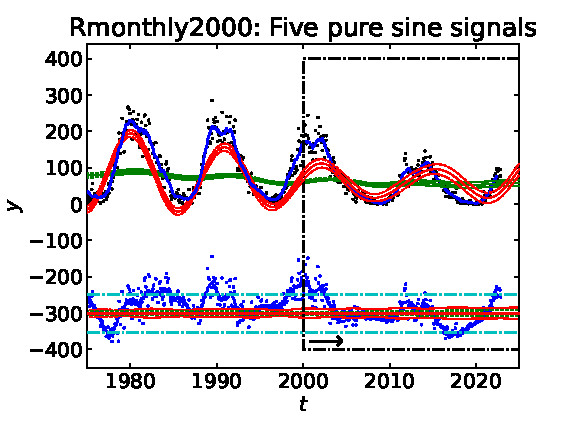} 
         \hspace*{-0.03\textwidth}
         \includegraphics[width=0.515\textwidth,clip=]{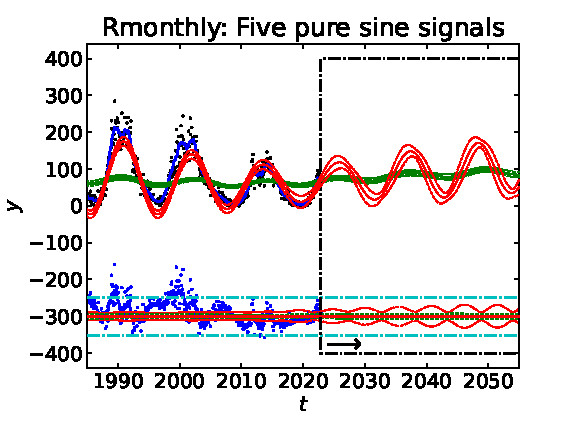}          
        }
\vspace{-0.33\textwidth}   
\centerline{\Large \bf     
\hspace{0.38 \textwidth}   \color{black}{(b)}
\hspace{0.43 \textwidth}   \color{black}{(e)}
   \hfill}
 \vspace{0.20\textwidth}    
\vspace{0.08\textwidth}    
\centerline{\hspace*{0.015\textwidth}
         \includegraphics[width=0.515\textwidth,clip=]{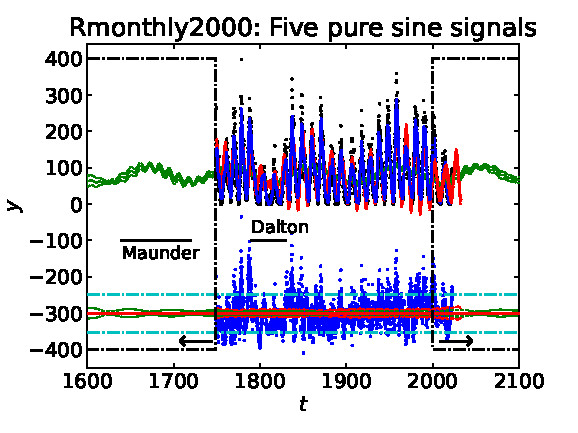} 
         \hspace*{-0.03\textwidth}
         \includegraphics[width=0.515\textwidth,clip=]{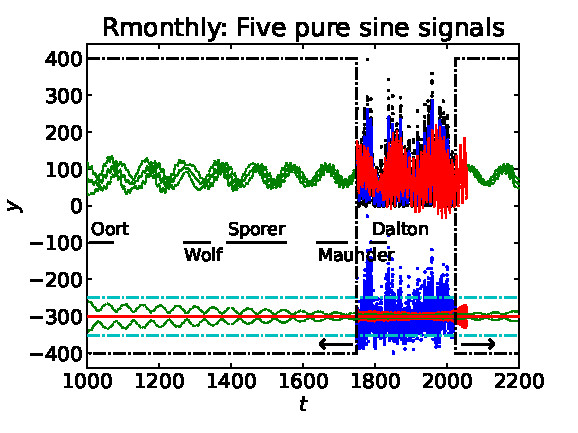}         
        }
\vspace{-0.33\textwidth}   
\centerline{\Large \bf     
\hspace{0.38 \textwidth}   \color{black}{(c)}
\hspace{0.43 \textwidth}   \color{black}{(f)}
   \hfill}
 \vspace{0.27\textwidth}    
 \caption{ (a) \RmonthlyTwo ~predictions for pure sines
   (\Supplementary Table   \ref{TableRmonthly2000K410R14}).
   (a) Predicted test statistic
           $z_{\mathrm{pred}}$
           (Equation \ref{EqPredz}) for pure sine models \M=1-8.
           Best model \M=5 minimises $z_{\mathrm{pred}}$.
           Units are x-axis [\M]=dimensionless and
           y-axis $[z_{\mathrm{pred}}]=$ dimensionless.
           (b) Black dots denote data $y_i$.
           Blue continuous line denotes
             13 month smoothed data.
            Red continuous line denotes  $g(t)$ model \M=5.
           Dotted red lines denote $\pm3\sigma$
           model error limits for $g(t)$.
           Green continuous line denotes model $g(t)$ 
           thirty years sliding mean.
           Dotted green lines show  $\pm3\sigma$
           error limits for this sliding mean.
             Predictive data ends at vertical dash-dotted
             black line.
             Predicted data, predicted model and
             predicted thirty years mean
             are surrounded by this
             black dash-dotted line.
             Black arrow shows prediction direction in time.
             Blue dots denote
           residuals $\epsilon_i$ offset
           to level -300, which is
           outlined with red continuous line.
           Dash-dotted cyan lines denote
           ISES prediction $s_{\mathrm{ISES}}$
           error limit (Equation \ref{EqISES}).
           Units are x-axis $[t]=$ years and
           y-axis $[y]=$ dimensionless.
           (c) Horizontal black continuous lines outline
           time intervals of
           Dalton and Maunder minima.
           Except for a longer time span,
           otherwise as in ``b''.
           (d) \RmonthlyOne ~predictions for pure sines
           (\Supplementary Table   \ref{TableRmonthlyK410R14}).
           Predicted mean level $m_{\mathrm{pred}}$
           (Equation \ref{EqPredMean})
           for pure sine \M=1-8 models during
           Maunder (red), Sporer (blue), Wolf
           (green) and Oort (yellow) minima.
           Dotted black line shows the mean $m=81.6$
           for all $y_i$ data.
           Units are x-axis [\M]=dimensionless and y-axis
           $[m_{\mathrm{pred}}]=$ dimensionless.
           (e) \M=5 model for modern times and
           its prediction after year 2022.
           Otherwise as in ``b''.
           (f) \M=5 model for long-term
           past and future. Otherwise as in ``c''
         }
         \label{FigRmonthly2000Sine}
       \end{figure}

\begin{figure}  
\centerline{\hspace*{0.015\textwidth}
         \includegraphics[width=0.515\textwidth,clip=]{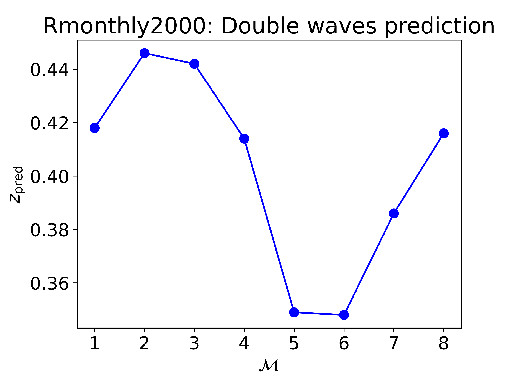} 
         \hspace*{-0.03\textwidth}
         \includegraphics[width=0.515\textwidth,clip=]{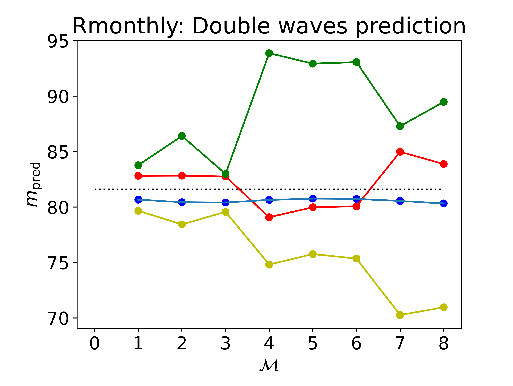}        
        }
\vspace{-0.33\textwidth}
\centerline{\Large \bf 
\hspace{0.38\textwidth}  \color{black}{(a)}
\hspace{0.43\textwidth}  \color{black}{(d)}
\hfill}
\vspace{0.29\textwidth}    
\centerline{\hspace*{0.015\textwidth}
         \includegraphics[width=0.515\textwidth,clip=]{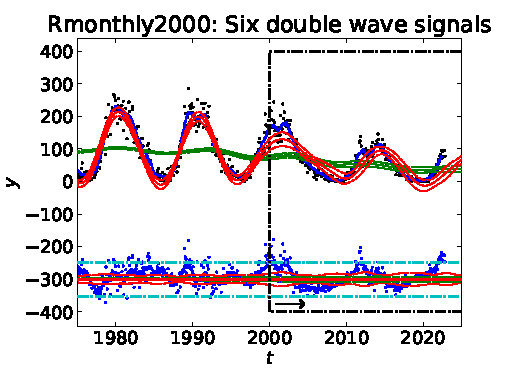} 
         \hspace*{-0.03\textwidth}
         \includegraphics[width=0.515\textwidth,clip=]{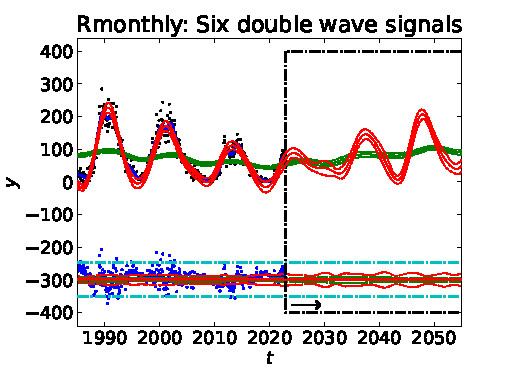}         
        }
\vspace{-0.33\textwidth}   
\centerline{\Large \bf     
\hspace{0.38 \textwidth}   \color{black}{(b)}
\hspace{0.43 \textwidth}   \color{black}{(e)}
   \hfill}
 \vspace{0.20\textwidth}    
\vspace{0.08\textwidth}    
\centerline{\hspace*{0.015\textwidth}
         \includegraphics[width=0.515\textwidth,clip=]{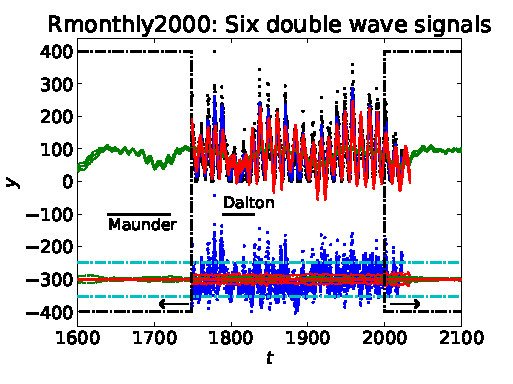} 
         \hspace*{-0.03\textwidth}
         \includegraphics[width=0.515\textwidth,clip=]{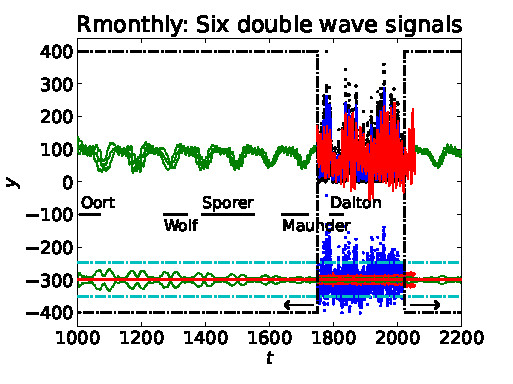}        
        }
\vspace{-0.33\textwidth}   
\centerline{\Large \bf     
\hspace{0.38 \textwidth}   \color{black}{(c)}
\hspace{0.43 \textwidth}   \color{black}{(f)}
   \hfill}
 \vspace{0.30\textwidth}    
 \caption{\RmonthlyTwo ~predictions for double waves
      (\Supplementary Table   \ref{TableRmonthly2000K420R14}).
   (a-c) Best double wave model \M=6 minimises
   $z_{\mathrm{pred}}$ (Equation \ref{EqPredz}). Otherwise as in
   Figs. \ref{FigRmonthly2000Sine}a-c.
   (d-f) \RmonthlyOne ~predictions for double waves
     (\Supplementary Table   \ref{TableRmonthlyK420R14}).   
   Otherwise as in  Figs. \ref{FigRmonthly2000Sine}d-f.
 } 
         \label{FigRmonthly2000Double}
       \end{figure}


{\bf Double waves.}       
The predictive data sample is \RmonthlyTwo.
All data sample is \RmonthlyOne.
The \RmonthlyOne ~sample values after the year 2000
are the predicted data.

The DCM double wave model results
for \RmonthlyTwo ~and 
\RmonthlyOne ~samples are given
in \Supplementary 
Tables  \ref{TableRmonthly2000K420R14}  and
  \ref{TableRmonthlyK420R14}, respectively.
For both samples,
the periods and amplitudes of the eight
strongest double wave signals
are summarised in Table \ref{TableCompare}
(Columns 4 and 5).

The prediction test statistic
$z_{\mathrm{pred}}$ (Equation \ref{EqPredz})
decreases until six signals are detected.
Then, $z_{\mathrm{pred}}$ begins to increase
for seven and eight signals
(Figure \ref{FigRmonthly2000Double}a).
This indicates that the first six detected signals
  are real, but the next seventh and eighth signals are not.

The prediction obtained from
the predictive data \RmonthlyTwo ~sample
begins after the black vertical dash-dotted
line at the year 2000  (Fig. \ref{FigRmonthly2000Double}b).
For twenty years,
  the majority of predictive model residuals (blue dots)
  are within the ISES limit
  (Equation \ref{EqISES}: cyan dash dotted lines).
The six signal double wave DCM model
can reproduce the Dalton minimum
inside \RmonthlyTwo.
The thirty years sliding mean
  shows a dip at the end of the Maunder minimum.
  This dip outside sample
  \RmonthlyTwo ~does not cover
  the whole Maunder minimum era (Fig. \ref{FigRmonthly2000Double}c:
green curve).

The predictive data \RmonthlyTwo ~sample
$z_{\mathrm{pred}}$ (Equation \ref{EqPredz})
values indicate the presence of six signals
(Fig. \ref{FigRmonthly2000Double}a).
The all data \RmonthlyOne ~sample
mean level prediction
$m_{\mathrm{pred}}$ (Equation \ref{EqPredMean})
for the Maunder minimum
supports this result.
The predicted mean level $m_{\mathrm{pred}}$
curve for the Maunder minimum
shows an expected dip
(Fig. \ref{FigRmonthly2000Double}d: red curve).
For the first three \M=1-3 signals,
this red $m_{\mathrm{pred}}$ curve
is above the all data mean level $m$
denoted with dotted black line. 
For the next three \M=4-6 signals,
the $m_{\mathrm{pred}}$ values
fall below the $m$ mean level.
Finally, these  $m_{\mathrm{pred}}$ values
for \M=7 and 8 signals
rise again above the mean level $m$.

The Maunder minimum is outside
all data sample \RmonthlyOne.
It is the closest activity minimum
before the beginning of this sample.
This could explain why
  only this particular
  deterministic activity minimum prediction succeeds.
The $m_{\mathrm{pred}}$ predictions fail
for the other activity minima, probably
because these minima are further
away in the past than the Maunder minimum.
The decreasing Oort minimum  $m_{\mathrm{mean}}$ trend deserves
to be mentioned (Fig. \ref{FigRmonthly2000Double}d: yellow curve).

The six double wave DCM model
for all data \RmonthlyOne ~sample 
predicts that the mean level of solar activity begins to rise
in the near future
(Fig. \ref{FigRmonthly2000Double}e: green curve).
This model predicts a turning
point at 2029, which does not necessarily represent
the beginning of a new activity cycle
(Fig. \ref{FigRmonthly2000Double}e: red curve).
The long-term thirty years sliding mean shows
a clear dip during the end of Maunder minimum
(Fig. \ref{FigRmonthly2000Double}f: green curve).

\begin{figure}  
\centerline{\hspace*{0.015\textwidth}
         \includegraphics[width=0.515\textwidth,clip=]{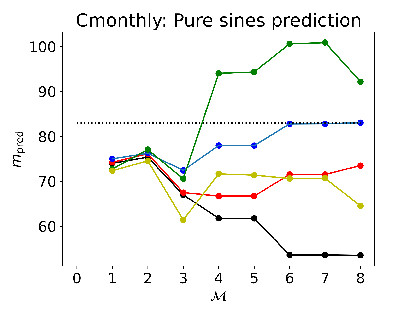} 
         \hspace*{-0.03\textwidth}
         \includegraphics[width=0.515\textwidth,clip=]{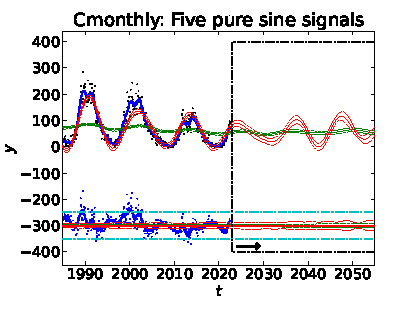}          
        }
\vspace{-0.33\textwidth}
\centerline{\Large \bf 
\hspace{0.38\textwidth}  \color{black}{(a)}
\hspace{0.43\textwidth}  \color{black}{(c)}
\hfill}
\vspace{0.29\textwidth}    
\centerline{\hspace*{0.015\textwidth}
         \includegraphics[width=0.515\textwidth,clip=]{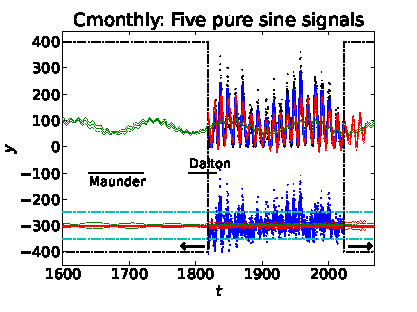} 
         \hspace*{-0.03\textwidth}
         \includegraphics[width=0.515\textwidth,clip=]{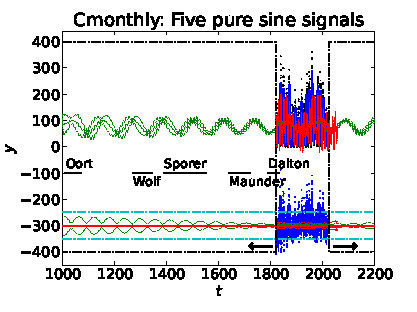} 
        }
\vspace{-0.33\textwidth}   
\centerline{\Large \bf     
\hspace{0.38 \textwidth}   \color{black}{(b)}
\hspace{0.43 \textwidth}   \color{black}{(d)}
   \hfill}
 \vspace{0.30\textwidth}    
 \caption{\CmonthlyOne ~predictions for pure sines
   (\Supplementary Table   \ref{TableCmonthly2000K410R14}).
   (a) Pure sine \M=1-8  model predictions for $m_{\mathrm{pred}}$
(Equation \ref{EqPredMean})
   during
           Dalton (black),
           Maunder (red), Sporer (blue), Wolf (green) and Oort(yellow) minima.
           Otherwise as in Fig. \ref{FigRmonthly2000Sine}d.
           (b) Model \M=5 pure sine prediction for near past.
           Otherwise as in Fig. \ref{FigRmonthly2000Sine}c.
           (c) Model \M=5 pure sine prediction for near future.
           Otherwise as in Fig. \ref{FigRmonthly2000Sine}e.
           (d) Model \M=5 pure sine long-term predictions.
           Otherwise as in Fig. \ref{FigRmonthly2000Sine}f.
         } 
         \label{FigCmonthlySine}
 \end{figure}

\subsubsection*{Weighted monthly sunspot data}

The \CmonthlyTwo ~sample contains the predictive data.
The \CmonthlyOne ~sample contains all data,
where the observations after the year
2000 are the predicted data.

For the predictive data \CmonthlyTwo ~sample,
both pure sine and double wave
models are unstable for more than two signals
(\Supplementary Tables
  \ref{TableCmonthly2000K410R14}
  and
  \ref{TableCmonthly2000K420R14}:``$\UM$'').
The detected signals are given in Table \ref{TableCompare}
(Columns 6 and 8). It is of no use
to study predictivity
for only two detected signals
in sample \CmonthlyTwo.

Only one double wave signal is detected in \CmonthlyOne
  ~(Table \ref{TableCompare}: Column 9).
  It would make no sense to study predictivity for
only one signal (\Supplementary Table  \ref{TableCmonthlyK420R14}).

Fortunately, we can detect five pure sine
signals in all data sample
(\Supplementary Table  \ref{TableCmonthlyK410R14}).
These signals are given in Table \ref{TableCompare} (Column 7).
The predicted mean level $m_{\mathrm{pred}}$ (Equation \ref{EqPredMean})
curves are promising (Fig. \ref{FigCmonthlySine}a).
The $m_{\mathrm{pred}}$  curves for all activity minima
in Table \ref{TableEvents}
show a dip at model \M=3.
This dip is clearly below the dotted black
line denoting the mean $m=81.6$  
of all monthly sunspot numbers.
Except for the green Wolf minimum curve,
all $m_{\mathrm{pred}}$ curves stay below this
mean $m$.
The predicted mean level $m_{\mathrm{pred}}$ curve for the
Dalton minimum is convincing
(Fig. \ref{FigCmonthlySine}a: black curve),
most probably because this activity minimum between
1790 and 1830
partly overlaps
the \CmonthlyOne ~sample between 1818 and 2022.
Furthermore, the predicted mean level  $m_{\mathrm{pred}}$ curve
for the second closest
Maunder minimum is good
(Fig. \ref{FigCmonthlySine}a: red curve).

The deterministic predictions for the Dalton and the Maunder
minima are illustrated in
Fig. \ref{FigCmonthlySine}b (green curve).
These curves can reproduce the duration of these
activity minima, but not the low values of sunspots.
The errors for the predicted model (red dotted lines)
  and the predicted sliding 30 years mean (green dotted lines) are
  far smaller than the ISES limit
  (Equation \ref{EqISES}: dashed dotted cyan lines).
This deterministic prediction is obtained
  for the two past centuries before the beginning of data. 
  The time span of this successful
  prediction is about twenty times longer than
  the one sunspot cycle limit of stochastic
  \Dhypothesis ~hypothesis predictions.
  Since this model can predict the time intervals
    of past weak activity, it should
also be able to predict the time intervals of future weak activity.
The same five pure sine signal prediction for
the next decades is shown in
Fig. \ref{FigCmonthlySine}c (red curve).
The long-term prediction confirms the result already
obtained earlier for other samples: the activity level of the Sun
increases during the next half a century
(Fig. \ref{FigCmonthlySine}d: green curve).

\begin{figure}  
\centerline{\hspace*{0.015\textwidth}
         \includegraphics[width=0.515\textwidth,clip=]{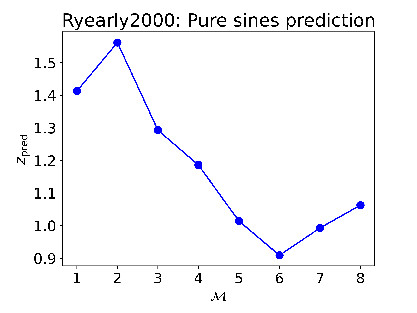} 
         \hspace*{-0.03\textwidth}
         \includegraphics[width=0.515\textwidth,clip=]{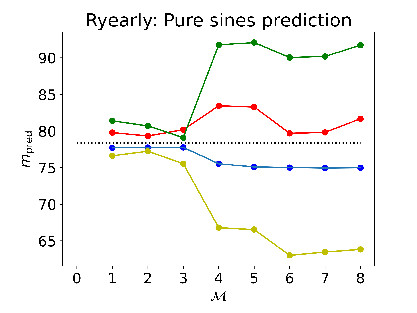}       
        }
\vspace{-0.33\textwidth}
\centerline{\Large \bf 
\hspace{0.38\textwidth}  \color{black}{(a)}
\hspace{0.43\textwidth}  \color{black}{(d)}
\hfill}
\vspace{0.29\textwidth}    
\centerline{\hspace*{0.015\textwidth}
         \includegraphics[width=0.515\textwidth,clip=]{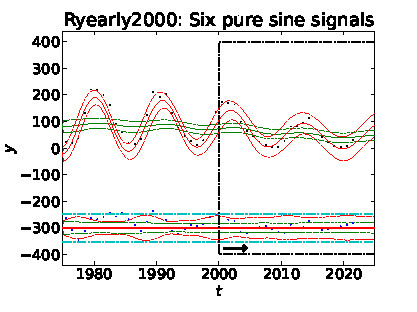} 
         \hspace*{-0.03\textwidth}
         \includegraphics[width=0.515\textwidth,clip=]{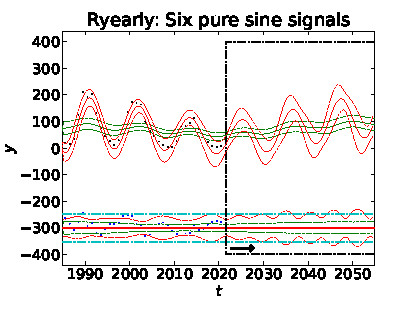}         
        }
\vspace{-0.33\textwidth}   
\centerline{\Large \bf     
\hspace{0.38 \textwidth}   \color{black}{(b)}
\hspace{0.43 \textwidth}   \color{black}{(e)}
   \hfill}
 \vspace{0.20\textwidth}    
\vspace{0.08\textwidth}    
\centerline{\hspace*{0.015\textwidth}
         \includegraphics[width=0.515\textwidth,clip=]{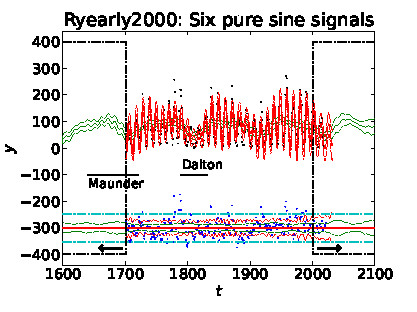} 
         \hspace*{-0.03\textwidth}
         \includegraphics[width=0.515\textwidth,clip=]{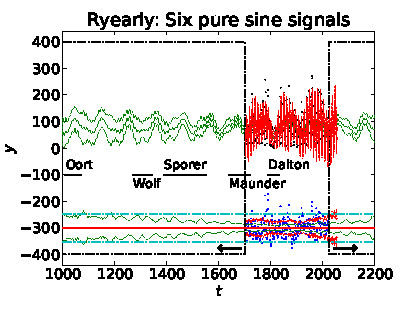}         
        }
\vspace{-0.33\textwidth}   
\centerline{\Large \bf     
\hspace{0.38 \textwidth}   \color{black}{(c)}
\hspace{0.43 \textwidth}   \color{black}{(f)}
   \hfill}
 \vspace{0.30\textwidth}    
 \caption{
(a-c) \RyearlyTwo ~predictions for pure sines
            (\Supplementary Table \ref{TableRyearly2000K410R14}).
            Otherwise as in Fig. \ref{FigRmonthly2000Sine}a-c.
            \RyearlyOne ~predictions for pure sines
                         (\Supplementary Table \ref{TableRyearlyK410R14}).
                       Otherwise as in Fig. \ref{FigRmonthly2000Sine}d-f.
                     } 
         \label{FigRyearlySine}
       \end{figure}

\begin{figure}  
\centerline{\hspace*{0.015\textwidth}
         \includegraphics[width=0.515\textwidth,clip=]{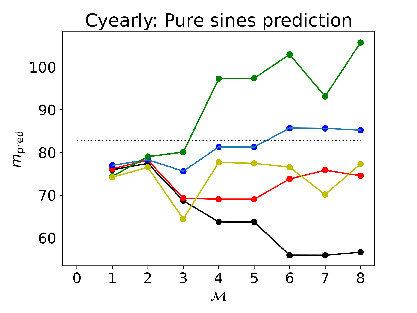} 
         \hspace*{-0.03\textwidth}
         \includegraphics[width=0.515\textwidth,clip=]{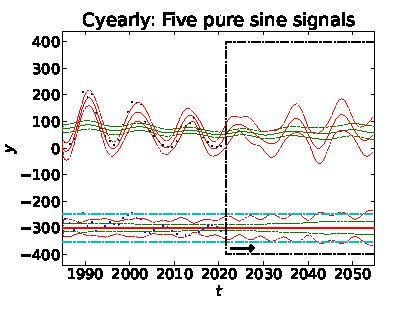}           
        }
\vspace{-0.33\textwidth}
\centerline{\Large \bf 
\hspace{0.38\textwidth}  \color{black}{(a)}
\hspace{0.43\textwidth}  \color{black}{(c)}
\hfill}
\vspace{0.29\textwidth}    
\centerline{\hspace*{0.015\textwidth}
         \includegraphics[width=0.515\textwidth,clip=]{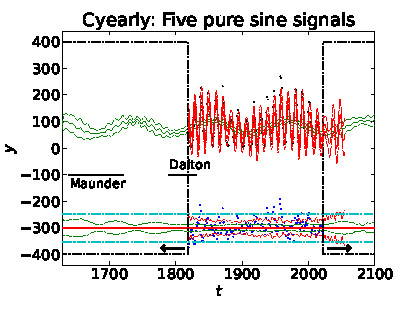}         
         \hspace*{-0.03\textwidth}
         \includegraphics[width=0.515\textwidth,clip=]{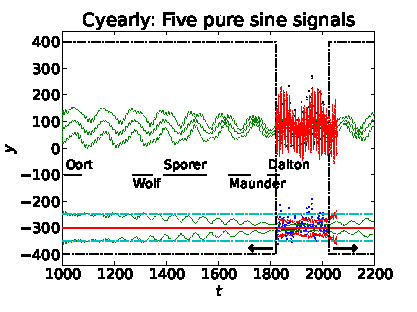}          
        }
\vspace{-0.33\textwidth}   
\centerline{\Large \bf     
\hspace{0.38 \textwidth}   \color{black}{(b)}
\hspace{0.43 \textwidth}   \color{black}{(d)}
   \hfill}
 \vspace{0.30\textwidth}    
 \caption{(a-d) \CyearlyOne ~predictions
              (\Supplementary Table \ref{TableCyearlyK410R14}).
            Otherwise as in Fig. \ref{FigCmonthlySine}a-d.
          } 
          \label{FigCyearlySine}
        \end{figure}

\subsubsection*{Non-weighted yearly sunspot data}

The \RyearlyOne ~sample contains all data,
where observations after 2000 are
the predicted data.
The \RyearlyTwo ~sample observations 
are the predictive data.

The four and eight
signal DCM double wave models
have $\eta=21$ and $\eta=21+20$ free
parameters, respectively.
The yearly sunspot data
sample \RyearlyTwo ~contains
only $n=300$ observations.
To avoid over-fitting, we analyse
these yearly data only with
the four and eight signal DCM pure sine models
having $\eta=13$ and $\eta=13+12$ free parameters.
The DCM resutls for the \RyearlyTwo ~and  \RyearlyOne 
  ~samples are given in \Supplementary  Tables
  \ref{TableRyearly2000K410R14}
  and
  \ref{TableRyearlyK410R14}.

The predictive data sample
\RyearlyTwo ~DCM analysis gives
the predicted test statistic
$z_{\mathrm{pred}}$ (Equation \ref{EqPredz})
values shown in Fig. \ref{FigRyearlySine}a.
The best six signal model gives the smallest
$z_{\mathrm{pred}}$ value.
The $z_{\mathrm{pred}}$ values for
the  seven and eight signal models
are larger.

This six signal model gives an amazingly accurate
prediction for the twenty years of predicted
data after the year 2000
(Fig. \ref{FigRyearlySine}b:
red curve after the vertical black dash-dotted line).
{\it All} predicted data values are between the
dotted red lines denoting $\pm3\sigma$ prediction
error limits. About half of these
black dots denoting
the predicted data
overlap the continuous red line
denoting the predicted model.
{\it All} residuals of predicted data (blue dots)
are significantly smaller
  than the ISES limit (Equation \ref{EqISES}: cyan dash dotted lines).
Had we been able to apply DCM to the \RyearlyTwo ~sample
in the year 2000, we could have
  predicted the yearly sunspots number for
  the next two decades!
  The time span of this successful deterministic
    prediction is about two
    times longer than the stochastic
    one sunspot cycle prediction limit
    of \Dhypothesis ~hypothesis.

For \RyearlyTwo ~sample, the six signal model 
can reproduce the Dalton minimum
inside the predictive data, but not the
Maunder minimum, which is outside this sample
(Fig. \ref{FigRyearlySine}c: green curve).

  The predicted mean levels $m_{\mathrm{pred}}$
  (Equation \ref{EqPredMean})
  for the all data sample \RyearlyOne ~are
  shown in Fig. \ref{FigRyearlySine}d.
Even the red curve for
the Maunder minimum, which is closest to the
beginning of this \RyearlyOne ~sample,
stays above the mean level
$m=78.4$ 
of all data.
These predicted mean level
$m_{\mathrm{pred}}$ estimates for the past activity
minima do not succeed.
The six pure sine signal model
for the next decades predicts rising
solar activity  (Fig. \ref{FigRyearlySine}e:
green line).

\subsubsection*{Weighted yearly sunspot data
\label{SectWeightedYearly}}

The predictive data sample is \CyearlyTwo.
Sample \CyearlyOne ~is all data, where the
observations after the year 2000 are the predicted data.
We study only the pure sine DCM models, because
the  \CyearlyOne ~and \CyearlyTwo ~samples are both
too small ($n=204$ and 182)
for the DCM double wave model analysis.

We detect only two pure
sine signals in the predictive data sample\CyearlyTwo .
Additional signals give unstable models
(\Supplementary Table \ref{TableCyearly2000K410R14}: ``$\UM$'').
Therefore, no prediction is made.

Five pure sine signals are detected in all data
sample \CyearlyOne ~(\Supplementary Table \ref{TableCyearlyK410R14}).
We give the periods and amplitudes of these five signals 
in Table \ref{TableCompare} (Column 13).

The pure sine signal DCM models \M=1-8 give the
predicted mean level $m_{\mathrm{pred}}$
(Equation \ref{EqPredMean}) values
shown in Fig. \ref{FigCyearlySine}a.
The black and red $m_{\mathrm{pred}}$
curves for the Dalton and the Maunder
minima fall clearly below the mean level
$m=82.8$ 
of all yearly sunspot data.
After this,
these two curves also stay below this $m$ level.
These results are not unexpected
because the Dalton and the Maunder minima are
the closest activity minima
before the beginning of the \CyearlyOne ~sample.

The all data sample \CyearlyOne ~prediction for
the time intervals of
the Dalton and the Maunder minima is excellent
(Fig. \ref{FigCyearlySine}b: green curve).
However, the prediction can not reproduce the low
number of sunspots during the Maunder minimum. 
The green dotted lines denoting the error of the predicted 
  30 years sliding mean stay well below the ISES limit denoted with
  cyan dash-dotted lines (Equation \ref{EqISES}).
This successful  deterministic
  prediction for the Maunder minimum time interval
  covers two centuries backwards,
  before the beginning of the analysed data sample \CyearlyOne.
  This time interval is about twenty times
  longer than the stochastic \Dhypothesis ~hypothesis
  predictability limit
  of only one solar cycle. It is actually
  quite unexpected
  that only about 200 years of data can predict so well
  the activity mean level for the past
    200 years 
    without any data. This can succeed only if the model
    is correct and deterministic.
    It also means that our prediction for the future
    can be considered reliable.
    This \CyearlyOne ~sample future prediction
    for the next decades
is shown in Fig. \ref{FigCyearlySine}c.
The long-term prediction indicates rising solar
activity during the next half a century
(Fig. \ref{FigCyearlySine}d).

Our best predictive models are very consistent
  because the number of signals in all models is either
  five (Figs. \ref{FigRmonthly2000Sine}a,
  \ref{FigCmonthlySine}a and \ref{FigCyearlySine}a)
  or six  (Figs. \ref{FigRmonthly2000Double}a
  and \ref{FigRyearlySine}a).
We conclude that these results
confirm that there is some predictability of sunspot numbers.

\begin{table}
  \caption{Predictions for next three sunspot cycles.
    (1) Sample: Fig. showing predictions. (2-3)
    Maximum and minimum $t$ and $y$ of cycle 25.
        (4-5) Sunspot cycle 26. Otherwise as in ``2-3''.
    (6-7) Sunspot cycle 27. Otherwise as in ``2-3''.
    Line below  
    each group of five estimates gives their weighted mean.
  } 
\label{TableMinimaMaxima}
 \begin{center}
  \begin{tabular}{ccccccc}
    \hline
  (1)             &
  (2)             &
  (3)             &
  (4)             &
  (5)             &
  (6)             &
  (7) \\
\hline
  Sample: Fig.                 &
  \multicolumn{2}{c}{Maximum 25}      &
  \multicolumn{2}{c}{Maximum 26}      &
  \multicolumn{2}{c}{Maximum 27}     \\
 & $t$ & $y$ 
 & $t$ & $y$ 
 & $t$ & $y$ \\                                   
 & (y) & (-) & (y) & (-) & (y) & (-) \\
  \hline
\RmonthlyOne: Fig.  \ref{FigRmonthly2000Sine}e   &$2025.86\pm0.16$&$118.4\pm5.9$&$2037.40\pm0.20$&$148.7\pm5.7$&$2048.41\pm0.22$&$170.9\pm5.4$\\ 
\RmonthlyOne: Fig.  \ref{FigRmonthly2000Double}e &$2024.42\pm0.20$&$88.6\pm4.8 $&$2037.6\pm4.9  $&$146\pm30   $&$2047.9\pm5.1  $&$207\pm18   $\\ 
\CmonthlyOne: Fig.  \ref{FigCmonthlySine}c       &$2024.07\pm0.19$&$89.4\pm4.1 $&$2037.39\pm0.13$&$102.8\pm5.3$&$2047.48\pm0.15$&$116.6\pm5.4$\\ 
\RyearlyOne:  Fig.  \ref{FigRyearlySine}e        &$2025.28\pm0.39$&$106\pm14   $&$2036.62\pm0.41$&$145\pm16   $&$2047.40\pm0.41$&$185\pm15   $\\ 
\CyearlyOne:  Fig.  \ref{FigCyearlySine}c        &$2024.36\pm0.78$&$90\pm10    $&$2037.4 \pm1.3 $&$112\pm22   $&$2047.5\pm2.0  $&$126\pm21   $\\ 
\hline
Weighted mean:                                        &$2024.94\pm0.79$&$95\pm12    $&$2037.34\pm0.19$&$125\pm22   $&$2047.72\pm0.41$&$148\pm30   $\\ 
      \hline
                                      &
  \multicolumn{2}{c}{Minimum 25}      &
  \multicolumn{2}{c}{Minimum 26}      &
                                        \multicolumn{2}{c}{Minimum 27}      \\
\hline 
 \RmonthlyOne:  Fig.  \ref{FigRmonthly2000Sine}e    &$2031.57\pm0.18$&$16.8\pm6.1  $&$2042.82\pm0.20$&$22.4\pm5.5$&$2053.76\pm0.23$&$34.0\pm5.0$ \\
\RmonthlyOne:   Fig.  \ref{FigRmonthly2000Double}e  &$2032.7\pm5.0  $&$25\pm10     $&$2043.4 \pm5.0 $&$6 \pm15    $&$2054.1 \pm5.4 $&$39.4\pm6.8$\\
\CmonthlyOne:   Fig.  \ref{FigCmonthlySine}c        &$2031.77\pm0.28$&$19.2\pm4.4  $&$2042.35\pm0.14$&$0.3\pm4.4 $&$2052.58\pm0.18$&$28.8\pm5.2$ \\
\RyearlyOne:    Fig.  \ref{FigRyearlySine}e         &$2030.77\pm0.38$&$-15\pm15    $&$2041.79\pm0.38$&$16\pm16   $&$2052.62\pm0.41$&$38\pm17   $ \\
\CyearlyOne:    Fig.  \ref{FigCyearlySine}c         &$2031.8\pm1.1  $&$16\pm13     $&$2042.3\pm2.0  $&$-1\pm19   $&$2052.7\pm2.2  $&$28\pm17   $ \\
\hline                        
Weighted mean:                                         &$2031.52\pm0.31$&$17.4\pm  7.4$&$2042.44\pm0.29$&$9 \pm 10  $&$2052.98\pm0.56$&$33.0\pm4.3$ \\
      \hline
    \end{tabular}
    \end{center}
    \end{table}
    
\subsubsection*{Future sunspot maxima
  and minima predictions}
  
We give the predictions for the next three sunspot cycle
maxima and minima in Table \ref{TableMinimaMaxima}.

The lowest possible real sunspot number is zero.
We can not prevent DCM from utilising
negative model $g(t)$ values.
For this reason,
two predicted cycle minimum sunspot
number estimates out of all fifteen estimates,
$y=-15\pm15$ and $y=-1\pm19$,
are negative (Table \ref{TableMinimaMaxima}).
However, in these two exceptional cases
the sunspot number
can be zero within $\pm 1 \sigma$.

The ISES prediction for the 
  cycle 25 maximum epoch was
  $t=$
  July 2025 $\pm$ 8.0 months =
  $2025.54\pm0.67$.
  Their smoothed maximum
  sunspot number prediction was
  $y=115$.
  The weighted mean values of all our predictions
    in Table \ref{TableMinimaMaxima},
  $t=$ December 2024 $\pm$ 9.5 months
  $=2024.94\pm0.79$
  and $y=95\pm12$,
  differ
  $0.76\sigma$ 
  and
  $1.67\sigma$ 
  from the ISES prediction
  (We made this prediction in the year 2023.
    It is known now that the cycle 25
    maximum took place in October 2024.).

  Our prediction error
  $\pm 0.^{\mathrm{y}}79$ for the cycle 25 maximum epoch
  is much larger than our prediction
  errors  $\pm 0.^{\mathrm{y}}19$
  and $\pm 0.^{\mathrm{y}}41$
  for the next two maximum epochs
  (Table \ref{TableMinimaMaxima}).
This is a real statistical effect.
The predicted peak of this next cycle 25 maximum
is lower than, and not so sharp as, the predicted
peaks of the next cycle 26 and 27 maxima
(Figs. 
\ref{FigRmonthly2000Sine}e,
\ref{FigRmonthly2000Double}e,
\ref{FigCmonthlySine}c,
\ref{FigRyearlySine}e,
and
\ref{FigCyearlySine}c).

The six double wave model predictions
for the largest \RmonthlyOne ~sample
are not very accurate in Table \ref{TableMinimaMaxima}
(\RmonthlyOne: Fig.  \ref{FigRmonthly2000Double}e).
The probable reason for this inaccuracy
is the large number $\eta=21+10$
of the free parameters
 ~(\Supplementary Table \ref{TableRmonthlyK420R14}, model \M=6).
This is also the only model that has
a turning point in the year 2029
(Fig. \ref{FigRmonthly2000Double}e).

We predict that
the mid-point of the next prolonged
grand sunspot minimum 
  is approximately in the year 2100
  (Figs.
  \ref{FigRmonthly2000Sine}f,
  \ref{FigRmonthly2000Double}f,
  \ref{FigCmonthlySine}d,
  \ref{FigRyearlySine}f
  and
  \ref{FigCyearlySine}d: green curve).
 


\begin{figure}  
\centerline{\hspace*{0.015\textwidth}
         \includegraphics[width=0.515\textwidth,clip=]{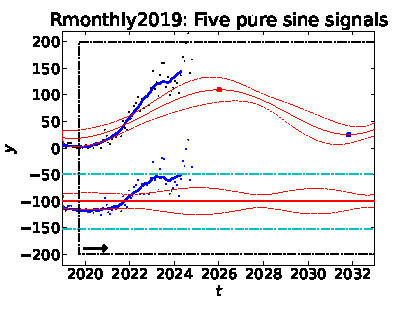} 
         \hspace*{-0.03\textwidth}
         \includegraphics[width=0.515\textwidth,clip=]{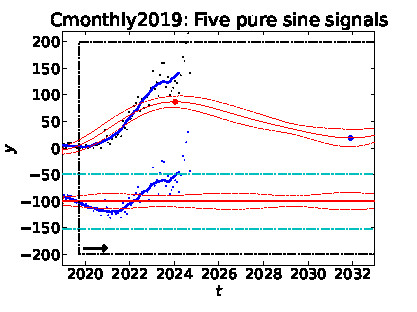}} 
\vspace{-0.33\textwidth}
\centerline{\Large \bf 
\hspace{0.38\textwidth}  \color{black}{(a)}
\hspace{0.43\textwidth}  \color{black}{(b)}
\hfill}
\vspace{0.28\textwidth}
 \caption{DCM predictions for cycle 25.
   (a) Nonweighted five pure sines prediction
   for sample \RmonthlyNew ~has
   maximum $y=109.6 \pm 7.7$ at  $2026.0 \pm 0.3$ (red circle) and
   minimum $y=25.3 \pm 6.1$ at  $2031.8 \pm 4.8$ (blue circle).
   Otherwise as in Fig. \ref{FigRmonthly2000Sine}.
   (b) Weighted five pure sines
   prediction for sample \CmonthlyNew ~has maximum
   $y=86.9 \pm 3.8$ at $t=2024.04 \pm 0.24$ and
   minimum $y = 18.9\pm 5.3$ at $t=2031.92 \pm 0.22$.
   Otherwise as in ``a''.
                     } \label{FigCycle}
\end{figure} 
  
\begin{figure}  
  \includegraphics[width=1.0\textwidth,clip=]{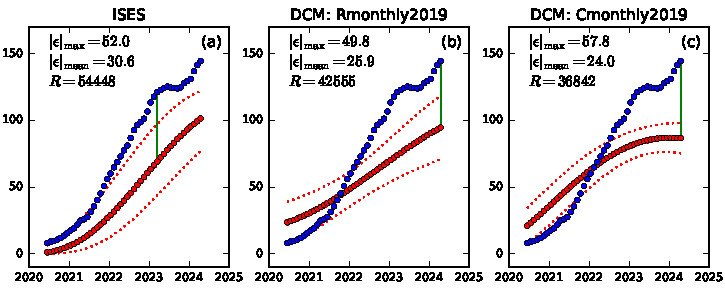} 
  \caption{Comparison of predictions.
    (a) ISES prediction: Predicted 
    $g_{13}$ (red circles), 
    observed $y_{13}$ (blue circles)
and
    prediction errors
    (red dotted curve).
    We give
    maximum of residual's absolute values
    ($|\epsilon|_{\mathrm{max}}$: green vertical line),
    mean of residual's absolute values $(|\epsilon|_{\mathrm{mean}})$
    and sum of squared residuals $(R)$.
    (b) Rmonthly2019 prediction from Fig. \ref{FigCycle}a,
    otherwise as in ``a''.
    (b) Cmonthly2019 prediction from Fig. \ref{FigCycle}b,
    otherwise as in ``a''.
  } 
                     \label{FigCycleComparison}
\end{figure} 

\subsubsection*{Prediction for current sunspot cycle 25}

  After the analysis of all eight data samples summarised
  in Table \ref{TableSamples} was already completed, 
  we were requested to
  give our own prediction for the ongoing cycle 25.
  We removed all data after September 2019
  from the samples \RmonthlyOne ~and
  \CmonthlyOne.
  This gave us all data before the beginning of cycle 25,
  samples \RmonthlyNew ~and \CmonthlyNew.
  The results of the DCM analysis of these samples are presented
  in \Supplementary Tables
    \ref{TableRmonthly2019K410R14}
    and 
   \ref{TableCmonthly2019K410R14}.
  
  Our DCM prediction for the \RmonthlyNew ~sample of non-weighted
  monthly sunspot numbers is shown in Fig. \ref{FigCycle}a.
  The predicted maximum and minimum epochs
  are January 2026 and
  September 2031, respectively.
  We show the DCM prediction
  for the weighted \CmonthlyNew ~sample
  in Fig. \ref{FigCycle}b.
  The  maximum 
  is at January 2024
  and  the minimum at
  November 2031.

  Cycle 25 began on October 2019. 
  The ISES panel gave no 
  $g_{13}$ prediction
  between October 2019 and May 2020. 
  Their prediction 
  begins from June 2020.
  The last available observed $y_{13}$ 
 value is from April 2024 (We accessed SILSO database
   on 21 November 2024).
 The ISES prediction for the cycle 25 maximum is
 July 2025.
 Their prediction for the cycle 25 minimum
  is vague
  because the forecast fades extremely slowly,
  $0 \le g_{13} \le 1 $ between July 2034 and April 2037
  (Fig. \ref{FigISES}).
  Their prediction is ``safe'' because the next
  minimum is bound to occur sooner or later.
  Their predicted long fading period begins three years after our
  predicted cycle 25 minimum at the end of the year 2031.

  In Figs. \ref{FigRmonthly2000Sine}-\ref{FigCyearlySine},
  we can compare our predictions only to the five year ISES limit
  $s_{\mathrm{ISES}}=|\epsilon|_{\mathrm{max}}=52$
  (Equation \ref{EqISES}).
  The ISES cycle 25 predictions
  between June 2020 and April 2024
  allow
  us to compare the three parameters
  $|\epsilon|_{\mathrm{max}}$,
  $|\epsilon|_{\mathrm{mean}}$ and
  $R$ (Equations
\ref{EqResiduals} and
\ref{EqR}).
The ISES prediction values are
$|\epsilon|_{\mathrm{max}}=52.0$,
$|\epsilon|_{\mathrm{mean}}=30.6$ and
$R=54448$
(Fig. \ref{FigCycleComparison}a).
Our DCM model prediction for
the \RmonthlyNew ~sample
gives smaller values
$|\epsilon|_{\mathrm{max}}=49.8$,
$|\epsilon|_{\mathrm{mean}}=25.9$ and
$R=42555$
(Fig. \ref{FigCycleComparison}b).
   All three parameters indicate that
  this prediction of ours is more accurate than
  the ISES prediction.
Our DCM model prediction for
the \CmonthlyNew ~sample gives
$|\epsilon|_{\mathrm{max}}=57.8$,
$|\epsilon|_{\mathrm{mean}}=24.0$ and
$R=36842$
(Fig. \ref{FigCycleComparison}c).
 The small $|\epsilon|_{\mathrm{mean}}$ and $R$
 values indicate that our prediction
 is better than the ISES prediction.
  However, our prediction $|\epsilon|_{\mathrm{max}}=57.8$
  is larger than the ISES prediction $|\epsilon|_{\mathrm{max}}=52.0$.
  For the last $y_{13}$ value in April 2024,
  our prediction for the \CmonthlyNew ~sample
  exceeds the five year ISES limit 
  (Figs. \ref{FigCycle}b and  \ref{FigCycleComparison}c).
  Statistically,
  this second prediction of ours is also better than the ISES prediction
  because
  only {\it one} $y_{13}$ value gives $|\epsilon|_{\mathrm{max}}$,
  but {\it all} $y_{13}$ values give
  $|\epsilon|_{\mathrm{mean}}$ and $R$.

%
%
%
%

\subsection*{Connections to planetary periods}

In time series analysis,
 the detected periods are usually considered real
 if they have an accepted physical cause,
like in our former DCM analyses\cite{Jet20,Jet21}.
Here, such a widely accepted cause for
the detected strictly
periodic, extremely significant and predictive signals
is missing.
  There must be an astrophysical mechanism
  that causes these signals in the sunspot data,
  unless both DCM and DFT give the same misleading results.
  Whatever this astrophysical mechanism may be,
  why would both of these
  statistical methods detect the same
  unreal signals?
What other solar system physical mechanism
can cause these signals,
except the planets?
This would mean that the solar cycle is
not generated only by the physical mechanisms
inside the Sun.
The only possible
  external cause can be
  the gravitational tidal forces of planets.

Parker  
formulated the dynamo model, where
the interaction of differential rotation
and convective motion drives the solar magnetic
field\cite{Par55}.
The first models for a dynamo generated sunspot cycle
were presented about a decade later\cite{Bab61,Lei69}.
All dynamo model physical
mechanisms causing the solar cycle
are not yet fully understood\cite{Cha10}.
As stated in our \Dhypothesis ~hypothesis,
it is widely accepted that the 
solar cycle is stochastic and unpredictable
beyond one solar cycle\cite{Pet20}.
For example, the ISES panel predictions
cover only the ongoing cycle
(Fig. \ref{FigISES}).
If these predictions were made for several cycles
in a row, the prediction errors would certainly
exceed the five year ISES limit (Equation \ref{EqISES}).

Recently, several attempts have been made to model
an internal solar dynamo influenced
by external gravitational tidal forces of planets. 
In one recent solar dynamo model,
the planetary tidal forcing causes periodic
11.07 and 22.14-year magnetic field
oscillations\cite{Ste19}. 
There are physical mechanisms
that could explain how
the weak planetary tidal forces can alter
the solar structure and cause the sunspot
cycle\cite{Sca22}.  
If the external planetary signal
is not constantly amplified,
the internal solar dynamo mechanism fluctuations
can mask this external signal\cite{Cha22}.
It has also been 
argued that the combined tidal
forcing of Venus, the Earth and Jupiter can not
cause the 11-years solar cycle\cite{Cio23}.

The mass $m_{\mathrm{E}}$ and the
major axis $a_{\mathrm{E}}$ of Earth
give the dimensionless relative tidal force
\begin{eqnarray}
  \TIDRel =
  {{m} \over {m_{\mathrm{E}}}}
  \left({{a_{\mathrm{E}}}\over{a}}\right)^3,
  \label{EqTidal}
\end{eqnarray}
exerted by each planet to the Sun,
where  the planet's mass is $m$ and its major axis is $a$.
The tidal forces of Venus and Jupiter are much stronger
than those of the Earth and Mercury (Table \ref{TableTidal}).
The tidal forces of Mars, Saturn, Neptune and Uranus
are negligible.
Therefore, we will search for the orbital periods of
 Mercury ($\Pmer$),
  Venus ($\Pven$),
  Earth ($\Pear$) and
  Jupiter ($\Pjup$),
  the synodic periods of two planets
  ($\PMV$,
  $\PME$,
  $\PMJ$,
  $\PVE$,
  $\PVJ$
  $\PEJ$),
  and the synodic periods of three planets
   ($\PMVE$,
    $\PMVJ$,
    $\PVEJ$).
    Only one of these periods, $\Pjup$, is inside the period
    interval between 5 and 200 years, where DCM detects periodicity
    (\Supplementary Section \ref{SectDCMone}). 
   All other orbital periods are shorter than 5 years.
   DCM detects no such shorter periods
   (\Supplementary Section \ref{SectDCMtwo}).

   \begin{table}
     \caption{Relative tidal
       forces (Equation \ref{EqTidal}).
       (1) Planet. (2) Major axis. (3) Mass.
       (4) Relative tidal force (Equation \ref{EqTidal}).
     } \label{TableTidal} 
     \begin{center}
\begin{tabular}{llll}
  \hline
  (1)    & (2)       & (3) & (4) \\
         & $a$ & $m$ & $\TIDRel$ \\
  Planet & $(a_{\mathrm{Earth}})$&$(m_{\mathrm{Earth}})$& (-) \\
  \hline
Mercury & 0.39  &        0.055  &         0.95  \\ 
Venus   & 0.72  &         0.82  &         2.16  \\ 
Earth   & 1.00  &         1.00  &         1.00  \\ 
Mars    & 1.52  &         0.11  &        0.030  \\ 
Jupiter & 5.20  &          318  &         2.26 \\ 
Saturn  & 9.53  &         95.3  &         0.11  \\ 
Uranus  & 19.2  &         14.5  &       0.0020  \\ 
Neptune & 30.1  &         17.3  &      0.00063  \\ 
  \hline
\end{tabular}
\end{center}
\end{table}

\begin{table}
  \caption{Three strongest signals in all data samples.
    (1) Sample: $K_2=1$ or 2 model.
    (2) Signal \SignalOne: Period $P_{\mathrm{11^y}}$ and amplitude $A_{\mathrm{11^y}}$.
    (3-4) Same for \SignalTwo ~and \SignalThree ~signals.
          Lowest line gives period and amplitude
     weighted means.
    Note that \SignalTwo ~signal
    in \RmonthlyOne: $K_2=2$
    is a double sinusoid\DW.}
  \label{TableRatio}
\begin{center}
  \begin{tabular}{lccc}
  \hline
  (1)                &     (2)              & (3)                  & (4)             \\
                     & \SignalOne           & \SignalTwo           & \SignalThree     \\
                     & $P_{\mathrm{11}}$ (y)  & $P_{\mathrm{10}}$ (y)  & $P_{\mathrm{11.86}}$ (y)  \\
  Sample: $K_2$    & $A_{\mathrm{11}}$ (-)  & $A_{\mathrm{10}}$ (-)  & $A_{\mathrm{11.86}}$ (-) \\
\hline
\RmonthlyOne: $K_2=1$& $11.0033\pm0.0064$   & $10.0001\pm0.0081$   & $11.807\pm0.012$ \\
                     &  $101.5\pm2.5$       & $65.5\pm2.1$         &   $56.4\pm2.4$  \\
\hline            
\RmonthlyOne: $K_2=2$& $10.9878\pm0.0051$   & $10.0031\pm0.0044$\DW& $11.770\pm0.011$\\
                     & $104.6\pm2.3$        & $75.2\pm2.7$         & $62.7\pm2.6$    \\
\hline
\CmonthlyOne: $K_2=1$& $10.8585\pm0.0048$   & $10.0658\pm0.0077$   & $11.863\pm0.021$\\
                     &  $117.5\pm2.9$       & $59.3\pm2.8$         & $43.3\pm2.0 $   \\
\hline
\RyearlyOne: $K_2=1$ & $10.981 \pm0.020$    & $9.975\pm0.017$      &$11.820\pm0.027$ \\
                     & $98.3\pm7.8  $       & $62.8\pm6.1$         & $40.5\pm3.3$   \\
\hline
\CyearlyOne:  $K_2=1$& $10.863\pm0.022$     & $10.058\pm0.026$     &$11.856\pm0.068$ \\  
                     &  $119.3\pm7.5 $      &  $61.0\pm9.6$        &$43.8\pm7.9$    \\
\hline
Weighted mean        &$10.938\pm0.066$      & $10.014\pm0.026$     &$11.799\pm0.031$  \\
                     &$107.0\pm6.8 $        & $66.3\pm5.8$         &$50.5\pm8.6$      \\
  \hline
\end{tabular}
\end{center}
\end{table}


\begin{table}
  \caption{Synodic period integer multiples.
    (1) Synodic period $P_{\mathrm{Syn}}$.
    (2-5) Signal \SignalOne:
    $\mathcal{P}$   (Equation \ref{EqPP}),
    $\mathcal{P}_0$ (Equation \ref{EqPPzero}),
    $\mathcal{P}_0P_{\mathrm{Syn}}$ and
    $\sigma =(\mathcal{P}_0P_{\mathrm{Syn}}-P)/\sigma_P$.
    (6-9) Signal \SignalTwo ~values.
    (10-13) Signal \SignalThree ~values.}
\label{TableSynodic}
\begin{center}
\addtolength{\tabcolsep}{-0.05cm}
\begin{tabular}{ccccccccccccccc}
\hline
(1) & (2) & (3) & (4) & (5) & & (6) & (7) & (8) & (9) & & (10) & (11) & (12) & (13) \\
  &  \multicolumn{4}{c}{\SignalOne$(P=10.938\pm0.066)$} & & 
     \multicolumn{4}{c}{\SignalTwo$(P=10.014\pm0.026)$} & &
     \multicolumn{4}{c}{\SignalThree$(P=11.799\pm0.031)$} \\
  \cline{2-5} \cline{7-10} \cline{12-15}
  $P_{\mathrm{Syn}}$ &
  $\mathcal{P}$ & $\mathcal{P}_0$ & $\mathcal{P}_0 P_{\mathrm{Syn}}$ &$\sigma$ & &
  $\mathcal{P}$ & $\mathcal{P}_0$ & $\mathcal{P}_0 P_{\mathrm{Syn}}$ &$\sigma$ & &
  $\mathcal{P}$ & $\mathcal{P}_0$ & $\mathcal{P}_0 P_{\mathrm{Syn}}$ &$\sigma$ \\
  (y)        &
  (-)        & (-)                & (y)                            & $(\sigma_P)$     & &
  (-)        & (-)                & (y)                            & $(\sigma_p)$     & &
  (-)        & (-)                & (y)                            & $(\sigma_P)$     \\
\hline
$\PMV= 0.3957$&$27.6$&$ 28$&$11.079$&$  2.1$&&$25.3$&$ 25$&$ 9.892$&$  4.7$&&$29.8$&$ 30$&$11.870$&$  2.3$\\ 
$\PME= 0.3172$&$34.5$&$ 34$&$10.784$&$  2.3$&&$31.6$&$ 32$&$10.150$&$  5.2$&&$37.2$&$ 37$&$11.736$&$  2.0$\\ 
$\PMJ= 0.2458$&$44.5$&$ 45$&$11.061$&$  1.9$&&$40.7$&$ 41$&$10.077$&$  2.4$&&$48.0$&$ 48$&$11.798$&$  0.0$\\ 
$\PVE= 1.5988$&$ 6.8$&$  7$&$11.191$&$  3.8$&&$ 6.3$&$  6$&$ 9.593$&$ 16.2$&&$ 7.4$&$  7$&$11.191$&$ 19.6$\\ 
$\PVJ= 0.6489$&$16.9$&$ 17$&$11.031$&$  1.4$&&$15.4$&$ 15$&$ 9.733$&$ 10.8$&&$18.2$&$ 18$&$11.680$&$  3.9$\\ 
$\PEJ= 1.0921$&$10.0$&$ 10$&$10.921$&$  0.3$&&$ 9.2$&$  9$&$ 9.829$&$  7.1$&&$10.8$&$ 11$&$12.013$&$  6.9$\\ 
\hline
\end{tabular} 
\addtolength{\tabcolsep}{+0.05cm}
\end{center}
\end{table}

The \SignalOne, \SignalTwo, \SignalThree, \SignalFour ~and
\SignalFive ~signals are the strongest ones
(Table \ref{TableCompare}).
Since these five 
signals are clearly
connected (Table \ref{TableConnections}),
their interference 
is repeated indefinitely,
as long as the orbital periods of the involved planets
remain constant.
For example, 
  the \RmonthlyOne ~sample $n=3287$ observations
  have $m=81.6$ and $s=67.7$.
  Only twenty $y_i$ values exceed  $m + 3s=285$.
  The best model for this sample,
  the sum of five pure sine signals,
  has the mean level $M_0=81.3\pm0.8$. 
  The sum of the peak to peak amplitudes
  for these five strongest \SignalOne, \SignalTwo, \SignalThree,
  \SignalFour ~and \SignalFive ~signals  
  is  $\AmpOne+\AmpTwo+\AmpThree+\AmpFour+\AmpFive=331$
  (Table \ref{TableCompare}: column 2). 
  These five signals alone can
  reproduce the whole range of observed $y_i$ changes.
  No additional signals are needed.
  The \SignalOne, \SignalTwo ~and \SignalThree
  ~signals may be the only real signals
   because 
   the interference of \SignalOne ~and \SignalTwo ~signals
  can cause the \SignalFour ~signal (Table \ref{TableConnections}).
  Then, the interference of
  \SignalThree ~and \SignalFour ~signals
  can cause the \SignalFive ~signal.
We give the weighted mean periods and amplitudes
of the three strongest signals in Table \ref{TableRatio}.

\subsubsection*{Integer multiples of one planet orbital period}

  The third
  strongest \SignalThree ~signal has
 $P_{\mathrm{11.^y86}}=11.799\pm0.031$ years weighted mean period.
 It differs 
 $1.9 \sigma$ 
 from the $1 \times \Pjup=11.857$ years period.
The orbital periods $\Pmer$,  $\Pven$ and  $\Pear$ are shorter
than 5 years. These periods can not be found
{\it directly} from the monthly sunspot data
(\Supplementary Section  \ref{SectDCMtwo}).
This does not, however, mean that we can not search
for multiples of these periods above the
five year limit $P_{\mathrm{max}}=5$.
These periods $i P_{\mathrm{Planet}}$
are called ``integer multiple orbital periods'',
where $P_{\mathrm{Planet}}$ is planet's orbital period 
and $i=1,2,3...$ is the ``integer multiplier''.
The aim is to find those integer multipliers
that give the detected signal periods.
The weighted mean $P_{\mathrm{11^y}}= 10.938\pm0.066$
years period
of the strongest \SignalOne ~signal differs
$0.94 \sigma$  
from $11\times\Pear$.
The weighted mean $P_{\mathrm{10^y}}=10.014\pm 0.026$ years period
of the second strongest \SignalTwo ~signal differs
$0.54 \sigma$ 
from $10\times \Pear$.

We use the known planet orbital period
$P_{\mathrm{Planet}}=\Pmer, \Pven, \Pear$ and $\Pjup$ 
in Earth years to compute
four dimensionless parameters
\begin{eqnarray}
\mathcal{P}    &=&P/P_{\mathrm{Planet}}  \label{EqPP}\\
  \sigma_{\mathcal{P}} &=&
  \sigma_P/P_{\mathrm{Planet}} \label{EqPPsigma}\\
  \Delta \mathcal{P} &=&
  \mathcal{P} - \mathcal{P}_0 \label{EqPPzero} \\
  \Delta \mathcal{P}_{\mathrm{rel}} &=&
  {{|\Delta \mathcal{P}|}
  \over {\sigma_{\mathcal{P}}}}, \label{EqRelativeError}
\end{eqnarray}
where $\mathcal{P}_0$ is the 
integer value closest to $\mathcal{P}$.
The two first parameters $\sigma_{\mathcal{P}}$ and
$\mathcal{P}$  are hereafter
called ``rounds error'' and ``rounds''
 (number of planet revolutions around the Sun).
The next two parameters
$\Delta \mathcal{P}$ 
and
$\Delta \mathcal{P}_{\mathrm{rel}}$  are called
``rounds deviation'' and ``relative rounds deviation''.
Note that we introduce a new letter ``$\mathcal{P}$''
to highlight the difference from the period letter ``$P$''.

In \Supplementary Section  \ref{SectIdentification},
  our detailed search for integer multiples
  of one planet orbital periods
 gives these two main results:

\begin{enumerate}

\item The $\Pear$ and $\Pjup$ integer multiples are 
  easier to detect than the  $\Pmer$ and $\Pven$ integer multiples.

\item The \SignalThree ~signal can be connected to Jupiter's  orbital motion
  (\Supplementary Section \ref{SectJupiter}).

  \end{enumerate}
  
\subsubsection*{Integer multiples of two planet synodic period}
\label{SectTwoPlanets}

The strongest \SignalOne ~signal period of
 $P_{\mathrm{11^y}}\pm \sigma_P=10.938\pm0.066$ years 
differs only
$0.3 \sigma$ 
from the synodic period multiple $10\times \PEJ=10.921$ years
(Table \ref{TableSynodic}).
The ratio $\PEJ/\sigma_P=16.5$ indicates that this $\sigma=0.3$
for $\PEJ$ is significant.
The ratio $\PVJ/\sigma_P=9.8$ indicates
that the $\sigma=1.4$ value for $17 \times \PVJ$
can be significant.
These two results indicate that $\PEJ$ and $\PVJ$ periods
can be connected to the strongest \SignalOne ~signal.
However, the ratios
$\PMV/\sigma_P=6.0$,
$\PME/\sigma_P=4.8$ and
$\PMJ/\sigma_P=3.7$
mean
that the respective
$\sigma=2.1, 2.3$ and 1.9 values
for $\PMV, \PME$ and $\PMJ$ multiples
are not significant.
This illustrates the effect that short period multiples
are more difficult to
detect (\Supplementary Equations \ref{EqProbP} and \ref{EqMigration}).

For the second strongest \SignalTwo ~signal,
the $P_{10^{\mathrm{y}}}\pm \sigma_P = 10.014 \pm 0.026$
years period differs $2.4\sigma$ from $41\times\PMJ$
(Table \ref{TableSynodic}).
The ratio $\PMJ/\sigma_P=9.4$ 
indicates that this result  may be significant.
All integer multiples for the other synodic
periods $\PMV, \PME, \PVE, \PVJ$ and $\PEJ$ 
differ more than $\pm3\sigma$ from
the \SignalTwo ~signal period.

The third strongest \SignalThree ~signal period of
 $P_{\mathrm{11.^y86}}\pm \sigma_P=10.938\pm0.031$ years 
 differs $2.3\sigma$, $2.0\sigma$ and $0.0\sigma$
 from
 $30\times\PMV$, $37\times \PME$ and $48\times\PMJ$,
 respectively.
 The ratios
 $\PMV/\sigma_P=12.8$,
 $\PME/\sigma_P=10.2$
 and
 $\PMJ/\sigma_P=7.9$
 indicate that all these the synodic period multiples
 may be connected to the \SignalThree ~signal period.

\begin{table}
  \caption[]{Three planet synodic periods. (1-2) Integers $i$ and $j$
    (Equations \ref{EqThreeOne} and \ref{EqThreeTwo}). (3-5) Two
    planet synodic periods  $P_{12}$, $P_{13}$ and $P_{23}$
    (Equations \ref{EqThreeOne} and \ref{EqThreeTwo}).
  (6) Three planet synodic period $P_{123}$ (Equation \ref{EqThreeTwo})}
  \label{TableThreeSynodic}
  \begin{center}
 \begin{tabular}{cccccc}
   \hline
   (1)&(2) & (3)        & (4)         & (5)        & (6)             \\
    i & j  & $P_{12}$    & $P_{13}$    & $P_{23}$    &$P_{\mathrm{123}}$ \\
   (-)&(-) & (y)        &(y)          & (y)        & (y)             \\
    \hline
    \multicolumn{6}{c}{Mercury, Venus, Earth} \\
    7 & 28 & $\PMV=0.396$&$\PME=0.317$&$\PVE=1.599$&  $\PMVE=P_{7,28}=11.101$ \\
    \multicolumn{6}{c}{Mercury, Venus, Jupiter} \\    
    8 & 13 & $\PMV=0.396$&$\PMJ=0.246$&$\PVJ=0.649$&  $\PMVJ=P_{8,13}=5.162 $ \\
   11 & 18 & $\PMV=0.396$&$\PMJ=0.246$&$\PVJ=0.649$&  $\PMVJ=P_{11,18}=7.128 $ \\
   14 & 23 & $\PMV=0.396$&$\PMJ=0.246$&$\PVJ=0.649$&  $\PMVJ=P_{14,23}=9.094 $ \\
   17 & 28 & $\PMV=0.396$&$\PMJ=0.246$&$\PVJ=0.649$&  $\PMVJ=P_{17,28}=11.061$ \\
   19 & 31 & $\PMV=0.396$&$\PMJ=0.246$&$\PVJ=0.649$&  $\PMVJ=P_{19,31}=12.290$ \\
    \multicolumn{6}{c}{Venus, Earth, Jupiter} \\    
    6 &  4 & $\PVE=1.599$&$\PVJ=0.649$&$\PEJ=1.092$&  $\PMVJ=P_{6,4}=6.489$ \\
    9 &  6 & $\PVE=1.599$&$\PVJ=0.649$&$\PEJ=1.092$&  $\PMVJ=P_{9,6}=9.733$ \\
   13 &  9 & $\PVE=1.599$&$\PVJ=0.649$&$\PEJ=1.092$&  $\PMVJ=P_{13,9}=14.275$ \\
   16 & 11 & $\PVE=1.599$&$\PVJ=0.649$&$\PEJ=1.092$&  $\PMVJ=P_{16,11}=17.519$ \\
   19 & 13 & $\PVE=1.599$&$\PVJ=0.649$&$\PEJ=1.092$&  $\PMVJ=P_{19,13}=20.764$ \\
    \hline
 \end{tabular}
 \end{center}
\end{table}

\subsubsection*{Synodic periods for three planets}
\label{SectThreePlanets}

There is an algorithm to
search for the synodic periods of three planets\cite{Bor12}.
The periods of these three planets are $P_1<P_2<P_3$.
Their synodic periods are $P_{12}$, $P_{13}$ and $P_{23}$.
The resonance of these three planets requires
\begin{eqnarray}
  {{P_{12}}\over{P_{23}}}={{i}\over{j}},
  \label{EqThreeOne}
\end{eqnarray}
where $i,j=1,2,3, ...$ are integers. The synodic
period of these three planets is
\begin{eqnarray}
  P_{i,j}= (i+j) P_{13}.
  \label{EqThreeTwo}
\end{eqnarray}
The algorithm\cite{Bor12} solves
the best values for the integers $i$ and $j$.
We compute the three planet synodic periods $P_{i,j}>5$ years
for $i\le20$ (Table \ref{TableThreeSynodic}).

The only suitable three planet synodic period
for Mercury, Venus and the Earth is
$\PMVE =P_{7,28}=11.101$  (Equation \ref{EqThreeTwo})
This value
differs $2.5 \sigma$ from $P=10.938\pm0.066$ of \SignalOne ~signal.

For Mercury, Venus and Jupiter,
the  $\PMVJ=P_{17,28}=11.061$ period differs
$1.9 \sigma$ from signal \SignalOne ~period
$P=10.938\pm0.066$.
None of the other three planet
synodic periods $\PMVJ=P_{8,13}$, $P_{11,18}$, $P_{14,23}$ and
$P_{19,31}$ is close to the \SignalOne, \SignalTwo ~or
\SignalThree ~signal periods.

For Venus, the Earth and Jupiter,
all five synodic periods
$\PVEJ=P_{6,4}, P_{9,6}$, $P_{13,3}$, $P_{16,11}$ and
$P_{19,13}$ are far from the \SignalOne, \SignalTwo ~and
\SignalThree ~signal periods.

In short, the \SignalOne ~signal connects to
  eight periods ($\Pear$,
  $\PMV$,
  $\PME$,
  $\PMJ$,
  $\PVJ$,
  $\PEJ$,
  $\PMVE$
  $\PMVJ$),
  the \SignalTwo ~signal connects to two periods
  ($\Pear$, $\PMJ$)
  and the \SignalThree ~signal
  connects to four periods
  ($\Pjup$,
  $\PMV$,
  $\PME$,
  $\PMJ$) periods.
  This would explain why
  the \SignalOne ~signal is
  stronger than the \SignalTwo ~and
  \SignalThree ~signals.

The two strongest \SignalOne ~and \SignalTwo ~signal periods
are close to $\Pjup$,
and the third strongest \SignalThree ~signal period is nearly equal to $\Pjup$.
This indicates that Jupiter's role is pronounced.

\section*{Discussion}

We present direct observational evidence that
the detected signals in the sunspot data can
be connected
to the planetary motions,
the \SignalThree ~signal
  case for Jupiter being the most obvious one
        (\Supplementary Figs. \ref{FigAnomaliesTwo}
        and  \ref{FigJupiterDistance}).
It is possible that
  the planetary motions can influence
  the solar cycle,
  but it is impossible that the solar magnetic field
  can influence the planetary motions.
  As far as we know, ours would be the first
  {\it direct} detection of planets
  from the sunspot data.

Astrophysical theories are used
  to construct models of astronomical phenomena.
  Observations are used to test these models.
  Statistical methods are not confined
  to any particular phenomenon.
  The same statistical method can
  be applied to different phenomena.
No phenomenon can fool statistics.
Our statistical
DCM time series analysis method
can detect combinations of phenomena
that are
repeated (Equation \ref{Eqharmonicone}: $h(t)$)
and not repeated (Equation \ref{EqPolyOne}: $p(t)$).
Phenomena are not stochastic only
because they appear unpredictable.
Phenomena may appear stochastic
because we can not yet predict them.
Both DCM and DFT detect the same extremely significant
and strictly periodic
signals that can predict sunspot data.
This does not support
the mainstream
dynamo-influence-theory \Dhypothesis ~hypothesis
of a stochastic sunspot cycle.
From the time series analysis
point of view, this is a clear-cut case.
Our results support
the deterministic
planetary-influence-theory
\Phypothesis ~hypothesis.

\renewcommand{\Ki}{}
No one has ever published a long-term
  sunspot cycle prediction that can be trusted,
  including the Solar Cycle Prediction Panel.
      The simulations of non-stationary stochastic datasets,
      which were claimed to prove that
      all our results can not be trusted,
      reproduced an exactly opposite result.
        The \SM ~can not
        reproduce the observed extremely significant long period
        \SignalFour ~and
  \SignalSeven ~signals
      (Figure \ref{FigDFTDistributions}).
     \Ki
  This model fails the \Detect.
     \Ki
     Furthermore, the \SM ~can not explain this interference quintet
      (Table \ref{TableConnections}: Equation \ref{EqConnections}).
  We promise to agree that our results  can not be trusted,
if the dynamo theorists can,
in the context of stochasticity,
answer our questions below. 
\Ki
The word ``why'' in these questions
refers to a statistical and/or an astrophysical cause. \Ki

\renewcommand{\Ki}{}

\begin{enumerate}

\item Why are the long period \SignalFour ~and
  \SignalSeven ~signals detected, if 
  the 11$^{\mathrm{y}}$ sunspot cycle is stochastic? \Ki

\item Why is there interference
  (Table \ref{TableConnections}: Equation \ref{EqConnections})
  between the
  five strongest detected signals? \Ki

\item Why can we predict the sunspot data? \Ki

\item Why are our
 DCM and DFT signal detections extremely significant?  \Ki

\item Why do we predict the current sunspot cycle 25 better
  than the Solar Cycle Prediction Panel? \Ki

\item Why do our predictions become better
  (Equations  \ref{EqPredz} and \ref{EqPredMean}) when
  new signals are detected? \Ki

\item Why is the third strongest \SignalThree ~signal
  period so close to Jupiter's period? \Ki

 \item Why does DCM perform better than DFT? \Ki

\end{enumerate}

\renewcommand{\Ki}{}
The mainstream dynamo models claim that
the sunspot cycle is non-stationary and 
stochastic.
\Ki
Our Discrete Chi-square Method (DCM)
and the Discrete Fourier Transform (DFT) 
detect the same extremely significant and
strictly periodic signals from the sunspot data.
\Ki
These signal detections are absolutely certain.
\Ki
The three strongest
detected \SignalOne, \SignalTwo ~and \SignalThree ~signals,
together with their interference
\SignalFour ~and \SignalFive ~signals,
can reproduce the full scale
of all observed sunspot number variations.
\Ki
We claim that the sunspot data are
stationary and multi-periodic,
and therefore deterministic, not stochastic.

\renewcommand{\Ki}{}
\Ki
Our DCM models give better sunspot cycle predictions than
the official Solar Cycle Panel. \Ki
Although the analysed sunspot data does not cover
the past prolonged activity minima,
the DCM models can predict some of these events,
like the Maunder minimum era.
\Ki
The predictability of solar activity can
help us to construct more accurate models for 
the climate change on the Earth and
perhaps allows us to prepare better for
the catastrophic geomagnetic storms,
like the Carrington event in the year 1859. \Ki

\renewcommand{\Ki}{}
The detected strictly periodic signals show
connections to the orbital motions of Mercury,
Venus, the Earth and Jupiter. \Ki
If the planets cause the sunspot cycle,
the starspot cycles observed in other chromospherically
active stars may represent indirect evidence
for the presence of exoplanets. \Ki

\section*{Methods}

\section*{Discrete Chi-Square Method (DCM)
  \label{SectMethod}} 

Jetsu formulated DCM\cite{Jet20}.
 He used DCM to discover the periods of
a third and a fourth body
from the O-C data of the eclipsing binary XZ~And.
An improved DCM version
revealed the presence of numerous
new companion candidates in the
eclipsing binary Algol\cite{Jet21}.
DCM is designed for detecting many signals superimposed
on an arbitrary trend.

We 
describe the sunspot  data in
\Supplementary Section \ref{SectData}.
The sunspot number data notations are observations
$y_i=y(t_i) \pm \sigma_i$,
where $t_i$ are the observing
times and $\sigma_i$ are the errors
$(i=1,2, ..., n)$. 
The units are $[y_i]=$ dimensionless, $[\sigma_i]=$ dimensionless
and $[t_i]=$~years.
The time span of data is $\Delta T=t_n-t_1$.
The mid point is $t_{\mathrm{mid}}=t_1+\Delta T/2$.
Our notations for 
the mean and the standard
deviation of all $y_i$ values are $m$ and $s$, respectively.

DCM model
\begin{eqnarray}
  g(t) = g(t,K_1,K_2,K_3)
   =  h(t) + p(t). 
\label{Eqmodel}
\end{eqnarray}
is a sum of a periodic function
\begin{eqnarray}
h(t)   & = & h(t,K_1,K_2) = \sum_{i=1}^{K_1} h_i(t) \label{Eqharmonicone} \\
h_i(t) & = & \sum_{j=1}^{K_2} 
             B_{i,j} \cos{(2 \pi j f_i t)} + C_{i,j} \sin{(2 \pi j f_i t)},
             \label{Eqharmonictwo}
\end{eqnarray}
and an aperiodic function
\begin{eqnarray}
  p(t)=p(t,K_3)=
  \begin{cases}
    0,                      & \text{if } K_3=-1           \\
   \sum_{k=0}^{K_3} p_k(t),  & \text{if } K_3=0, 1, 2, ...
 \end{cases}
\label{EqPolyOne}                               
\end{eqnarray}
where
\begin{eqnarray}
p_k(t) & =  & M_k \left[
                    {{2(t-t_{\mathrm{mid}}}) \over {\Delta T}}
                    \right]^k.
\label{EqPolyTwo}
\end{eqnarray} 
The periodic $h(t)$ function is a sum of 
$K_1$ harmonic $h_i(t)$
signals having 
frequencies $f_i$.

The signal order is $K_2$.
For simplicity, we refer to $K_2=1$ order
models as ``pure sine'' models, and to 
$K_2=2$ order models as ``double wave'' models.
The former category of signal curves are pure sines.
The latter category of signal curves can deviate from a pure sine.

The $h_i(t)$ signals
are superimposed
on the aperiodic $K_3$ 
order polynomial trend $p(t)$.
Function $h(t)$ repeats itself in time.
  Function $p(t)$ does not repeat itself,
  except when $p(t)=0$ for $K_3=-1$ or $p(t)=M_0=$ constant
  for $K_3=0$.

DCM model residuals are
\begin{eqnarray}
  \epsilon_i=y(t_i)-g(t_i)= y_i-g_i.
  \label{EqResiduals}
\end{eqnarray}
Our notations for the mean and standard deviation
of the residuals
are $m_{\epsilon}$ and $s_{\epsilon}$.
The notations for the mean and maximum
  values of the absolute values of residuals are
$|\epsilon|_{\mathrm{mean}}$ and $|\epsilon|_{\mathrm{max}}$.
The residuals 
give the sum of squared residuals
\begin{eqnarray}
R=\sum_{i=1}^n \epsilon_i^2,
\label{EqR}
\end{eqnarray}
and also the Chi-square
\begin{eqnarray}
  \chi^2=\sum_{i=1}^n {{\epsilon_i^2} \over {\sigma_i^2}}.
  \label{EqChi}
\end{eqnarray}
If the data errors $\sigma_i$ are known,
we use $\chi^2$ to estimate the goodness of our model.
For unknown errors $\sigma_i$, we use $R$.

Our notation
for $K_1$, $K_2$ and $K_3$ order DCM model $g(t)$
is 
``\RModel{K_1,K_2,K_3}'' or
``\CModel{K_1,K_2,K_3}''.
The last subscripts ``$R$" or ``$\chi^2$'' refer
to the use of Equation \ref{EqR} or \ref{EqChi} in
estimating the goodness of our model,
respectively.

The free parameters of model $g(t)$  are
\begin{eqnarray}
\bm{\beta}=[\beta_1, \beta_2, ..., 
\beta_{\eta}] = 
[B_{1,1},C_{1,1},f_1, ..., 
B_{K_1,K_2}, 
C_{K_1,K_2}, f_{K_1},
  M_0, ..., M_{K_3}],
\end{eqnarray}
where
\begin{eqnarray}
\eta = K_1 \times (2K_2+1) + K_3+1
\label{EqEta}
\end{eqnarray}
 is the number of free parameters.
We divide the free parameters $\bm{\beta}$ into
two groups
\begin{eqnarray}
\bm{\beta}_{I}  & = & [f_1, ..., f_{K_1}] \label{EqBetaOne} \\
\bm{\beta}_{II} & = & [B_{1,1}C_{1,1}, ..., 
B_{K_1,K_2}, C_{K_1,K_2},M_0, ..., M_{K_3}] \label{EqBetaTwo}
\end{eqnarray}
The first frequency group $\bm{\beta}_{I}$ 
makes the $g(t)$ model non-linear,
because all free parameters are not
eliminated from 
all partial derivatives $\partial g / \partial \beta_i$.
If these $\bm{\beta}_{I}$ frequencies are fixed to the constant
known tested numerical values,
none of the partial derivatives
$\partial g / \partial \beta_i$ 
contain any free parameters.
In this case, the model becomes {\it linear},
and the solution for the second group of free parameters,
$\bm{\beta}_{II}$ is {\it unambiguous}.
Our concepts like ``linear model'' and ``unambiguous result''
refer to this type of models
and their free parameter solutions.

For every tested frequency
  combination $\bm{\beta}_{I}=[f_1, f_2, ..., f_{K_1}]$,
   we   compute the DCM test statistic
  \begin{eqnarray}
    z & = &z(f_1, f_2, ..., f_{K_1})=\sqrt{R/n}
    \label{EqzR} \\
    z & = &z(f_1, f_2, ..., f_{K_1})=\sqrt{\chi^2/n}
    \label{EqzChi}
  \end{eqnarray}
\noindent
from a linear model least squares fit.
We use Equation \ref{EqzR} or Equation \ref{EqzChi},
for unknown or known $\sigma_i$ errors,
respectively.

For two signal $K_1 =2$ model,
the sum $h(t)$ of signals $h_1(t)$ and $h_2(t)$ does not depend
on the order in which these signals are added.
This causes the symmetry $z(f_1,f_2)=z(f_2,f_1)$
for all tested frequency pairs $f_1$ and $f_2$.
The same symmetry applies to any other
$K_1$ number of signals.
Therefore, we compute $z$ test statistic
only for all combinations
\begin{eqnarray}
  f_{\mathrm{max}} \ge f_1 > f_2 > ... > f_{K_1} \ge f_{\mathrm{min}},
\label{EqCombinations}
\end{eqnarray}
where $f_{\mathrm{min}}$ and  $f_{\mathrm{max}}$ are the minimum
and maximum tested frequencies, respectively.
In the long search, we test an evenly spaced
grid of $n_{\mathrm{L}}$ frequencies between
$f_{\mathrm{min}}$ and  $f_{\mathrm{max}}$.
This gives us the best frequency candidates
$f_{\mathrm{1,mid}}, ... f_{\mathrm{K_1,mid}}$.
In the short search, we test a denser grid
of $n_{\mathrm{S}}$ frequencies within an interval
\begin{eqnarray}
[f_{\mathrm{i,mid}}-a,f_{\mathrm{i,mid}}+a],
\label{Eqc}
\end{eqnarray}
where $a = c (f_{\mathrm{min}}-f_{\mathrm{max}})/2$ and $i=1,..,K_1$.
In this study, we use $c=0.05$ which means that
the tested short search frequency interval
represents 5\% of the tested long search interval.

We search for periods between $P_{\mathrm{min}}=1/f_{\mathrm{max}}=5$ years
and  $P_{\mathrm{max}}=1/f_{\mathrm{min}}=200$ years.
The reasons for not detecting periods below
$P_{\mathrm{min}}=5$ years are discussed in
Section  \ref{SectIdentification}.

The best linear model for the data gives
the global periodogram minimum 
\begin{eqnarray}
  z_{\mathrm{min}}=
  z(f_{\mathrm{1,best}},f_{\mathrm{2,best}},...,
  f_{\mathrm{K_1,best}})
  \label{Eqzmin}
\end{eqnarray}
at the tested frequencies
$f_{\mathrm{1,best}},f_{\mathrm{2,best}},...,
f_{\mathrm{K_1,best}}$. 

The scalar $z$ periodogram values
are computed from $K_1$ frequency values.
For example, $K_1=2$ two signal periodogram $z(f_1,f_2)$
could be plotted like a map,
where $f_1$ and $f_2$ are the coordinates,
and $z=z(f_1,f_2)$ is the height.
For three or more signals, such a direct graphical presentation 
becomes impossible, because it requires more than three
dimensions.
We solve this
problem by
presenting only the
following one-dimensional
slices of the full periodogram
\begin{eqnarray}
  z_1(f_1) & = & z(f_1,f_{\mathrm{2,best}}, ...,f_{\mathrm{K_1,best}}) \nonumber \\
  z_2(f_2) & = & z(f_{\mathrm{1,best}},f_2, f_{\mathrm{3,best}}, ...,f_{\mathrm{K_1,best}}) \nonumber \\
  z_3(f_3) & = & z(f_{\mathrm{1,best}},f_{\mathrm{2,best}},f_3, f_{\mathrm{4,best}},...,f_{\mathrm{K_1,best}}) ~\nonumber \\
  z_4(f_4) & = & z(f_{\mathrm{1,best}},f_{\mathrm{2,best}},f_{\mathrm{3,best}},f_4, f_{\mathrm{5,best}},f_{\mathrm{K_1,best}}) \label{EqPeriodograms} \\
  z_5(f_5) & = & z(f_{\mathrm{1,best}},f_{\mathrm{2,best}},f_{\mathrm{3,best}},f_{\mathrm{4,best}},f_5,f_{\mathrm{K_1,best}}) \nonumber \\
  z_6(f_6) & = & z(f_{\mathrm{1,best}},f_{\mathrm{2,best}},f_{\mathrm{3,best}},f_{\mathrm{4,best}},f_{\mathrm{5,best}},f_6). \nonumber 
\end{eqnarray}
Using the above $K_1=2$ map analogy,
the slice $z_1(f_1)$ represents the height $z$ at a $f_1$ coordinate
when moving along the constant line $f_2=f_{\mathrm{2,best}}$
that crosses the global minimum $z_{\mathrm{min}}$
(Equation \ref{Eqzmin})
at the coordinate point $(f_{\mathrm{1,best}},f_{\mathrm{2,best}})$.

The best frequencies detected in the short search
give the initial values for the first group of free parameters
$\bm{\beta}_{\mathrm{I,initial}}=[f_{\mathrm{1,best}}, ..., f_\mathrm{K_1,best}]$
(Equation \ref{EqBetaOne}).
The linear model with these constant
$[f_{\mathrm{1,best}}, ..., f_\mathrm{K_1,best}]$
frequency values gives the unambiguous
initial values for
the second group 
$\bm{\beta}_{\mathrm{II,initial}}$ of free parameters
(Equation \ref{EqBetaTwo}).
The final non-linear iteration is performed from
\begin{eqnarray}
  \bm{\beta}_{\mathrm{initial}}=[\bm{\beta}_{\mathrm{I,initial}},
  \bm{\beta}_{\mathrm{II,initial}}] \rightarrow
   \bm{\beta}_{\mathrm{final}}.
\label{EqIteration}
\end{eqnarray}

DCM determines the following parameters for
$h_i(t)$ signals
\begin{itemize}

\item[] $P_i = 1/f_i = $ Period
\item[] $A_i = $ Peak to peak amplitude
\item[] $t_{\mathrm{i,min,1}} = $ Deeper primary minimum epoch 
\item[] $t_{\mathrm{i,min,2}} = $ Secondary minimum epoch (if present)
\item[] $t_{\mathrm{i,max,1}} = $ Higher primary maximum epoch
\item[] $t_{\mathrm{i,max,2}} = $ Secondary maximum epoch (if present),
\end{itemize}
as well the $M_k$ parameters 
of the $p(t)$ trend.
For the sunspots,
the most interesting
parameters
are 
the signal periods $P_i$,
the signal  amplitudes $A_i$,
and signal primary minimum
epochs $t_{\mathrm{min,1}}$ 
given in \Supplementary Tables
\ref{TableRmonthly2000K410R14}
-
\ref{TableCyearlyK410R14}.

The subtraction
\begin{eqnarray}
  y_{i,j}=y_i-[g(t_i)-h_j(t_i)]
\label{EqSignals}
\end{eqnarray}
shows how well all observations $y_i$
are connected to any signal $h_j(t)$, where $j=1,...,K_1$.
In other words, the full
  model $g(t_i)$ is subtracted from
  the data $y_i$, except for the $h_j(t_i)$ signal itself.
  We show an example of this subtraction 
  in our  \Supplementary Fig. \ref{Rmonthly2000K410R14Signals}.
  This subtraction resembles the 
  Discrete Fourier Transform de-trending procedure \Rr{\cite{Mur12}},
  but here we remove both the trend and all other signals,
  except for the $h_j(t_i)$ signal.

The DCM model
parameter errors are
determined
with
the bootstrap procedure
\Rr{\cite{Efr86}}.
We have previously used this same
bootstrap procedure
in our TSPA- and CPS-methods
\Rr{\cite{Jet99,Leh11}}.
A random sample $\bm{\epsilon}^*$
is selected from
the residuals $\bm{\epsilon}$
of the DCM model (Equation \ref{EqResiduals}).
In this reshuffling of $\epsilon_i$ residuals,
any $\epsilon_i$ 
can be chosen
as many times as
the random selection 
happens to favour it.
This random sample of residuals gives the {\it artificial} 
bootstrap data sample
\begin{eqnarray}
y_i^*=g_i+\epsilon_i^*.
\label{EqBoot}
\end{eqnarray}
We create numerous such $\bm{y}^*$  random samples.
The DCM model for each 
$\bm{y}^*$ sample gives
one estimate for
every model parameter.
The error estimate
for each particular 
model parameter is
the standard deviation
of all estimates obtained
from all $\bm{y}^*$
bootstrap samples.

DCM models are nested. For example,
a one signal model is a special case of a two signal model.
DCM uses the Fisher-test to compare
any pair of simple $g_1(t)$ and complex $g_2(t)$ models.
Their number of free parameters are $\eta_1 < \eta_2$.
Their sums of squared residuals $(R_1,R_2)$
and Chi-squares $(\chi_1,\chi_2)$
give the Fisher-test test statistic 
\begin{eqnarray}
F_R & = &
\left(
{
{R_1}
\over
{R_2}
}
-1
\right)
\left(
{
{n-\eta_2-1}
\over
{\eta_2-\eta_1}
}
\right)
          \label{EqFR} \\
 F_{\chi} & = &
\left(
{
{\chi_1^2}
\over
{\chi_2^2}
}
-1
\right)
\left(
{
{n-\eta_2-1}
\over
{\eta_2-\eta_1}
}
\right).
          \label{EqFChi}          
\end{eqnarray}
The Fisher-test null hypothesis is
\begin{itemize}

\item[] $H_{\mathrm{0}}$: {\it ``The complex model $g_2(t)$ 
does not provide
a significantly better fit to the data
than the simple
model $g_{\mathrm{1}}(t)$.'' }

\end{itemize}
\noindent
Under $H_0$, both test statistic parameters 
$F_R$ and $F_{\chi}$ have an $F$ distribution with 
$(\nu_1,\nu_2)$ degrees of freedom,
where $\nu_1=\eta_2-\eta_1$ and $\nu_2=n-\eta_2$
\Rr{\cite{Dra98}}. 
The probability for 
$F_R$ or $F_{\chi}$ reaching values higher than $F$
is called the critical level
$Q_{F} = P(F_R \ge F)$ or $Q_{F} = P(F_{\chi} \ge F)$.
We reject the $H_0$ hypothesis,
if
\begin{eqnarray}
  Q_F < \gamma_F=0.001,
\label{EqFisher}
\end{eqnarray}
where 
$\gamma_F$ is the pre-assigned significance level.
This  $\gamma_F$ represents the probability of falsely rejecting
$H_0$ hypothesis when it is in fact true.
If $H_0$ is rejected, we rate the complex $g_2(t)$ model 
better than the simple $g_1(t)$ model.

The basic idea of the Fisher-test is simple.
The $H_0$ hypothesis
rejection probability increases for larger $F_R$ values
having smaller $Q_F$ critical levels.
The complex model $R_2$ or $\chi^2_2$ values decrease
when the $\eta_2$ number of free parameters increases.
This increases the first $(R_1/R_2-1)$ and
$(\chi^2_1/\chi^2_2-1)$ terms in Equations \ref{EqFR} and \ref{EqFChi}.
However, the second $(n-\eta_2-1)/(\eta_2-\eta_1)$ penalty 
term decreases at the same time.
This second penalty term prevents over-fitting,
i.e. accepting complex models having
too many $\eta_2$ free parameters.

The key ideas of DCM are based on the following robust,
thoroughly tested statistical approaches

  \begin{enumerate}

  \item 
    The DCM model $g(t)$ is {\it non-linear}
    (Equation \ref{Eqmodel}).
    This model becomes {\it linear} when the frequencies
    $\bm{\beta}_{I}=[f_1, ..., f_{K_1}]$
    (Equation \ref{EqBetaOne})
    are fixed to their tested
    numerical values. This
    linear
    model
    gives {\it unambiguous} results for the other
    free parameters, 
    $\bm{\beta}_{II}$
    (Equation \ref{EqBetaTwo}).

 \item 
    The short search
    $f_{\mathrm{1,best}} > f_{\mathrm{2,best}} > ... >f_{\mathrm{K_1,best}}$
    grid combination that 
    minimises the $z$
    test statistic gives the best initial 
    $\bm{\beta}_{\mathrm{initial}}$ values
    for the non-linear iteration of Equation \ref{EqIteration}.

  \item 
    DCM tests a dense grid of all possible
    frequency combinations
    $f_{\mathrm{max}} \ge f_1 \!> \!f_2 \!>\!... \!>\!f_{K_1} \ge f_{\mathrm{min}}$
    (Equation \ref{EqCombinations}).
    For every 
    frequency combination, the
    {\it unambiguous linear} model least squares fit
    gives the test statistic 
    $z=\sqrt{\chi^2/n}$
    (errors $\sigma_i$ known)
     or
     $z=\sqrt{R/n }$
     (errors $\sigma_i$ unknown).

  \item  All model
    parameter error estimates are determined with the bootstrap method
    (Equation \ref{EqBoot}).

  \item  \label{IdeaFive}
    The best model is identified using the Fisher-test,
    which compares all different tested 
    $K_1$, $K_2$ and $K_3$
    order combinations for nested models
    \Rr{\cite{Dra98,All04}}.

  \end{enumerate}

  DCM has the following restrictions

  \begin{enumerate}

  \item
    If the frequency grid 
    $f_1> f_2 > ... > f_{K_1}$
    contains $n_L$ values in the long search, then
    the total number of all tested
    frequency combinations is
    \begin{eqnarray}
      \binom{n_L}{K_1}={{n_L!}\over{K_1! (n_L-K_1)!}}.
      \label{EqCPU}
    \end{eqnarray}
    For example, it took
    about two days 
    for
    a cluster of processors to
    compute the four signal DCM model
    for the \RmonthlyTwo ~sample
    (\Supplementary Table \ref{TableRmonthly2000K410R14} model \M=4,
    Figs. \ref{Rmonthly2000K410R14z}
    -
    \ref{Rmonthly2000K410R14fA}).

  \item An adequately dense tested frequency
    grid eliminates the possibility 
    that the best frequency combination is missed.
    The restriction is that denser grids
    require more computation time.
    If the tested frequency grid is sufficiently dense,
    no abrupt periodogram jumps occur,
    because the $z$ values for all
    close tested frequencies correlate.
    Hence, the frequencies of the minima
    of these periodograms
    are accurately determined.
    There is no need to test
    an even denser grid,
    because this would not alter
    the final result of the non-linear
    iteration of Equation \ref{EqIteration}.

  \item 
    Some DCM models are unstable.
    They are simply wrong
    models for the data,
    like a wrong $p(t)$
    trend order $K_3$ 
    or a search for too few or too many $K_1$
    signals (e.g.,  Figs. 5-10 in Ref\cite{Jet20}).
    This causes model instability. 
    We denote such unstable
    models with ``$\UM$''
    in \Supplementary Tables
\ref{TableRmonthly2000K410R14}
-
\ref{TableCyearlyK410R14}.
     The signatures of such unstable models are

  \begin{itemize}
  \item[] ``$\IF$'' = Intersecting frequencies
  \item[] ``$\AD$'' = Dispersing amplitudes
  \item[] ``$\LP$'' = Leaking periods   
  \end{itemize}

\item[] Intersecting
  frequencies ``$\IF$''  occur when the frequencies
   of two signals are very close to each other.
   For example, if frequency $f_1$ approaches frequency $f_2$,
   the $h_1(t)$ and $h_2(t)$ signals become essentially one and
   the same signal. This ruins the least squares fit.
   It makes no sense to add the same signal twice.

 \item[] Dispersing amplitudes ``$\AD$'' also occur when two signal
   frequencies are close to each other, like
   in the above-mentioned ``$\IF$'' cases. The least squares fit uses 
   two high amplitude signals which 
   cancel out. Hence, the  sum of these two high amplitude signals
   is one low amplitude signal that fits to the data.

 \item[] We take extra care to identify the suspected
   ``$\IF$'' and $"\AD$'' unstable model cases. 
   The reshuffling of bootstrap $\bm{\epsilon}^{\star}$ residuals
   provides a good test for identifying such unstable models
from the DCM analysis of numerous artificial
bootstrap data samples $\bm{y}^{*}$ (Equation \ref{EqBoot}).
If we encounter any signs of instability,
      we test combinations
      $n_{\mathrm{L}}=100~ \& ~c=0.05$,
      $n_{\mathrm{L}}=100~ \& ~c=0.10$,
      $n_{\mathrm{L}}=120~ \& ~c=0.05$ and
      $n_{\mathrm{L}}=120~ \& ~c=0.10$.
      If there is instability in any of these combinations,
      we reject the model as unstable $("\UM")$.
      If there is no instability,
      we take the $n_{\mathrm{L}}$ and $c$ combination
      that gives the lowest value
      for $R$ (non-weighted data)
      or $\chi^2$ (weighted data).
      
    \item[] Leaking periods ``$\LP$'' instability
      refers to the cases, where
      the detected frequency $f$ is
      outside the tested frequency interval
      between $f_{\mathrm{min}}$ and  $f_{\mathrm{max}}$,
      or the period $P=1/f$ of this frequency is
      longer than the time span $\Delta T$ of the data.
   
 \end{enumerate}

 We use the DCM model $g(t_i,\bm{\beta})$
 to predict the future and past data.
 The three samples used in these predictions
 are the predictive data, the predicted data and all data.
 We use the following notations for
the time points, the observations
and the errors of these three samples.

\begin{itemize}

  \item[-] Predictive data:  $n$ values of $t_i$, $y_i$ and $\sigma_i$
  \item[-] Predicted  data: $n'$ values of $t_i'$, $y_i'$ and $\sigma_i'$
  \item[-] All data: $n''=n+n'$ values of
    $t''_i$, $y''_i$ and $\sigma''_i$ from predictive
    and predicted data

  \end{itemize}

  We also compute the monthly or yearly time points during
  past activity minima of Table \ref{TableEvents}.
  Our notations for these time points are
  
  \begin{itemize}
  \item[-] Activity minimum data:
    $n'''$ values of $t'''_i$ 
\end{itemize}

\noindent
DCM gives the best free parameter $\bm{\beta}$ values
for the predictive data model  $g_i=g(t_i,\bm{\beta})$,
where the ``old'' $t_{\mathrm{mid}}$ and $\Delta T$ values are
computed from the predictive data time
points $t_i$.
These ``old'' predictive data $t_{\mathrm{mid}}$ and $\Delta T$ values
are used to compute the predicted data model values
\begin{eqnarray}
g'_i & = & g(t'_i,\bm{\beta})   \nonumber 
\end{eqnarray}
In other words, we do not compute ``new''
$t_{\mathrm{mid}}$ and $\Delta T$ values
from the predicted data time points $t'_i$.
There is an obvious reason for this.
The predictive data free parameter values
$\bm{\beta}$ give the correct
$g'_i$ values
only for the  ``old'' predictive data
$t_{\mathrm{mid}}$ and $\Delta T$ values.
The residuals
\begin{eqnarray}
\epsilon'_i  =  y'_i-g'_i   \nonumber
\end{eqnarray}
of the predicted data give the {\it predicted test statistic}
\begin{eqnarray}
  z_{\mathrm{pred}}  =
  \textnormal{z test statistic for predicted data 
  (Equation \ref{EqzR} or \ref{EqzChi}),} \label{EqPredz} 
\end{eqnarray}
\noindent
which measures how well the prediction obtained
from the predictive data works.

The best DCM model $g(t''_i,\bm{\beta})$ for all data
uses the ``old'' $t_{\mathrm{mid}}$ and $\Delta T$ values
computed from $t''_i$ time points of all data.
We use these  ``old'' $t_{\mathrm{mid}}$ and $\Delta T$
values to compute
the predicted model values
\begin{eqnarray}
  g'''_i=g(t'''_i,\bm{\beta}) 
\label{EqGthree}
\end{eqnarray}
for the monthly or yearly time points $t'''_i$ during the
chosen activity minimum of Table \ref{TableEvents}.
These $g'''_i$ values give the
{\it predicted mean level}
\begin{eqnarray}
  m_{\mathrm{pred}}  = {{1}\over{n'''}}\sum_{i=1}^{n'''}g'''_i
  \label{EqPredMean}
\end{eqnarray}
during the selected activity minimum of Table \ref{TableEvents}.

It is essential to understand how and
why the above two predictive parameters
of Equations \ref{EqPredz} and \ref{EqPredMean} are computed.
The former parameter, $z_{\mathrm{pred}}$, is computed from
the {\it known}
$t'_i$, $y'_i$ and $\sigma'_i$ values of the predicted data.
If our predictions are correct,
the following regularities should occur
\begin{itemize}
\item[] $z_{\mathrm{pred}}$ decreases,
  if DCM detects real new signals in the predictive data
\item[] $z_{\mathrm{pred}}$ increases,
  if DCM detects unreal new signals in the predictive data
\end{itemize}

\noindent
For the latter $m_{\mathrm{pred}}$ parameter,
the $t'''_i$, $y'''_i$ or $\sigma'''_i$ data values
during the past activity minima 
are {\it unknown},
but we can create the $n'''$ monthly or yearly $t'''_i$ time points
during these minima  of Table \ref{TableEvents}.
We can then compute the $g_i'''$ values from these
$t'''_i$ values (Equation \ref{EqGthree}).
This gives us the mean
$m_{\mathrm{pred}}$ value (Equation \ref{EqPredMean})
that can be compared to the mean level $m$ for all data.
If our predictions for the past activity minima of
Table \ref{TableEvents} are correct,
the following regularities should occur
\begin{itemize}
\item[] $m_{\mathrm{pred}}$ falls below  $m$,
  if DCM detects real new signals
  in all data
\item[] $m_{\mathrm{pred}}$ decreases,
  if DCM detects real new signals in all data 
\item[] $m_{\mathrm{pred}}$ increases,
  if DCM detects unreal new signals in all data
\end{itemize}

Our model is non-linear
(Equations~ \ref{Eqmodel} - \ref{EqPolyTwo}).
We find a unique solution for this ill-posed problem.
The stages of finding this solution are

\begin{enumerate}

\item We fix the numerical values of the
  tested frequencies $\bm{\beta}_{I}$ (Equation \ref{EqBetaOne}).
  The model becomes linear and the solution
  for the other free parameters 
  $\bm{\beta}_{II}$ is unique (Equation \ref{EqBetaTwo}).

\item We test all possible  $\bm{\beta}_{I}$
  frequency combinations
  (Equation \ref{EqCombinations}).
  The linear model for
  every tested frequency combination $\bm{\beta}_{I}$
  gives a unique value for
  the test statistic $z$ (Equations~ \ref{EqzR} and \ref{EqzChi}).
  We select the best frequency combination  $\bm{\beta}_{\mathrm{I,best}}$
  that minimises $z$.
  The model for this best frequency combination
  is unique
  and the free parameter values
  $\bm{\beta}_{\mathrm{initial}}=
  [\bm{\beta}_{\mathrm{I,best}},\bm{\beta}_{\mathrm{II,best}}]$ of this model
  are unique.

\item
  These unique initial free parameter values
  $\bm{\beta}_{\mathrm{initial}}$ are used in the non-linear
  iteration (Equation \ref{EqIteration})
  that gives the unique final free parameter
  values  $\bm{\beta}_{\mathrm{final}}$.
  This solution is unique.

\item We  use the Fisher-test
  to compare many different non-linear
  models against each other.
  The selection criterion
  for the best model is unique (Equation \ref{EqFisher}).
  
\end{enumerate}

\noindent
After the selection of the tested non-linear models,
these four stages give a unique
the solution for this ill-posed problem.


\section*{Acknowledgements}

  We thank the Finnish Computing
  Competence Infrastructure (FCCI)
  for supporting this project
  with computational resources.
  Juha Helin,
  Jani Jaakkola,
  Sami Maisala and
  Pasi Vettenranta
  have
  helped us utilise
  parallel computation resources in
  the High Performance Computing (HPC) platform.
  We thank professor Alexis Finoguenov
  for his insightful comments,
  as well as for recommending the use
  of grammar packages.
This work has made use of NASA's 
Astrophysics Data System (ADS) services.

\section*{Author contributions statement}
L.J. analysed the data, prepared the figures and the tables,
  and wrote the manuscript.

\section*{Additional information}

\noindent
{\bf \Supplementary information} The online version contains
supplementary material available at https:...  

\section*{Data availability}

All data files, DCM Python code, 
DCM control files and Fisher-test Python code are stored to the
\Link{https://zenodo.org/uploads/11503698}
{Zenodo database in https://zenodo.org/uploads/11503698}.
Our \Supplementary  material 
      contains instructions for repeating
      the whole DCM analysis. 

\section*{Declarations}

  \section*{Competing interests}
  The author declares no competing interests. \\

\noindent
{\bf Correspondence} and requests for materials should be
addressed to L.J.

\clearpage


\pagestyle{empty}
\setcounter{linenumber}{1}


\setcounter{section}{0}
\setcounter{figure}{0}
\setcounter{table}{0}
\setcounter{equation}{0}

\renewcommand{\thesection}{S\arabic{section}}
\renewcommand{\thefigure}{S\arabic{figure}}
\renewcommand{\thetable}{S\arabic{table}}
\renewcommand{\theequation}{S\arabic{equation}}

\begin{center}
  {\bf \Large Supplementary material \\
    Do the planets cause the sunspot cycle? } \\
  L. Jetsu \\
  Department of Physics, P.O. Box 64,
FI-00014 University of Helsinki, Finland

\end{center}

\section{Data} \label{SectData}

We retrieved the sunspot data
from the
Solar Influences Data Analysis
Center (Source: WDC SILSO,
Royal Observatory of Belgium, Brussels
in December 2022). The monthly mean total sunspot number
data begin in January 1749 and 
end to November 2022 (Table \ref{TableMonthly}: $n=3287$).
After January 1818,
the monthly mean standard deviation $(S_i)$
of the input sunspot numbers
from individual stations are available, as well as
the total number of observations $(N_i)$.

The yearly mean total sunspot number
data begin in the year
1700 and end to 2021 (Table \ref{TableYearly}: $n=322$).
The $S_i$ and $N_i$ estimates are available
after 1818.

The  standard errors 
\begin{eqnarray}
  \sigma_i =  S_i/\sqrt{N_i}   \label{EqE} 
\end{eqnarray}
for these data 
give the weights $w_i=\sigma_i^{-2}$.
The normalised weights are
\begin{eqnarray}
  w_{\mathrm{nor,i}} = n w_i/\sum_{i=1}^n  w_i.   \label{EqWnor}
\end{eqnarray}

If all errors were equal,
the normalised weight $w_{\mathrm{nor,i}}$
for every observation
would be one.
For all monthly data,
the normalised weights show that
one observation out of $n=2458$ observations
has the largest weight $w_{\mathrm{nor,max}}=381$
(Table \ref{TableWeights}, Line 1).
The four most accurate, $N_{1/2}=4$,
observations influence
the modeling more than the
remaining $n-4=2454$ observations.
These monthly data statistics improve slightly
if the more accurate data after the year
2000 are removed, which gives
$N_{1/2}=64$ for sample size $n=2183$.
All yearly data show an extreme case,
$N_{1/2}=1$,
where the weight $w_{\mathrm{nor,max}}=135$
of one observation
exceeds the weight of all other $n-1=203$ observations.
Again, the statistics improve slightly
if the more accurate data after
the year 2000 are removed.

For these biased normalised weights,
the period and the amplitude error estimates
for the weighted data signals would be dramatically
larger than the respective errors
for the non-weighted data signals.
For example, the bootstrap reshuffling
(see Equation \ref{EqBoot})
of the four $N_{1/2}=4$ most accurate
monthly values, or the one $N_{1/2}=1$
most accurate yearly value,
resembles Russian roulette,
where the majority of
remaining other data values do not
seem to need to fit to
the model at all.
We emphasise that the weighted data itself causes this bias,
not our period analysis method.

%

\begin{table}[h] 
\caption{Times $t_i$ of monthly mean total sunspot numbers $y_i$
   $(n=3287)$.
   The standard deviation $(S_i)$
   and the total number of estimates $(N_i)$
  are available from January 1818 onward.
  Values -1 indicate cases having no 
  $S_i$ and $N_i$ estimate.
  This table shows only the first
  and the last lines,   because all data is available in electronic form.}
  \label{TableMonthly} 
 \begin{center}
  \begin{tabular}{rrrr}
\hline
  $t_i$ & $y_i$ & $S_i$  & $N_i$ \\
   (y)  &  (-)  &  (-)  &  (-)  \\
\hline
         1749.042 &            96.7     &  -1   &  -1    \\
         ...      &           ...       &  ...  &  ...  \\
         2022.873 &            77.6     & 14.1  &  881  \\
\hline
  \end{tabular}
   \end{center}
\end{table}

\begin{table}[h]  
  \caption{Yearly mean total sunspot numbers $(n=322)$, 
    otherwise as in Table \ref{TableMonthly}}
  \label{TableYearly} 
 \begin{center}
\begin{tabular}{rrrr}
  \hline
  $t_i$ & $y_i$ & $S_i$ & $N_i$ \\
   (y)  &  (-)  &  (-)  &  (-)  \\ 
  \hline
         1700.5 &            8.3     &  -1   &  -1    \\
         ...    &           ...      &  ...  &  ...  \\
         2021.5 &           29.6     & 7.9   &  15233 \\
  \hline
\end{tabular}
\end{center}
\end{table}

\begin{table}[h] 
  \caption{Normalised weights $w_{\mathrm{nor,i}}$
    for Equation \ref{EqWnor}.
    (1) Sample.
    (2) Sample size.
    (3,4) Maximum and minimum  $w_{\mathrm{nor,i}}$.
    (5) Number of largest $w_{\mathrm{nor,i}}$ having a sum exceeding  $n/2$.}
    \label{TableWeights} 
 \begin{center}
    \begin{tabular}{lcccr}
    \hline
         (1)      &
         (2)      &
         (3)      &
         (4)      &
         (5)     \\
  Sample               &
  $n$                  &
  $w_{\mathrm{nor,max}}$ &
  $w_{\mathrm{nor,min}}$ & 
                           $N_{\mathrm{1/2}}$   \\
                  &
         (-)        &
         (-)        &
         (-)        &
         (-)        \\                               
\hline
\multicolumn{5}{c}{Monthly data} \\
\hline
All          & 2458 & 381  &$2.02\times10^{-5}$ & 4  \\
Before 2000  & 2183 & 48.9 &$1.83\times10^{-3}$& 64  \\
 \hline
  \multicolumn{5}{c}{Yearly data} \\
\hline
    All          & 204 & 135  &$1.08\times10^{-3}$ & 1  \\
    Before 2000  & 182 & 21.6 &$1.11\times10^{-2}$ & 6  \\
\hline
    \end{tabular}
     \end{center}
\end{table}

We solve this statistical bias of errors $\sigma_i$ by
using the Sigma-cutoff weights
\begin{eqnarray}
    \begin{cases}
w_i'=1,                           & \text{if $\sigma_i \le K \bar{\sigma}$} \\
w_i'=(K \bar{\sigma}/\sigma_i)^x, & \text{if $\sigma_i >  K \bar{\sigma}$,}
    \end{cases}       
\end{eqnarray}
where $\bar{\sigma}$ is the mean of all $\sigma_i$,
and the $K$ and $x$ values can be freely
chosen\cite{Han03,Bre02}.
We use the same $K=1$ and $x=2$
values as in one former study\cite{Rod03},
which gives
\begin{equation}
  \begin{cases}
    \sigma_i'=1 ,& \text{if $\sigma_i \le \bar{\sigma}$} \\
    \sigma_i'=\sigma_i/\bar{\sigma} > 1, & \text{if $\sigma_i > \bar{\sigma}$}.
    \label{EqSigma}
    \end{cases}       
 \end{equation}
  Our chosen $K$ and $x$ values are reasonable,
  because they give 55 percent of monthly
  data having full weight 1,
  and 25 percent having weight below 1/2.
  The respective values for the yearly data
  are 53 and 24 percent.

There are four exceptional cases.
We use the sigma cutoff weights $w_i'=1$
for the three exceptional cases $y_i=0$, $S_i=0$ and $N_i>0$,
which solves the infinite weight
$w_i=\sigma_i^{-2}=0^{-2}=\infty$ problem.
Finally, we compute no error estimate
for the one exceptional case, $S_i>0$ and $N_i=0$.

We analyse eight samples.
The first four samples are monthly sunspot data
drawn from Table \ref{TableMonthly}.
The last four samples
are yearly sunspot data drawn from Table \ref{TableYearly}.
All eight samples are published only in electronic form.
We summarise the sample contents and sample
naming abbreviations 
in Table \ref{TableSamples}.

\begin{table}[h]  
  \caption{Eight sunspot data
    samples. (1) Table from where sample is drawn. (2) Sample name.
    (3-4) Sample first and last observing time. (5) Sample time span.
    (6) Sample size. (7) Sample electronic file name.}
  \label{TableSamples} 
 \begin{center}
  \begin{tabular}{llccccl}
  \hline
  (1)        &
  (2)        &
  (3)        &
  (4)        &
  (5)        &
  (6)        &
  (7)        \\
  Table      &
  Name       &
  $t_1$      &
  $t_n$      &
  $\Delta T$ &
  $n$        &
  Data file      \\
             &
             &
   (y)         &
   (y)         &
   (y)         &
   (-)         &
            \\
\hline
Table \ref{TableMonthly}  &  \RmonthlyOne    & 1749.0 & 2022.8 & 273.8    & 3287  & \PRtext{Rmontly.dat}      \\
Table \ref{TableMonthly}  &  \RmonthlyTwo    & 1749.0 & 1999.9 & 250.9    & 3012  & \PRtext{Rmonthly2000.dat} \\
Table \ref{TableMonthly}  &  \CmonthlyOne    & 1818.0 & 2022.8 & 204.8    & 2458  & \PRtext{Cmonthly.dat}     \\
Table \ref{TableMonthly}  &  \CmonthlyTwo    & 1818.0 & 1999.9 & 181.9    & 2182  & \PRtext{Cmonthly2000.dat} \\
Table \ref{TableYearly}   &  \RyearlyOne     & 1700.5 & 2021.5 & 321.0    & 322   & \PRtext{Ryearly.dat}      \\
Table \ref{TableYearly}   &  \RyearlyTwo     & 1700.5 & 1999.5 & 299.0    & 300   & \PRtext{Ryearly2000.dat}  \\
Table \ref{TableYearly}   &  \CyearlyOne     & 1818.5 & 2021.5 & 203.0    & 204   & \PRtext{Cyearly.dat}      \\
Table \ref{TableYearly}   &  \CyearlyTwo     & 1818.5 & 1995.5 & 181.0    & 182   & \PRtext{Cyearly2000.dat}  \\
  \hline
\end{tabular}
\end{center}
\end{table}

\begin{figure}  
\vspace{0.02\textwidth}
\centerline{\hspace*{0.005\textwidth}
 \includegraphics[width=0.245\textwidth,clip=]{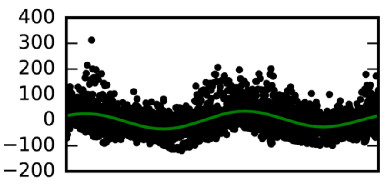} 
 \hspace*{-0.01\textwidth}
 \includegraphics[width=0.24\textwidth,clip=]{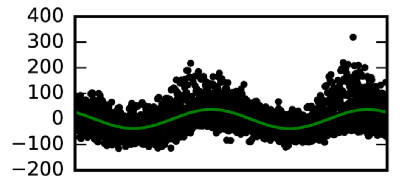}   
 \hspace*{-0.01\textwidth}
 \includegraphics[width=0.245\textwidth,clip=]{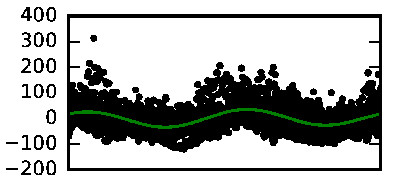} 
 \hspace*{-0.01\textwidth}
 \includegraphics[width=0.245\textwidth,clip=]{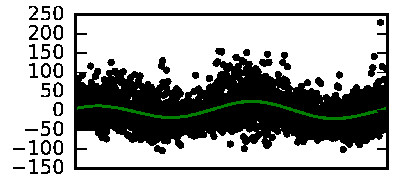}  
        }
\vspace{-0.15\textwidth}
\centerline{\Large \bf 
\hspace{0.11\textwidth}   \color{black}{(a)}
\hspace{0.165\textwidth}  \color{black}{(b)}
\hspace{0.175\textwidth}  \color{black}{(c)}
\hspace{0.175\textwidth}  \color{black}{(d)}
\hfill}
\vspace{0.15\textwidth}
\centerline{\hspace*{0.005\textwidth}
 \includegraphics[width=0.245\textwidth,clip=]{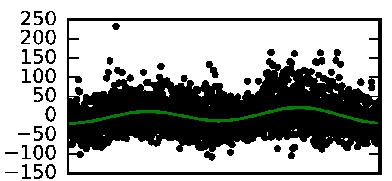}  
 \hspace*{-0.01\textwidth}
 \includegraphics[width=0.24\textwidth,clip=]{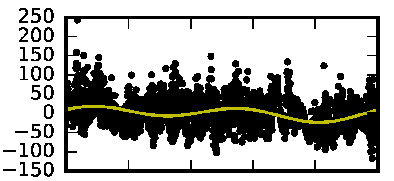}    
 \hspace*{-0.01\textwidth}
 \includegraphics[width=0.245\textwidth,clip=]{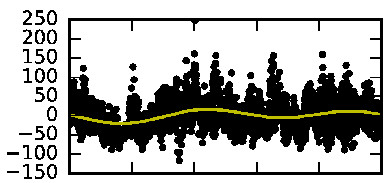}  
        }
\vspace{-0.15\textwidth}
\centerline{\Large \bf 
\hspace{0.23\textwidth}  \color{black}{(e)}
\hspace{0.175\textwidth}  \color{black}{(f)}
\hspace{0.1805\textwidth}  \color{black}{(g)}
\hfill}
\vspace{0.12\textwidth}
\caption{Double sinusoids\DW
(Table \ref{TableCompare}).
      (a) \CmonthlyOne: \SignalOne ~signal period
         $21.667\pm0.014  $ $= 2\times (10.834  \pm 0.007)$.
    (b) \RmonthlyOne: \SignalTwo ~signal period
         $20.0062\pm0.0088$ $= 2\times (10.0031 \pm 0.0044)$.
    (c) \RmonthlyTwo: \SignalThree ~signal period
         $23.686\pm0.018  $ $= 2\times (11.843  \pm 0.009)$.
    (d) \RmonthlyTwo: \SignalSix ~signal period
         $16.781\pm0.034  $ $= 2\times (8.390   \pm 0.017)$.
    (e) \RmonthlyOne: \SignalSix ~signal period
         $16.753\pm0.027  $ $= 2\times (8.376   \pm 0.014)$.
    (f) \RmonthlyTwo: \SignalNine ~signal period
         $136.43\pm0.99  $ $= 2\times (68.22   \pm 0.50)$.
    (g) \RmonthlyOne: \SignalNine ~signal period
         $143.39\pm0.99  $ $= 2\times (71.70   \pm 0.50)$.
            Numbers and labels
            for x-axis phases $\phi$
            between 0 and 1 are missing,
            because
            these figures are extracts from larger figures,
            like Fig. \ref{Rmonthly2000K410R14Signals}.
            Period units are years.
            Fig. units are x-axis $[\phi]=$
            dimensionless and y-axis $[y]=$ dimensionless.}
\label{FigDw}
\end{figure}

\subsection{\RmonthlyOne ~and \RmonthlyTwo}

\RmonthlyOne ~is
our largest sample ($n=3287$, $\Delta T=274^{\mathrm{y}}$).
It contains all $t_i$ and $y_i$ values from Table \ref{TableMonthly}.
Since the error estimates $\sigma_i$ are unknown
before January 1818,
we use an arbitrary error $\sigma_i=1$
for all data in this sample.
We perform a non-weighted period analysis,
which is based on the
sum of squared residuals (Equation \ref{EqR}: $R$).
Hence, the chosen $\sigma_i=1$
value has no effect to our analysis results,
because every observation has an equal weight.
The particular file name \RmonthlyOne ~is used
because the
period analysis is based on a non-weighted
$R$ test statistic. 

\RmonthlyTwo ~contains observations
from \RmonthlyOne,
which were made before the year 2000
$(n=3012, \Delta T=251^{\mathrm{y}})$.
Number ``2000'' 
refers to the omitted data after the year 2000.

\subsection{\CmonthlyOne ~and \CmonthlyTwo}

\CmonthlyOne ~contains all $t_i$ and $y_i$ observations
having an error estimate $\sigma_i'$ computed from
Equation \ref{EqSigma} $(n=2458,~ \Delta T=205^{\mathrm{y}})$.
For this sample, we apply the weighted period analysis, which
utilises the error $\sigma_i'$ information (Equation \ref{EqChi}: $\chi^2$).
The sample name begins with letter ``C'',
because our analysis
is based on the weighted Chi-square test statistic
(Equation \ref{EqzChi}).

\CmonthlyTwo ~contains those
\CmonthlyOne ~  observations,
which were made before the year 2000
$(n=2182, ~\Delta T=182^{\mathrm{y}})$.
We refer to the omitted data after the year 2000 
by using the number ``2000''.

\subsection{\RyearlyOne ~and \RyearlyTwo}

\RyearlyOne ~is our longest sample $(\Delta T = 321^{\mathrm{y}})$
containing $n=322$ yearly mean total sunspot number
observations over more than three centuries.
Since no error estimates are available for observations before 1818,
the value $\sigma_i=1$ is used for all data. 
We perform a non-weighted period analysis
based on the $R$ test statistic (Equation \ref{EqR}).
Therefore,
the sample name begins with the letter ``R''.

\RyearlyTwo ~contains observations from \RyearlyOne ~that
were made before the year 2000.

\subsection{\CyearlyOne ~and \CyearlyTwo}

\CyearlyOne ~is the smallest sample $(n=204, \Delta T=204^{\mathrm{y}})$.
It contains all yearly mean total sunspot numbers
having an error estimate $\sigma_i'$ (Equation \ref{EqSigma}).
We perform a weighted period analysis
based on the $\chi^2$ test statistic (Equation \ref{EqChi}).
Therefore, the sample name begins with the letter ``C''.

\CyearlyTwo ~contains all \CyearlyOne ~data before the year 2000.

\section{DCM period detections}

\subsection{Periods
  between 5 and 200 years}
\label{SectDCMone}

We must introduce the ``double sinusoid'' concept.
This concept is necessary when comparing the consistency of our
DCM period detections
in different samples.
  The shape of these double sinusoid signals
  resembles that of a pure sinusoid
(Figs. \ref{FigDw}a-g).
All seven double sinusoids 
are highlighted with the notation ``\DW'' in Table \ref{TableCompare}.
For example, we detect only one signal for the
  double wave $K_2=2$ model of \CmonthlyOne ~sample
  ~(Table \ref{TableCompare}: Column 9, Fig. \ref{FigDw}a).
  The period of this 
  signal
  is $P=21.667\pm0.014=2 \times (10.834\pm0.007)$ years.
The period of this double sinusoid signal is about
two times longer than the 
$\sim11$ years signal periods detected in all other eleven
samples.
In other words,
 including the \CmonthlyOne ~sample,
DCM detects this same $\sim 11$ years signal
in all twelve samples.

In Tables
\ref{TableRmonthly2000K410R14}-\ref{TableCyearlyK410R14},
  the four first \M=1-4 models,
  are computed for the original
  $y_i$ data.
  The next four \M=5-8 models 
  are computed for the $\epsilon_i$
  residuals of the \M=4 model.
 The total number of tested DCM models
 in these twelve tables is 80.
 All signal detections are consistent.
 If a new signal is detected,
the previously detected
 signal periods, amplitudes
and phases are also re-detected.
 Without this consistency,
  our analysis could be considered unreliable.
  Inside each table, 
    we use the Fisher-test to compare
    all models in an orderly fashion
  against each other.
  The extremely high
    Fisher-test critical level estimates,
  $Q_F$ \REJ ~$\ll \gamma = 0.001$
  (Equation \ref{EqFisher}),
  confirm that these signal detections are very significant.

  \begin{figure} 
  \centerline{\includegraphics[width=0.7\textwidth,clip=]{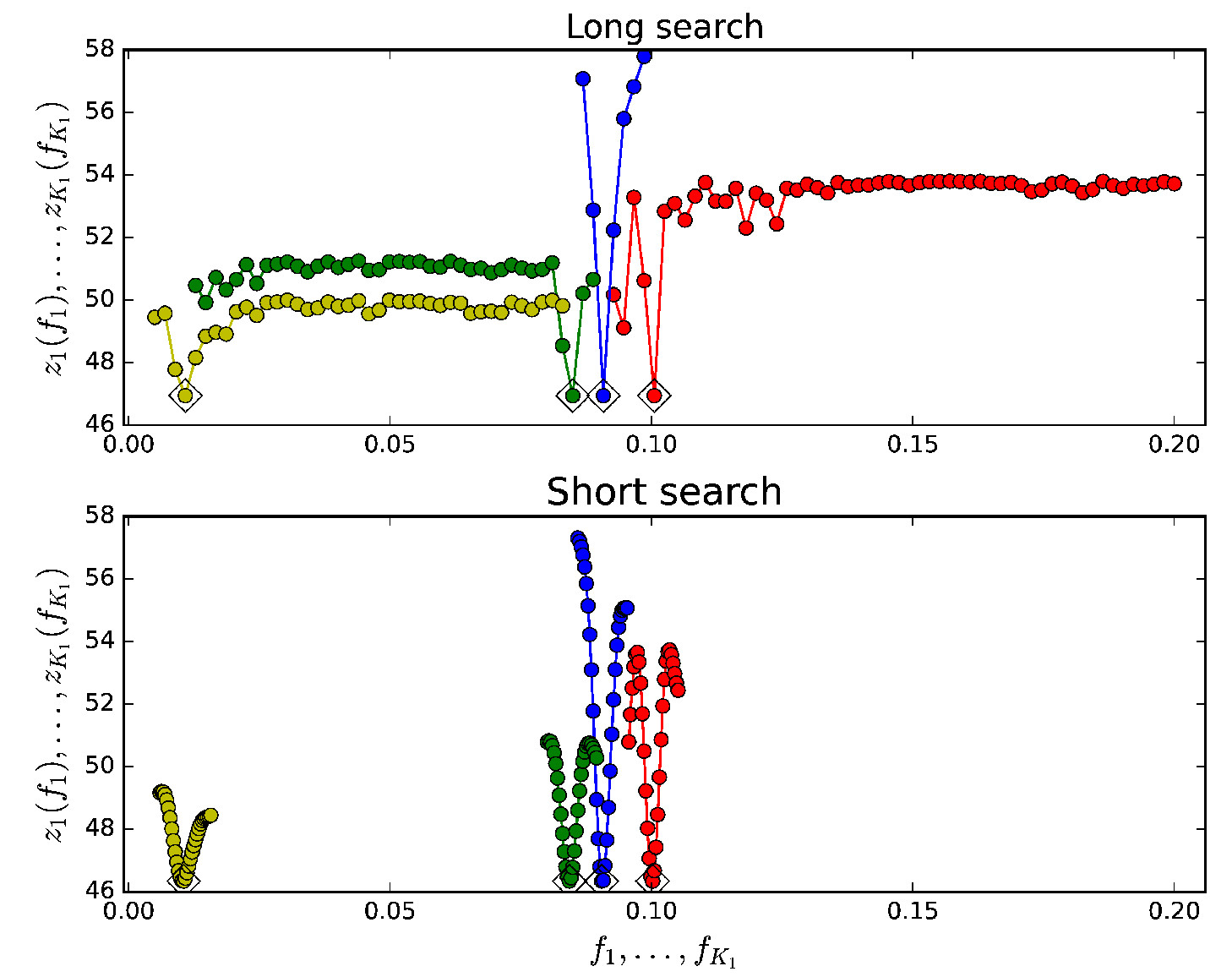}} 
  \caption{\RmonthlyTwo ~sample four pure sine signal model
    \M=4 (Table \ref{TableRmonthly2000K410R14}).
  Periodograms are computed between $P_{\mathrm{min}}=5$
  and $P_{\mathrm{max}}=200$ years
  for grids having
   $n_{\mathrm{L}}=100$ and $n_{\mathrm{S}}=30$ tested frequencies.
  Upper panel shows long search 
  $z_1(f_1)$ (red),
  $z_2(f_2)$ (blue),
  $z_3(f_3)$ (green),
  and
  $z_4(f_4)$ (yellow)
  periodograms (Equation \ref{EqPeriodograms}).
  Best frequencies are marked with diamonds.
  Lower panel shows short search periodograms for
  a denser frequency grid
  having an interval width $c=0.05\equiv$ 5\% (Equation \ref{Eqc}).
  Otherwise as in upper panel.
  Units in both panels
  are x-axis $[f]=$ 1/years and y-axis $[z]=$ dimensionless.} 
\label{Rmonthly2000K410R14z}
\end{figure}

\begin{figure} 
\centerline{\includegraphics[width=0.7\textwidth,clip=]{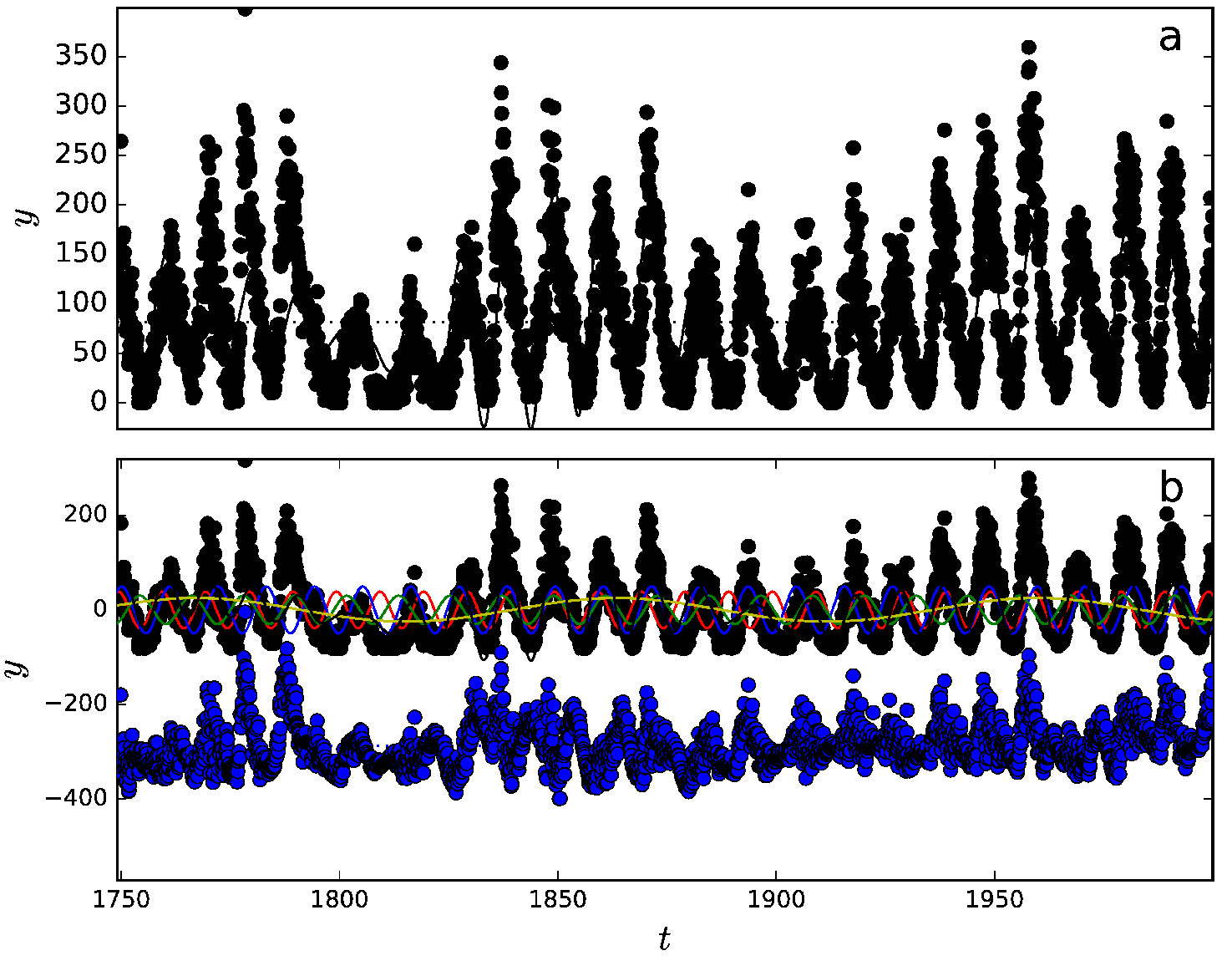}} 
  \caption{\RmonthlyTwo ~sample four pure sine signal model
    \M=4 (Table \ref{TableRmonthly2000K410R14}).
  Upper panel shows data $y_i$ (black circles),
  $g(t)$ (continuous black curve) and
  constant trend $p(t)=M_0=81.3$ (dotted black curve).
  Lower panel shows data minus
  trend $y(t_i)-p(t_i)$ values (black circles),
  and $g(t)-p(t)$ (black curve), $h_1(t)$ (red curve),
  $h_2(t)$ (blue curve), $h_3(t)$ (green curve)
  and $h_4(t)$ (yellow curve).
  Residuals (blue circles) are offset to $y=-300$ level.
  Signal periods are $P_1=9.9842$, $P_2=11.0234$, $P_3=11.846$
  and $P_4=96.2$ years.
 Units in both panels are x-axis $[t]=$ years and y-axis $[y]=$ dimensionless.} 
\label{Rmonthly2000K410R14gdet} 
\end{figure}

\begin{figure}  
\centerline{\includegraphics[width=0.7\textwidth,clip=]{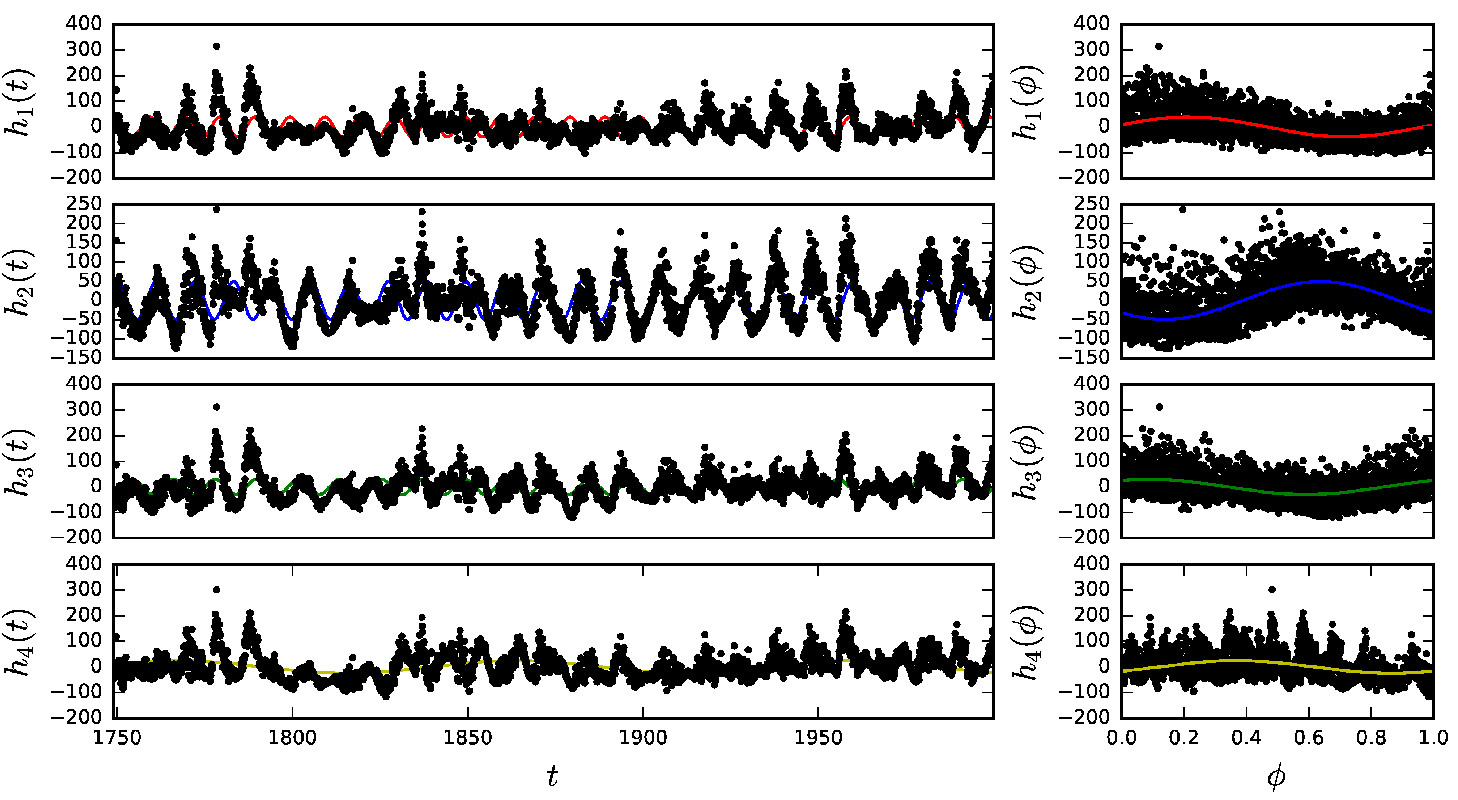}} 
\caption{\RmonthlyTwo ~sample four pure sine signal model
    \M=4 (Table \ref{TableRmonthly2000K410R14}).
    Left-hand panels.
    Values of $h_j(t)$ and 
    $y_{i,j}$ (Equation \ref{EqSignals}) for four signals
    as a function of time.
    Signal periods in years are $P_1=9.9842$ (red curve),
    $P_2=11.0234$ (blue curve),
    $P_3=11.846$ (green curve),
    and $P_4=96.2$ (yellow curve).
 Units are x-axis $[t]=$ years and y-axis $[h_j(t)]=[y_{i,j}]=$ dimensionless.
 Right-hand panels.
 Same signals as a function of
 phase $\phi$.
  Units are x-axis $[\phi]=$ dimensionless
  and y-axis $[h_j(t)]=[y_{i,j}]=$ dimensionless.}
\label{Rmonthly2000K410R14Signals} 
\end{figure}

\begin{figure}  
\centerline{\includegraphics[width=0.7\textwidth,clip=]{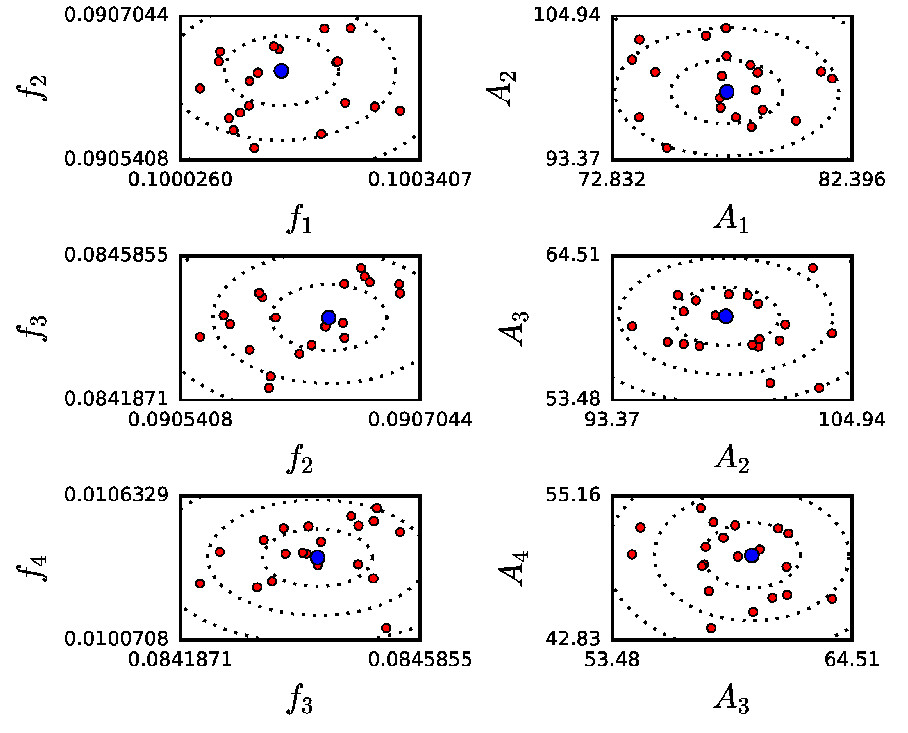}} 
\caption{\RmonthlyTwo ~sample four pure sine signal model
    \M=4 (Table \ref{TableRmonthly2000K410R14}).
  Left-hand panels show bootstrap signal frequency estimates
  $(f_1,f_2,f_3,f_4)$.
 Units are $[f_i]=$ 1/years.
  Right-hand panels show respective peak to peak
  amplitude estimates
  $(A_1,A_2,A_3,A_4)$.
 Units are $[A_i]=$ dimensionless.
  Larger blue circles denote values for
  original data.
  Smaller red circles denote values for
  bootstrap random samples $y_i^{\star}$ (Equation \ref{EqBoot}).
  Dotted lines denote one, two and three
  sigma error limits.} 
\label{Rmonthly2000K410R14fA}
\end{figure}

  We present a detailed full DCM analysis
  description {\it only} for
  sample \RmonthlyTwo.
  It is explained thoroughly
  how to read all the
  results given in Table \ref{TableRmonthly2000K410R14}.
  We explain the DCM periodogram
  (Fig. \ref{Rmonthly2000K410R14z}),
  the DCM model (Fig. \ref{Rmonthly2000K410R14gdet}),
  the DCM signals (Fig.  \ref{Rmonthly2000K410R14Signals}),
  and the DCM bootstrap results
  (Fig.  \ref{Rmonthly2000K410R14fA}).
  Our aim is to make it as easy as possible
to read and understand the results
presented in the other
eleven Tables
\ref{TableRmonthlyK410R14}-\ref{TableCyearlyK410R14}.
We do not describe the
  details of the analyses for other samples,
  because we analyse these other samples
  exactly same way as sample \RmonthlyTwo.

In Table \ref{TableRmonthly2000K410R14},
the first one signal \M=1 model is 
\RModel{1,1,0}, where
the model orders are $K_1=1$, $K_2=1$ and $K_3=0$,
and the subscript ``$R$'' refers
to a non-weighted DCM analysis based on $R$
test statistic (Equation \ref{EqR}).
This model has
$\eta=4$ free parameters (Equation \ref{EqEta}).
The sum of squared residuals of this model
is $R=1.09\times10^7$ (Equation \ref{EqR}).
The first signal period is $P_1=11.0336\pm0.0074$ years.
The dimensionless peak to peak amplitude
of this one signal is $A_1=92.2\pm2.8$.
The primary minimum epoch of this one signal 
is at year $t_{\mathrm{min,1}}=1755.30\pm0.12$.

Every new model is more complex when
 going down in our Table \ref{TableRmonthly2000K410R14},
 because
    the number $\eta$ of free parameters increases. 
    The Fisher-test between the simple
    \M=1 model and the complex \M=2
    model gives a test statistic
    value $F_R=255$ (Equation \ref{EqFR})
    reaching a critical level
    $Q_F<10^{-16}$.
    The Fisher-test null hypothesis $H_0$ is rejected because
    the criterion $Q_F<\gamma=0.001$ is fulfilled
    (Equation \ref{EqFisher}).
    Therefore, an upward arrow ``$\uparrow$'' indicates that
    the \M=2 model is a better model than the \M=1 model.
    A leftward arrow ``$\leftarrow$'' would have been used
    if model \M=1 were better.
    The electronic control file \PRtext{Rmonthly2000K110R14.dat}
    can be used to repeat
    whole DCM
    analysis of model \M=1, as explained later in the end of this
    Section \ref{SectDCMone}.
    The four first signals are detected from the original data
    in electronic file \PRtext{Rmontly2000.dat}.
    The next four \M=5-8 model signals are detected
    from the residuals
    of model \M=4, which are stored to electronic file
    \PRtext{RmonthlyK410R14Residuals.dat}.
    The best model for the \RmonthlyTwo ~sample
    is the six signal \M=6 model,
    because the \M=7 and 8 models are unstable $("\UM")$.
    These two unstable models
    suffer from the amplitude dispersion $("\AD")$.
    No  intersecting frequencies $("\IF")$
   or
    leaking periods $("\LP")$
    are encountered in any of the eight \M=1-8 models in
    Table \ref{TableRmonthly2000K410R14}.
    
    The best model for the \RmonthlyOne ~sample,
    the six signal model in Table \ref{TableRmonthly2000K410R14},
    is the $g(t)$ model \M=4 added to the $g(t)$ model \M=6.
     We refer to this six signal model simply as the \M=6 model.

We show the four pure sine signal \M=4 model
periodograms between $P_{\mathrm{min}}=5$ and $P_{\mathrm{max}}=200$ years
in Fig. \ref{Rmonthly2000K410R14z}.
All four 
    $z_1$,
    $z_2$,
    $z_3$
    and
    $z_4$
    periodograms
    are continuous.
    Since the close-by periodogram values correlate,
    there are no abrupt
    deviations from the smooth periodogram changes,
   and a denser grid of
    tested frequencies would give exactly the same
    initial values $\bm{\beta}_{\mathrm{initial}}$
    for the non-linear iteration (Equation \ref{EqIteration}).
    The minima denoted with diamonds are well defined.
    The narrow, sharp shapes of these periodogram minima
    strongly support the idea that
    DCM detects real periods.

The four pure sine signal $g(t)$ model for sample \RmonthlyTwo ~is
shown in Fig. \ref{Rmonthly2000K410R14gdet}.
The remaining additional \M=5 and 6 model signals,
which are still undetected at this stage,
cause some of the regularities
seen in the blue circles denoting the $\epsilon_i$ residuals.
The four signals of \M=4 model are shown
separately in Fig. \ref{Rmonthly2000K410R14Signals}.
When we plot these signals as a function of
phase  in the right-hand panels,
they appear strong, clear and convincing.
The time span of these
  four signals covers about 25 $(\sim 274/11)$
  sunspot cycles, and their changes as a function
  of phase,
  certainly do not appear stochastic
  (Fig.  \ref{Rmonthly2000K410R14Signals}:
  Right-hand panels).
Finally, we emphasise that
the bootstrap estimates for the frequencies
and amplitudes of these
four signals are stable (Fig. \ref{Rmonthly2000K410R14fA}).

\begin{center}
  {\bf How to repeat the analysis of this
    Section \ref{SectDCMone}.
}
\end{center}
The model \M=1
analysis in Table 
\ref{TableRmonthly2000K410R14} can be repeated
with two python commands. \\
\PRtext{cp Rmonthly2000K110R14.dat dcm.dat} \\
\PRtext{python dcm.py} \\
The model \M=2 analysis requires the commands \\
\PRtext{cp Rmonthly2000K210R14.dat dcm.dat} \\
\PRtext{python dcm.py} \\
All analysis in Tables
\ref{TableRmonthly2000K410R14}-\ref{TableCyearlyK410R14}
can be repeated in this fashion.

\begin{table}
  \caption{\RmonthlyTwo ~analysis for pure sines $(K_2=1)$.
    (1) Model number \M (2) Model \RModel{K_1,K_2,K_3},
    number of free parameters $\eta$, and sum of squared
    residuals $R$.
    (3-6) Periods $P$, amplitudes $A$ and primary minima $t_{\mathrm{min,1}}$
    for one, two, three and four signal models.
    (7-9) Fisher-test comparison between models \M=1, 2, 3 and 4.
    An upward arrow $\uparrow$ indicates that
    complex model is better than simple model.
    A leftward arrow $\leftarrow$ would be used,
    if simple model were the better one.
    (10) Electronic control file.
    Four first signals are detected from original data
    (file \PRtext{Rmontly2000.dat}).
    Next four model \M=5-8 signals are detected from residuals
    of model \M=4 (electronic file
    \PRtext{RmonthlyK410R14Residuals.dat}).
    Models \M=7 and 8 are unstable $(\UM)$ because they
    suffer from amplitude dispersion $(\AD)$.
    There are no models having
    intersecting frequencies $(\IF)$ or leaking periods $(\LP)$
    in any of these eight \M=1-8 models.
    Models \M=4 and \M=6 give final results.}
  \label{TableRmonthly2000K410R14}
  \begin{scriptsize}
     \begin{center}
    \begin{adjustbox}{angle=90}
\begin{tabular}{lcccccccccc}
  \hline
 & & \multicolumn{4}{c}{Period analysis} & & \multicolumn{3}{c}{Fisher-test} &  \\
\cline{3-6} \cline{8-10}
\multicolumn{11}{c}{Data: Original non-weighted data ($n=3012, \Delta T=250.9$: \PR{Rmonthly2000.dat})} \\
\M                    &
\RModel{K_1,K_2,K_3}  &
$P_1$ (y)             &
$P_2$ (y)             &
$P_3$ (y)             &
$P_4$ (y)             &
                      &
\M=2                  &
\M=3                  &
\M=4                  &
                      \\ 
             &
$\eta$ (-)   &
$A_1$ (-)    &
$A_2$ (-)    &
$A_3$ (-)    &
$A_4$ (-)    &
             &
$F_R$ (-)    &
$F_R$ (-)    &
$F_R$ (-)    &
Control file \\ 
                       &
$R$ (-)                &
$t_{\mathrm{min,1}}$ (y) &
$t_{\mathrm{min,1}}$ (y) &
$t_{\mathrm{min,1}}$ (y) &
$t_{\mathrm{min,1}}$ (y) &
                       &
$Q_F$ (-)              &
$Q_F$ (-)              &
$Q_F$ (-)              &
       \\ 
(1)    & 
(2)    & 
(3)    & 
(4)    & 
(5)    & 
(6)    & 
       & 
(7)    & 
(8)    & 
(9)    & 
(10)   \\
\hline 
\M=1 &\RModel{1,1,0}   &$11.0336\PM0.0074$&     -            & -             & -                     &&$\uparrow    $&$\uparrow     $&$\uparrow    $&                                \\
        &       4         &$92.2\PM2.8      $&     -            & -             & -                     &&   255        & 254          &  231          &  \PR{Rmonthly2000K110R14.dat}  \\
\hline
\end{tabular}
\end{adjustbox}
\end{center}
\end{scriptsize}
\addtolength{\tabcolsep}{+0.05cm}
\end{table}

\begin{table}
  \caption{\RmonthlyOne ~analysis for pure sines $(K_2=1)$. 
    Otherwise as in Table \ref{TableRmonthly2000K410R14}.}
    \label{TableRmonthlyK410R14} 
  \begin{scriptsize}
     \begin{center}
    \begin{adjustbox}{angle=90}
      \begin{tabular}{lcccccccccc}
   \hline
\multicolumn{11}{c}{Data: Non-weighted original data ($n=3287, \Delta T=273.8$: \PR{Rmonthly.dat})} \\
 & & \multicolumn{4}{c}{Period analysis} & & \multicolumn{3}{c}{Fisher-test} &  \\
\cline{3-6} \cline{8-10}
\M                    &
\RModel{K_1,K_2,K_3}  &
$P_1$ (y)             &
$P_2$ (y)             &
$P_3$ (y)             &
$P_4$ (y)             &
                      &
\M=2                  &
\M=3                  &
\M=4                  &
                      \\ 
             &
$\eta$ (-)   &
$A_1$ (-)    &
$A_2$ (-)    &
$A_3$ (-)    &
$A_4$ (-)    &
             &
$F_R$ (-)    &
$F_R$ (-)    &
$F_R$ (-)    &
Control file \\ 
                       &
$R$ (-)                &
$t_{\mathrm{min,1}}$ (y) &
$t_{\mathrm{min,1}}$ (y) &
$t_{\mathrm{min,1}}$ (y) &
$t_{\mathrm{min,1}}$ (y) &
                       &
$Q_F$ (-)              &
$Q_F$ (-)              &
$Q_F$ (-)              &
                               \\ 
(1)    & 
(2)    & 
(3)    & 
(4)    & 
(5)    & 
(6)    & 
       & 
(7)    & 
(8)    & 
(9)    & 
(10)   \\
\hline
\M=1 &\RModel{1,1,0}   &$11.0046\PM0.0076$&     -            & -              & -                     &&$\uparrow    $&$\uparrow     $&$\uparrow    $&                                  \\
        &       4         &$93.8\PM2.3      $&     -            & -              & -                     &&   217        &  243          & 237          &  \PR{RmonthlyK110R14.dat}        \\
\hline
\end{tabular}
\end{adjustbox}
\end{center}
\end{scriptsize}
\addtolength{\tabcolsep}{+0.05cm}
\end{table}

\begin{table}
  \caption{\RmonthlyTwo ~analysis for double waves $(K_2=2)$. 
    Otherwise as in Table \ref{TableRmonthly2000K410R14}.}
    \label{TableRmonthly2000K420R14} 
  \begin{scriptsize}
     \begin{center}
    \begin{adjustbox}{angle=90}
      \begin{tabular}{lcccccccccc}
    \hline
                  &  & \multicolumn{4}{c}{Period analysis} & & \multicolumn{3}{c}{Fisher-test} &  \\
\cline{3-6} \cline{8-10}
\multicolumn{11}{c}{Data: Non-weighted original data  ($n=3012, \Delta T=250.9$: \PR{Rmonthly2000.dat})} \\
\M                  &
\RModel{K_1,K_2,K_3} &
$P_1$ (y)       &
$P_2$ (y)       &
$P_3$ (y)       &
$P_4$ (y)       &
                &
\M=2            &
\M=3            &
\M=4            &
                \\ 
                &
$\eta$ (-)      &
$A_1$ (-)       &
$A_2$ (-)       &
$A_3$ (-)       &
$A_4$ (-)       &
                &
$F_R$ (-)       &
$F_R$ (-)       &
$F_R$ (-)       &
Control file    \\ 
                       &
$R$ (-)                &
$t_{\mathrm{min,1}}$ (y) &
$t_{\mathrm{min,1}}$ (y) &
$t_{\mathrm{min,1}}$ (y) &
$t_{\mathrm{min,1}}$ (y) &
                       &
$Q_F$ (-)              &
$Q_F$ (-)              &
$Q_F$ (-)              &
                       \\ 
(1) & 
(2) & 
(3) & 
(4) & 
(5) & 
(6) & 
    & 
(7) & 
(8) & 
(9) & 
(10)\\
\hline
\M=1 &\RModel{1,2,0}   &$11.0242\PM0.0084 $&     -          & -             & -                     &&$\uparrow    $&$\uparrow    $&$\uparrow     $&                                \\
        &       6         &$94.7\PM3.6       $&     -          & -             & -                     &&   154        & 149          & 164           &  \PR{Rmonthly2000K120R14.dat}  \\
\hline
\end{tabular}
\end{adjustbox}
\end{center}
\end{scriptsize}
\addtolength{\tabcolsep}{+0.05cm}
\end{table}

\begin{table}
  \caption{\RmonthlyOne ~analysis for double waves $(K_2=2)$. 
    Otherwise as in Table \ref{TableRmonthly2000K410R14}.} 
  \label{TableRmonthlyK420R14} 
  \begin{scriptsize}
     \begin{center}
    \begin{adjustbox}{angle=90}
      \begin{tabular}{lcccccccccc}  
    \hline
\multicolumn{11}{c}{Data: Original non-weighted data ($n=3287, \Delta T=273.8$: \PR{Rmonthly.dat})}\\
 &  & \multicolumn{4}{c}{Period analysis} & & \multicolumn{3}{c}{Fisher-test} &  \\
\cline{3-6} \cline{8-10}
\M                 &
\RModel{K_1,K_2,K_3}&
$P_1$ (y)          &
$P_2$ (y)          &
$P_3$ (y)          &
$P_4$ (y)          &
                   &
\M=2               &
\M=3               &
\M=4               &
                     \\ 
             &
$\eta$ (-)   &
$A_1$ (-)    &
$A_2$ (-)    &
$A_3$ (-)    &
$A_4$ (-)    &
             &
$F_R$ (-)    &
$F_R$ (-)    &
$F_R$ (-)    &
Control file \\ 
                       &
$R$ (-)                &
$t_{\mathrm{min,1}}$ (y) &
$t_{\mathrm{min,1}}$ (y) &
$t_{\mathrm{min,1}}$ (y) &
$t_{\mathrm{min,1}}$ (y) &
                       &
$Q_F$ (-)              &
$Q_F$ (-)              &
$Q_F$ (-)              &
                       \\ 
(1)    & 
(2)    & 
(3)    & 
(4)    & 
(5)    & 
(6)    & 
       & 
(7)    & 
(8)    & 
(9)    & 
(10)   \\
\hline
\M=1 &\RModel{1,2,0}  &$10.9922\PM0.0064   $&     -           & -                & -                    &&$\uparrow    $&$\uparrow     $&$\uparrow    $&                                  \\
        &       6        &$97.5\PM2.8         $&     -           & -                & -                    &&     157      &  155          &    177       &  \PR{RmonthlyK120R14.dat}        \\
\hline
\end{tabular}
\end{adjustbox}
\end{center}
\end{scriptsize}
\addtolength{\tabcolsep}{+0.05cm}
\end{table}

\begin{table}
  \caption{\CmonthlyTwo ~analysis for pure sines $(K_2=1)$. 
    Otherwise as in Table \ref{TableRmonthly2000K410R14}. }
    \label{TableCmonthly2000K410R14} 
 \begin{scriptsize}
     \begin{center}
    \begin{adjustbox}{angle=90}
      \begin{tabular}{lcccccccccc}  
    \hline
 &  & \multicolumn{4}{c}{Period analysis} & & \multicolumn{3}{c}{Fisher-test} &  \\
\cline{3-6} \cline{8-10}
\multicolumn{11}{c}{Data: Weighted original data before 2000 ($n=2182, \Delta T=181.9$: \PR{Cmonthly2000.dat})} \\
\M                 &
\CModel{K_1,K_2,K_3}&
$P_1$ (y)          &
$P_2$ (y)          &
$P_3$ (y)          &
$P_4$ (y)          &
                   &
\M=2               &
\M=3               &
\M=4               &
                   \\ 
             &
$\eta$ (-)   &
$A_1$ (-)    &
$A_2$ (-)    &
$A_3$ (-)    &
$A_4$ (-)    &
             &
$F_R$ (-)    &
$F_R$ (-)    &
$F_R$ (-)    &
Control file \\ 
                       &
$\chi^2$ (-)           &
$t_{\mathrm{min,1}}$ (y) &
$t_{\mathrm{min,1}}$ (y) &
$t_{\mathrm{min,1}}$ (y) &
$t_{\mathrm{min,1}}$ (y) &
                       &
$Q_F$ (-)              &
$Q_F$ (-)              &
$Q_F$ (-)              &
                       \\ 
(1)    & 
(2)    & 
(3)    & 
(4)    & 
(5)    & 
(6)    & 
       & 
(7)    & 
(8)    & 
(9)    & 
(10)   \\
\hline
\M=1  &\CModel{1,1,0}   &$10.8338\PM0.0094$&     -            & -             & -                     &&$\uparrow    $&$\uparrow    $&$\uparrow     $&                                 \\
         &       4         &$121.9\PM3.6     $&     -            & -             & -                     && 203          & 230          &  196          &  \PR{Cmonthly2000K110R14.dat}   \\
\hline
\end{tabular}
\end{adjustbox}
\end{center}
\end{scriptsize}
\addtolength{\tabcolsep}{+0.05cm}
\end{table}

\begin{table}
  \caption{\CmonthlyOne ~analysis for pure sines $(K_2=1)$. 
    Otherwise as in Table \ref{TableRmonthly2000K410R14}.}
  \label{TableCmonthlyK410R14} 
 \begin{scriptsize}
     \begin{center}
    \begin{adjustbox}{angle=90}
      \begin{tabular}{lcccccccccc}  
      \hline
 &  & \multicolumn{4}{c}{Period analysis} & & \multicolumn{3}{c}{Fisher-test} &  \\
\cline{3-6} \cline{8-10}
\multicolumn{11}{c}{Data: Weighted original data ($n=2458, \Delta T=204.8$: \PR{Cmonthly.dat})} \\
\M                 &
\CModel{K_1,K_2,K_3}&
$P_1$ (y)          &
$P_2$ (y)          &
$P_3$ (y)          &
$P_4$ (y)          &
                   &
\M=2               &
\M=3               &
\M=4               &
                     \\ 
             &
$\eta$ (-)   &
$A_1$ (-)    &
$A_2$ (-)    &
$A_3$ (-)    &
$A_4$ (-)    &
             &
$F_R$ (-)    &
$F_R$ (-)    &
$F_R$ (-)    &
Control file \\ 
                       &
$\chi^2$ (-)           &
$t_{\mathrm{min,1}}$ (y) &
$t_{\mathrm{min,1}}$ (y) &
$t_{\mathrm{min,1}}$ (y) &
$t_{\mathrm{min,1}}$ (y) &
                       &
$Q_F$ (-)              &
$Q_F$ (-)              &
$Q_F$ (-)              &
                       \\ 
(1)    & 
(2)    & 
(3)    & 
(4)    & 
(5)    & 
(6)    & 
       & 
(7)    & 
(8)    & 
(9)    & 
(10)   \\
\hline
\M=1 &\CModel{1,1,0}   &$10.8350\PM0.0089$&     -          & -               & -                    &&$\uparrow      $&$\uparrow    $&$\uparrow     $&                                \\
        &       4         &$121.6\PM2.9     $&     -          & -               & -                    &&   173         & 197          &  210          &  \PR{CmonthlyK110R14.dat}      \\
\hline
\end{tabular}
\end{adjustbox}
\end{center}
\end{scriptsize}
\addtolength{\tabcolsep}{+0.05cm}
\end{table}

\begin{table}
  \caption{\CmonthlyTwo ~analysis for double waves $(K_2=2)$. 
    Otherwise as in Table \ref{TableRmonthly2000K410R14}.}
    \label{TableCmonthly2000K420R14} 
 \begin{scriptsize}
     \begin{center}
    \begin{adjustbox}{angle=90}
      \begin{tabular}{lcccccccccc}   
        \hline
 & & \multicolumn{4}{c}{Period analysis} & & \multicolumn{3}{c}{Fisher-test} &  \\
\cline{3-6} \cline{8-10}
\multicolumn{11}{c}{Data: Weighted original data before 2000 ($n=2182, \Delta T=181.9$: \PR{Cmonthly2000.dat})} \\
\M                 &
\CModel{K_1,K_2,K_3}&
$P_1$ (y)          &
$P_2$ (y)          &
$P_3$ (y)          &
$P_4$ (y)          &
                   &
\M=2               &
\M=3               &
\M=4               &
                   \\ 
             &
$\eta$ (-)   &
$A_1$ (-)    &
$A_2$ (-)    &
$A_3$ (-)    &
$A_4$ (-)    &
             &
$F_R$ (-)    &
$F_R$ (-)    &
$F_R$ (-)     &
Control file \\ 
                       &
$\chi^2$ (-)           &
$t_{\mathrm{min,1}}$ (y) &
$t_{\mathrm{min,1}}$ (y) &
$t_{\mathrm{min,1}}$ (y) &
$t_{\mathrm{min,1}}$ (y) &
                       &
$Q_F$ (-)              &
$Q_F$ (-)              &
$Q_F$ (-)              &
                       \\ 
(1)    & 
(2)    & 
(3)    & 
(4)    & 
(5)    & 
(6)    & 
       & 
(7)    & 
(8)    & 
(9)    & 
(10)   \\
\hline
\M=1 &\CModel{1,2,0}   &$10.830\PM0.010  $&     -          & -                  & -                      &&$\uparrow$&$\uparrow$&$\uparrow$&                              \\
        &       6         &$126.1\PM4.2     $&     -          & -                  & -                      && 125      &  180     & 160      &  \PR{Cmonthly2000K120R14.dat}\\
\hline
\end{tabular}
\end{adjustbox}
\end{center}
\end{scriptsize}
\addtolength{\tabcolsep}{+0.05cm}
\end{table}

\begin{table}
  \caption{\CmonthlyOne ~analysis for double waves $(K_2=2)$. 
    Otherwise as in Table \ref{TableRmonthly2000K410R14}.}
    \label{TableCmonthlyK420R14} 
 \begin{scriptsize}
     \begin{center}
    \begin{adjustbox}{angle=90}
      \begin{tabular}{lcccccccccc}       
  \hline
 &  & \multicolumn{4}{c}{Period analysis} & & \multicolumn{3}{c}{Fisher-test} &  \\
\cline{3-6} \cline{8-10}
\multicolumn{11}{c}{Data: Weighted original data ($n=2458, \Delta T=204.8$: \PR{Cmonthly.dat})} \\
\M                  &
\CModel{K_1,K_2,K_3}&
$P_1$ (y)          &
$P_2$ (y)          &
$P_3$ (y)          &
$P_4$ (y)          &
                   &
\M=2               &
\M=3               &
\M=4               &
                     \\ 
             &
$\eta$ (-)   &
$A_1$ (-)    &
$A_2$ (-)    &
$A_3$ (-)    &
$A_4$ (-)    &
             &
$F_R$ (-)    &
$F_R$ (-)    &
$F_R$ (-)    &
Control file \\ 
                       &
$\chi^2$ (-)           &
$t_{\mathrm{min,1}}$ (y) &
$t_{\mathrm{min,1}}$ (y) &
$t_{\mathrm{min,1}}$ (y) &
$t_{\mathrm{min,1}}$ (y) &
                       &
$Q_F$ (-)              &
$Q_F$ (-)              &
$Q_F$ (-)              &
                       \\ 
(1)    & 
(2)    & 
(3)    & 
(4)    & 
(5)    & 
(6)    & 
       & 
(7)    & 
(8)    & 
(9)    & 
(10)   \\
\hline
\M=1 &\CModel{1,2,0}   &$21.667\PM0.014  $&     -          & -               & -              &&$\uparrow    $&$\uparrow    $&$\uparrow     $&                              \\
        &       6         &$132.6\PM3.4     $&     -          & -               & -              && 140          & 171          &  170          &  \PR{CmonthlyK120R14.dat}    \\
\hline
\end{tabular}
\end{adjustbox}
\end{center}
\end{scriptsize}
\addtolength{\tabcolsep}{+0.05cm}
\end{table}

\begin{table}
  \caption{\RyearlyTwo ~analysis for pure sines $(K_2=1)$. 
    Otherwise as in Table \ref{TableRmonthly2000K410R14}}.
    \label{TableRyearly2000K410R14} 
 \begin{scriptsize}
     \begin{center}
    \begin{adjustbox}{angle=90}
      \begin{tabular}{lcccccccccc}  
\hline
 &  & \multicolumn{4}{c}{Period analysis} & & \multicolumn{3}{c}{Fisher-test} & \\
\cline{3-6} \cline{8-10}
\multicolumn{11}{c}{Data: Non-weighted original data ($n=300, \Delta T=299$: \PR{Ryearly2000.dat})} \\
\M                 &
\RModel{K_1,K_2,K_3}&
$P_1$ (y)          &
$P_2$ (y)          &
$P_3$ (y)          &
$P_4$ (y)          &
                   &
\M=2               &
\M=3               &
\M=4               &
                     \\ 
             &
$\eta$ (-)   &
$A_1$ (-)    &
$A_2$ (-)    &
$A_3$ (-)    &
$A_4$ (-)    &
             &
$F_R$ (-)    &
$F_R$ (-)    &
$F_R$ (-)    &
Control file \\ 
                       &
$R$ (-)                &
$t_{\mathrm{min,1}}$ (y) &
$t_{\mathrm{min,1}}$ (y) &
$t_{\mathrm{min,1}}$ (y)&
$t_{\mathrm{min,1}}$ (y) &
                       &
$Q_F$ (-)              &
$Q_F$ (-)              &
$Q_F$ (-)              &
                       \\ 
(1)    & 
(2)    & 
(3)    & 
(4)    & 
(5)    & 
(6)    & 
       & 
(7)    & 
(8)    & 
(9)    & 
(10)   \\
\hline
\M=1 &\RModel{1,1,0}   &$11.018\PM0.018  $&     -          & -              & -              &&$\uparrow    $&$\uparrow    $&$\uparrow     $&                             \\
        &       4         &$94.7\PM7.9      $&     -          & -              & -              &&   28         &      24      &   26          &  \PR{Ryearly2000K110R14.dat}    \\
\hline
\end{tabular}
\end{adjustbox}
\end{center}
\end{scriptsize}
\addtolength{\tabcolsep}{+0.05cm}
\end{table}

\begin{table}
  \caption{\RyearlyOne ~analysis for pure sines $(K_2=1)$. 
    Otherwise as in Table \ref{TableRmonthly2000K410R14}.}
    \label{TableRyearlyK410R14} 
 \begin{scriptsize}
     \begin{center}
    \begin{adjustbox}{angle=90}
      \begin{tabular}{lcccccccccc}
    \hline
 &  & \multicolumn{4}{c}{Period analysis} & & \multicolumn{3}{c}{Fisher-test} & \\
\cline{3-6} \cline{8-10}
\multicolumn{11}{c}{Data: Non-weighted original data ($n=322, \Delta T=321$: \PR{Ryearly.dat})} \\
\M                    &
\RModel{K_1,K_2,K_3}  &
$P_1$ (y)             &
$P_2$ (y)             &
$P_3$ (y)             &
$P_4$ (y)             &
                      &
\M=2                  &
\M=3                  &
\M=4                  &
                     \\ 
             &
$\eta$ (-)   &
$A_1$ (-)    &
$A_2$ (-)    &
$A_3$ (-)    &
$A_4$ (-)    &
             &
$F_R$ (-)    &
$F_R$ (-)    &
$F_R$ (-)    &
Control file \\ 
                      &
$R$ (-)               &
$t_{\mathrm{min,1}}$ (y)&
$t_{\mathrm{min,1}}$ (y)&
$t_{\mathrm{min,1}}$ (y)&
$t_{\mathrm{min,1}}$ (y) &
                       &
$Q_F$ (-)              &
$Q_F$ (-)              &
$Q_F$ (-)              &
                       \\ 
(1)    & 
(2)    & 
(3)    & 
(4)    & 
(5)    & 
(6)    & 
       & 
(7)    & 
(8)    &
(9)    & 
(10)  \\
\hline
\M=1 &\RModel{1,1,0}   &$11.003\PM0.017  $&     -          & -              & -              &&$\uparrow    $&$\uparrow    $&$\uparrow     $&                             \\
        &       4         &$95.9\PM7.4      $&     -          & -              & -              &&  26          &      26      &     29        &  \PR{RyearlyK110R14.dat}    \\
\hline
\end{tabular}
\end{adjustbox}
\end{center}
\end{scriptsize}
\addtolength{\tabcolsep}{+0.05cm}
\end{table}

\begin{table}
  \caption{\CyearlyTwo ~analysis for pure sines $(K_2=1)$. 
    Otherwise as in Table \ref{TableRmonthly2000K410R14}.}
  \label{TableCyearly2000K410R14} 
 \addtolength{\tabcolsep}{-0.03cm}
 \begin{scriptsize}
     \begin{center}
    \begin{adjustbox}{angle=90}
      \begin{tabular}{lcccccccccc}
      \hline
 &  & \multicolumn{4}{c}{Period analysis} & & \multicolumn{3}{c}{Fisher-test} &  \\
\cline{3-6} \cline{8-10}
\multicolumn{11}{c}{Data: Weighted original data ($n=182, \Delta T=181$: \PR{Cyearly2000.dat})} \\
\M                    &
\CModel{K_1,K_2,K_3}  &
$P_1$ (y)             &
$P_2$ (y)             &
$P_3$ (y)             &
$P_4$ (y)             &
                      &
\M=2                  &
\M=3                  &
\M=4                  &
                      \\ 
                      &
$\eta$ (-)   &
$A_1$ (-)    &
$A_2$ (-)    &
$A_3$ (-)    &
$A_4$ (-)    &
             &
$F_R$ (-)    &
$F_R$ (-)    &
$F_R$ (-)    &
Control file \\ 
                       &
$\chi^2$ (-)           &
$t_{\mathrm{min,1}}$ (y) &
$t_{\mathrm{min,1}}$ (y) &
$t_{\mathrm{min,1}}$ (y) &
$t_{\mathrm{min,1}}$ (y) &
                       &
$Q_F$ (-)              &
$Q_F$ (-)              &
$Q_F$ (-)              &
                       \\ 
(1)    & 
(2)    & 
(3)    & 
(4)    & 
(5)    & 
(6)    & 
       & 
(7)    & 
(8)    & 
(9)    & 
(10)   \\
      \hline
\M=1 &\CModel{1,1,0}   &$10.835\PM0.030  $&     -          & -              & -              &&$\uparrow    $&$\uparrow    $&$\uparrow     $&                             \\
        &       4         &$123.5\PM8.7     $&     -          & -              & -              &&  21          &  24          &    21         &  \PR{Cyearly2000K110R14.dat}    \\
\hline
\end{tabular}
\end{adjustbox}
\end{center}
\end{scriptsize}
\addtolength{\tabcolsep}{+0.05cm}
\end{table}

\begin{table}
  \caption{\CyearlyOne ~analysis for pure sinusoids $(K_2=1)$. 
    Otherwise as in Table \ref{TableRmonthly2000K410R14}.}
    \label{TableCyearlyK410R14} 
\begin{scriptsize}
     \begin{center}
    \begin{adjustbox}{angle=90}
      \begin{tabular}{lcccccccccc}
      \hline
     &  & \multicolumn{4}{c}{Period analysis} & & \multicolumn{3}{c}{Fisher-test} &  \\
\cline{3-6} \cline{8-10}
\multicolumn{11}{c}{Data: Weighted original data ($n=204, \Delta T=203$: \PR{Cyearly.dat})} \\
\M                    &
\CModel{K_1,K_2,K_3}  &
$P_1$ (y)             &
$P_2$ (y)             &
$P_3$ (y)             &
$P_4$ (y)             &
                      &
\M=2                  &
\M=3                  &
\M=4                  &
                     \\ 
             &
$\eta$ (-)   &
$A_1$ (-)    &
$A_2$ (-)    &
$A_3$ (-)    &
$A_4$ (-)    &
             &
$F_R$ (-)    &
$F_R$ (-)    &
$F_R$ (-)    &
Control file \\ 
                       &
$\chi^2$ (-)           &
$t_{\mathrm{min,1}}$ (y) &
$t_{\mathrm{min,1}}$ (y) &
$t_{\mathrm{min,1}}$ (y) &
$t_{\mathrm{min,1}}$ (y) &
                       &
$Q_F$ (-)              &
$Q_F$ (-)              &
$Q_F$ (-)              &
                       \\ 
(1)    & 
(2)    & 
(3)    & 
(4)    & 
(5)    & 
(6)    & 
       & 
(7)    & 
(8)    & 
(9)    & 
(10) \\
      \hline
\M=1 &\CModel{1,1,0}      &$10.832\PM0.022  $&     -          & -              & -              &&$\uparrow    $&$\uparrow    $&$\uparrow     $&                             \\
        &       4         &$123\PM11        $&     -          & -              & -              &&     18       &   22         &   25          &  \PR{CyearlyK110R14.dat}    \\
\hline
\end{tabular}
\end{adjustbox}
\end{center}
\end{scriptsize}
\addtolength{\tabcolsep}{+0.05cm}
\end{table}

\clearpage

\begin{figure}
  \centerline{
     \includegraphics[width=0.60\textwidth,clip=]{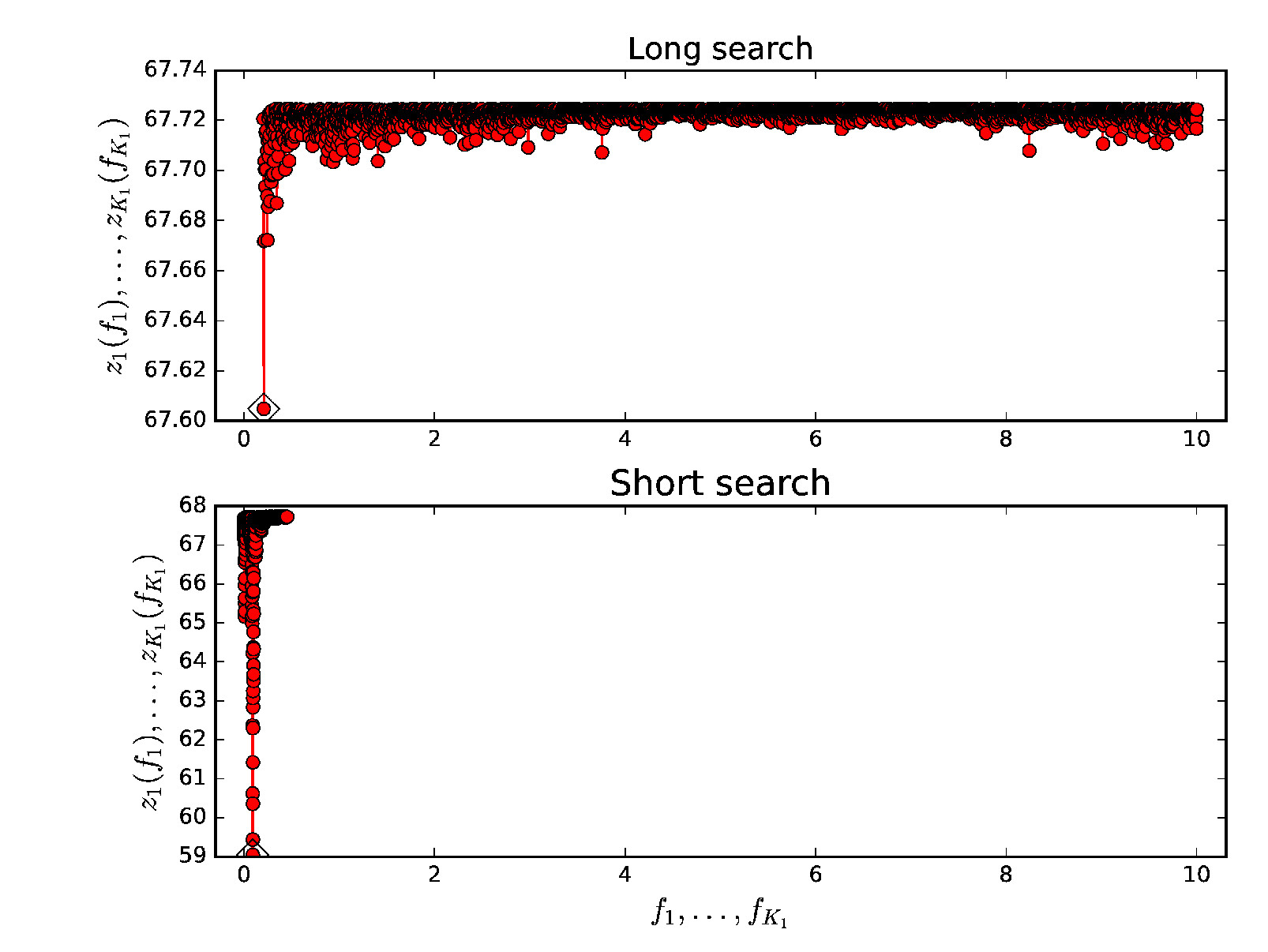}} 
  \caption{\RmonthlyOne ~one pure sine signal model.
    (Upper panel) Long search
    periodogram between
    $P_{\mathrm{min}}=0.1$ and $P_{\mathrm{max}}=5$ years.
    Tested grid contains $n_L=3000$ frequencies. Diamond denotes
    best 4.84 year period. (Lower panel) Short search
    periodogram for $n_S=1000$ and $c=0.05$. Diamond denotes
    best 11.00 year period.} 
\label{ShortRmonthlyK110R14z}
\end{figure}

\subsection{Periods between 0.1 and 5 years}
\label{SectDCMtwo}

Here,
we apply DCM to search for periods
between $P_{\mathrm{min}}=0.1$  and   $P_{\mathrm{max}}=5$ years
from the non-weighted monthly
sunspot  \RmonthlyOne ~sample.
This sample is our largest one.
The one pure sine signal DCM
model periodogram is shown in
Fig. \ref{ShortRmonthlyK110R14z}.
The long search grid contains
an extremely large number of
$n_L=3000$ tested frequencies,
because we do not want to miss
any of the inner planet periods.
For all shorter periods,
the long search $z_1$ periodogram
continuum is nearly
constant, between 67.70 and 67.73
(Fig. \ref{ShortRmonthlyK110R14z}: upper panel).
The diamond denoting the best detected long search
4.84 year period is at the edge of the
tested  period interval.
The short search grid also contains
a large number of
$n_S=1000$
tested frequencies
(Fig. \ref{ShortRmonthlyK110R14z}: lower panel).
The tested frequency interval is 5\% 
of the full tested long search
frequency interval $(c=0.05)$.
For the best 11.00 years period
denoted with a diamond,
the short search periodogram shows
a dramatic dip to the level $z_1=59$,
which is far below the $\sim 67.7$
continuum of the long search periodogram.
This best 11 year ``leaking period'' is outside
the tested period interval upper limit
$P_{\mathrm{max}}=5$ years (``$\LP$").
What we have here is a clearly unstable model case (``$\UM$'').

For the same 
$P_{\mathrm{min}}=0.1$  and   $P_{\mathrm{max}}=5$ year interval,
the one pure sine signal DCM model
for the weighted monthly sunspots
\CmonthlyOne ~sample also gives an unstable model result
(``$\LP,\UM$'').

In our earlier DCM search
for periods between 5 and 200 years,
we classified the ``$\LP$"
cases as unstable ``$\UM$'' models,
and stopped searching for additional signals.
For the same reason, we can conclude that
the orbital and synodic periods
of inner planets 
can not be found
{\it directly} from the monthly sunspot data.

\begin{center}
  {\bf How to repeat the analysis of this
    Section \ref{SectDCMtwo}}
\end{center}

\noindent
The DCM analysis in Fig. \ref{ShortRmonthlyK110R14z}
can be repeated with two
commands \\ 
\PRtext{cp ShortRmonthlyK110R14.dat dcm.dat} \\
\PRtext{python dcm.py}

\section{Discrete Fourier Transform
  (DFT) cross-check} \label{SectDFT} 

The Discrete Fourier Transform (DFT) has been applied
to the daily sunspot number data from 
WDC SILSO\cite{Zhu18}
(Royal Observatory of Belgium, Brussels).
The applied DFT version was formulated in Ref\cite{Hor86}.
DFT searches for the best pure sine model for the data.
Here, we cross-check if a similar DFT
analysis of our monthly sunspot number sample \RmonthlyOne ~gives
the same results as our pure sine model DCM analysis
(Table \ref{TableCompare}: Column 3).
Before our cross-check, we perform DCM and DFT
analysis of simulated data.

\subsection{Simulated two pure
  sine signal data for DCM and DFT analysis}
\label{SectDFTdata} 

We create simulated data for
comparing the
performance of DFT and DCM.
Let us assume that the data contains the
sum signal $s(t)=s_1(t)+s_2(t)$ of two pure sine signals
\begin{eqnarray}
  s_1(t) & = & a_1 \sin{(2 \pi f_1 t )} \label{EqSignalOne} \\
  s_2(t) & = & a_2 \sin{(2 \pi f_2 t )}, \label{EqSignalTwo}
\end{eqnarray}
where the frequencies fulfil $f_1 \approx f_2$.
DFT can detect both of the $s_1(t)$ and $s_2(t)$ signals, if 
\begin{eqnarray}
  \Delta f = |f_2 - f_1| \gtrsim f_0,
  \label{EqPrevent}
\end{eqnarray}
where $f_0=1/\Delta T$ is the distance
between independent frequencies\cite{Lou78}.
We give the $f/f_0=\Delta T/P$ ratios for the
\SignalOne, \SignalTwo, \SignalFive ~and
\SignalThree ~signals in Table \ref{TableIndependent}.
The \SignalTwo ~and \SignalFive ~signal frequencies
fulfil the criterion of Equation \ref{EqPrevent}
only in the two longest
\RyearlyOne ~and \RyearlyTwo ~samples.
This criterion is
not fulfilled in our DFT cross-check
\RmonthlyOne ~sample (Table \ref{TableIndependent}).
  Furthermore, many
  pairs of \SignalOne, \SignalTwo, \SignalFive ~and
  \SignalThree ~signals detected in this sample
  \RmonthlyOne ~have
frequency differences $\Delta f$
  close to $1\times f_0$.  
\begin{table}[h]
  \caption{Independent frequency limitation (Equation \ref{EqPrevent}).
    Ratios $f/f0=\Delta T/P$ in order of decreasing
    sample time interval $\Delta T$.
    Arrows above numerical values indicate cases
    where $\Delta f < f_0$.}
  \label{TableIndependent}
  \begin{center}
  \begin{tabular}{lccccc}
    \hline
            &             & \SignalTwo   &\SignalFive  & \SignalOne & \SignalThree \\
Sample      &  $\Delta T$ &$\Delta T/P  $&$\Delta T/P $&$\Delta T/P$&$\Delta T/P$  \\
            &   (y)       & (-)         & (-)          & (-)        & (-)          \\
    \hline
\RyearlyOne &  321.0      & 32.10       & 30.28        & 29.18      & 27.07        \\
\RyearlyTwo &  299.0      & 29.90       & 28.21        & 27.18      & 25.21        \\
\RmonthlyOne&  273.8      & 27.38       & \Rar{25.83}  & \Lar{24.89}& 23.09        \\
\RmonthlyTwo&  250.9      & 25.09       & \Rar{23.67}  & \Lar{22.81}& 21.16        \\
\CmonthlyOne&  204.8      & 20.48       & \Rar{19.32}  & \Lar{18.62}& 17.27        \\
\CyearlyOne &  203.0      & 20.30       & \Rar{19.15}  & \Lar{18.45}& 17.12        \\
\CmonthlyTwo&  181.9      & 18.19       & \Rar{17.16}  & \Lar{16.54}& 15.34        \\
\CyearlyTwo &  181.0      & 18.10       & \Rar{17.07}  & \Lar{16.45}& 15.25        \\
   \hline
\end{tabular}
\end{center}
\end{table}

  The interference pattern
  of the sum $s(t)$ of the two signals
$s_1(t)$ and $s_2(t)$ 
is
repeated during the beat period
$P_{\mathrm{beat}}=|P_1^{-1}-P_2^{-1}|^{-1}=|f_1-f_2|^{-1}$.
The solutions for the frequency, the amplitude and the
phase of this sum signal $s(t)$  are quite
complex\cite{Fet74,Agu12,Rut14,Sch23}.
For the purposes of our DFT cross-check,
it suffices to discuss the frequency $f(t)$ changes and
the abrupt $\phi$ phase shifts
of the sum signal $s(t)$.

The frequency $f(t)$ of the $s(t)$ sum signal is constant
\begin{eqnarray}
f(t) = (f_1+f_2)/2, \textnormal{~if $a_1=a_2$.}
\label{EqMisleadOne}
\end{eqnarray}
This frequency $f(t)$
varies between the lower and upper limits given below
\begin{eqnarray}
  f_2-{{a_1(f_1-f_2)}\over {a_2+a_1}}
  \le
  f(t)
  \le
  f_2+{{a_1(f_1-f_2)}\over {a_2+a_1}},
  \textnormal{~if $a_2>a_1$.}
\label{EqMisleadTwo}
\end{eqnarray}
In this case,
the $f(t)$ frequency changes are dominated by 
 the stronger $f_2$ signal.

The sum signal $s(t)$ amplitude $a_s$ is constantly changing.
An abrupt instantaneous phase shift
\begin{eqnarray}
  \Delta \phi = [{\mathrm{arc~sin(a_1/a_2)}}]/\pi
  \label{EqAbrupt}
\end{eqnarray}
occurs at the epoch of $a_s$ amplitude minimum.
This abrupt phase shift is the largest, $\Delta \phi=0.5$,
for equal amplitudes $a_1=a_2$.

The limitation of Equation \ref{EqPrevent} can
{\it prevent} the DFT detection of both $f_1$ and $f_2$ values.
The effects of Equations \ref{EqMisleadOne}, \ref{EqMisleadTwo}
and \ref{EqAbrupt}
can {\it mislead} the DFT detection of the correct
$f_1$ and $f_2$ values.

The units of time intervals and periods in
all our simulations are years. The amplitude unit is dimensionless.
These units are therefore no longer mentioned in the simulations below.
We simulate $y_i^{\star}$ data,
where the monthly sunspot number time points $t_i^{\star}$
are computed for four alternative time intervals of
$\Delta T=70$, 90, 110 and 274.
These time intervals begin from
the year 1749 at the beginning of sample \RmonthlyOne.
The longest $\Delta T=274$
simulation has the same time span
and the same time points
as \RmonthlyOne.
The $s_1(t_i^{\star})$ and $s_2(t_i^{\star})$ signal periods are
$P_1=1/f_1=10$,  $P_2=1/f_2=11$.
They are equal to the
periods of the two strongest \SignalTwo ~and
\SignalOne ~signals detected in sample \RmonthlyOne.
This gives
the beat period $P_{\mathrm{beat}}=110$.
We simulate data for two alternative amplitude combinations
$a_1=a_2=50 \Rightarrow f(t) =$ constant (Equation \ref{EqMisleadOne})
and $a_1=25 < a_2=50 \Rightarrow f(t)=$ variable
(Equation \ref{EqMisleadTwo}).
Hence,
the total number of simulated samples is $4 \times 2=8$.
The tested period interval between 5 and 200 years is the same
as in our earlier DCM analysis of \RmonthlyOne ~sample.

For the \RmonthlyOne ~sample, the DCM five pure sine
signal \M=5 model
gives the residuals $\epsilon_i$,
which have a standard deviation $s_{\epsilon}=42$.
We assume that after the detection of these five pure sine signals,
these remaining $\epsilon_i$ residuals represent the noise in
the original data. Therefore,
the errors $\sigma_i^{\star}$ for the simulated data are drawn from
a Gaussian distribution having a mean $m=0$ and a standard
deviation $s=s_{\mathrm{\epsilon}}=42$.
The simulated
data are
\begin{eqnarray}
  y_i^{\star}= m^{\star}+
  s_1(t_i^{\star})+ s_2(t_i^{\star}) + \sigma_i^{\star},
\label{EqSimData}
\end{eqnarray}
where $m^{\star}=81.6$ is the mean of all $y_i$ values in
  \RmonthlyOne. The selected $m^{\star}$ value has no impact on
  our results, because it is subtracted from the
  simulated
  $y_i^{\star}$ data before the computation of DFT periodogram.
  The DCM two sine wave model \RModel{2,1,0} 
  ~solves this $M_0=m^{\star}$ value.

We analyse the $2\times 4=8$ simulated data samples
using the same DFT approach as \cite{Zhu18}.
The DFT pre-whitening technique,
detects one signal at the
time\cite{Sca82,Ree07}.
We apply DFT to the original simulated data.
This gives us the first $P_1$ period.
The least squares fit is then made to the original simulated data,
where the pure sine model has the $P_1$ period.
This ``first DFT model'' gives the first sample of residuals.
DFT is applied to this first sample of residuals,
which gives the second $P_2$ period.
The next least squares fit is made to the first sample of
residuals using a pure sine model having a period $P_2$.
The model for this sample is called
``the second DFT model''.
This gives the second sample of residuals.

The DFT pre-whitening technique has its limitations.
The DFT {\it can only} be used to cross-check the DCM pure sine model
analysis results for non-weighted data. 
DFT {\it can not} be used
to cross-check the DCM double wave model analysis results,
nor the DCM weighted data analysis results.

In the next Sections \ref{SectDFTEqual} and \ref{SectDFTUnequal},
  we perform the DCM and DFT analyses
  of the $2\times4=8$ simulated data samples,
  and compare the results.
  The results for each individual simulated sample
  are presented in one vertical column of
  Figs. \ref{FigDFTDCMOne} and
  \ref{FigDFTDCMTwo}, where
  the dashed vertical lines
  intersecting the
  periodograms denote
  the {\it simulated frequencies} $1/P_1=1/10$ and $1/P_2=1/11$.
  The diamonds at the DFT periodogram maxima and the DCM
  periodogram minima 
  denote the {\it detected frequencies}.
Hence, the vertical dashed lines intersect the diamonds only
if the method detects the correct frequencies.

\begin{figure}  
\vspace{-0.10\textwidth}
\centerline{\hspace*{0.005\textwidth}
 \includegraphics[width=0.25\textwidth,clip=]{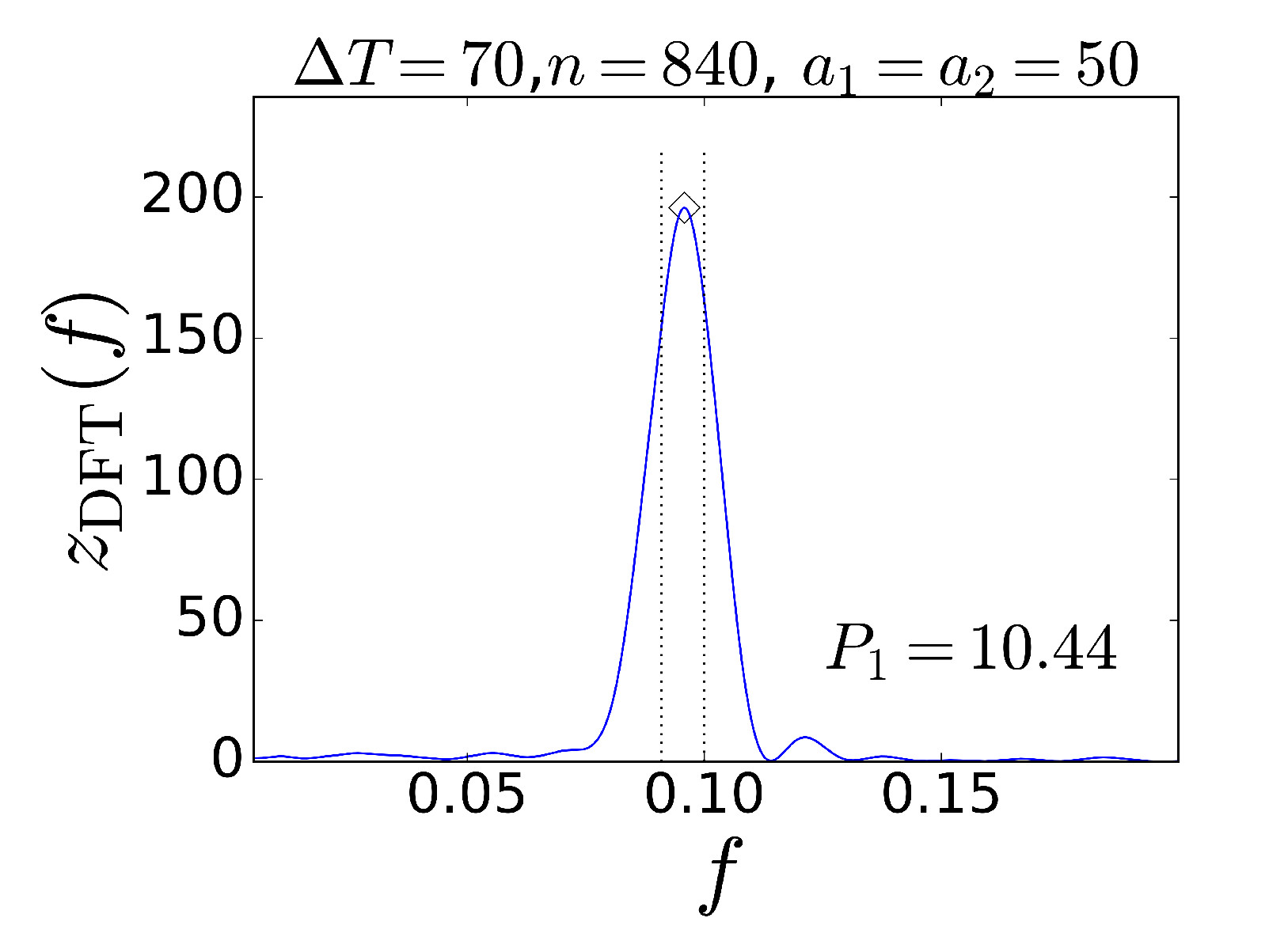} 
 \hspace*{-0.01\textwidth}
 \includegraphics[width=0.25\textwidth,clip=]{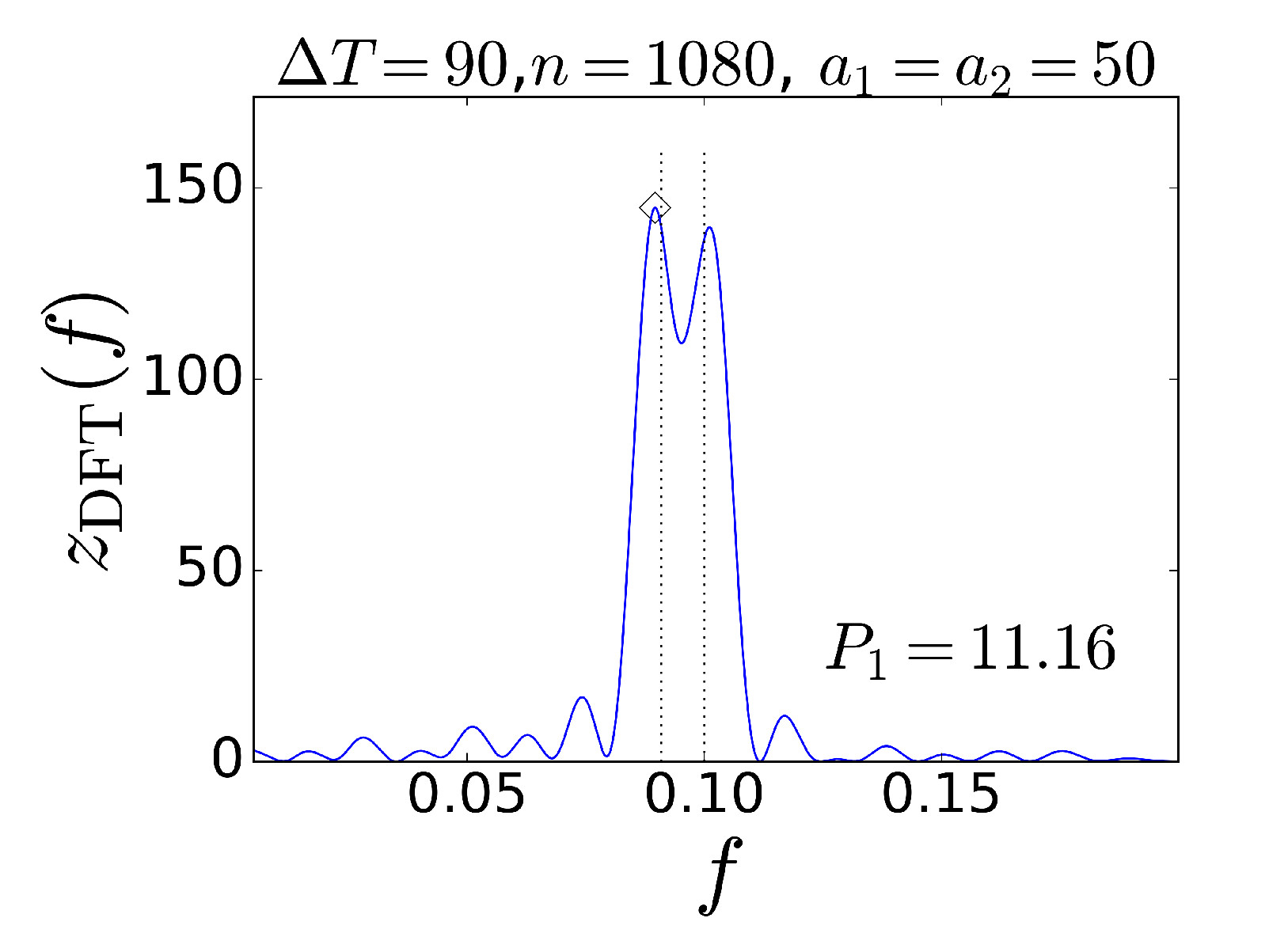} 
 \hspace*{-0.01\textwidth}
 \includegraphics[width=0.25\textwidth,clip=]{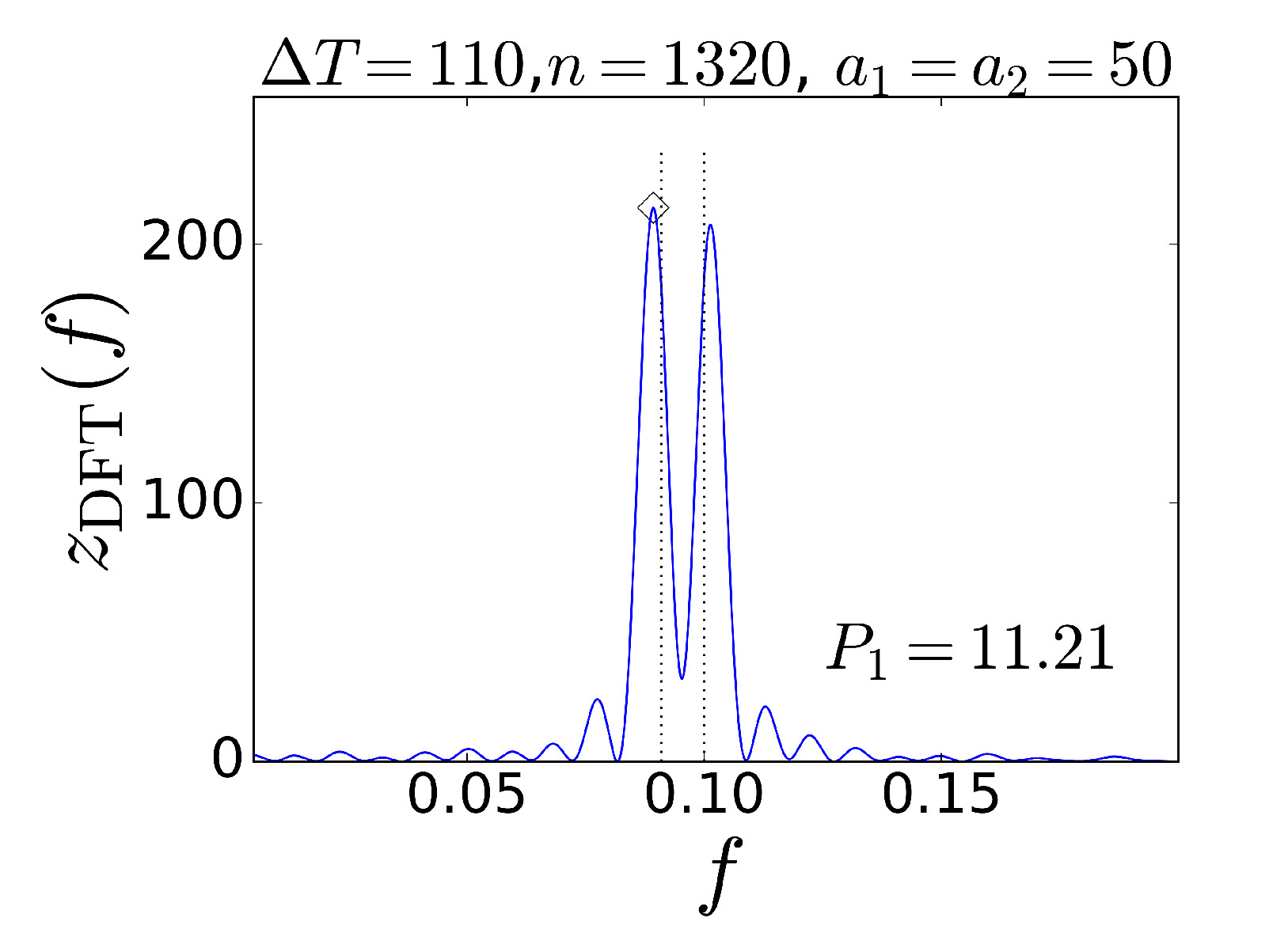} 
 \hspace*{-0.01\textwidth}
 \includegraphics[width=0.25\textwidth,clip=]{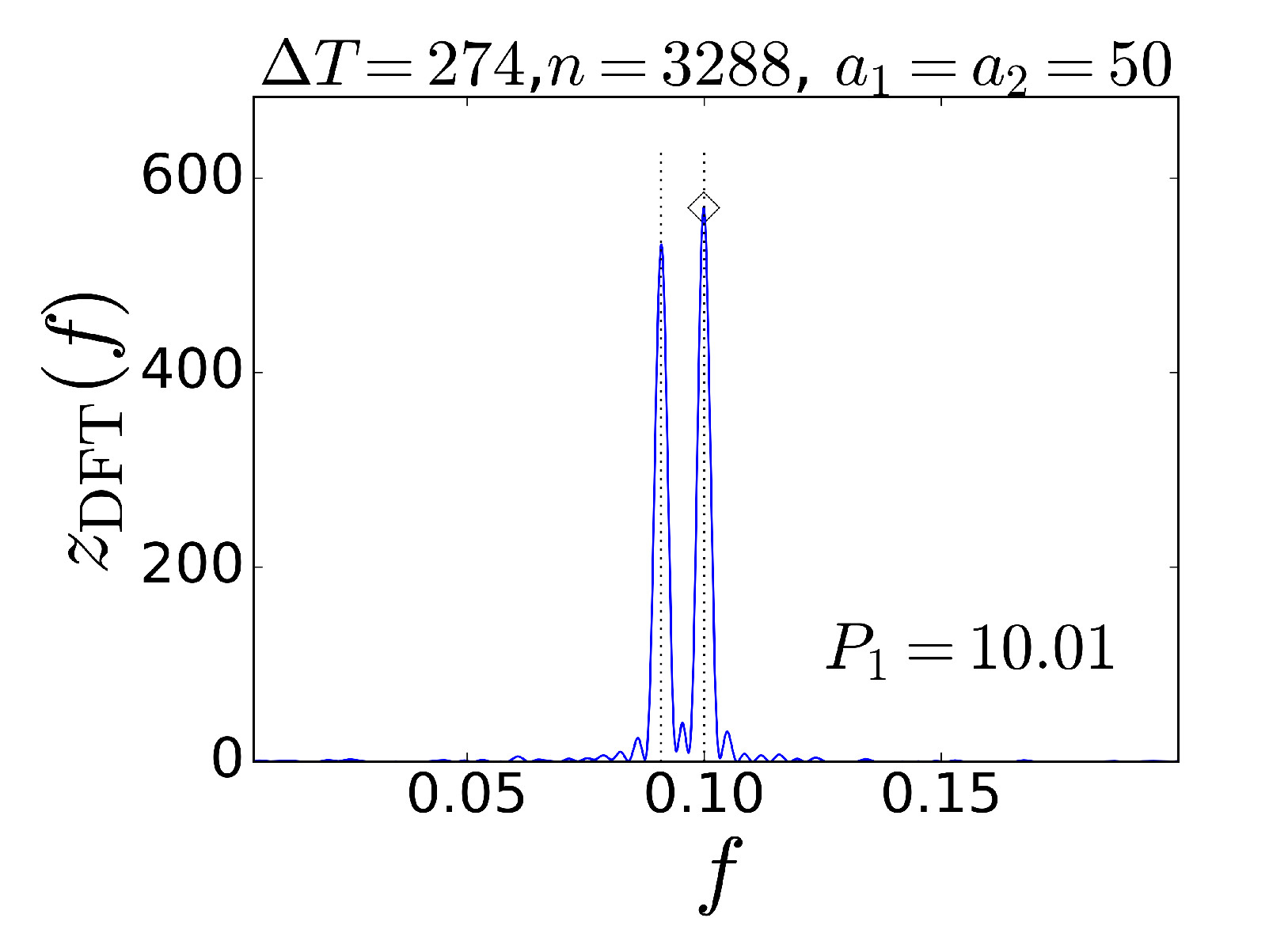} 
 }
\vspace{-0.22\textwidth}
\centerline{\normalsize \bf 
\hspace{0.11\textwidth}   \color{black}{(a)}
\hspace{0.23\textwidth}   \color{black}{(g)}
\hspace{0.22\textwidth}  \color{black}{(m)}
\hspace{0.22\textwidth}  \color{black}{(s)}
\hfill}
\vspace{0.21\textwidth}
\centerline{\hspace*{0.005\textwidth}
 \includegraphics[width=0.25\textwidth,clip=]{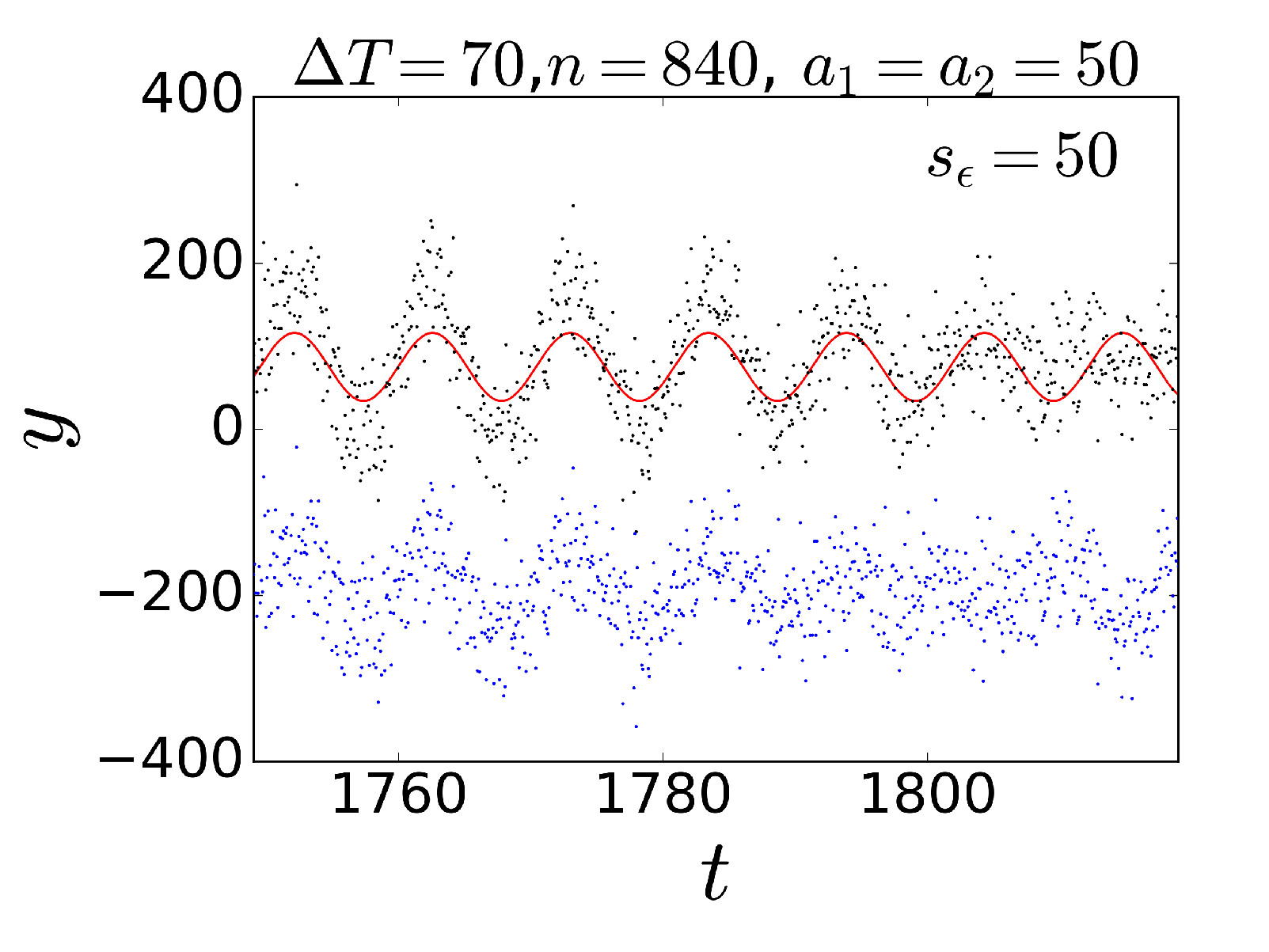} 
 \hspace*{-0.01\textwidth}
 \includegraphics[width=0.25\textwidth,clip=]{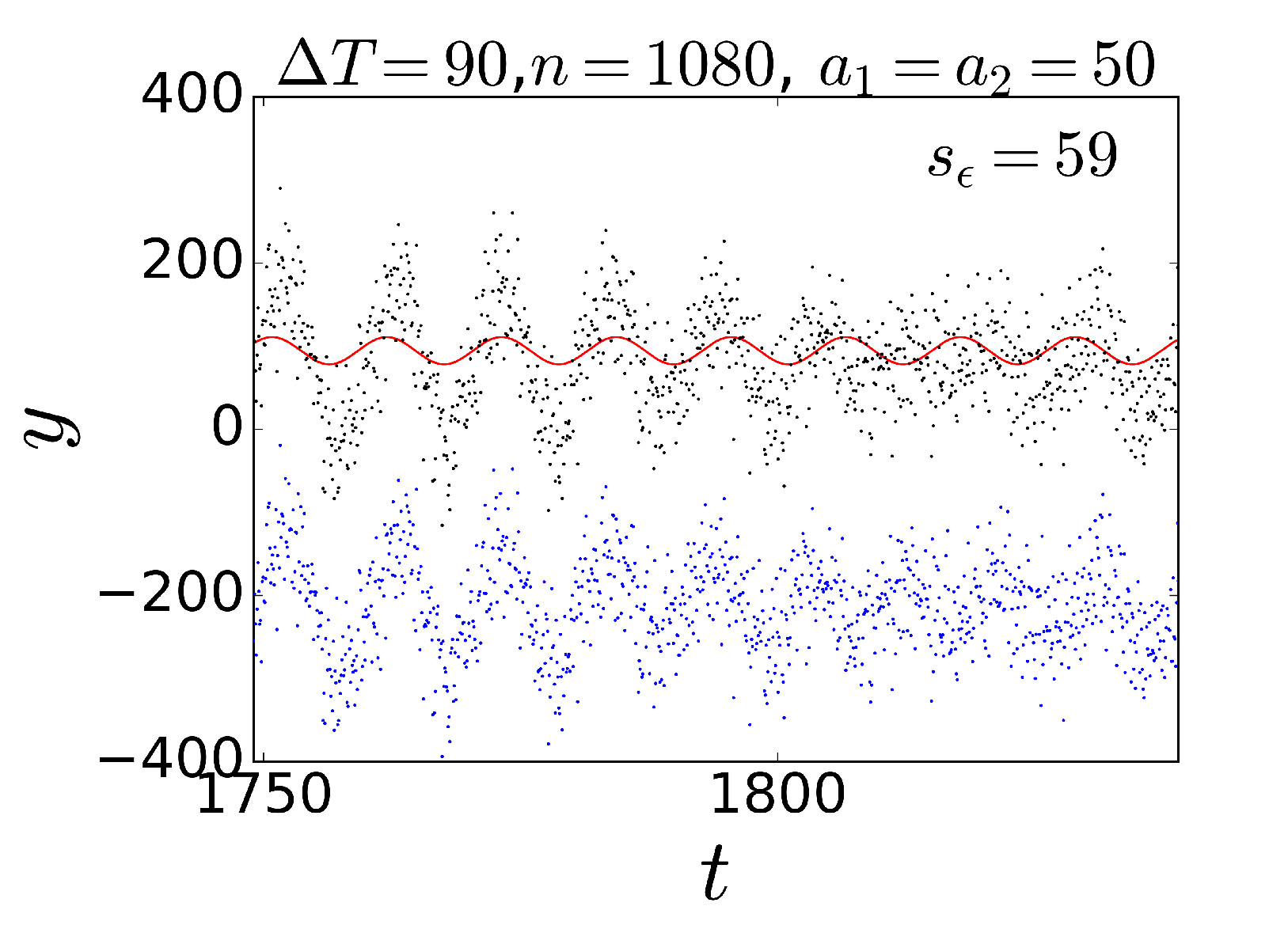} 
 \hspace*{-0.01\textwidth}
 \includegraphics[width=0.25\textwidth,clip=]{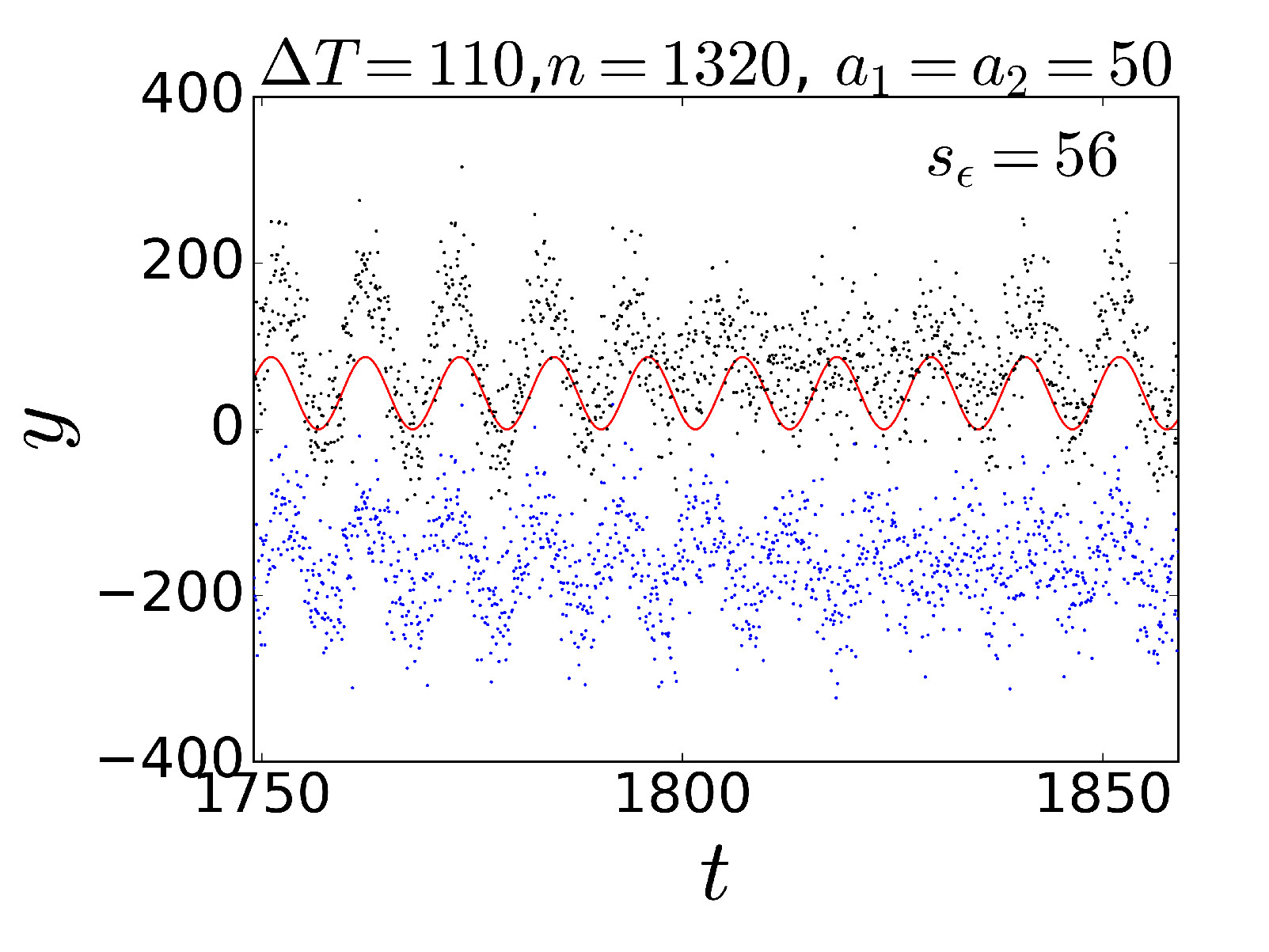} 
 \hspace*{-0.01\textwidth}
 \includegraphics[width=0.25\textwidth,clip=]{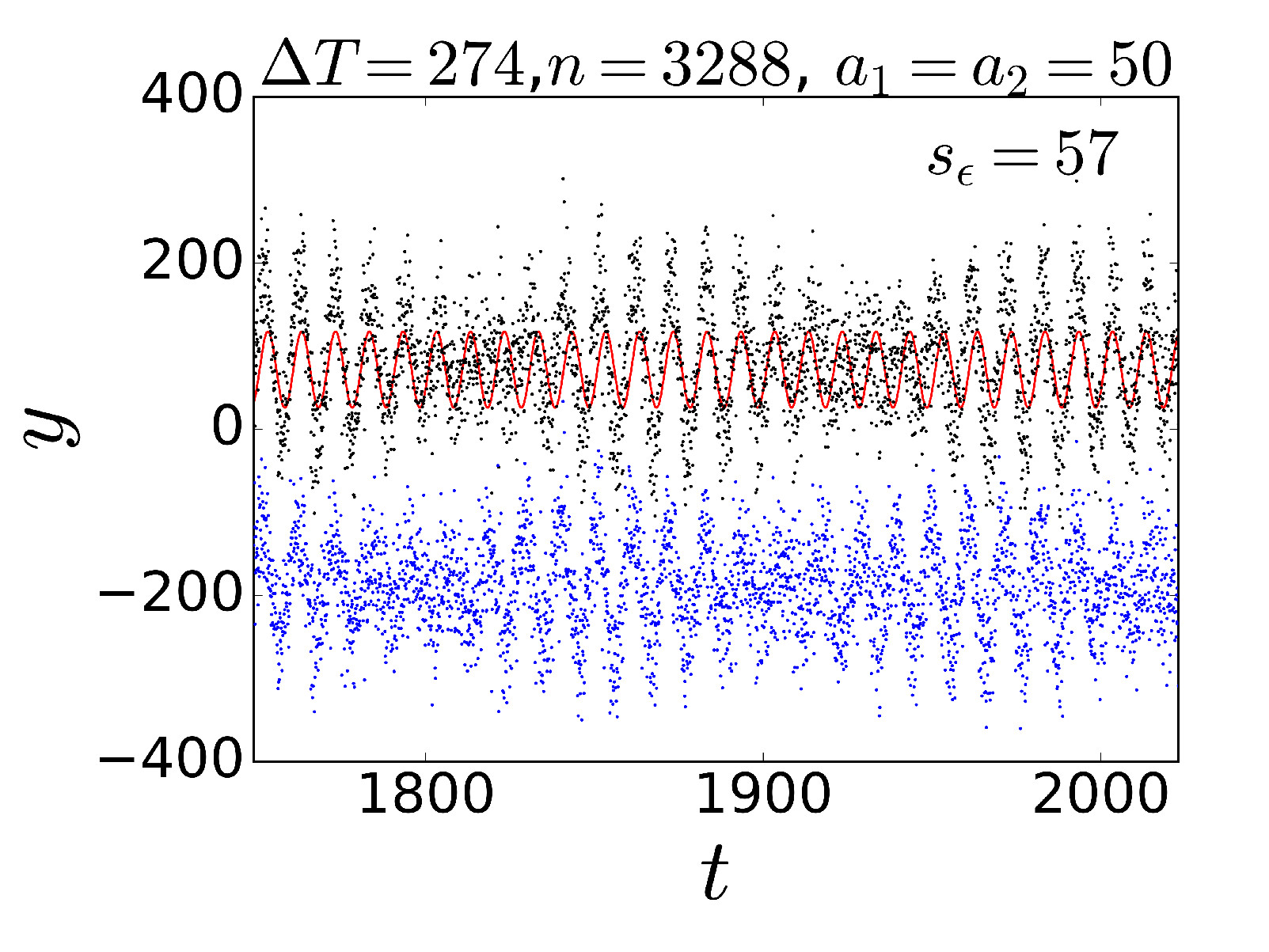} 
  }
\vspace{-0.22\textwidth}
\centerline{\normalsize \bf 
\hspace{0.11\textwidth}   \color{black}{(b)}
\hspace{0.23\textwidth}  \color{black}{(h)}
\hspace{0.23\textwidth}  \color{black}{(n)}
\hspace{0.22\textwidth}  \color{black}{(t)}
  \hfill} 
\vspace{0.21\textwidth}
\centerline{\hspace*{0.005\textwidth}
 \includegraphics[width=0.25\textwidth,clip=]{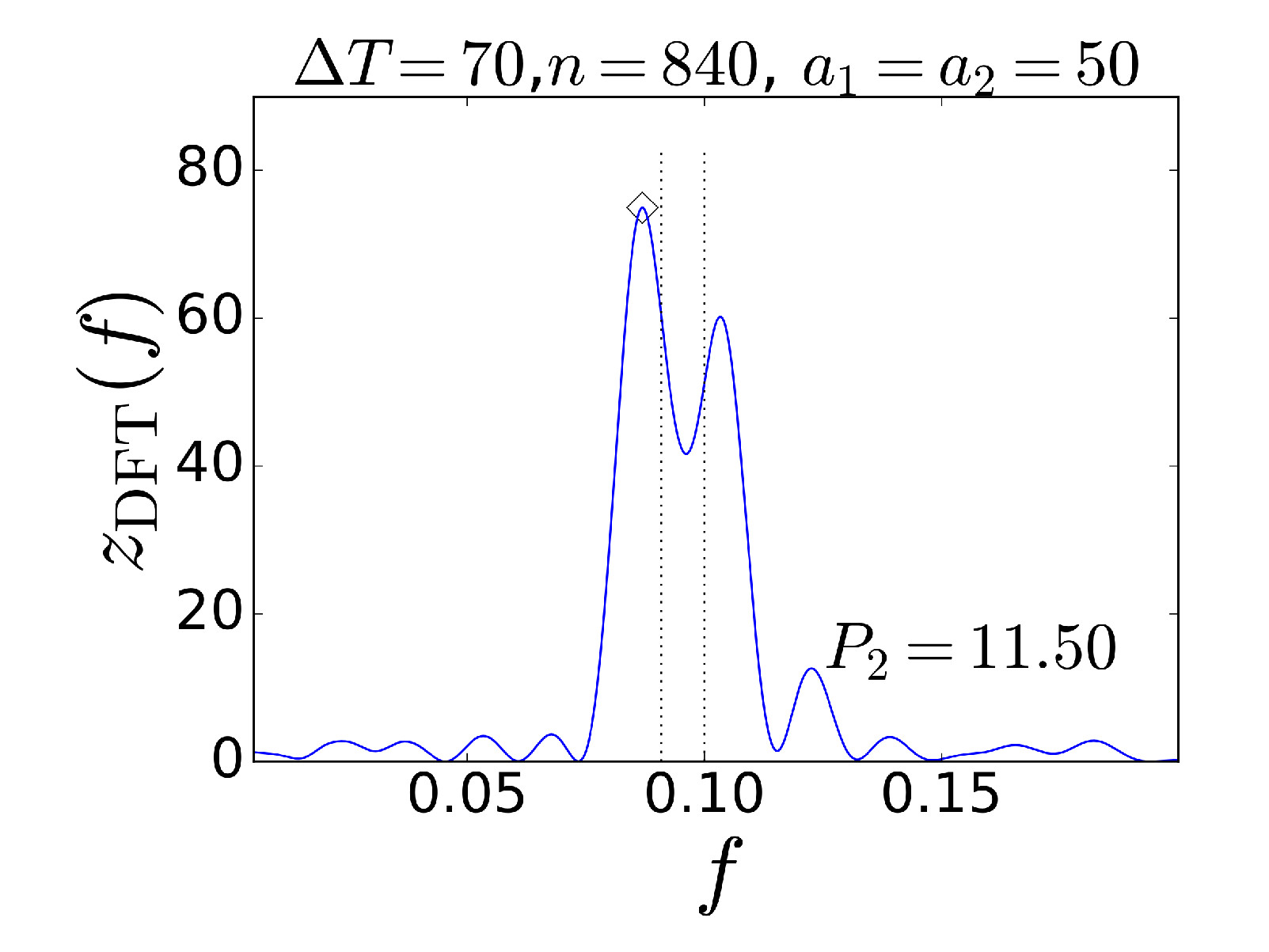} 
 \hspace*{-0.01\textwidth}
 \includegraphics[width=0.25\textwidth,clip=]{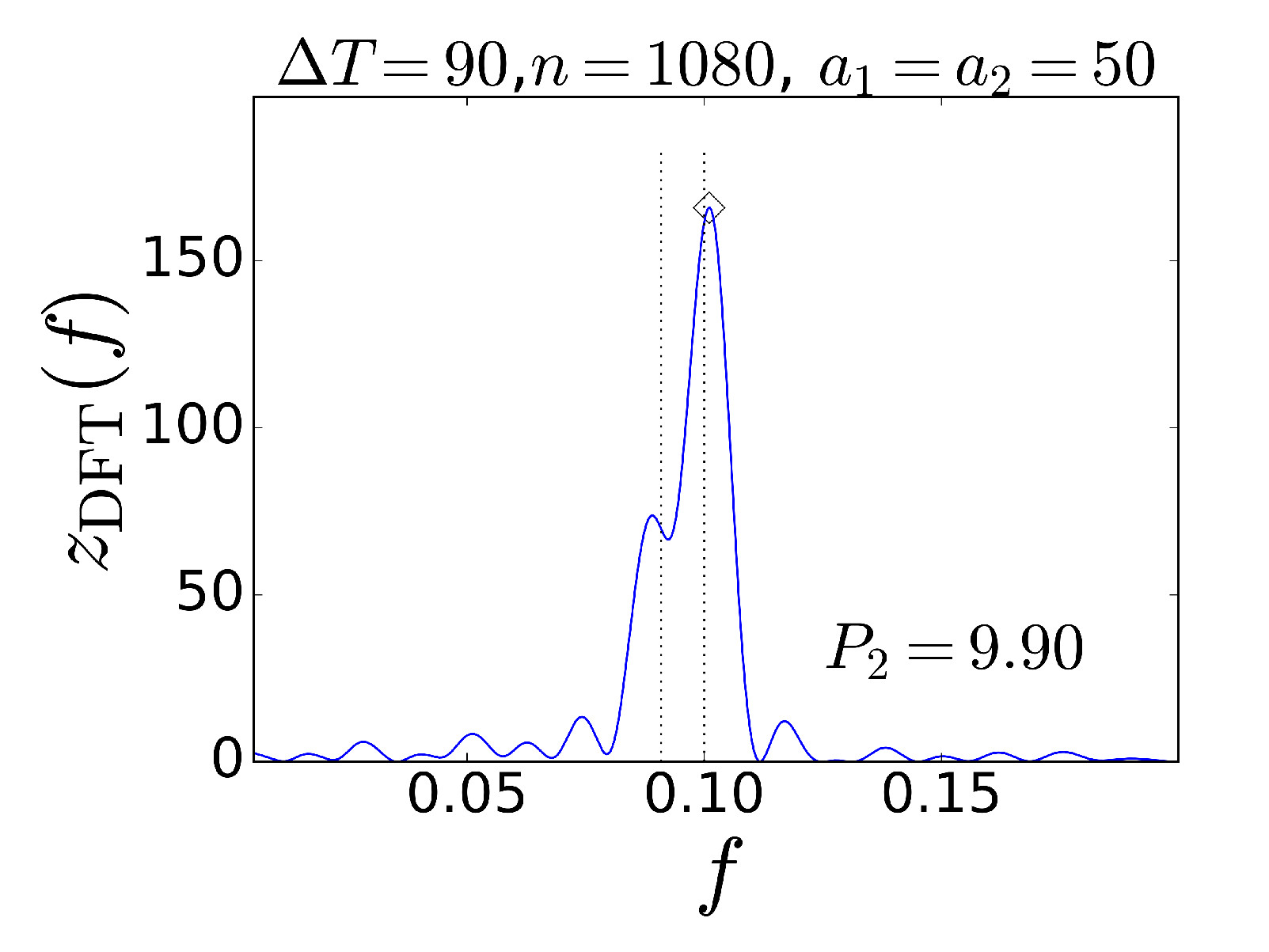} 
 \hspace*{-0.01\textwidth}
 \includegraphics[width=0.25\textwidth,clip=]{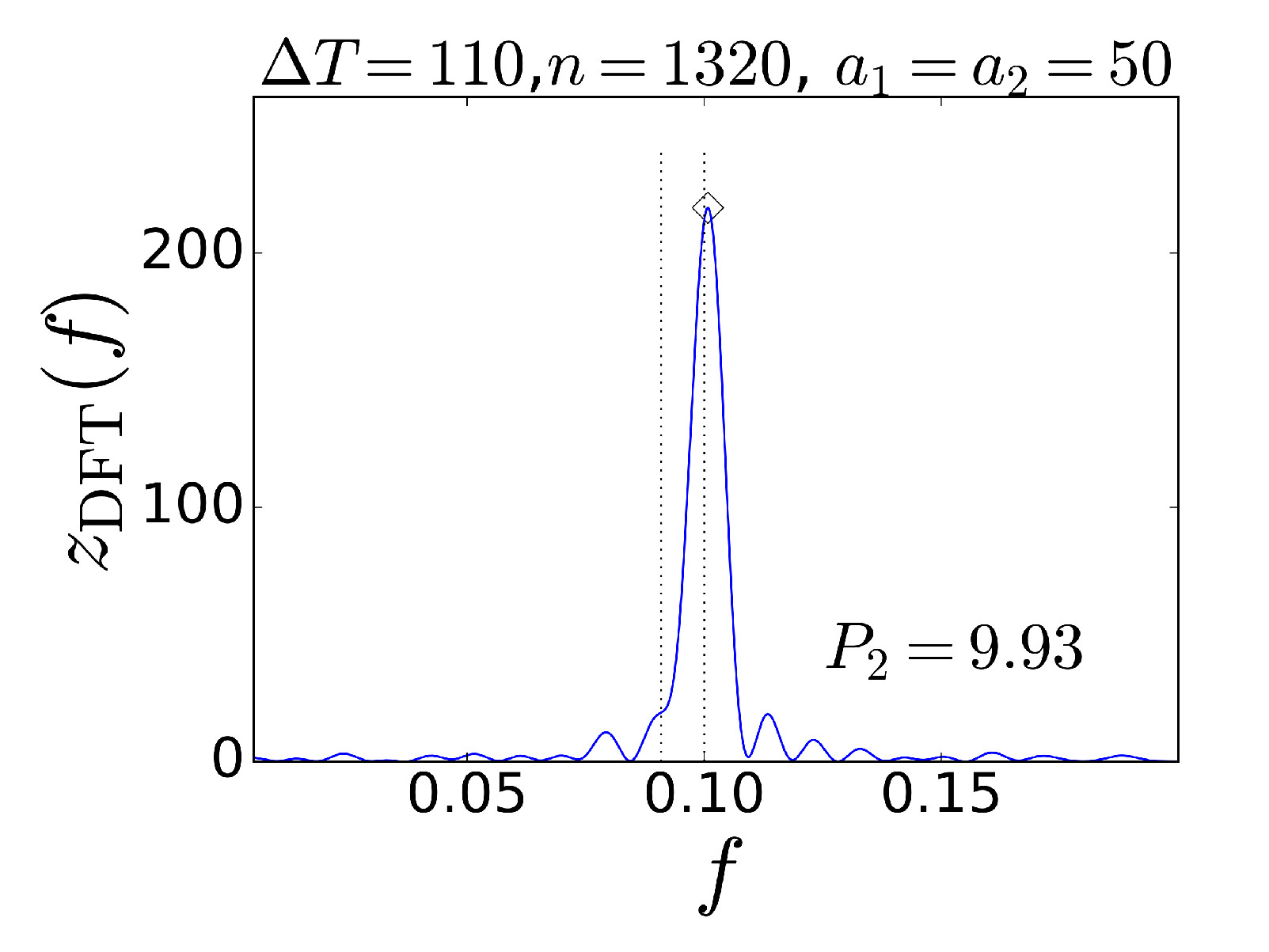} 
 \hspace*{-0.01\textwidth}
 \includegraphics[width=0.25\textwidth,clip=]{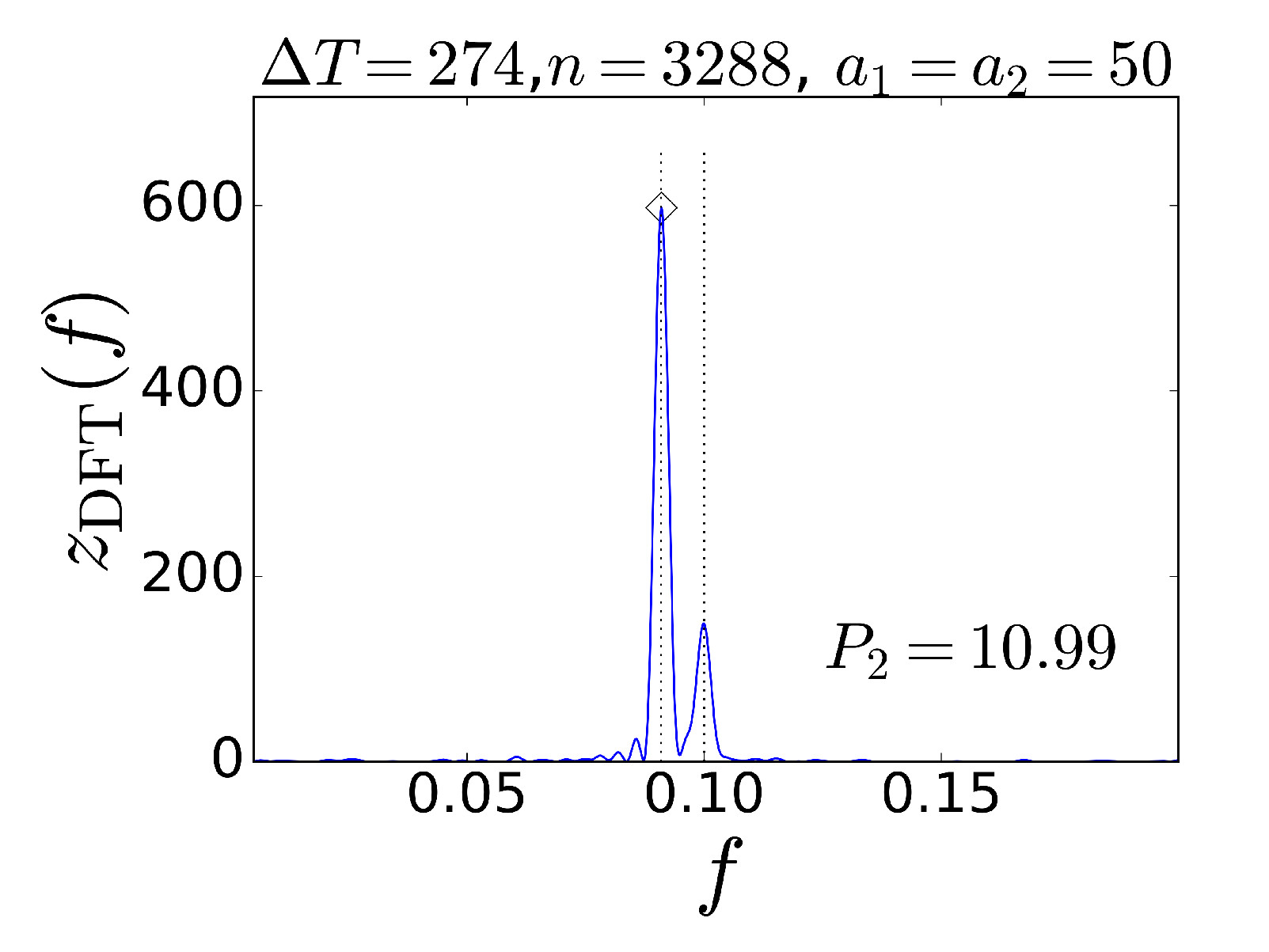} 
}
\vspace{-0.22\textwidth}
\centerline{\normalsize \bf 
\hspace{0.11\textwidth}   \color{black}{(c)}
\hspace{0.23\textwidth}  \color{black}{(i)}
\hspace{0.22\textwidth}  \color{black}{(o)}
\hspace{0.22\textwidth}  \color{black}{(u)}
  \hfill}
\vspace{0.21\textwidth}
\centerline{\hspace*{0.005\textwidth}
 \includegraphics[width=0.25\textwidth,clip=]{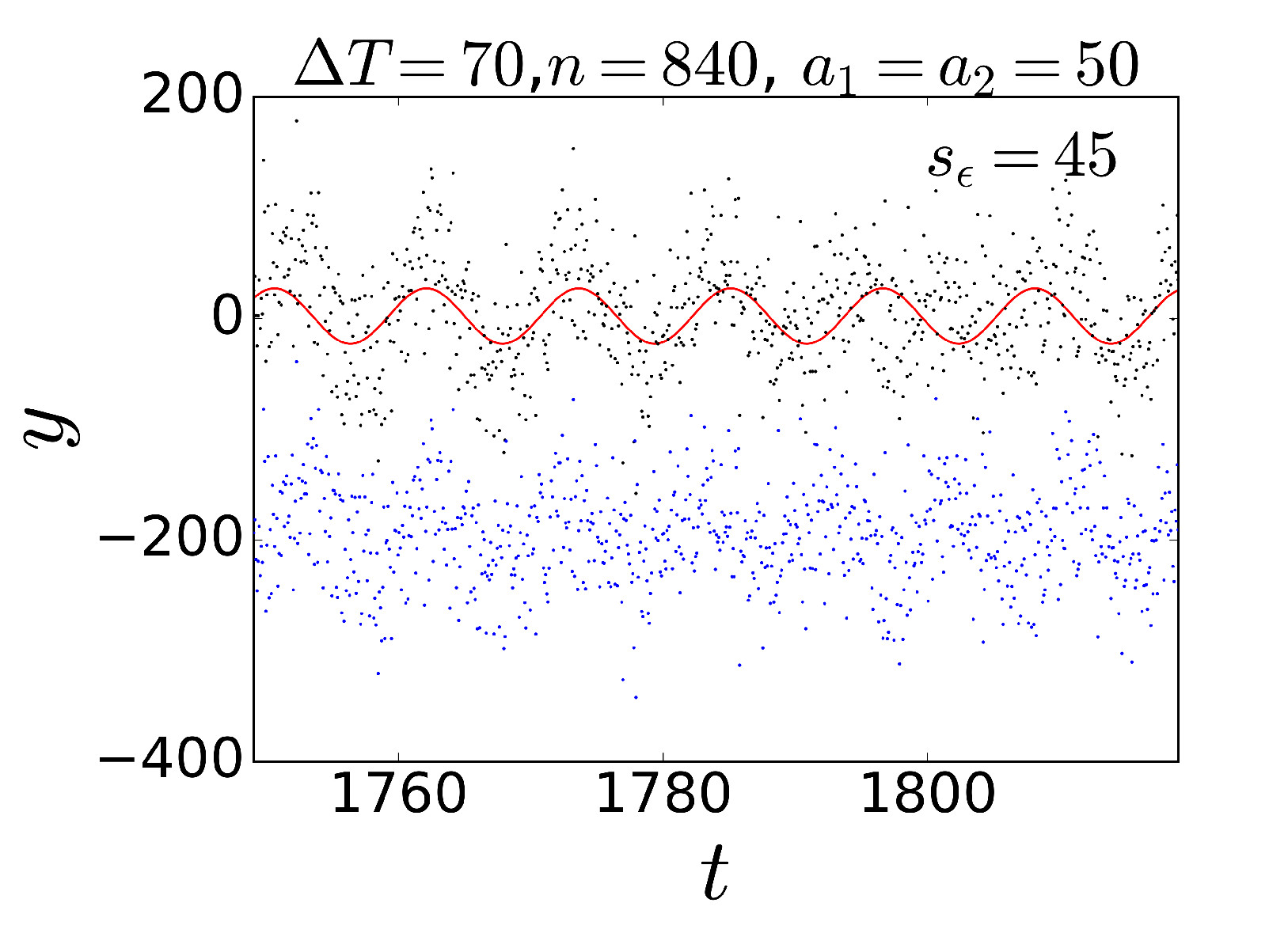} 
 \hspace*{-0.01\textwidth}
 \includegraphics[width=0.25\textwidth,clip=]{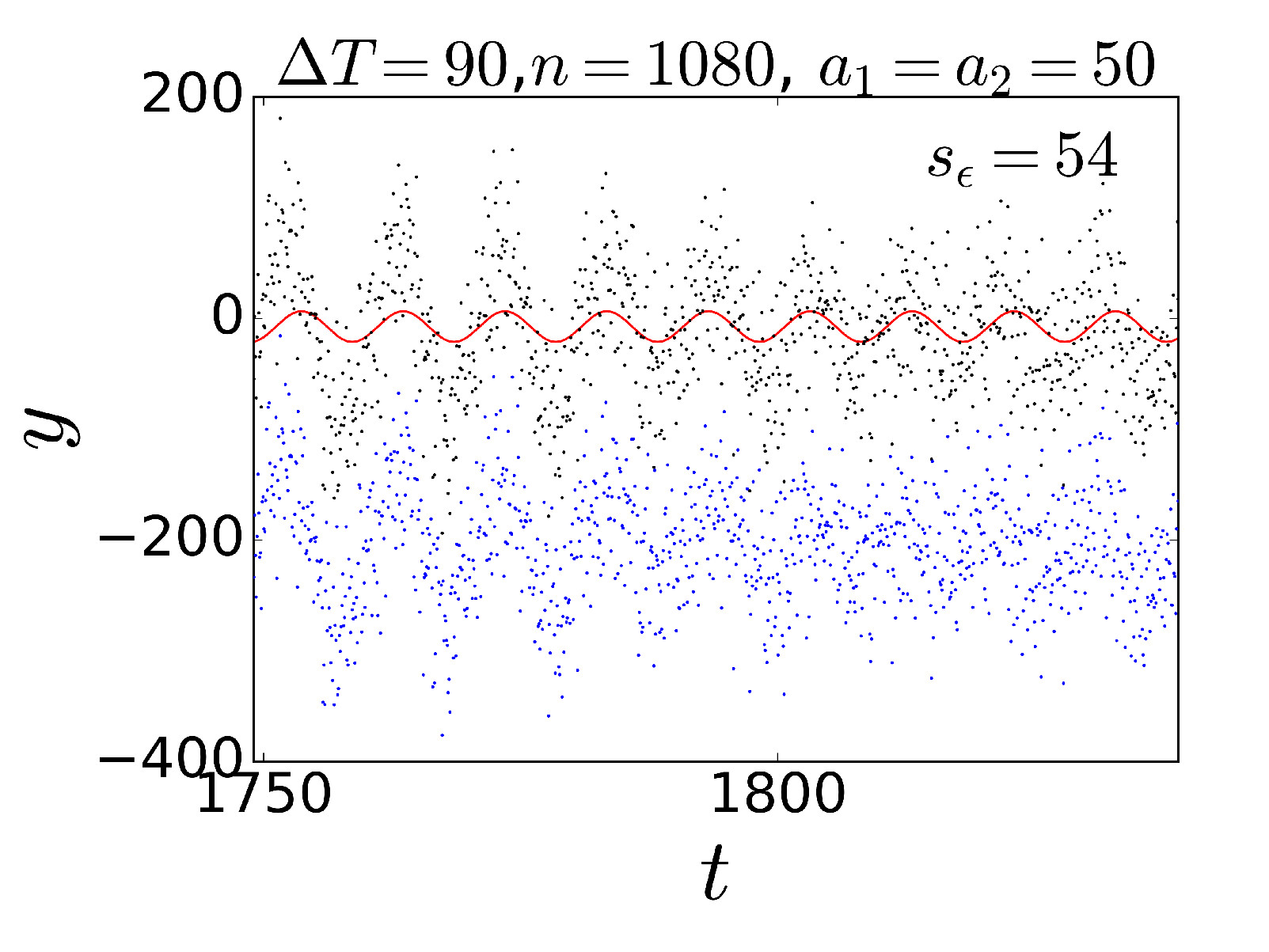} 
 \hspace*{-0.01\textwidth}
 \includegraphics[width=0.25\textwidth,clip=]{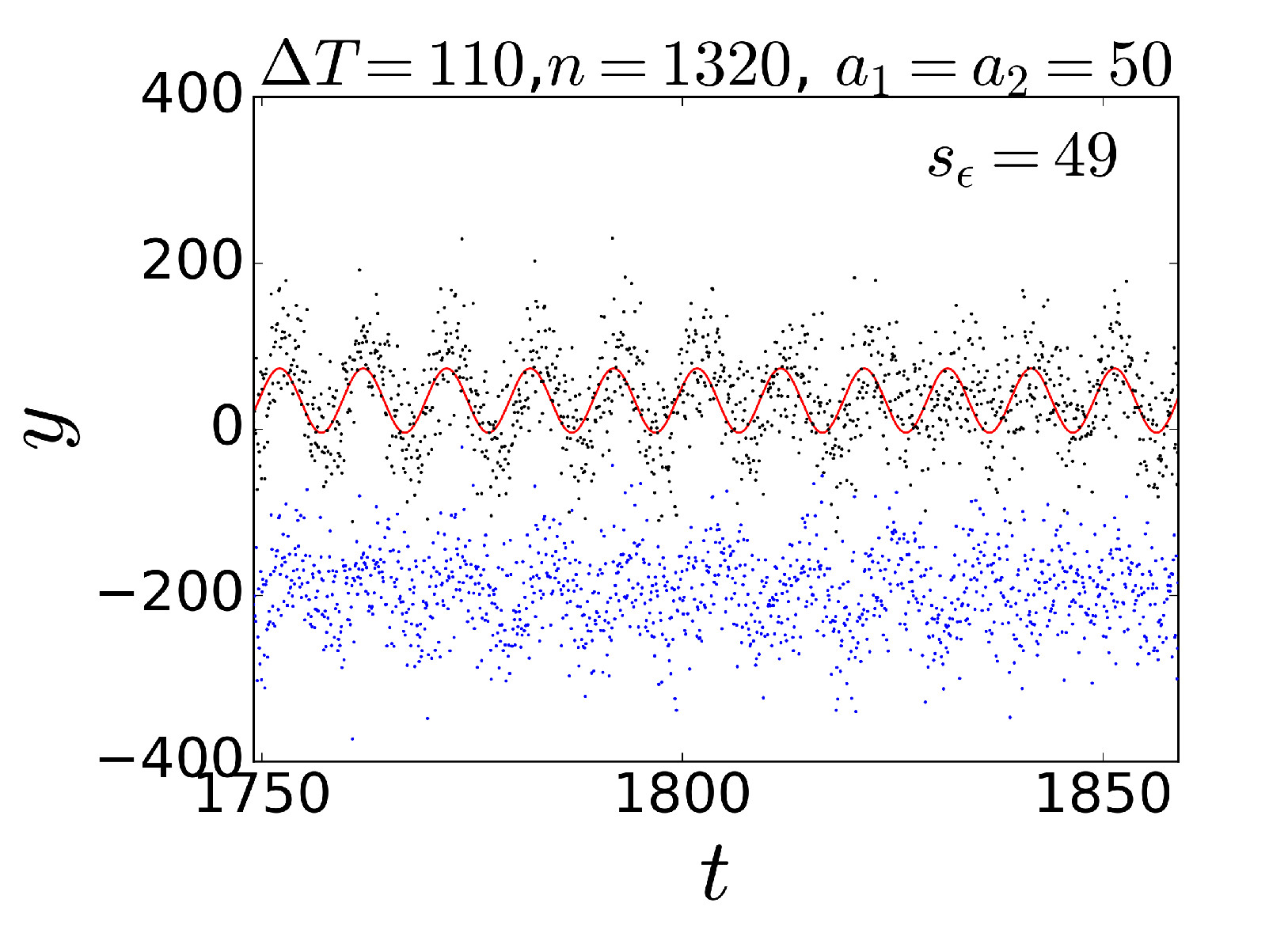} 
 \hspace*{-0.01\textwidth}
 \includegraphics[width=0.25\textwidth,clip=]{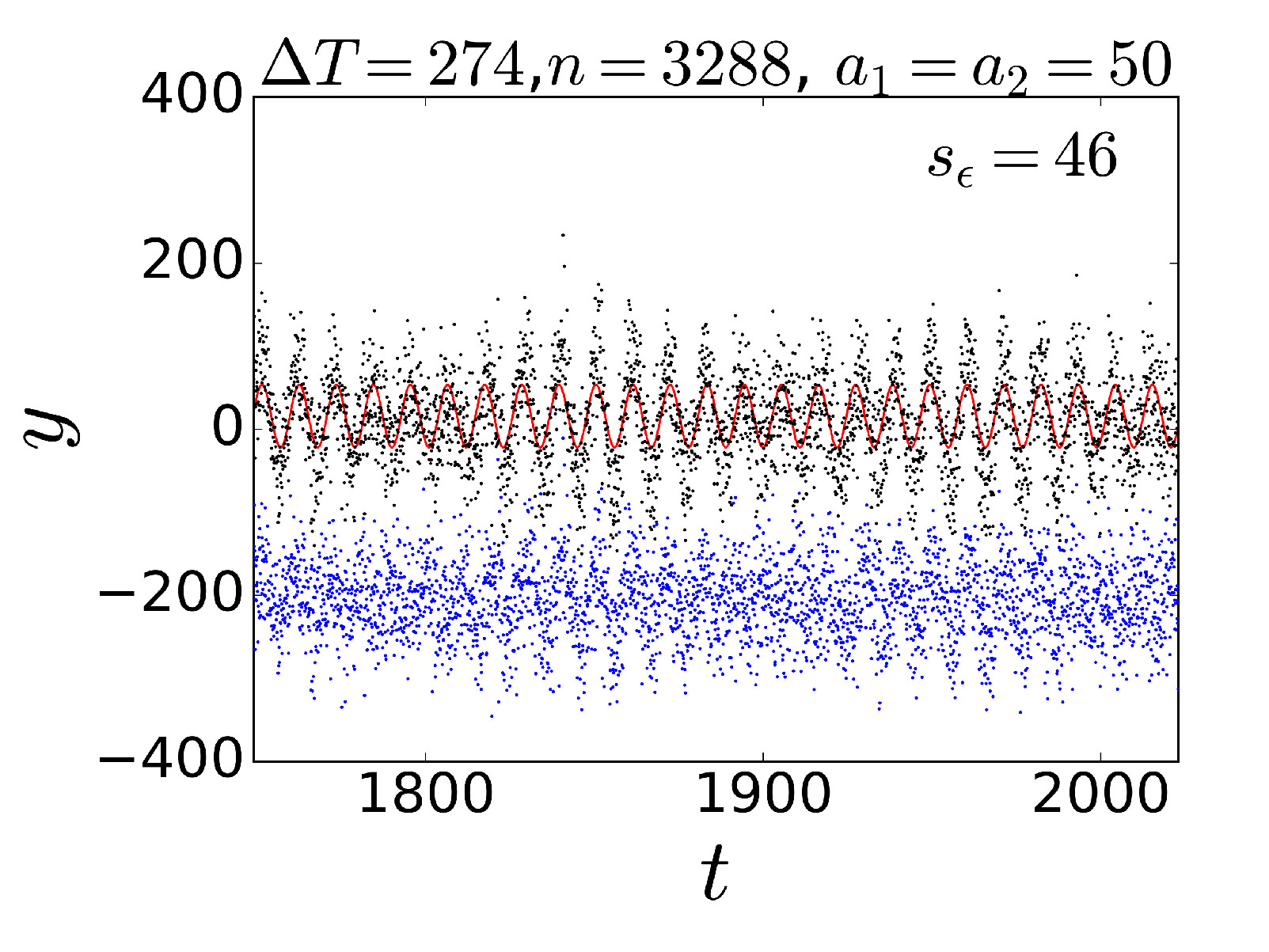} 
}
\vspace{-0.22\textwidth}
\centerline{\normalsize \bf 
\hspace{0.11\textwidth}   \color{black}{(d)}
\hspace{0.23\textwidth}  \color{black}{(j)}
\hspace{0.22\textwidth}  \color{black}{(p)}
\hspace{0.22\textwidth}  \color{black}{(v)}
  \hfill}
\vspace{0.21\textwidth}
\centerline{\hspace*{0.005\textwidth}
 \includegraphics[width=0.25\textwidth,clip=]{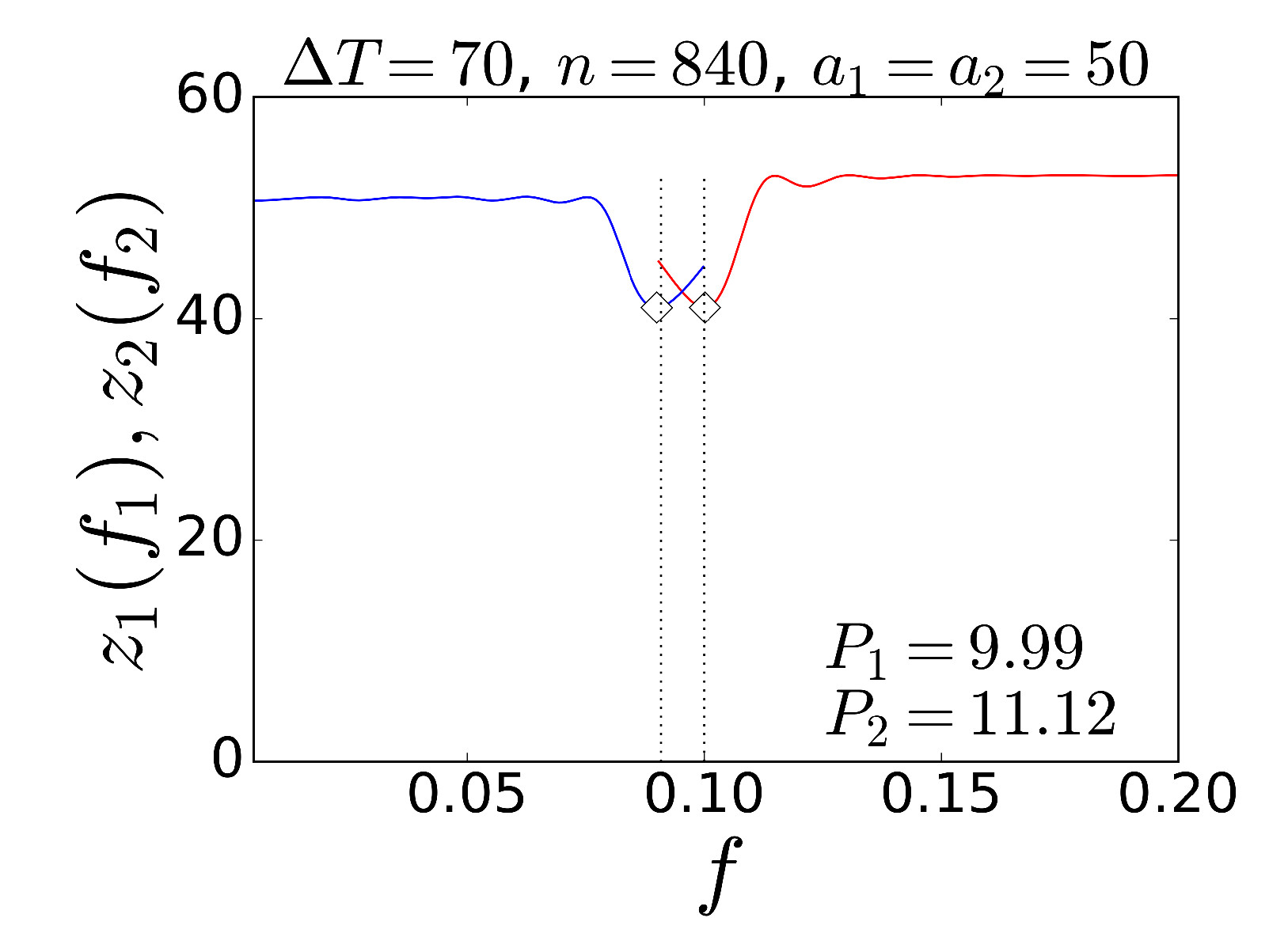} 
 \hspace*{-0.01\textwidth}
 \includegraphics[width=0.25\textwidth,clip=]{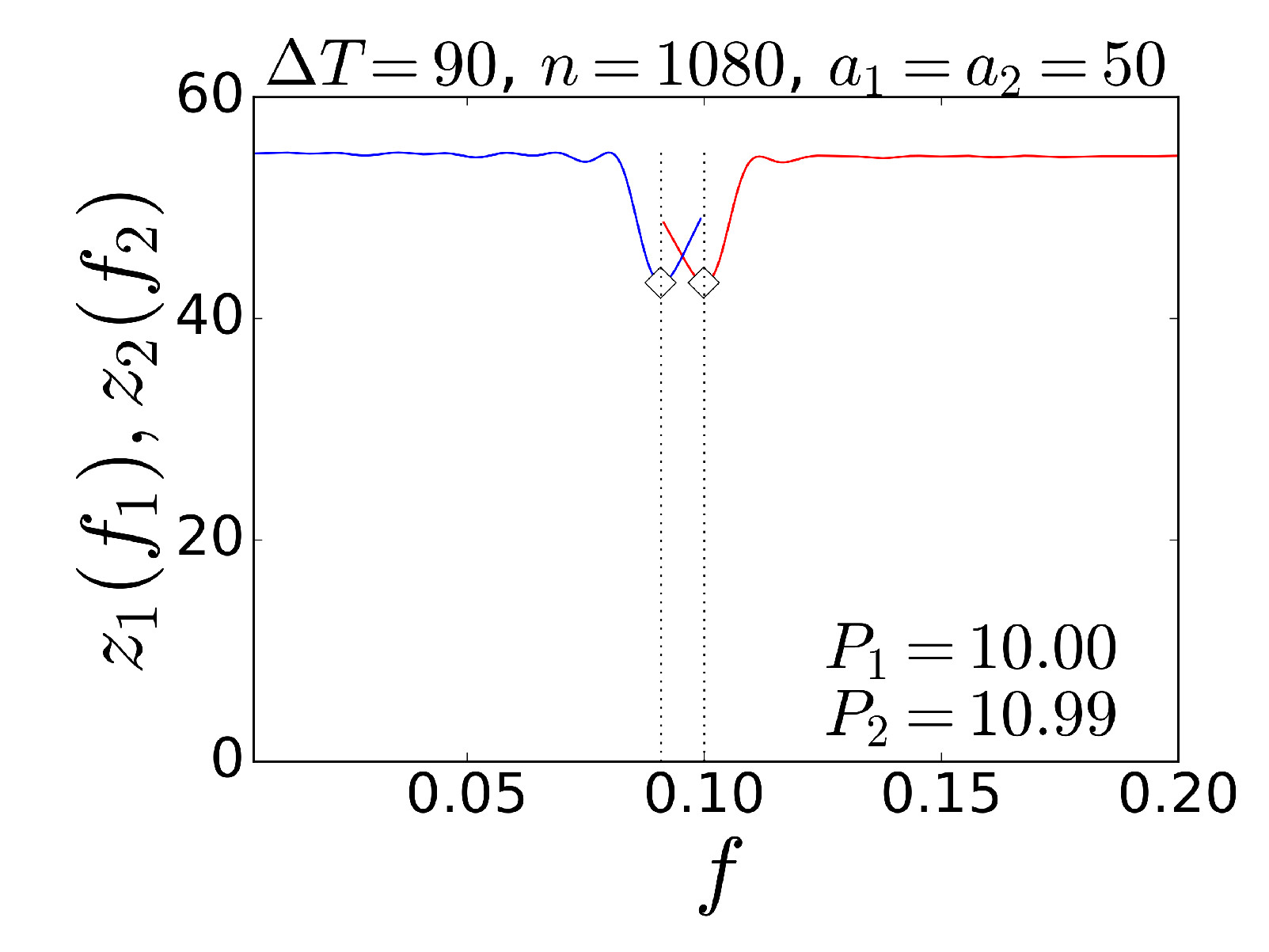} 
 \hspace*{-0.01\textwidth}
 \includegraphics[width=0.25\textwidth,clip=]{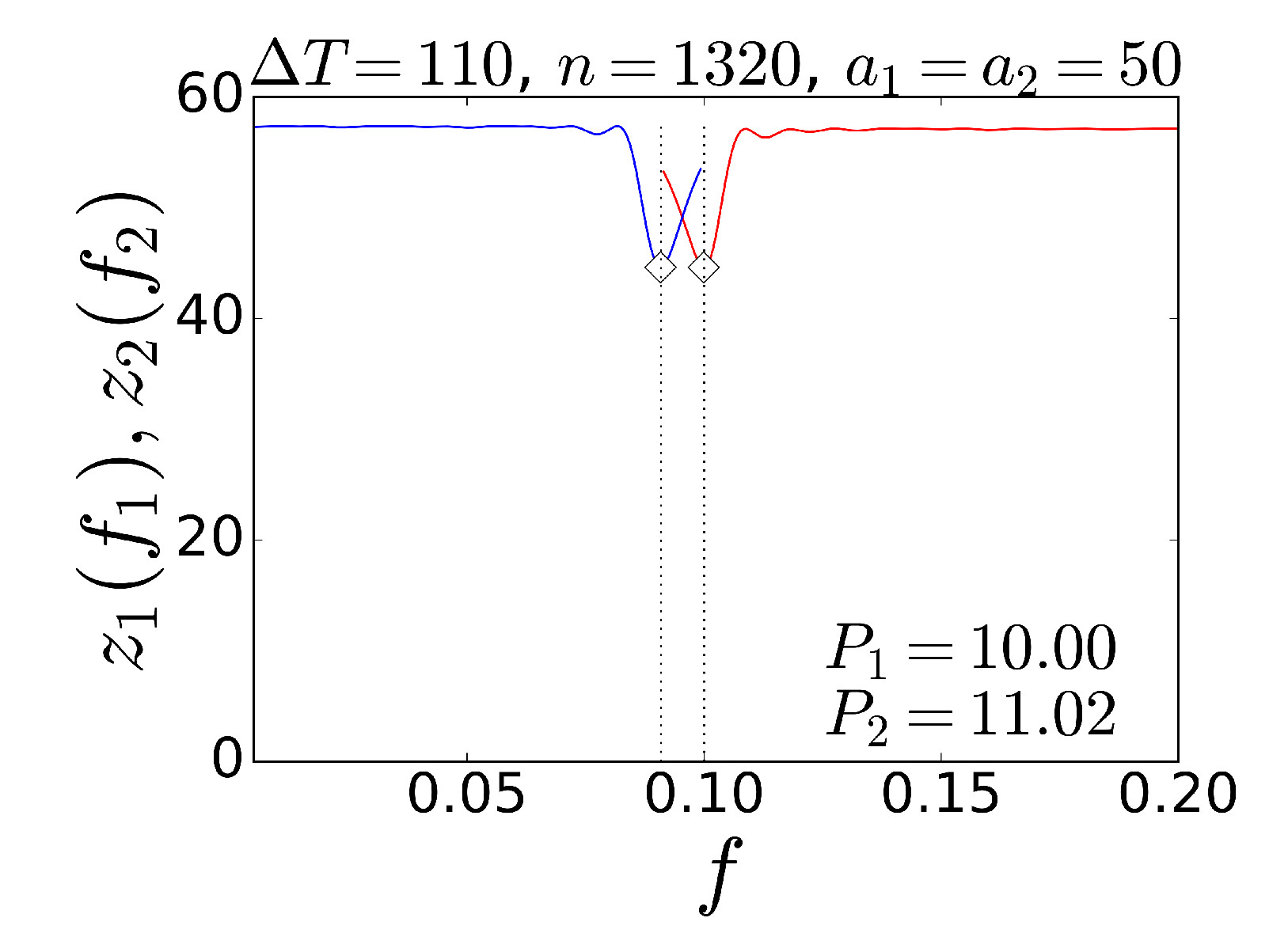} 
 \hspace*{-0.01\textwidth}
 \includegraphics[width=0.25\textwidth,clip=]{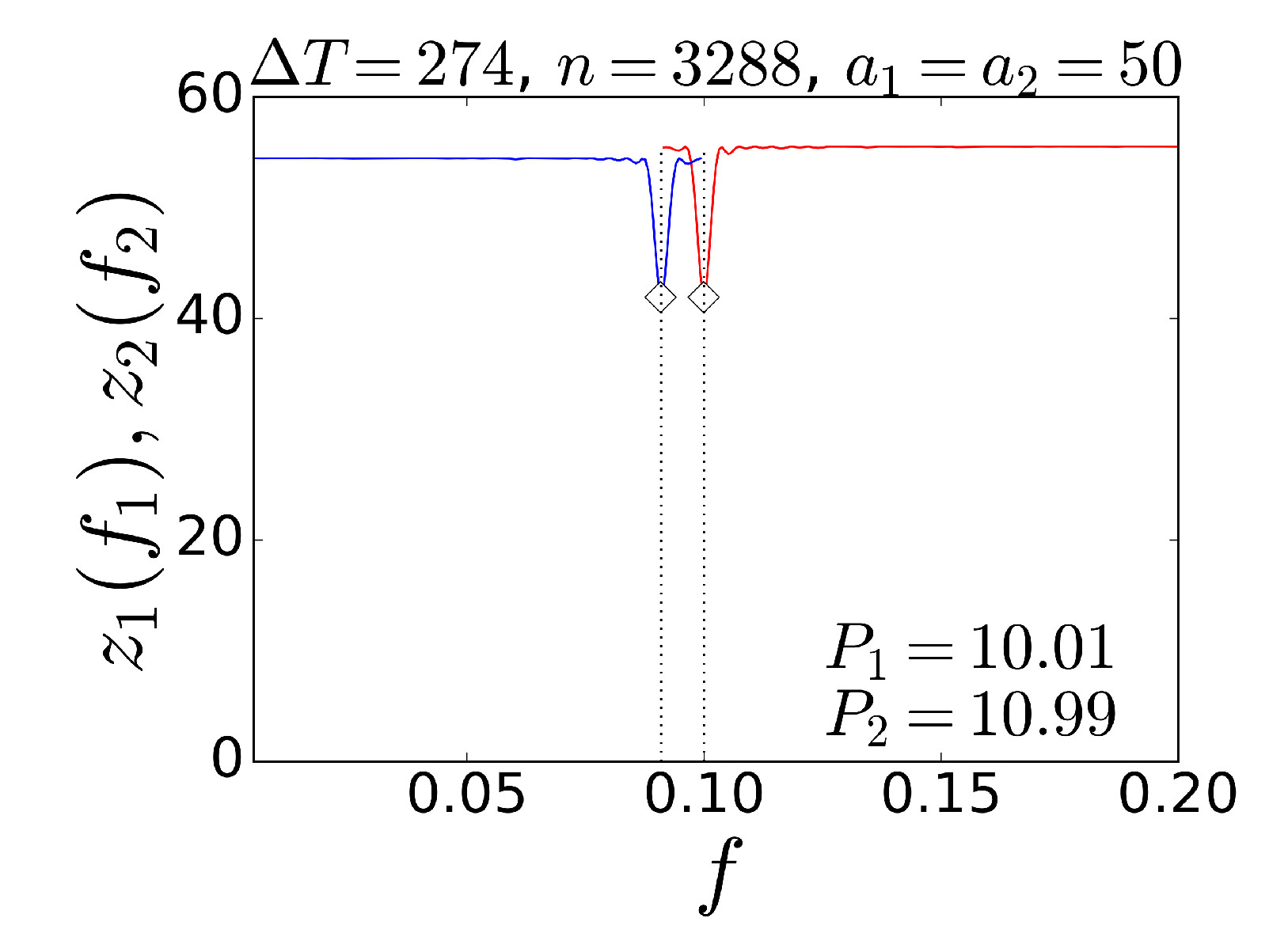} 
 \hspace*{-0.01\textwidth}
}
\vspace{-0.22\textwidth}
\centerline{\normalsize \bf 
\hspace{0.11\textwidth}   \color{black}{(e)}
\hspace{0.23\textwidth}  \color{black}{(k)}
\hspace{0.22\textwidth}  \color{black}{(q)}
\hspace{0.22\textwidth}  \color{black}{(w)}
  \hfill}
\vspace{0.21\textwidth}
\centerline{\hspace*{0.005\textwidth}
 \includegraphics[width=0.25\textwidth,clip=]{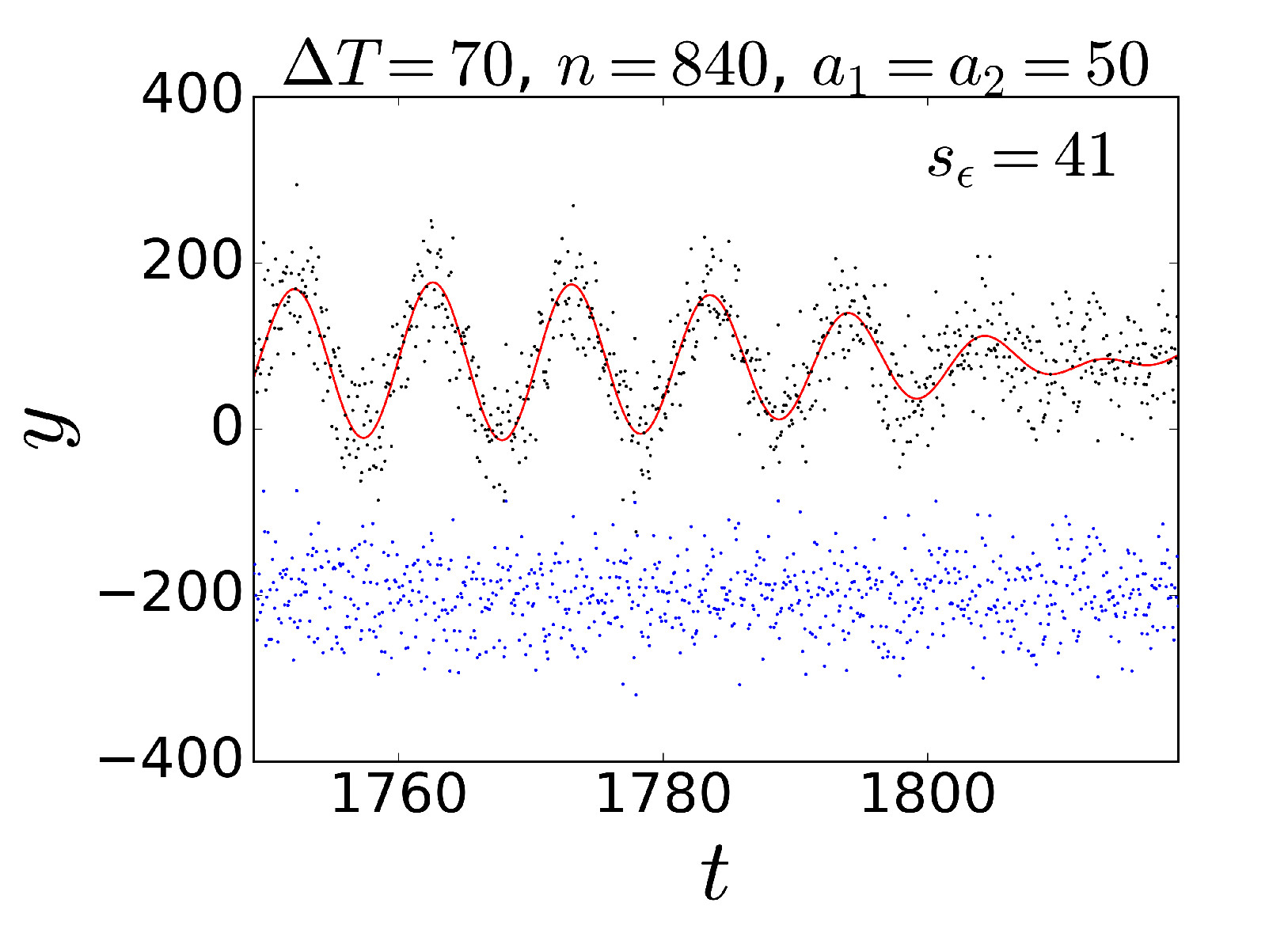} 
 \hspace*{-0.01\textwidth}
 \includegraphics[width=0.25\textwidth,clip=]{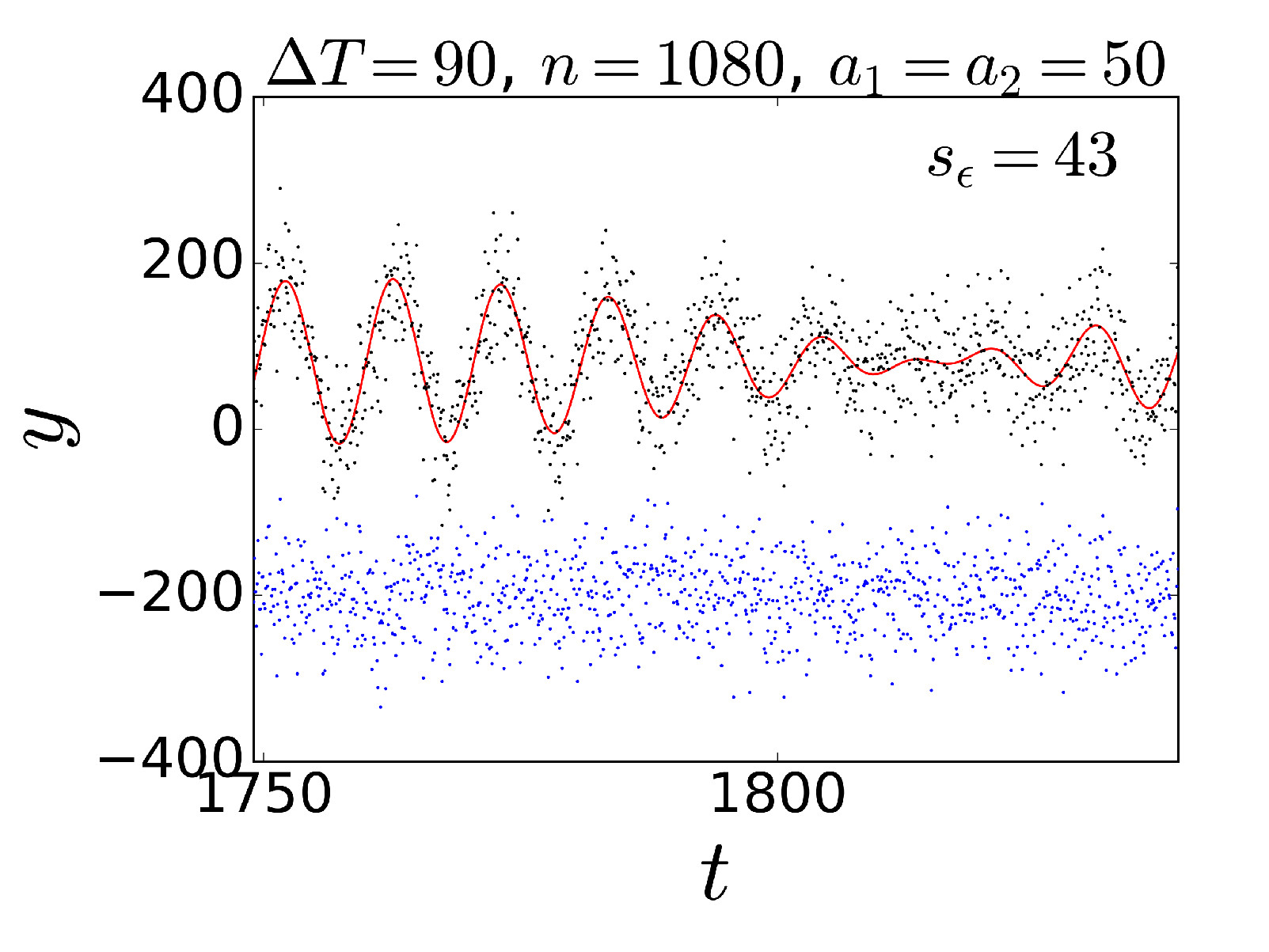} 
 \hspace*{-0.01\textwidth}
 \includegraphics[width=0.25\textwidth,clip=]{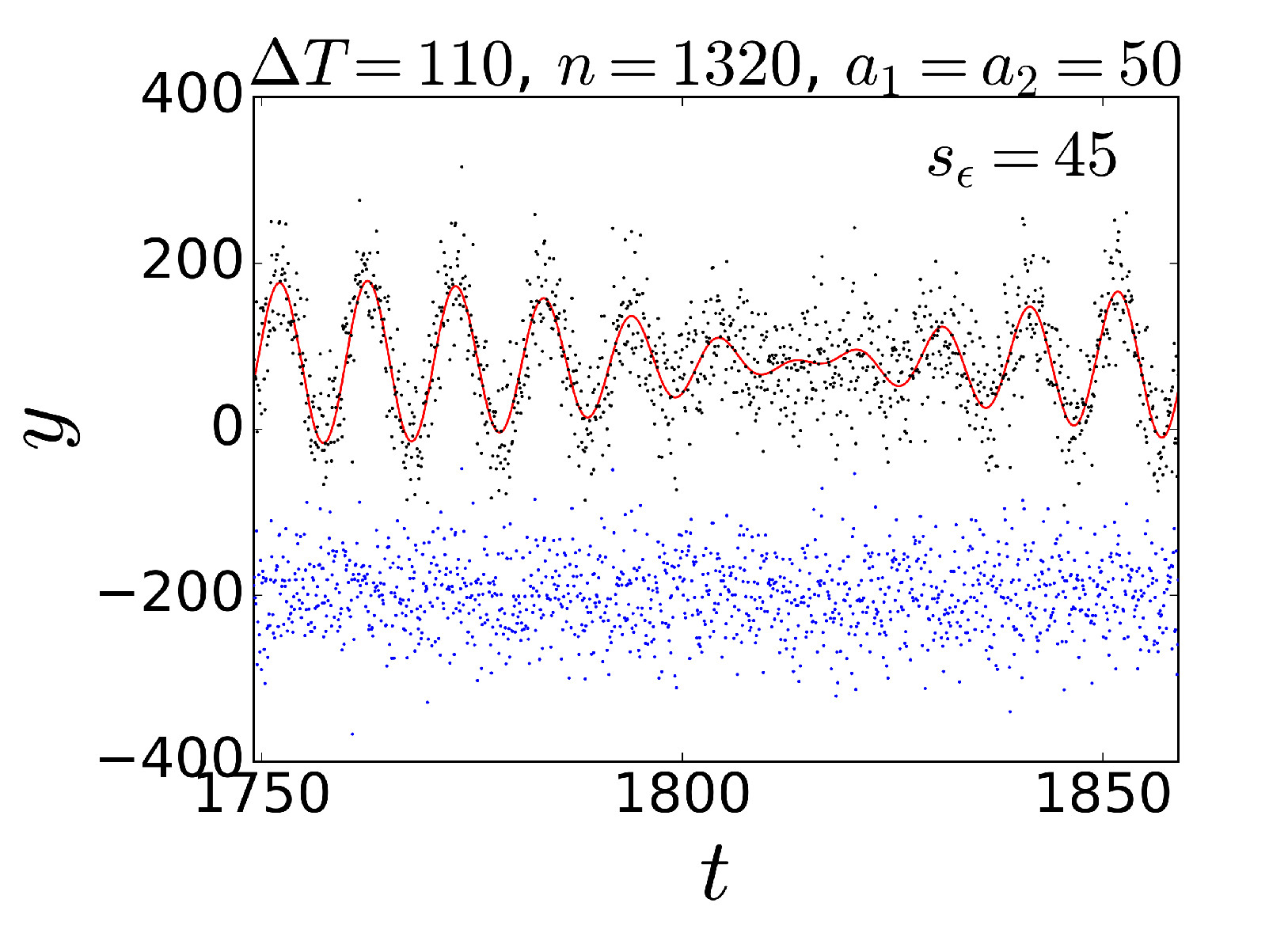} 
 \hspace*{-0.01\textwidth}
 \includegraphics[width=0.25\textwidth,clip=]{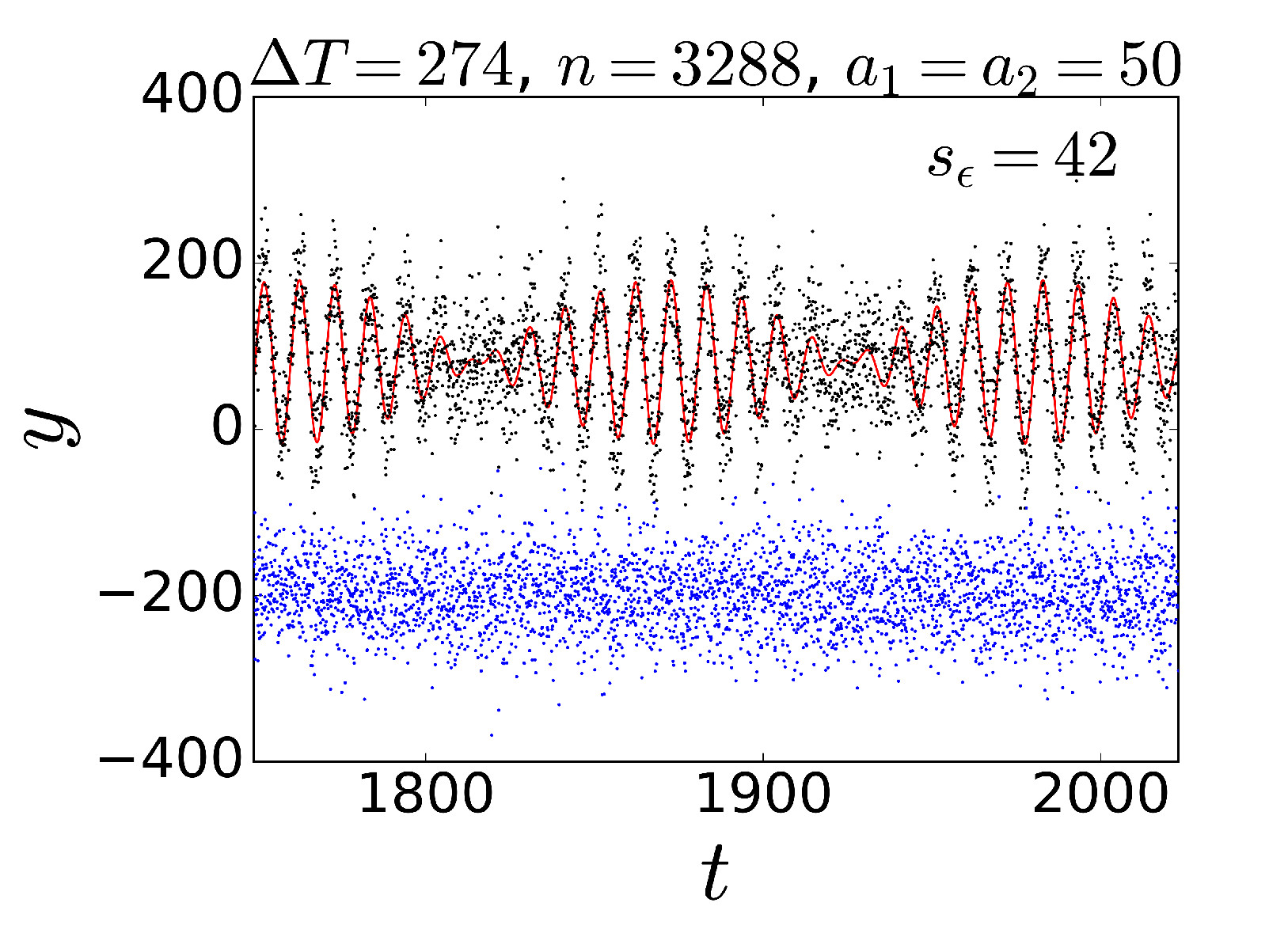} 
        }
\vspace{-0.22\textwidth}
\centerline{\normalsize \bf 
\hspace{0.11\textwidth}   \color{black}{(f)}
\hspace{0.23\textwidth}  \color{black}{(l)}
\hspace{0.22\textwidth}  \color{black}{(r)}
\hspace{0.22\textwidth}  \color{black}{(x)}
  \hfill}
\vspace{0.168\textwidth}
\caption{Equal amplitude $a_1=a_2=50$ simulations
  for signal periods $P_1=10$ and $P_2=11$
  (Equations \ref{EqSignalOne} and \ref{EqSignalTwo}).
  (a) DFT periodogram $z_{\mathrm{DFT}}(f)$.
  Period $P_1=10.44$ is detected from $y_i^{\star}$
  data simulated with $n=840$, $\Delta T=70$,
  $P_1=10$, $a_1=50$, $P_2=11$
  and $a_2=50$. Units are $[f]=$ 1/y and
  $[z_{\mathrm{DFT}}]=$ dimensionless.
  Vertical dashed black lines denote locations of
  simulated frequencies $f_1=1/P_1$ and $f_2=1/P_2$.
  Diamond denotes location of detected frequency.
  (b) Model for simulated data.
  Black dots denote simulated $y_{i}^{\star}$ data.
  Red curve denotes pure sine model having period $P_1=10.44$.
  Black dots
  denote $\epsilon_i$ residuals offset to level -200.
  Standard deviation of
  residuals is $s_{\epsilon}=50$.
  Units are $[t]=$ year and $[y]=[\epsilon]=$ dimensionless.
  (c-d) DFT results for first sample of residuals.
  Otherwise as in a-b.
  (e) DCM periodograms $z_1(f_1)$ and $z_2(f_2)$.
  Periods $P_1=9.99$ and $P_=11.12$ are
  detected from $y_i^{\star}$
  data simulated with $n=840$, $\Delta T=70$,
  $P_1=10$, $a_1=50$, $P_2=11$
  and $a_2=50$.
  Diamonds denote locations of detected frequencies.
  Units are $[f]=$ 1/y and
  $[z_1]=[z_2]=$ dimensionless.
  (f) Model for simulated data.
  Black dots denote simulated $y_{i}^{\star}$ data.
  Red curve shows two pure sine signal DCM model.
  Blue dots denote $\epsilon_i$ residuals offset to level -200.
  Standard deviation of
  residuals is $s_{\epsilon}=41$.
  Units are $[t]=$ year and $[y]=[\epsilon]=$ dimensionless.
  (g-l) $\Delta T=90$ simulations.
 (m-r) $\Delta T=110$ simulations.
 (s-x) $\Delta T=274$ simulations.}
\label{FigDFTDCMOne}
\end{figure}

\subsection{Results for two equal amplitude signals
\label{SectDFTEqual} }

We present the results
for equal amplitudes $a_1=a_2=50$ 
in Fig. \ref{FigDFTDCMOne}.

The results for the shortest simulated time interval,
$\Delta T = 70$, are shown in
Figs. \ref{FigDFTDCMOne}a-\ref{FigDFTDCMOne}f.
The DFT periodogram $z_{\mathrm{DFT}}(f)$ shows only
one peak at $P_1=10.44$ (Fig. \ref{FigDFTDCMOne}a).
The limitation of Equation \ref{EqPrevent} prevents
the detection
of two peaks.
The effect $[(f_1+f_2)/2]^{-1}=10.48$
(Equation \ref{EqMisleadOne})
explains the detected wrong period $P_1=10.44$ value.
The first DFT model residuals have a standard deviation
$s_{\epsilon}=50$, and show regular periodic
variation (Fig. \ref{FigDFTDCMOne}b).
The DFT periodogram for these residuals shows the highest
peak at $P_2=11.50$ (Fig. \ref{FigDFTDCMOne}c).
The second DFT model residuals have a standard
deviation of $s_{\epsilon}=45$
and show some regular periodic variation
(Fig. \ref{FigDFTDCMOne}d).
DCM detects the $P_1=9.99$ and $P_2=11.12$ periods,
which are close to, but not exactly equal to
the simulated 10 and 11 values 
(Fig. \ref{FigDFTDCMOne}e).
The DCM model residuals show no periodic variation
and follow a straight line (Fig. \ref{FigDFTDCMOne}f).
This residual distribution resembles a random  distribution.
The DCM model residuals
$s_{\epsilon}=41$ standard deviation
is much smaller than the DFT model residuals
$s_{\epsilon}=45$ standard deviation.
In this $\Delta T=70$ simulation,
DFT fails and DCM succeeds.

Both DFT periodograms are double-peaked
in the longer $\Delta T=90$ simulation
(Figs.  \ref{FigDFTDCMOne}g and \ref{FigDFTDCMOne}i).
DFT detects the $P_1=11.16$ and $P_2=9.90$ periods,
which are closer to the simulated values.
The residuals of both DFT models
show periodic variability
(Figs.  \ref{FigDFTDCMOne}h and \ref{FigDFTDCMOne}j).
The mean residual $s_{\epsilon}=59$ and 54
values
are larger than in the earlier
shorter $\Delta T=70$ simulation,
because there is a $\Delta \phi=0.5$ phase shift
(Equation \ref{EqAbrupt}) in the
simulated $y_i^{\star}$ data at the year $1804$.
DCM detects the correct $P_1=10.00$ and $P_2=10.99$ periods
(Fig.  \ref{FigDFTDCMOne}k).
The distribution of DCM model residuals
resembles that of a random distribution,
and there is no periodic variability
(Fig.  \ref{FigDFTDCMOne}l).
The $s_{\epsilon}=43$ value of DCM model
is much smaller than
the $s_{\epsilon}=54$ value of DFT model,
because the DCM can handle
the $\Delta \phi=0.5$ phase shift
(Equation \ref{EqAbrupt}).
Again, DCM performs better than DFT.

In the next $\Delta T=110$ simulation,
the criterion of Equation \ref{EqPrevent}
is fulfilled, and therefore
the first DFT periodogram is clearly double-peaked
(Fig. \ref{FigDFTDCMOne}m).
DFT detects the $P_1=11.21$ and $P_2=9.93$ periods,
which differ from the simulated periods.
Both DFT model $s_{\epsilon}=56$ and 49
mean residual values are large
(Figs. \ref{FigDFTDCMOne}n and \ref{FigDFTDCMOne}p).
These DFT residuals
show periodic variability.
DCM detects the correct
$P_1=10.00$ and $P_2=11.02$ periods
(Fig. \ref{FigDFTDCMOne}q).
The distribution of DCM model residuals appears random,
and its standard deviation is $s_{\epsilon}=45$
(Fig. \ref{FigDFTDCMOne}r).
The abrupt phase shift in the simulated $y_i^{\star}$ data
(Equation \ref{EqAbrupt}) at the year
1804 misleads DFT, but it does not mislead DCM.

Two abrupt phase shifts, in the years 1804 and 1914,
occur in the last $\Delta T=274$ simulation
(Equation \ref{EqAbrupt}).
Finally, DFT succeeds in finding the correct
$P_1=10.01$ and $P_2=10.99$ periods
(Figs. \ref{FigDFTDCMOne}s and \ref{FigDFTDCMOne}u).
However, the residuals of both DFT models still
show periodic variability
(Figs. \ref{FigDFTDCMOne}t and \ref{FigDFTDCMOne}v).
DCM detects the correct 
$P_1=10.01$ and $P_2=10.99$ periods
(Figs. \ref{FigDFTDCMOne}w).
The DCM model residuals show no periodic variability,
and their distribution resembles a random distribution
(Figs. \ref{FigDFTDCMOne}x).
The standard deviation
$s_{\epsilon}=46$ of the DFT model clearly
exceeds the $s_{\epsilon}=42$ value of DCM model.
In short, DCM performs better than DFT.

\begin{figure}  
\vspace{0.02\textwidth}
\centerline{\hspace*{0.005\textwidth}
 \includegraphics[width=0.25\textwidth,clip=]{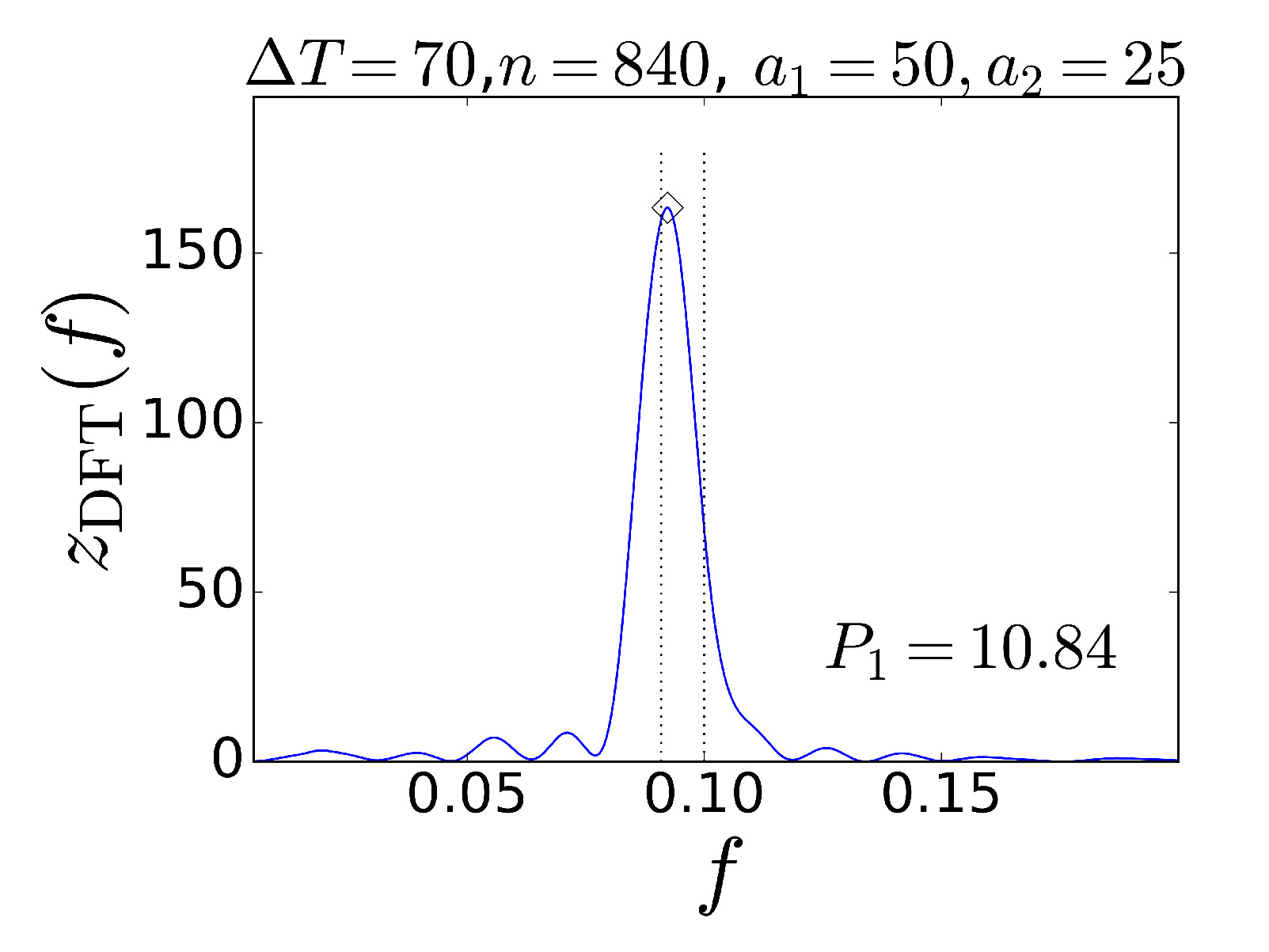} 
 \hspace*{-0.01\textwidth}
 \includegraphics[width=0.25\textwidth,clip=]{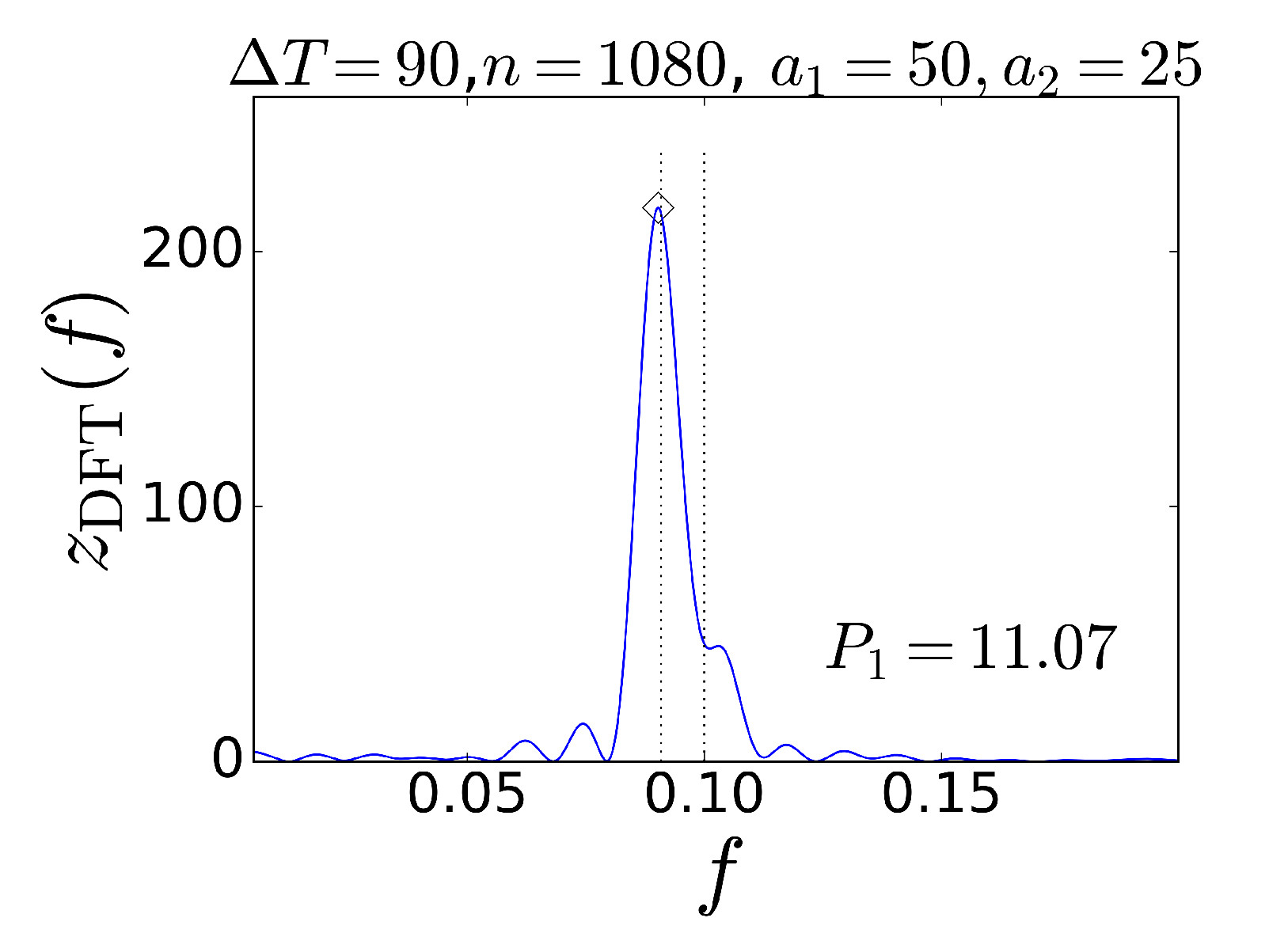} 
 \hspace*{-0.01\textwidth}
 \includegraphics[width=0.25\textwidth,clip=]{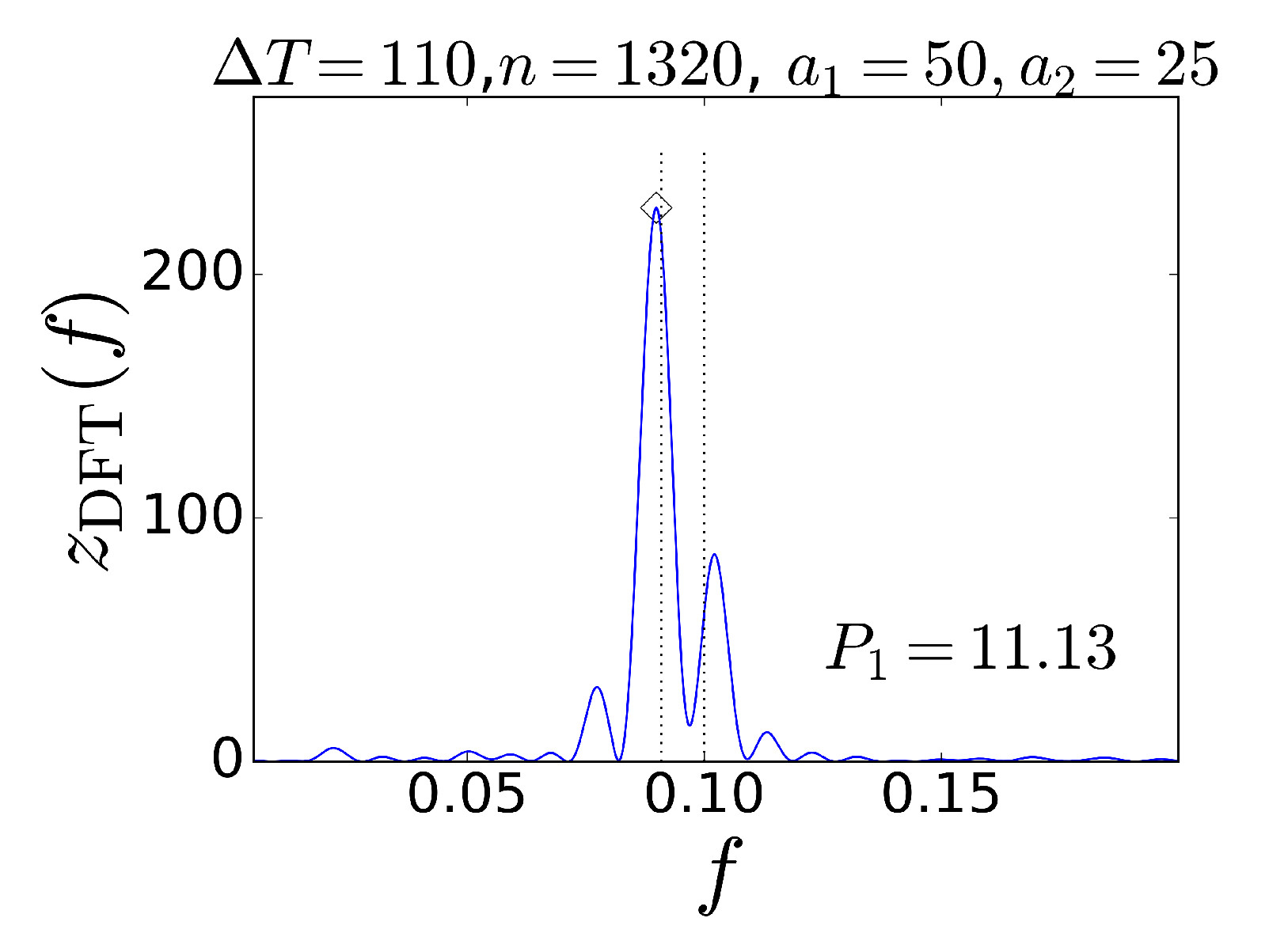} 
 \hspace*{-0.01\textwidth}
 \includegraphics[width=0.25\textwidth,clip=]{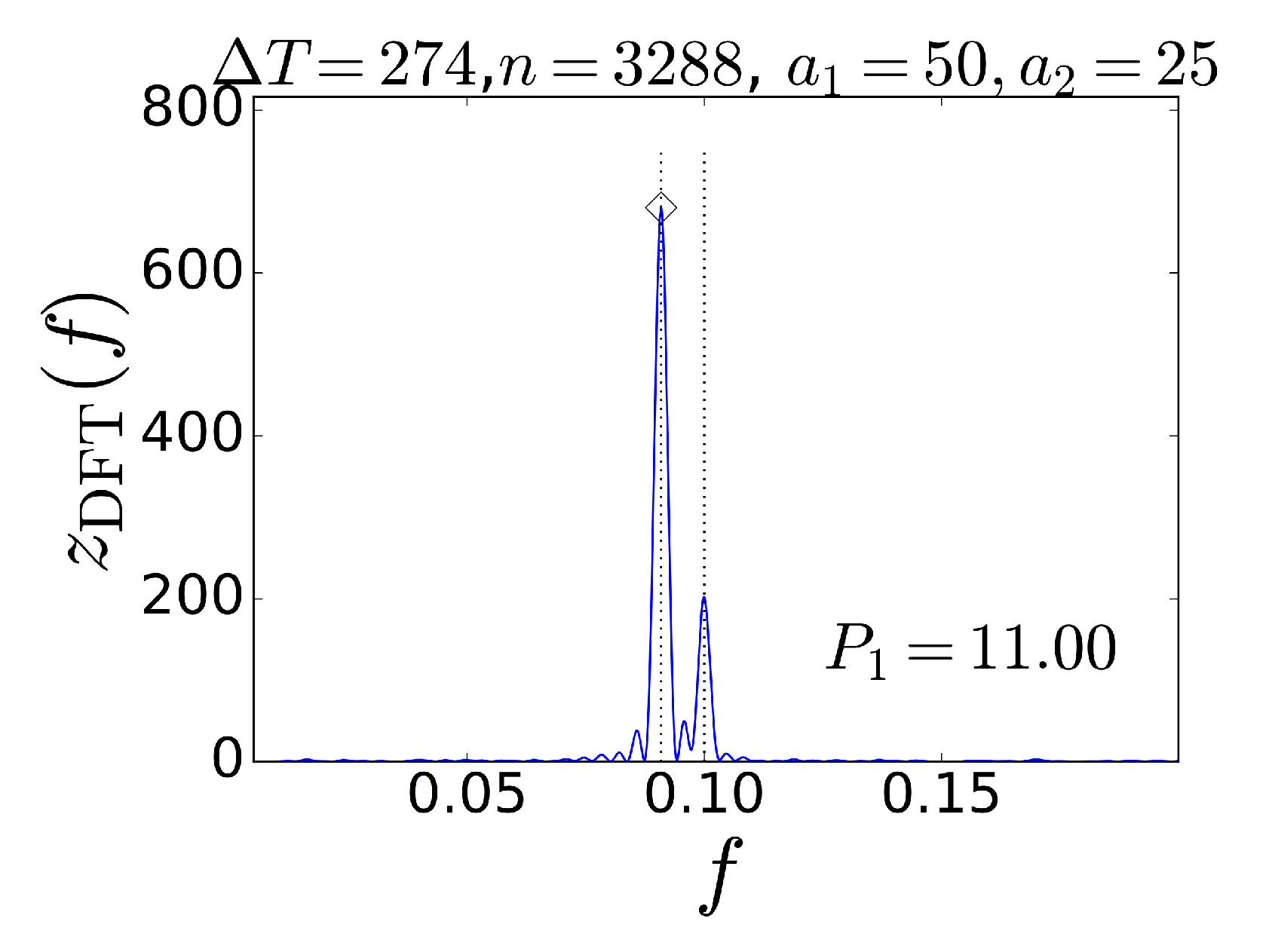} 
 }
\vspace{-0.22\textwidth}
\centerline{\normalsize \bf 
\hspace{0.11\textwidth}   \color{black}{(a)}
\hspace{0.23\textwidth}  \color{black}{(g)}
\hspace{0.22\textwidth}  \color{black}{(m)}
\hspace{0.22\textwidth}  \color{black}{(s)}
\hfill}
\vspace{0.21\textwidth}
\centerline{\hspace*{0.005\textwidth}
\includegraphics[width=0.25\textwidth,clip=]{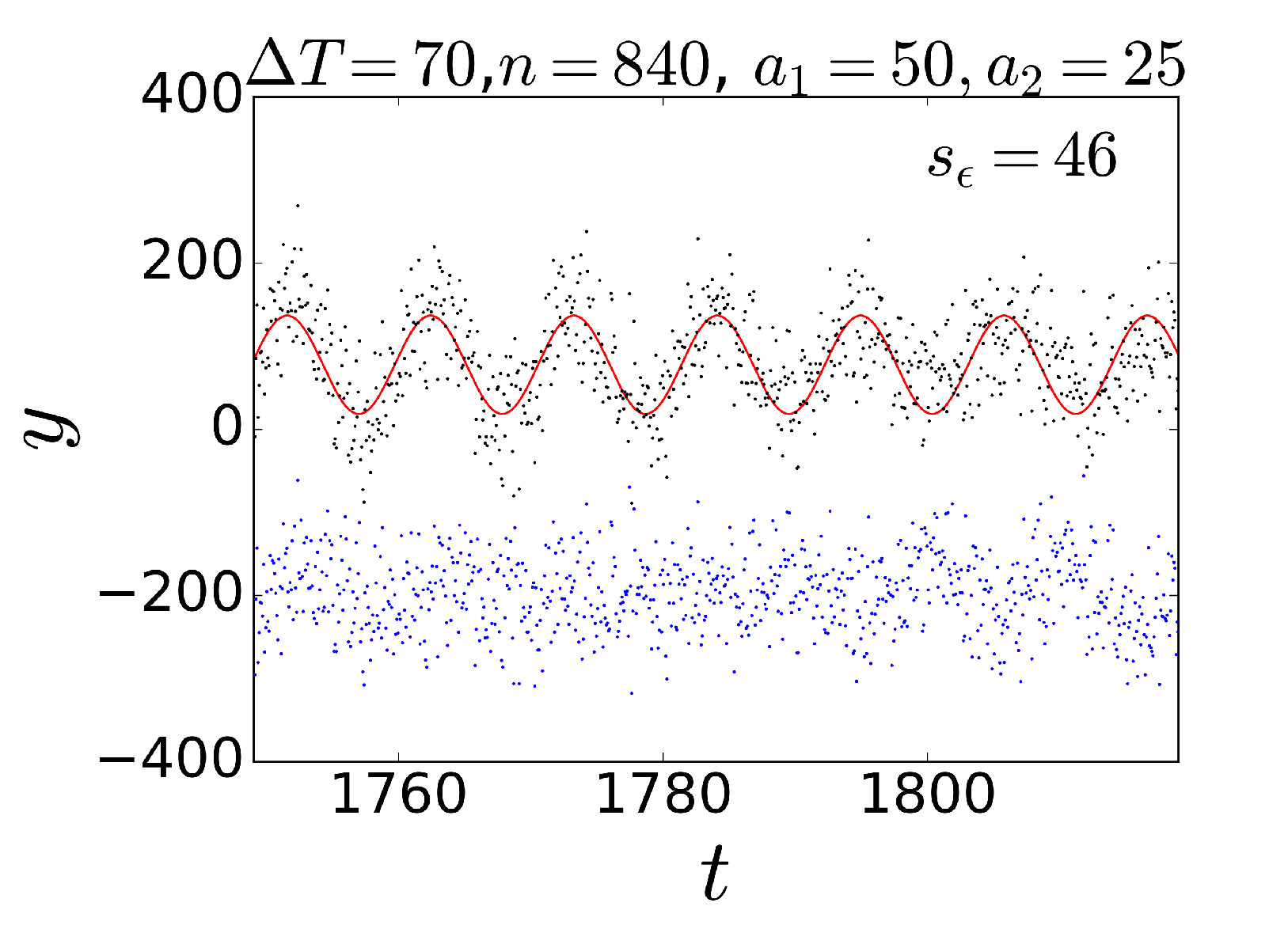}  
 \hspace*{-0.01\textwidth}
 \includegraphics[width=0.25\textwidth,clip=]{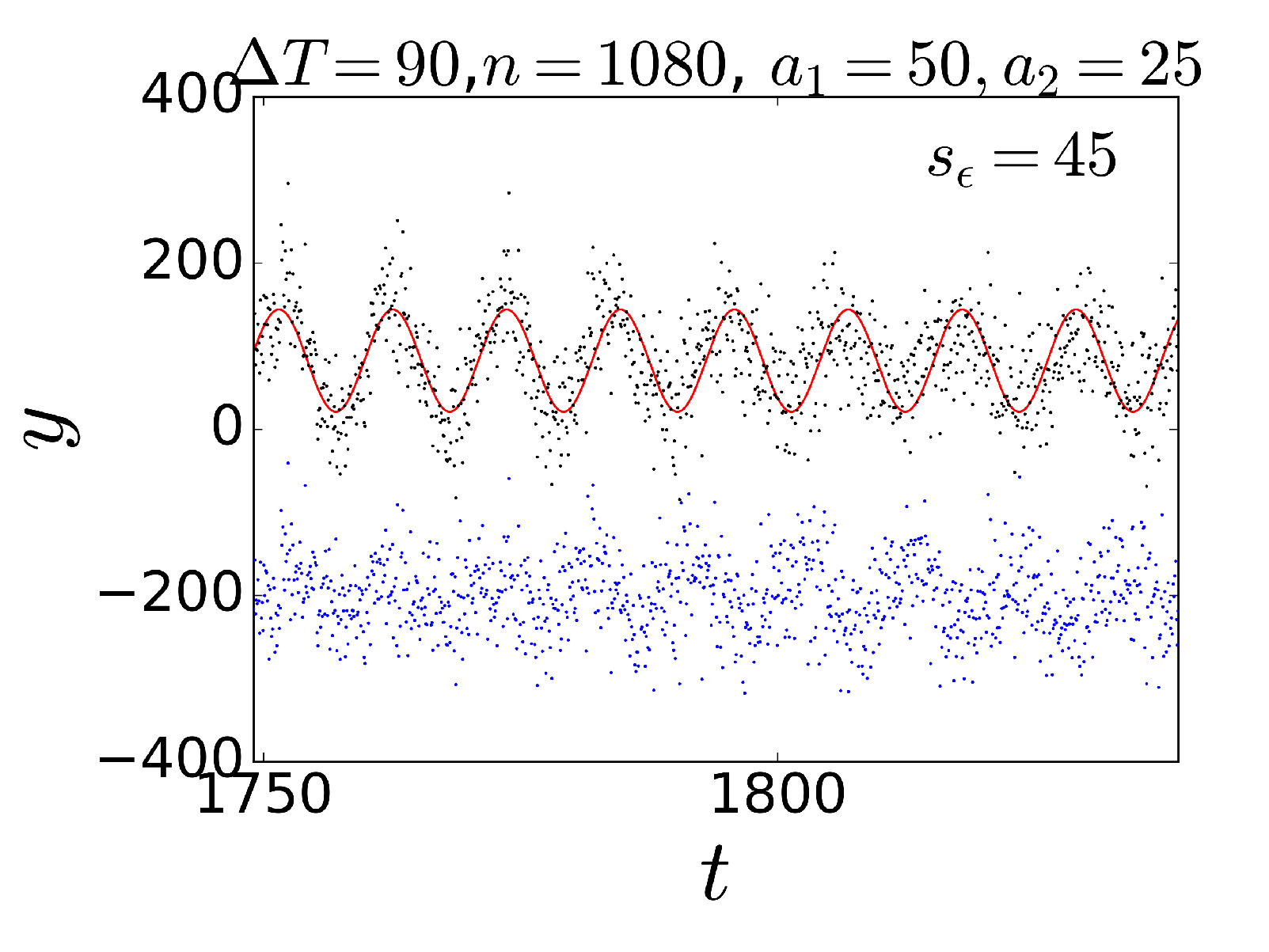} 
 \hspace*{-0.01\textwidth}
 \includegraphics[width=0.25\textwidth,clip=]{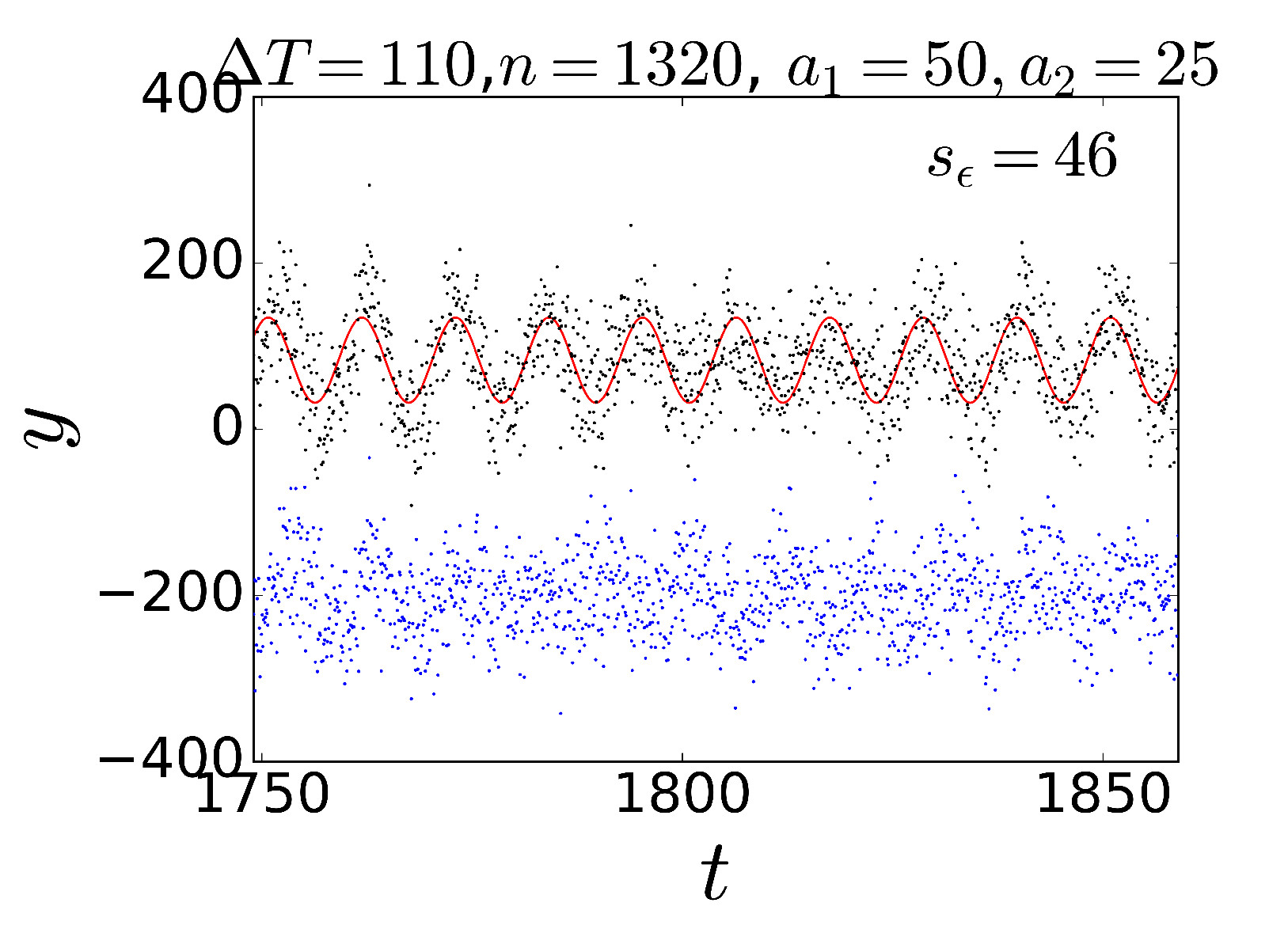} 
 \hspace*{-0.01\textwidth}
 \includegraphics[width=0.25\textwidth,clip=]{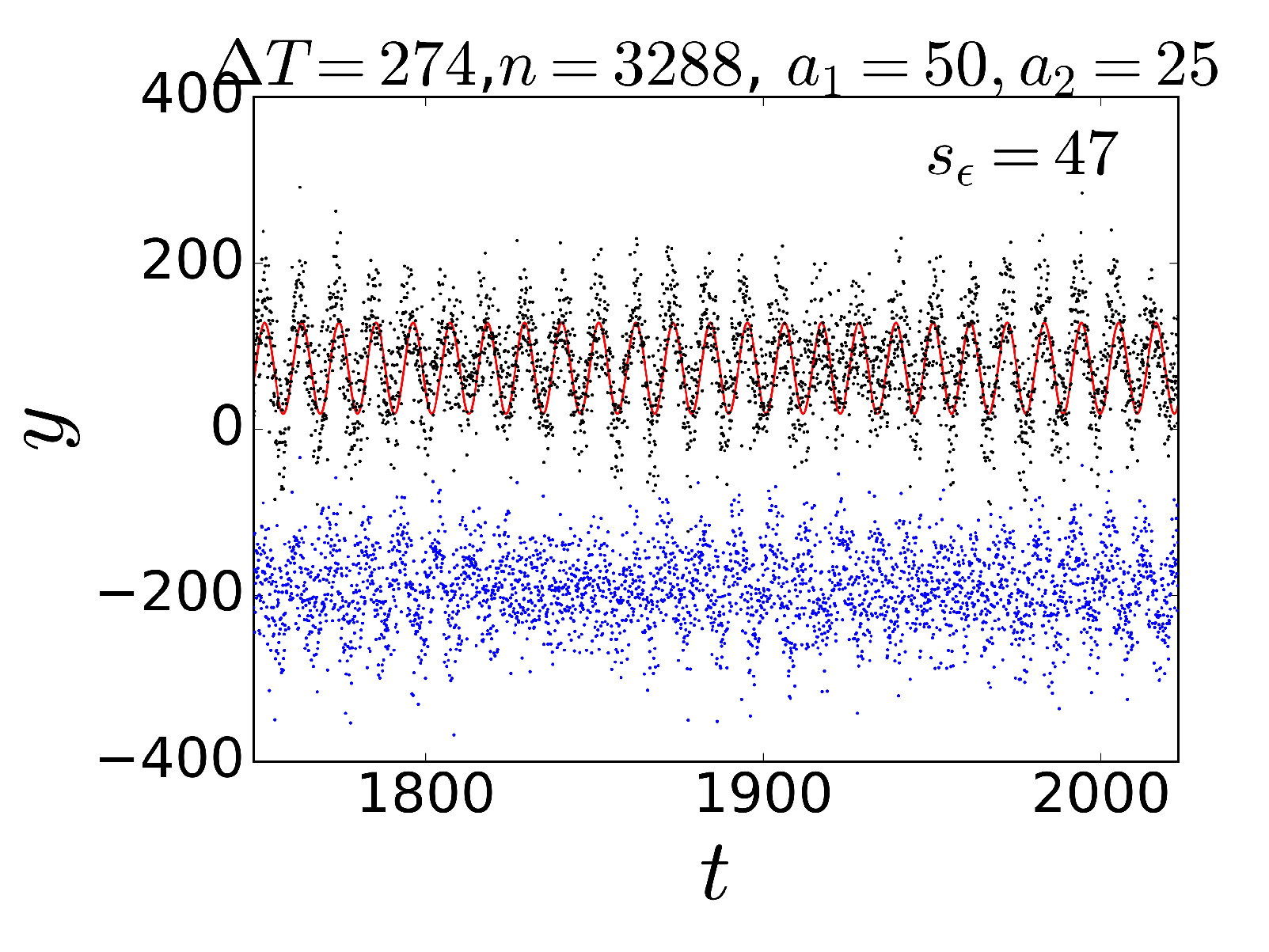} 
  }
\vspace{-0.22\textwidth}
\centerline{\normalsize \bf 
\hspace{0.11\textwidth}   \color{black}{(b)}
\hspace{0.23\textwidth}  \color{black}{(h)}
\hspace{0.22\textwidth}  \color{black}{(n)}
\hspace{0.22\textwidth}  \color{black}{(t)}
  \hfill}
\vspace{0.21\textwidth}
\centerline{\hspace*{0.005\textwidth}
 \includegraphics[width=0.25\textwidth,clip=]{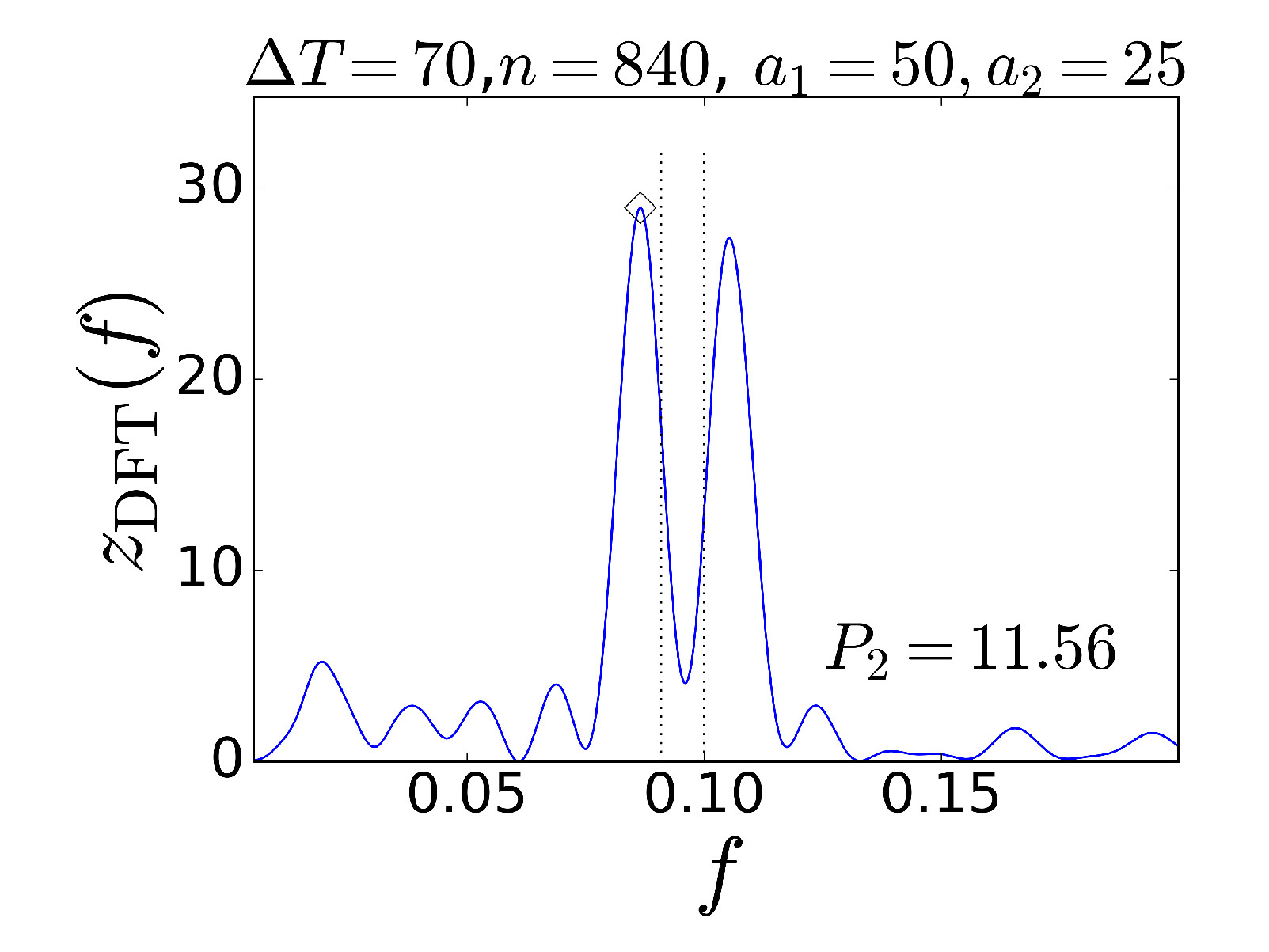} 
 \hspace*{-0.01\textwidth}
 \includegraphics[width=0.25\textwidth,clip=]{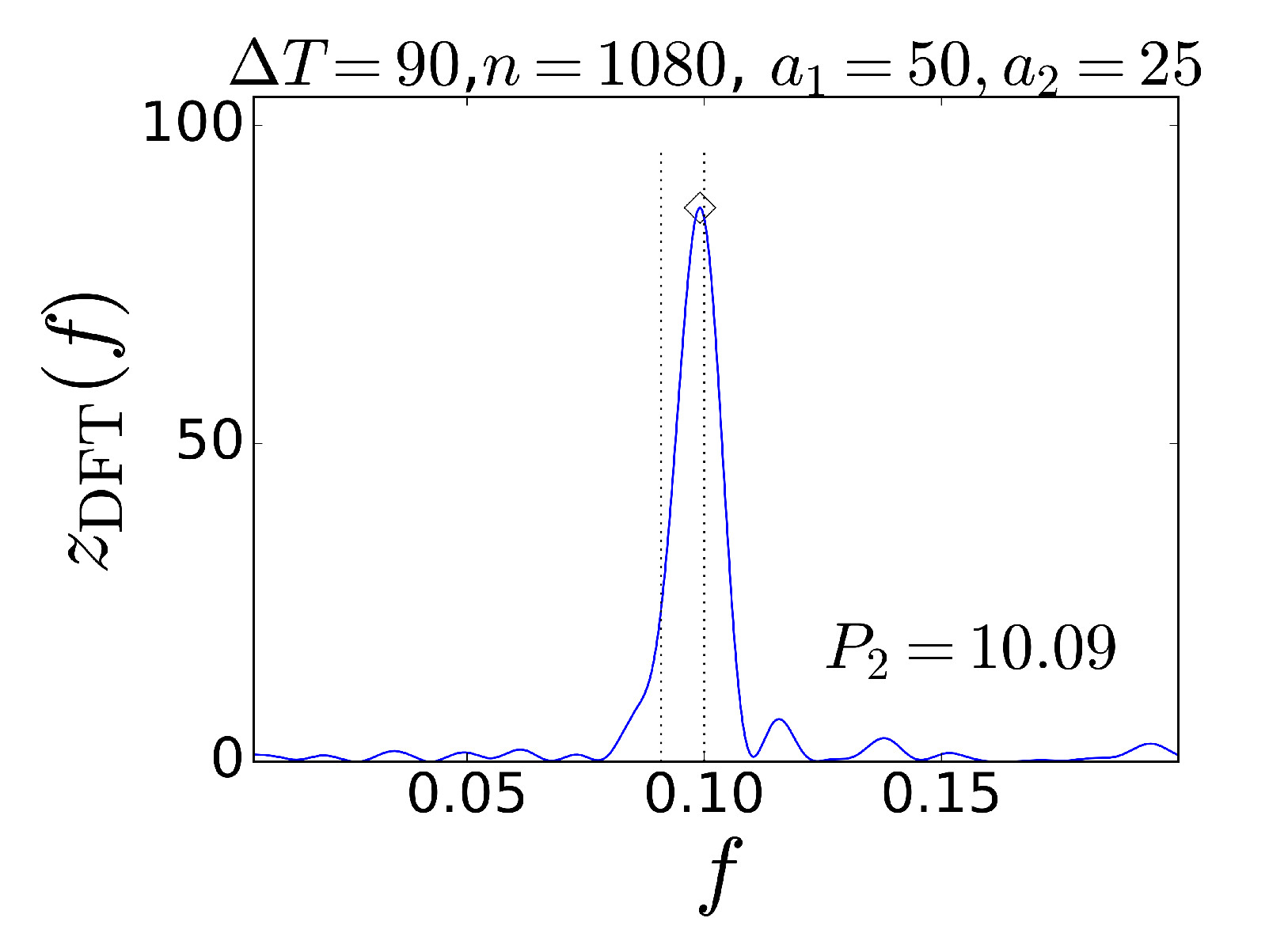} 
 \hspace*{-0.01\textwidth}
 \includegraphics[width=0.25\textwidth,clip=]{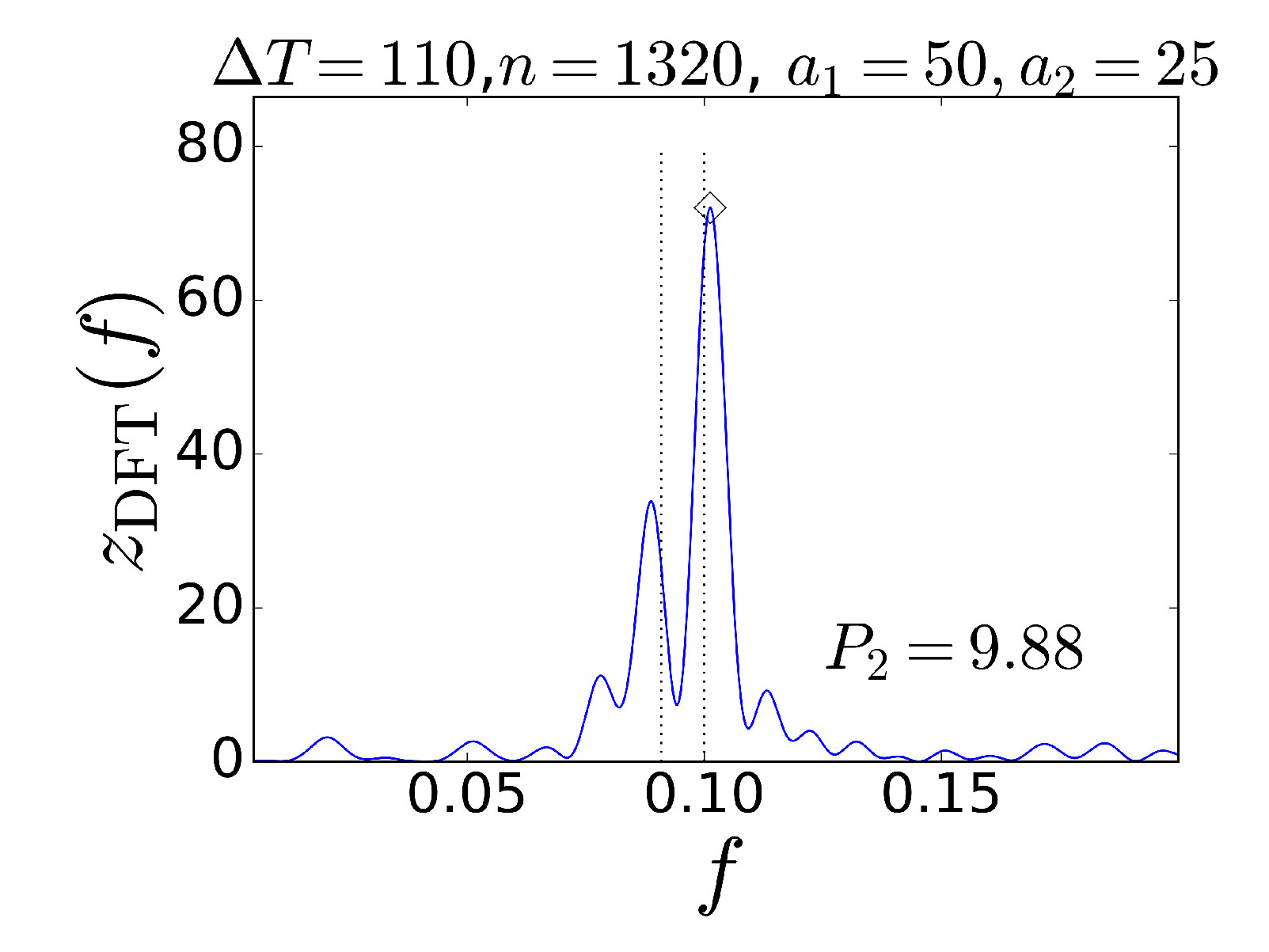} 
 \hspace*{-0.01\textwidth}
 \includegraphics[width=0.25\textwidth,clip=]{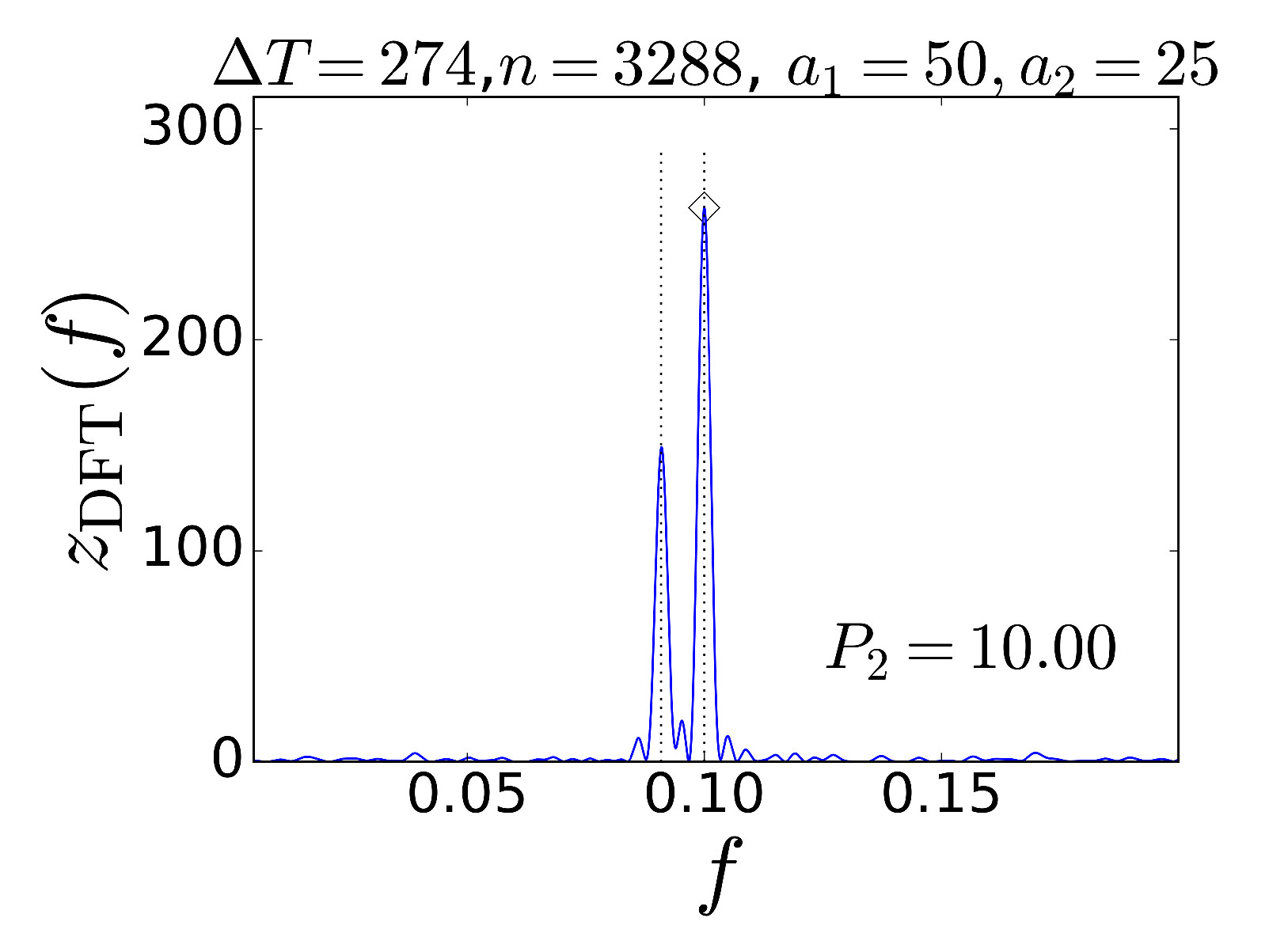} 
}
\vspace{-0.22\textwidth}
\centerline{\normalsize \bf 
\hspace{0.11\textwidth}   \color{black}{(c)}
\hspace{0.23\textwidth}  \color{black}{(i)}
\hspace{0.22\textwidth}  \color{black}{(o)}
\hspace{0.22\textwidth}  \color{black}{(u)}
  \hfill}
\vspace{0.21\textwidth}
\centerline{\hspace*{0.005\textwidth}
 \includegraphics[width=0.25\textwidth,clip=]{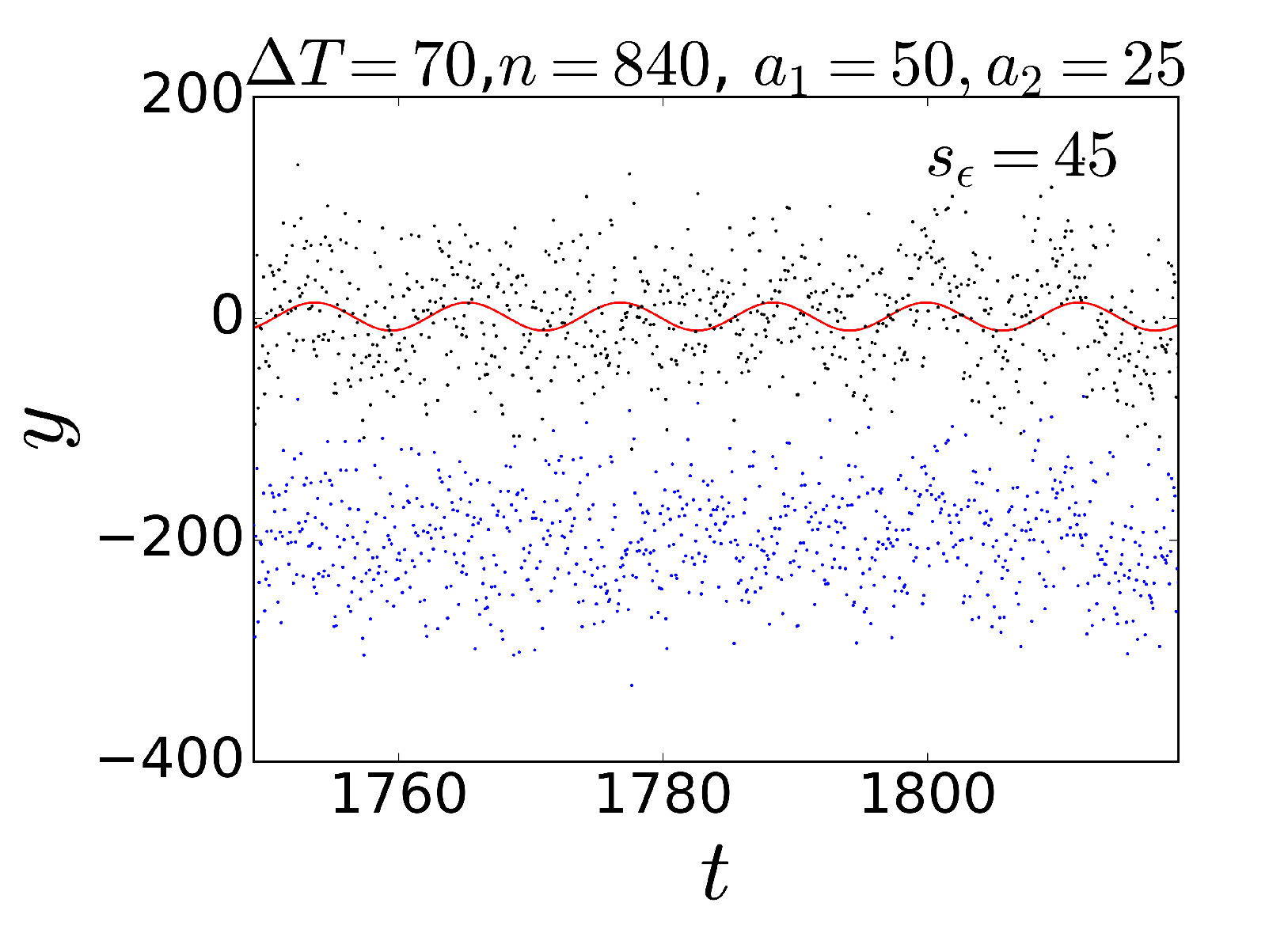}  
 \hspace*{-0.01\textwidth}
 \includegraphics[width=0.25\textwidth,clip=]{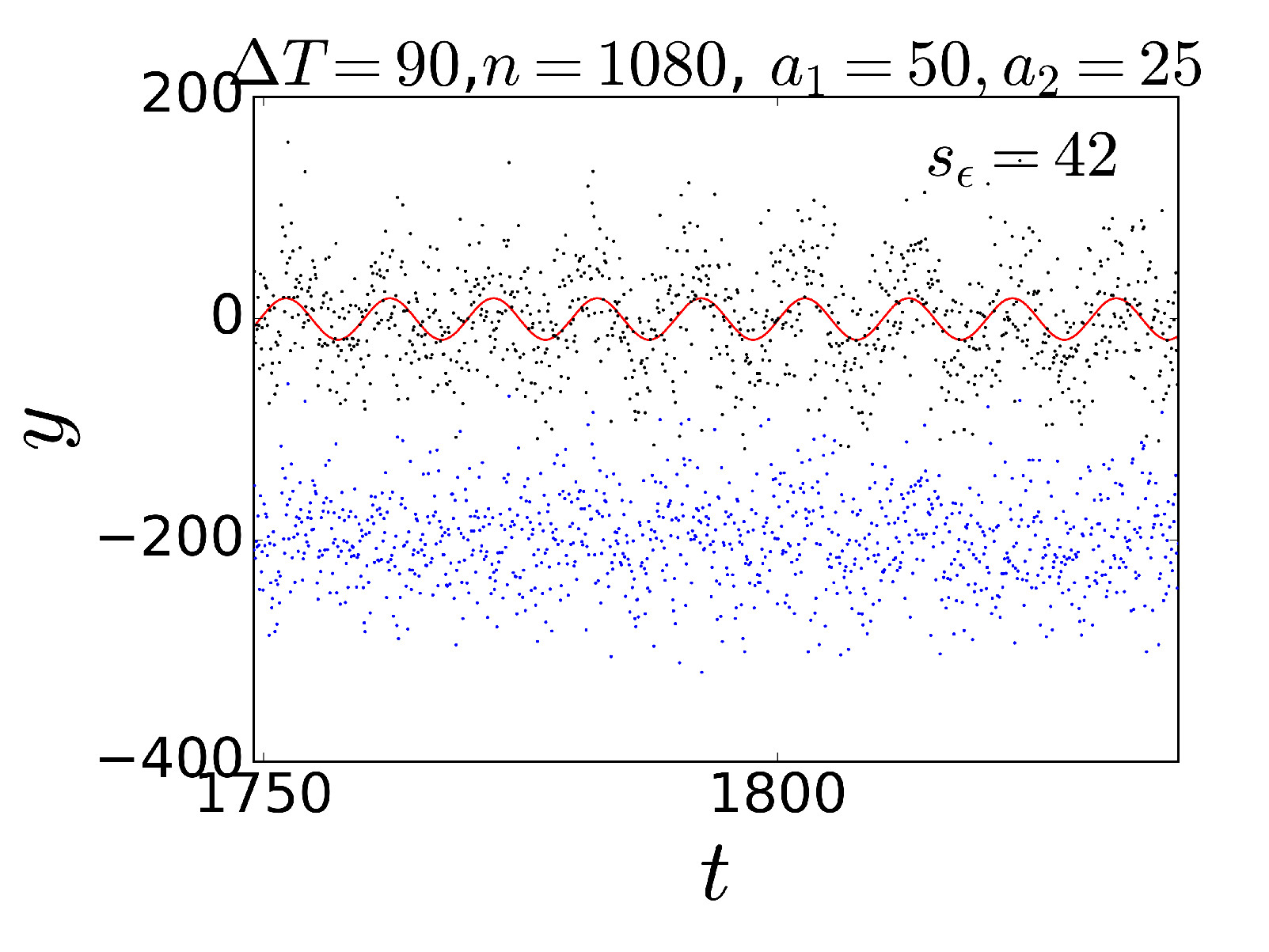}   
 \hspace*{-0.01\textwidth}
 \includegraphics[width=0.25\textwidth,clip=]{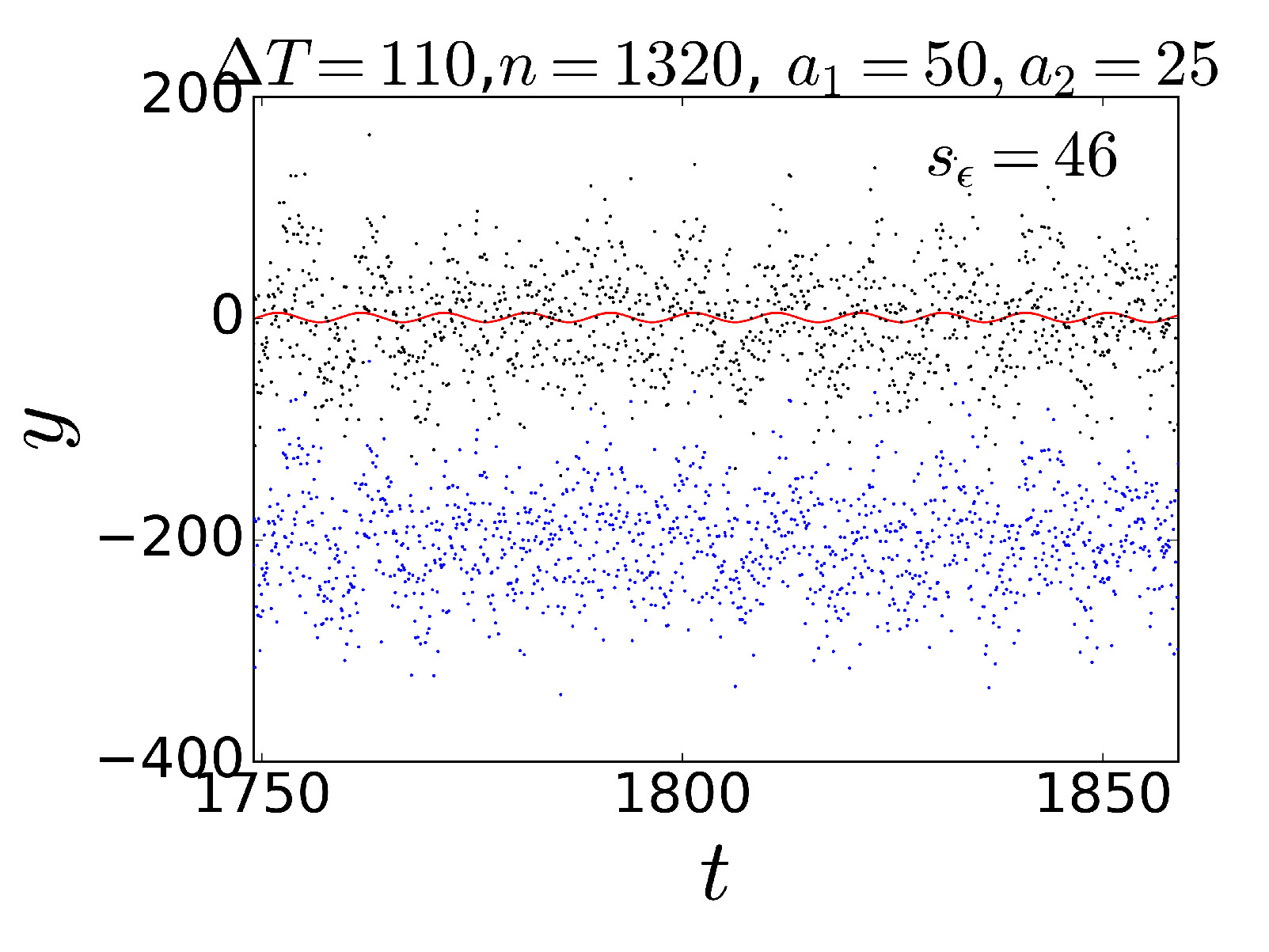} 
 \hspace*{-0.01\textwidth}
 \includegraphics[width=0.25\textwidth,clip=]{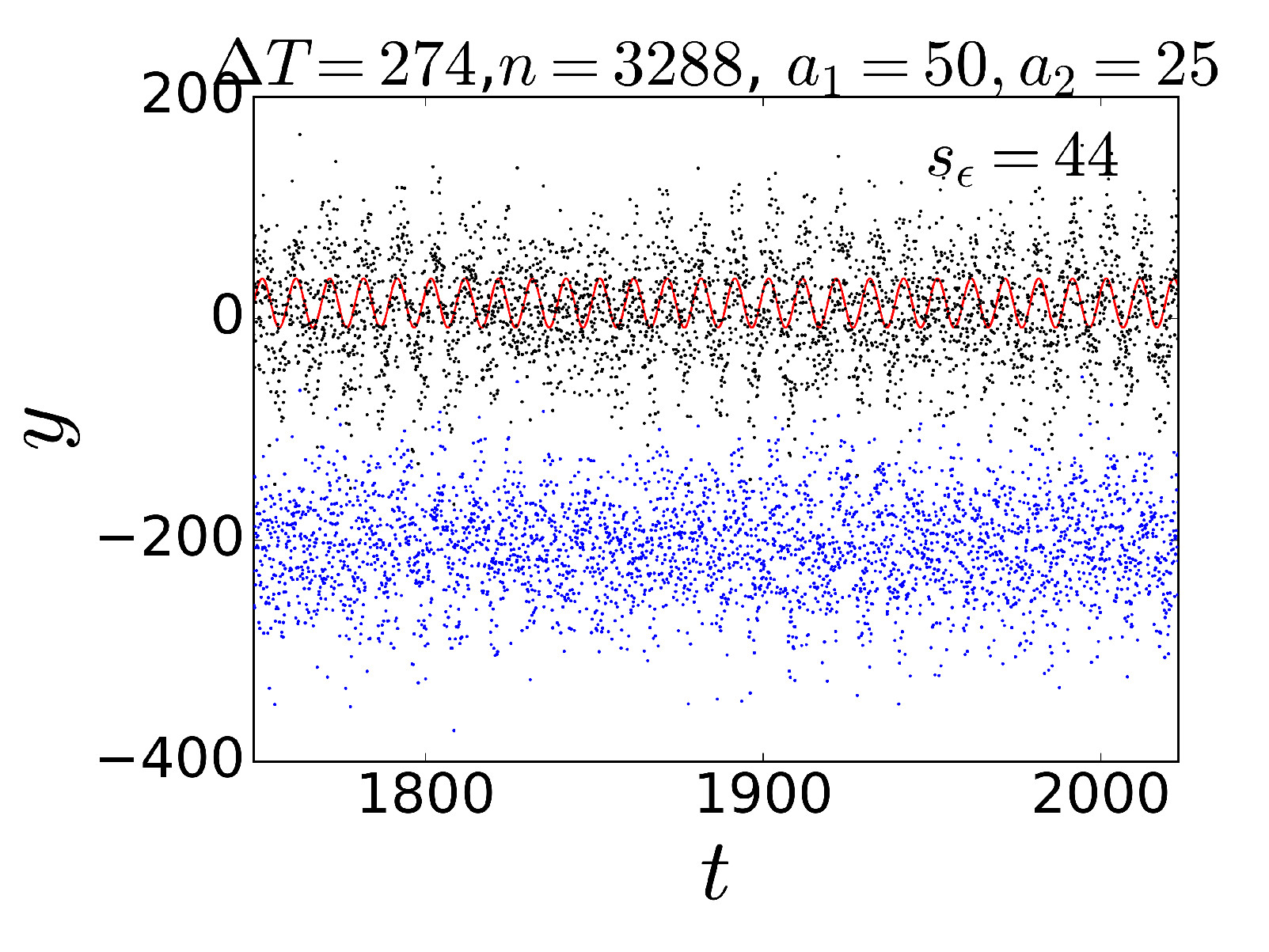} 
}
\vspace{-0.22\textwidth}
\centerline{\normalsize \bf 
\hspace{0.11\textwidth}   \color{black}{(d)}
\hspace{0.23\textwidth}  \color{black}{(j)}
\hspace{0.22\textwidth}  \color{black}{(p)}
\hspace{0.22\textwidth}  \color{black}{(v)}
  \hfill}
\vspace{0.21\textwidth}
\centerline{\hspace*{0.005\textwidth}
 \includegraphics[width=0.25\textwidth,clip=]{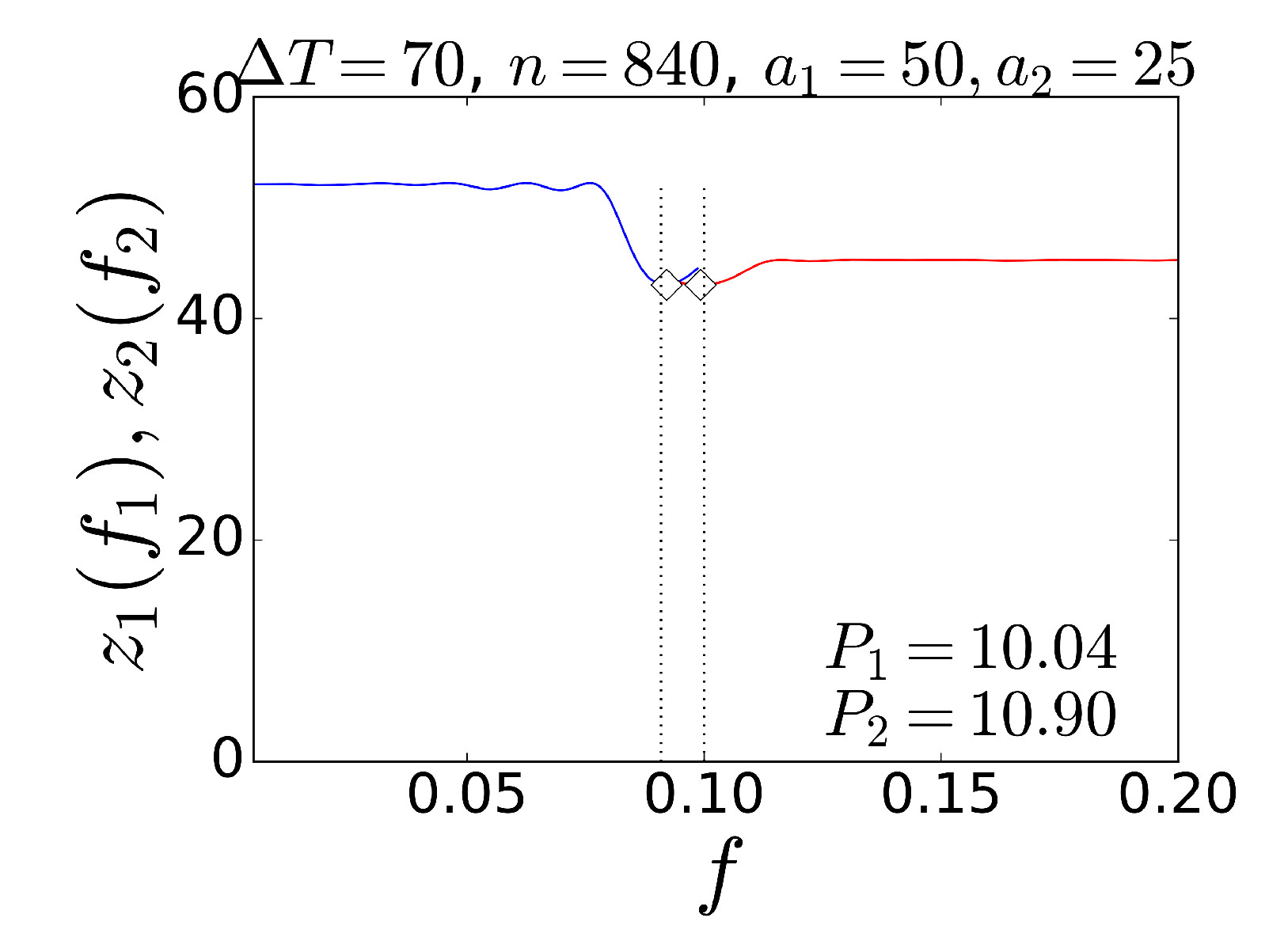}  
 \hspace*{-0.01\textwidth}
 \includegraphics[width=0.25\textwidth,clip=]{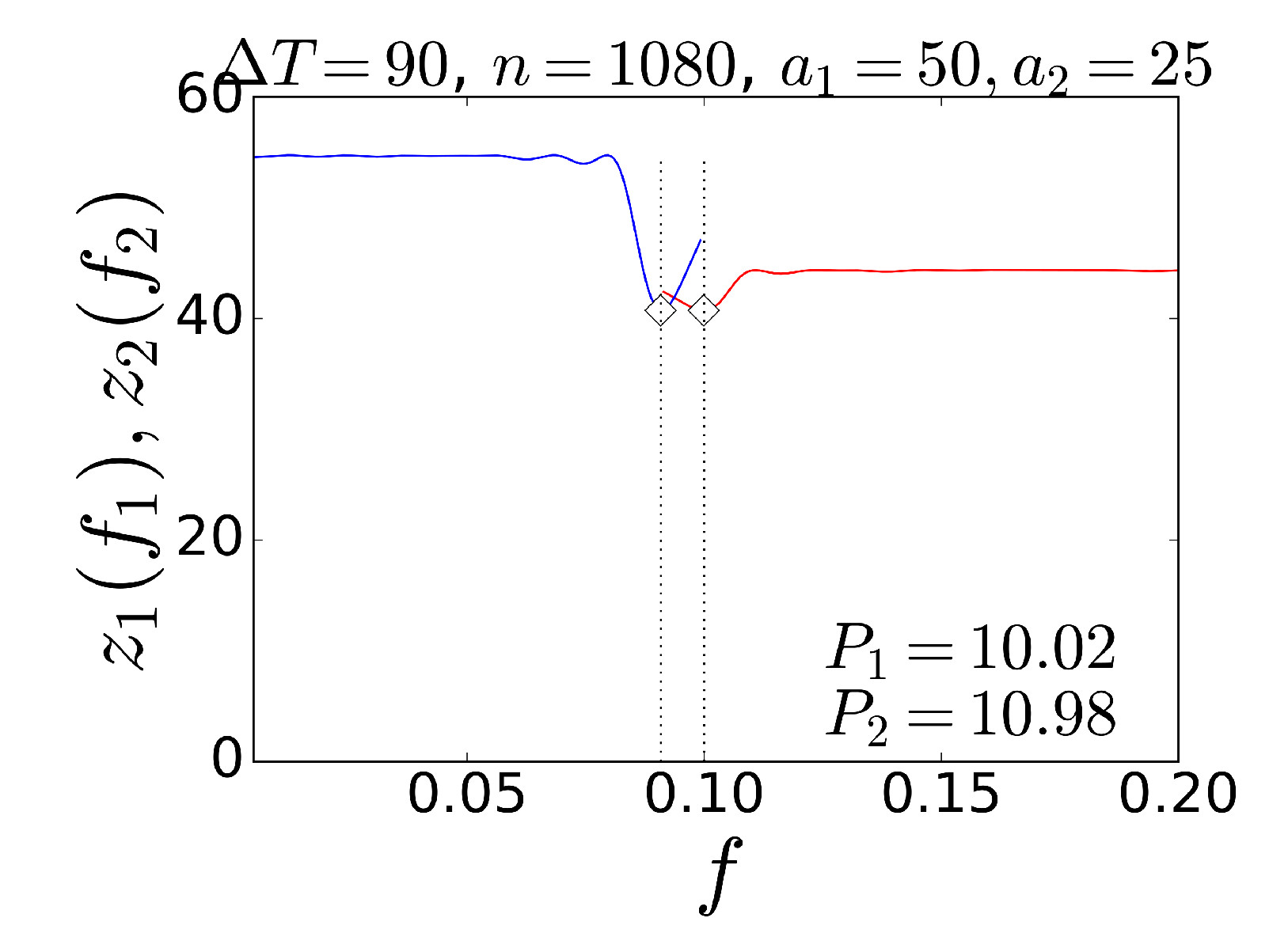}   
 \hspace*{-0.01\textwidth}
 \includegraphics[width=0.25\textwidth,clip=]{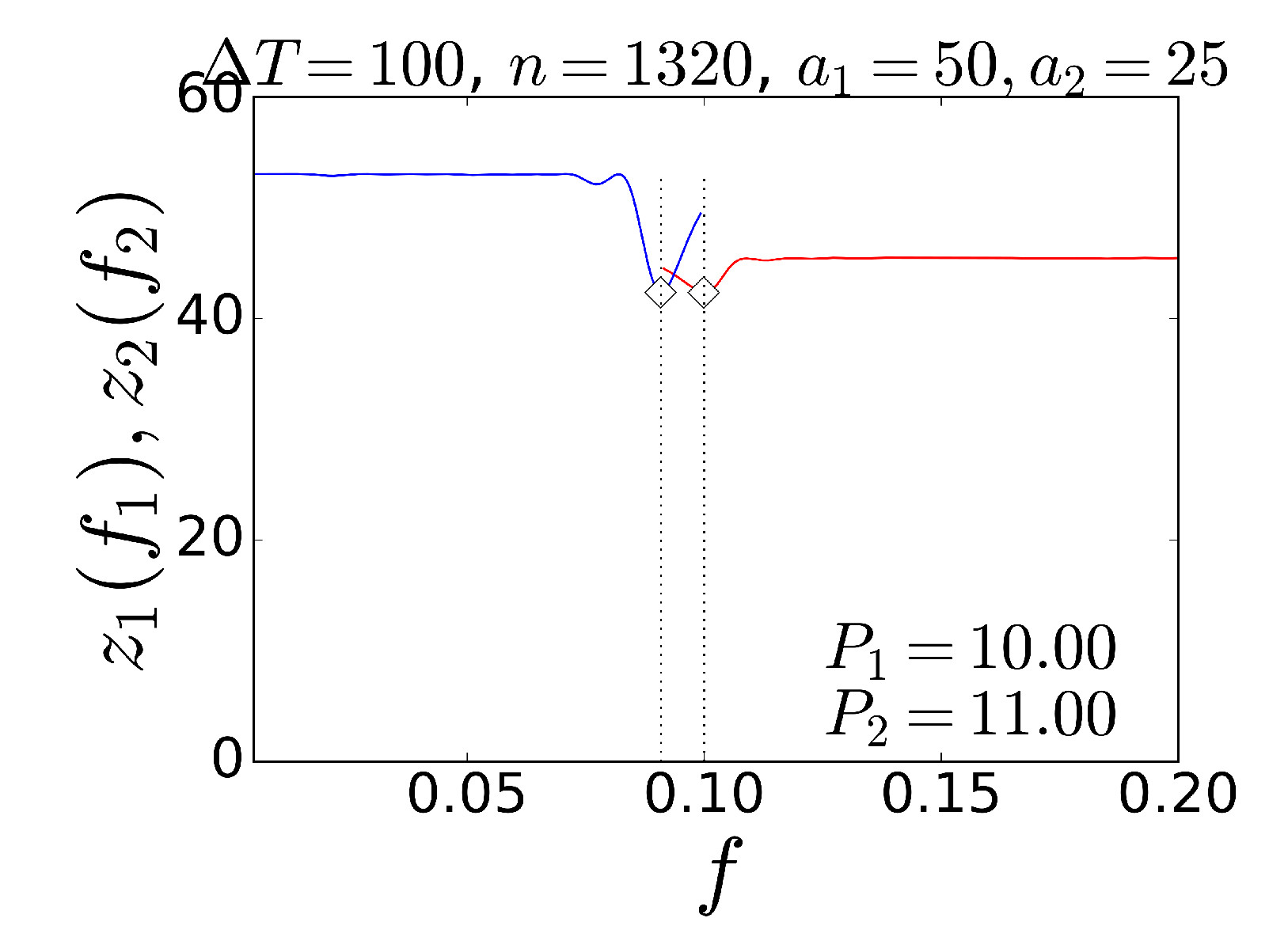} 
 \hspace*{-0.01\textwidth}
 \includegraphics[width=0.25\textwidth,clip=]{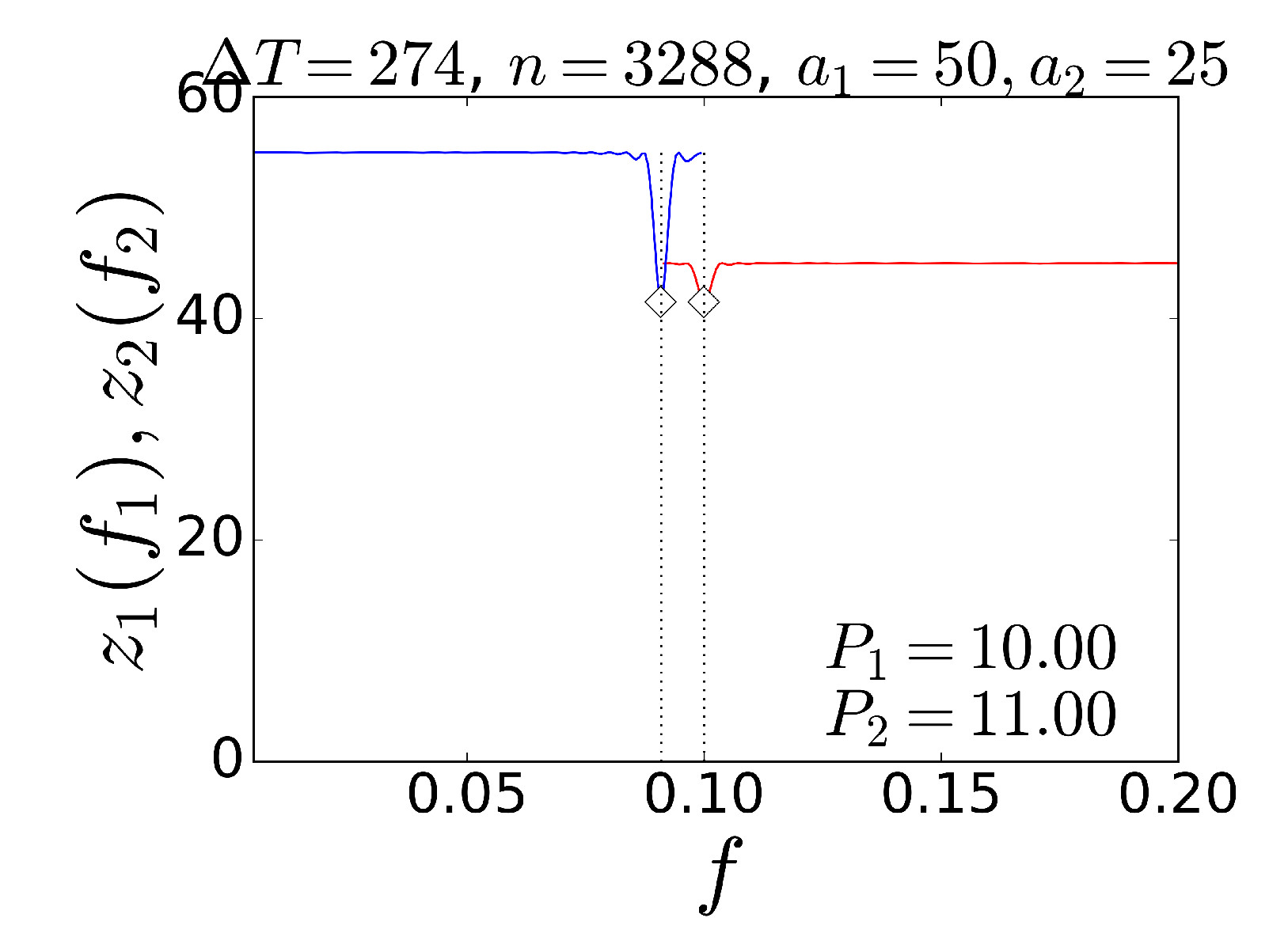} 
 \hspace*{-0.01\textwidth}
}
\vspace{-0.22\textwidth}
\centerline{\normalsize \bf 
\hspace{0.11\textwidth}   \color{black}{(e)}
\hspace{0.23\textwidth}  \color{black}{(k)}
\hspace{0.22\textwidth}  \color{black}{(q)}
\hspace{0.22\textwidth}  \color{black}{(w)}
  \hfill}
\vspace{0.21\textwidth}
\centerline{\hspace*{0.005\textwidth}
 \includegraphics[width=0.25\textwidth,clip=]{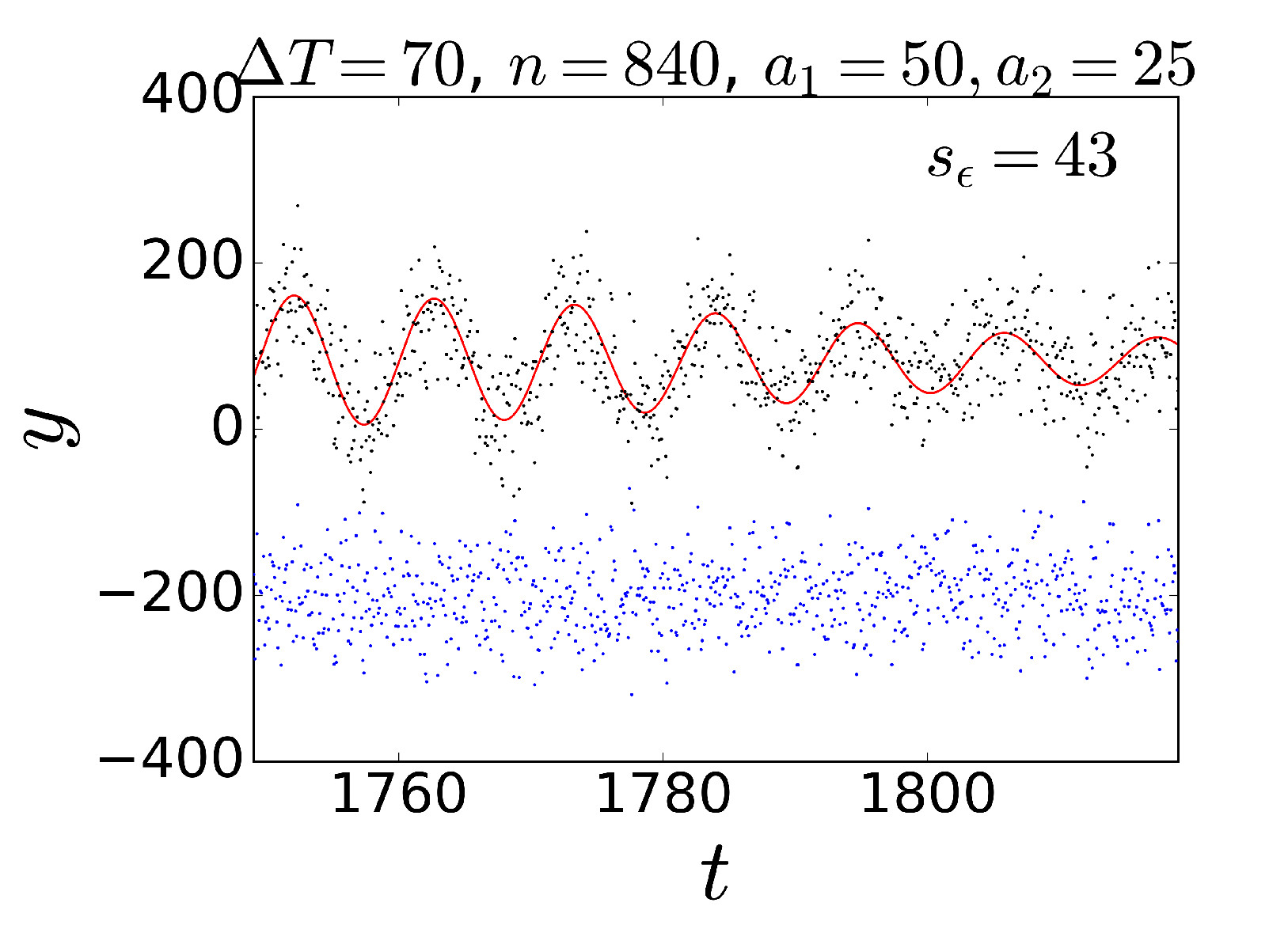}   
 \hspace*{-0.01\textwidth}
 \includegraphics[width=0.25\textwidth,clip=]{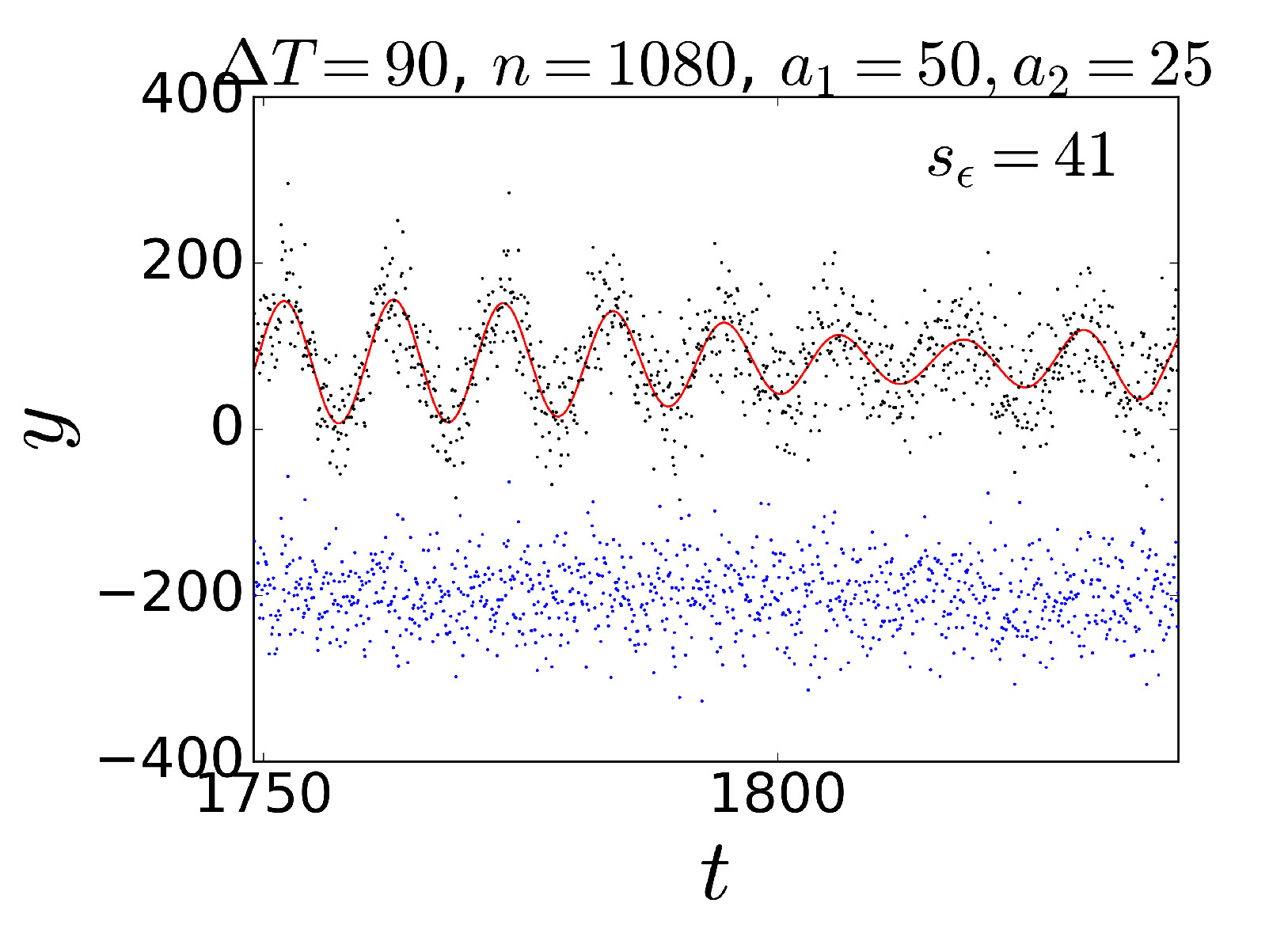}   
 \hspace*{-0.01\textwidth}
 \includegraphics[width=0.25\textwidth,clip=]{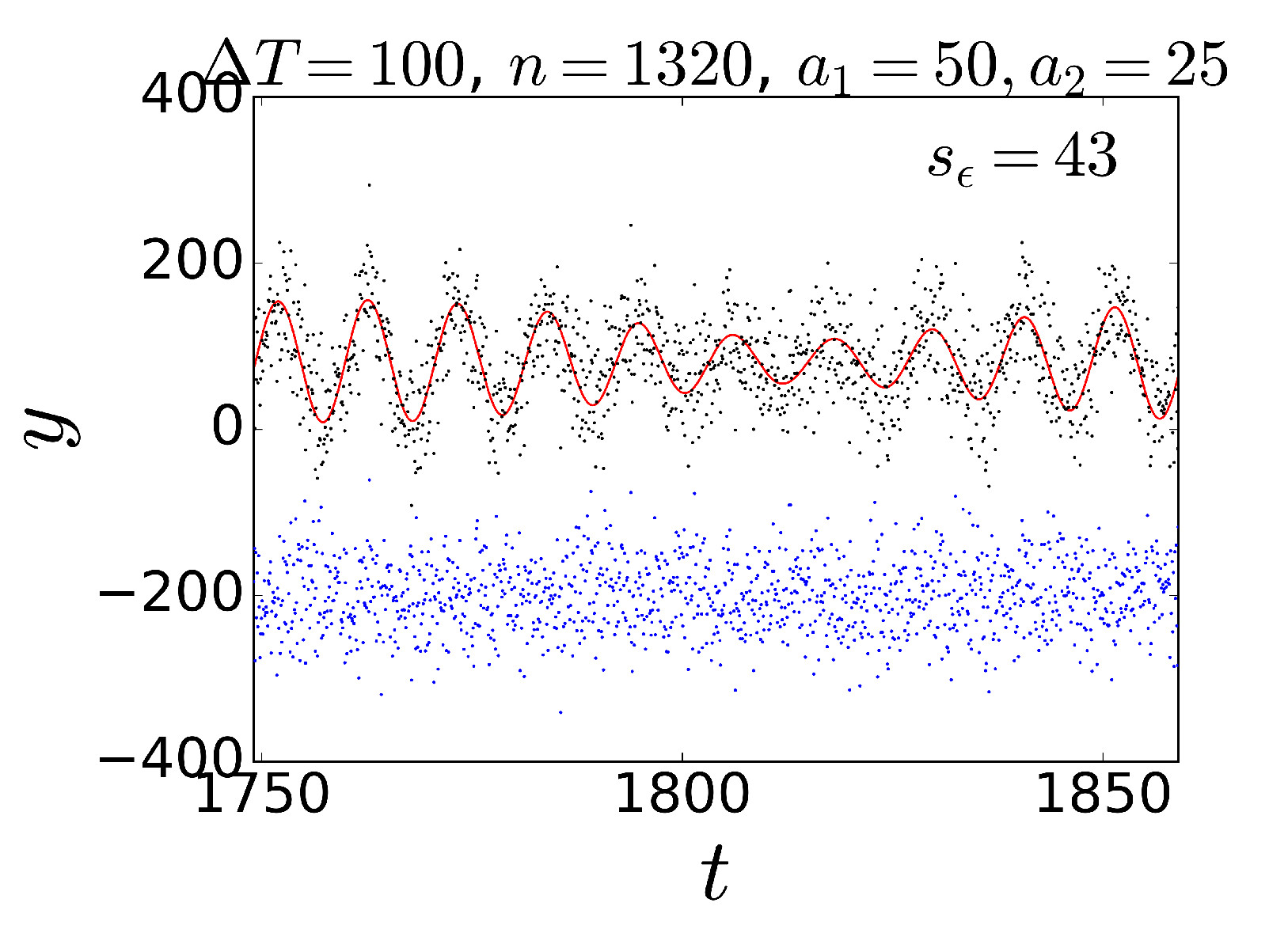} 
 \hspace*{-0.01\textwidth}
 \includegraphics[width=0.25\textwidth,clip=]{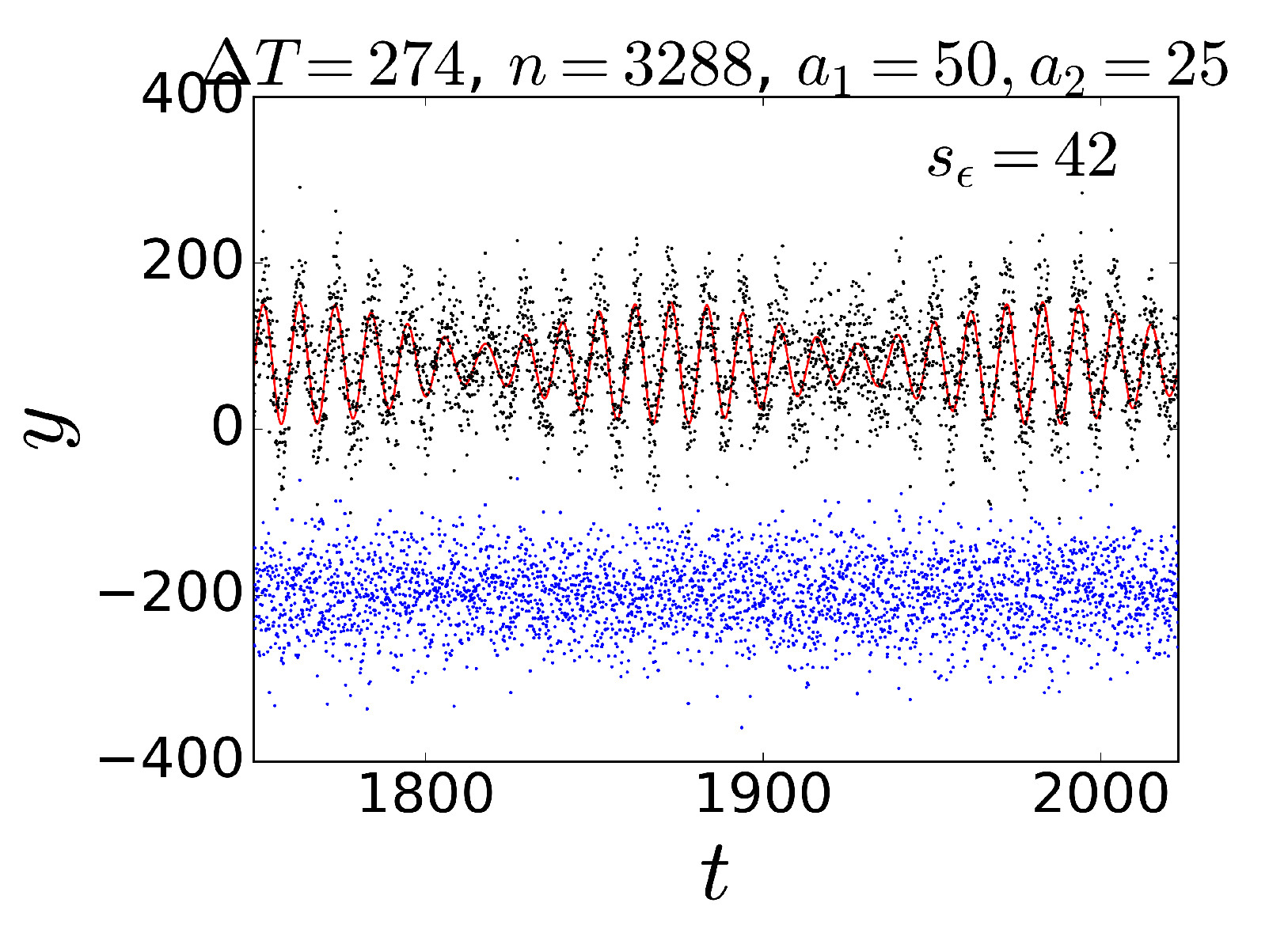} 
        }
\vspace{-0.22\textwidth}
\centerline{\normalsize \bf 
\hspace{0.11\textwidth}   \color{black}{(f)}
\hspace{0.23\textwidth}  \color{black}{(l)}
\hspace{0.22\textwidth}  \color{black}{(r)}
\hspace{0.22\textwidth}  \color{black}{(x)}
  \hfill}
\vspace{0.18\textwidth}
\caption{Unequal amplitude $a_1=25$ and $a_2=50$ simulations
  for signal periods $P_1=10$ and $P_2=11$
    (Equations \ref{EqSignalOne} and \ref{EqSignalTwo}).
  Otherwise as in Fig. \ref{FigDFTDCMOne}. }
\label{FigDFTDCMTwo}
\end{figure}

\subsection{Results for
  two unequal amplitude signals
  \label{SectDFTUnequal} } 

We present the results  for
unequal amplitudes $a_1=25$ and $a_2=50$
in Fig. \ref{FigDFTDCMTwo}.

All details are not discussed,
as in the case of Fig. \ref{FigDFTDCMOne}.
We concentrate only on the general results
in Fig. \ref{FigDFTDCMTwo},
because they are the same for all
$\Delta T=70$, 90, 110 and 274 simulations.

If the relation of Equation \ref{EqPrevent} is not fulfilled,
DFT can not detect the correct period values,
but DCM can. If this relation is fulfilled,
DFT can also succeed.

The relation of Equation \ref{EqMisleadTwo} predicts
that the frequency $f_2=1/P_2=1/11$ of the stronger
signal dominates in the frequency $f(t)$ of the $s(t)$.
Therefore, DFT always detects this
$P_2=11$ period signal first.
DCM detects both signals simultaneously,
neither one of them being ``the first''.

DFT can not model the amplitude changes or
  the abrupt phase shifts  (Equation \ref{EqAbrupt}).
For this reason,
the DFT model residuals always show periodic
variability, while the DCM model residuals do not.
DCM residuals have a random distribution concentrated
on a straight line, which is a signature of a good model.
DFT model $s_{\epsilon}$ values are always
larger than DCM model $s_{\epsilon}$ values.
This means that some information is always lost at every DFT
pre-whitening stage, where one pure sine signal is detected
at the time.
Both methods reach $s_{\epsilon}$ levels that
are close to the standard deviation $s=42$ of the
random errors $\sigma_i^{\star}$,
which are added as noise
to the simulated data (Equation \ref{EqSimData}).
This means that both methods can also detect
the second, weaker $P_1=10$ signal, which has the peak
to peak amplitude $A_1=2a_1=50$ that has
the same order of magnitude as $s=42$.
However, only
DCM can reach this $\sigma_{\epsilon} \approx s=42$
level for
all $2\times 4=8$ simulations.

\subsection{Discrete Fourier Transform
  (DFT) analysis of \RmonthlyOne}
  \label{SectDFTResults}

We simulate data 
arising from
the interference between two sine waves
having frequencies $f_1 \approx f_2$
(Equation \ref{EqSimData}).
Our DFT and DCM analyses of these 
simulated
data
reveal the performance differences
between these two methods. 
We encounter the same five DFT
drawbacks for equal and unequal
amplitude sine
waves (Sections \ref{SectDFTEqual} and \ref{SectDFTUnequal}).

\begin{enumerate}

\item Detection of close frequencies:
  DFT suffers from the limitation of Equation \ref{EqPrevent},
  but DCM does not.
  
\item Wrong frequency detection:
  DFT suffers from the
  misleading effects of Equations \ref{EqMisleadOne}
  and \ref{EqMisleadTwo}, but DCM does not.

\item Abrupt phase shifts: DFT can not model the abrupt
  phase shifts of Equation \ref{EqAbrupt}, but DCM can.

\item Biased residuals: DFT pre-whitening model
  residuals do not have a random distribution, but DCM
  model residuals do. 
  The mean residual
  $s_{\epsilon}$  values of DFT models 
  are always larger those of
  DFT models,
  which  means that some information
  is lost at every DFT pre-whitening stage.

\item Detection accuracy: The difference between
  the detected and the simulated period values is
  larger for DFT than for DCM.
  This difference decreases as the samples become longer.
  For real data, this means that DCM detects the correct
  period values more probably than DFT.
  
\end{enumerate}

\noindent
When DFT analysis fails,
DCM analysis can succeed.
We can safely state that DCM performs better than DFT.
The two strongest \SignalOne ~and \SignalTwo
~signals in the real data sample \RmonthlyOne ~are
exactly the same as in our simulations.
When we use DFT pre-whitening
to search for more than two signals
in this real data sample \RmonthlyOne,
it is logical to assume that the above-mentioned five
DFT drawbacks become more pronounced.

\begin{figure}  
\vspace{0.02\textwidth}
\centerline{\hspace*{0.005\textwidth}
 \includegraphics[width=0.25\textwidth,clip=]{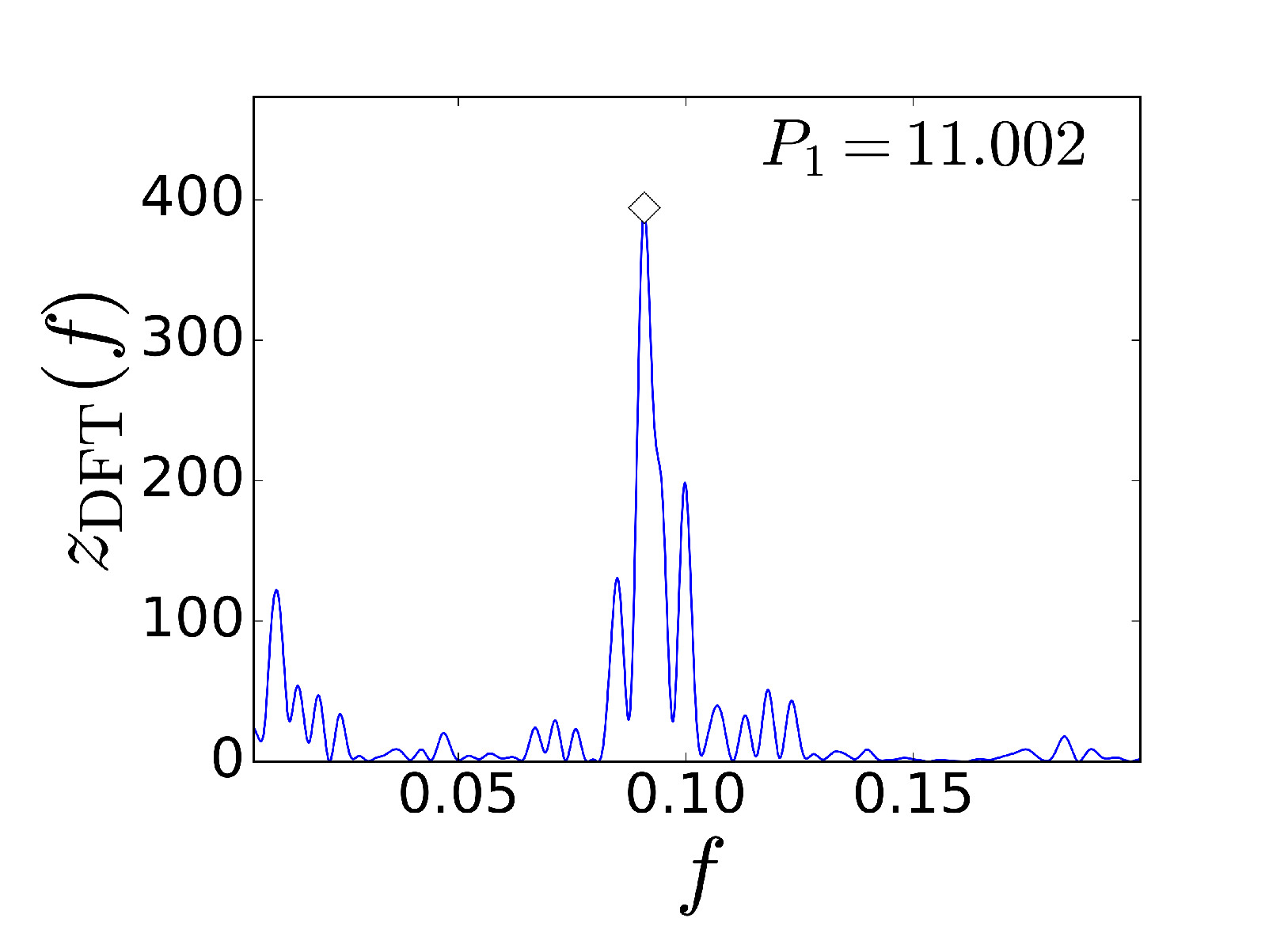} 
 \hspace*{-0.01\textwidth}
 \includegraphics[width=0.25\textwidth,clip=]{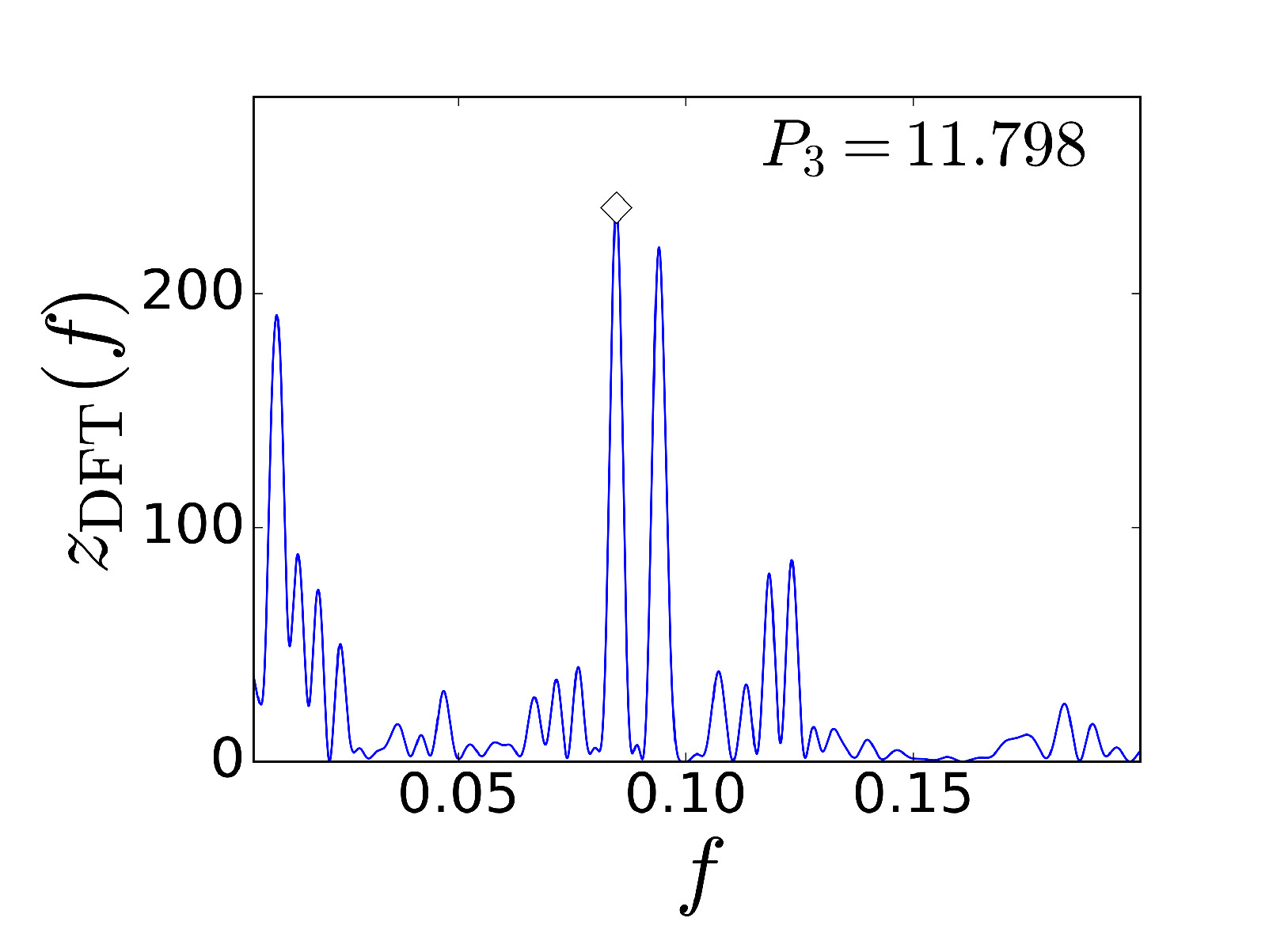} 
 \hspace*{-0.01\textwidth}
 \includegraphics[width=0.25\textwidth,clip=]{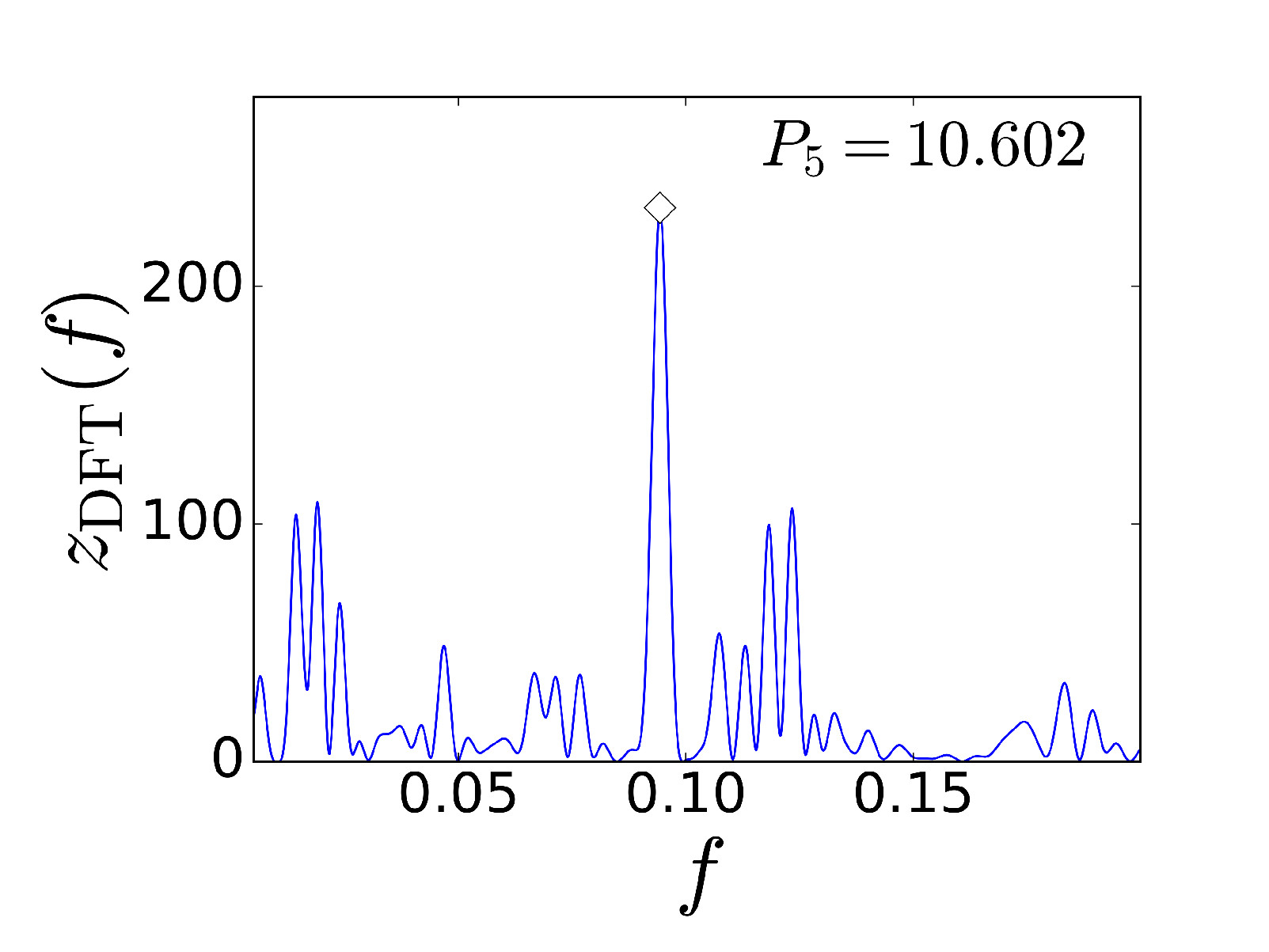} 
 \hspace*{-0.01\textwidth}
 \includegraphics[width=0.25\textwidth,clip=]{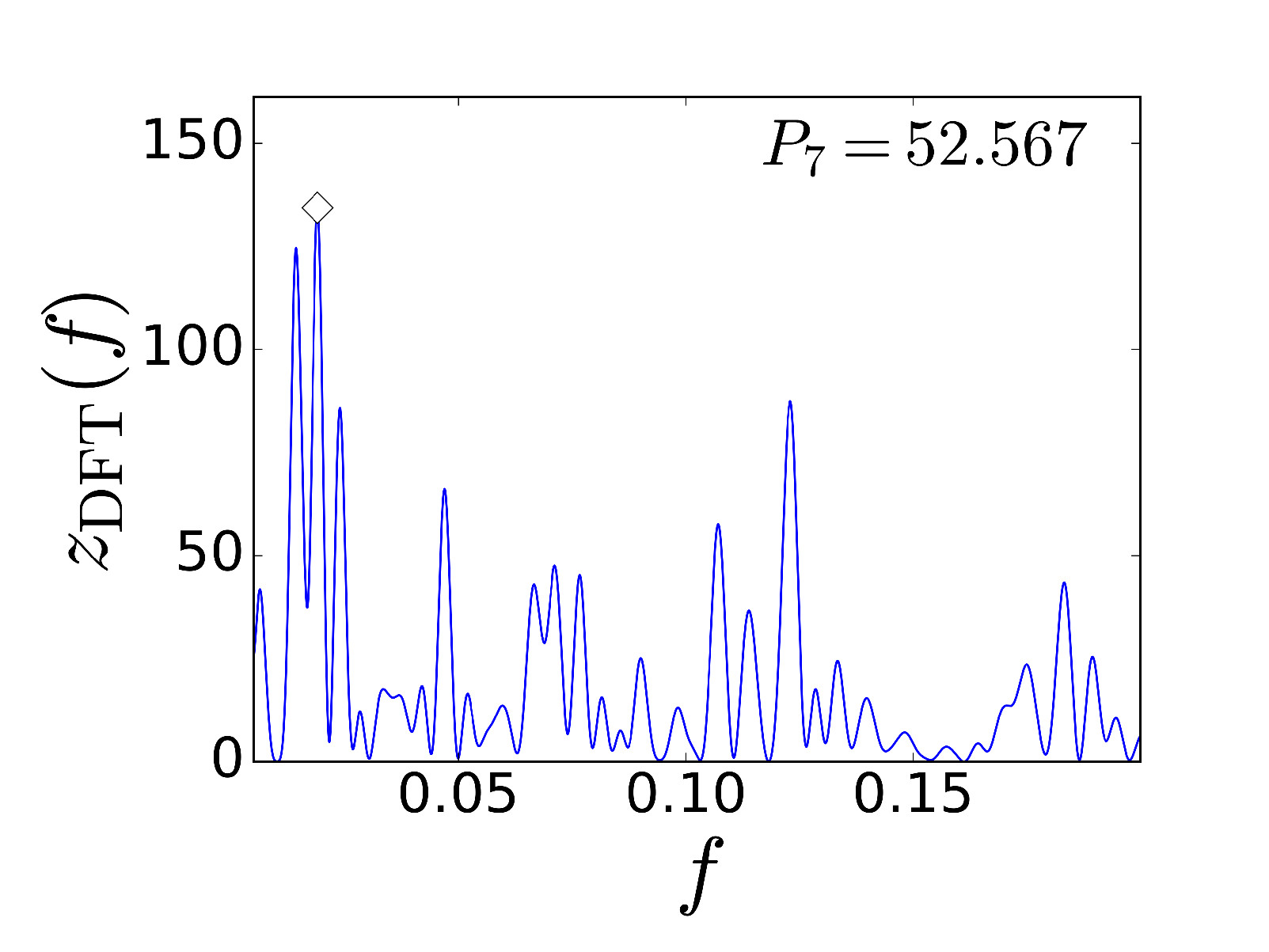} 
 }
\vspace{-0.22\textwidth}
\centerline{\normalsize \bf 
\hspace{0.11\textwidth}   \color{black}{(a)}
\hspace{0.23\textwidth}  \color{black}{(e)}
\hspace{0.22\textwidth}  \color{black}{(i)}
\hspace{0.22\textwidth}  \color{black}{(m)}
\hfill}
\vspace{0.21\textwidth}
\centerline{\hspace*{0.005\textwidth}
 \includegraphics[width=0.25\textwidth,clip=]{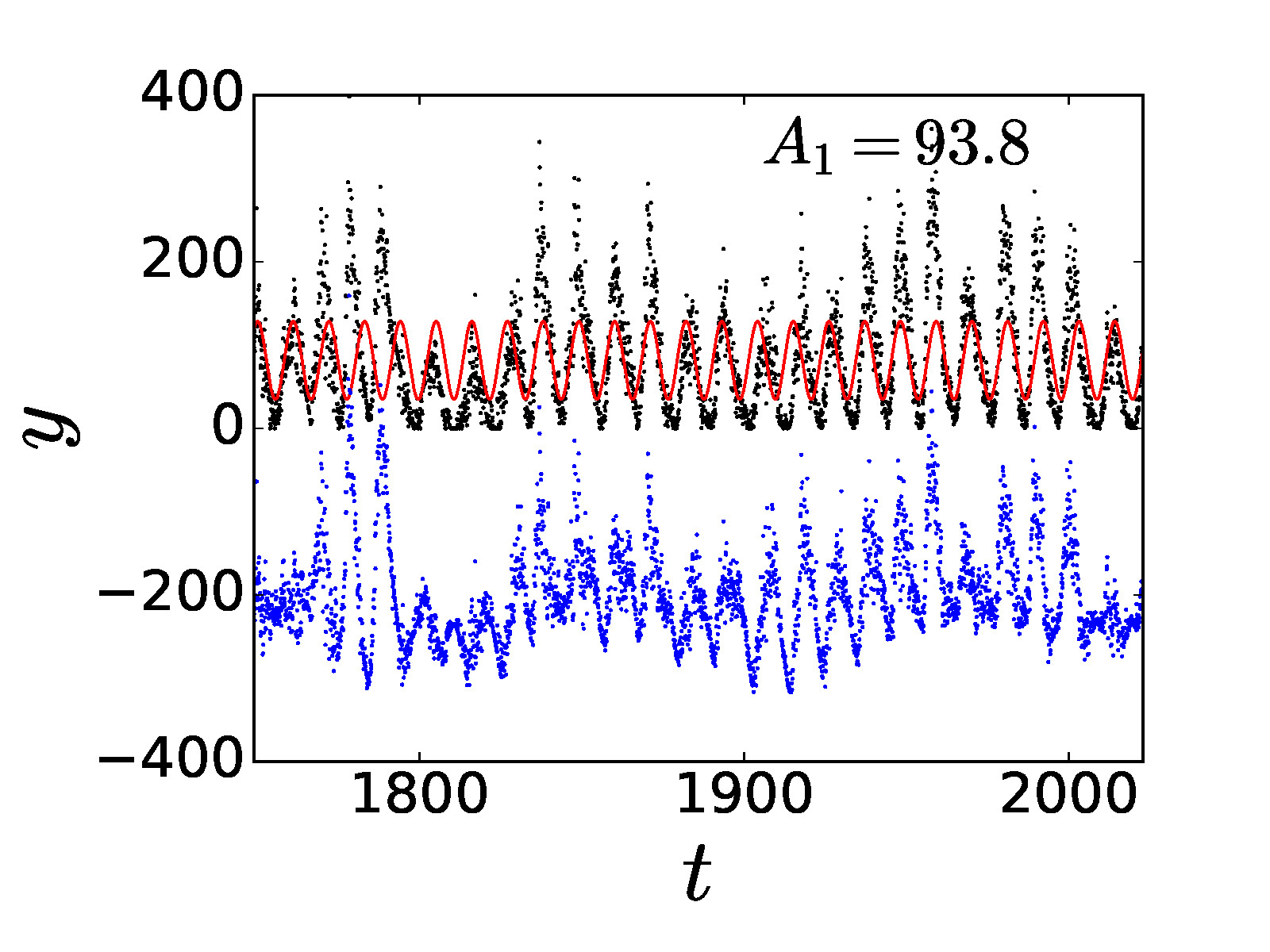} 
 \hspace*{-0.01\textwidth}
 \includegraphics[width=0.25\textwidth,clip=]{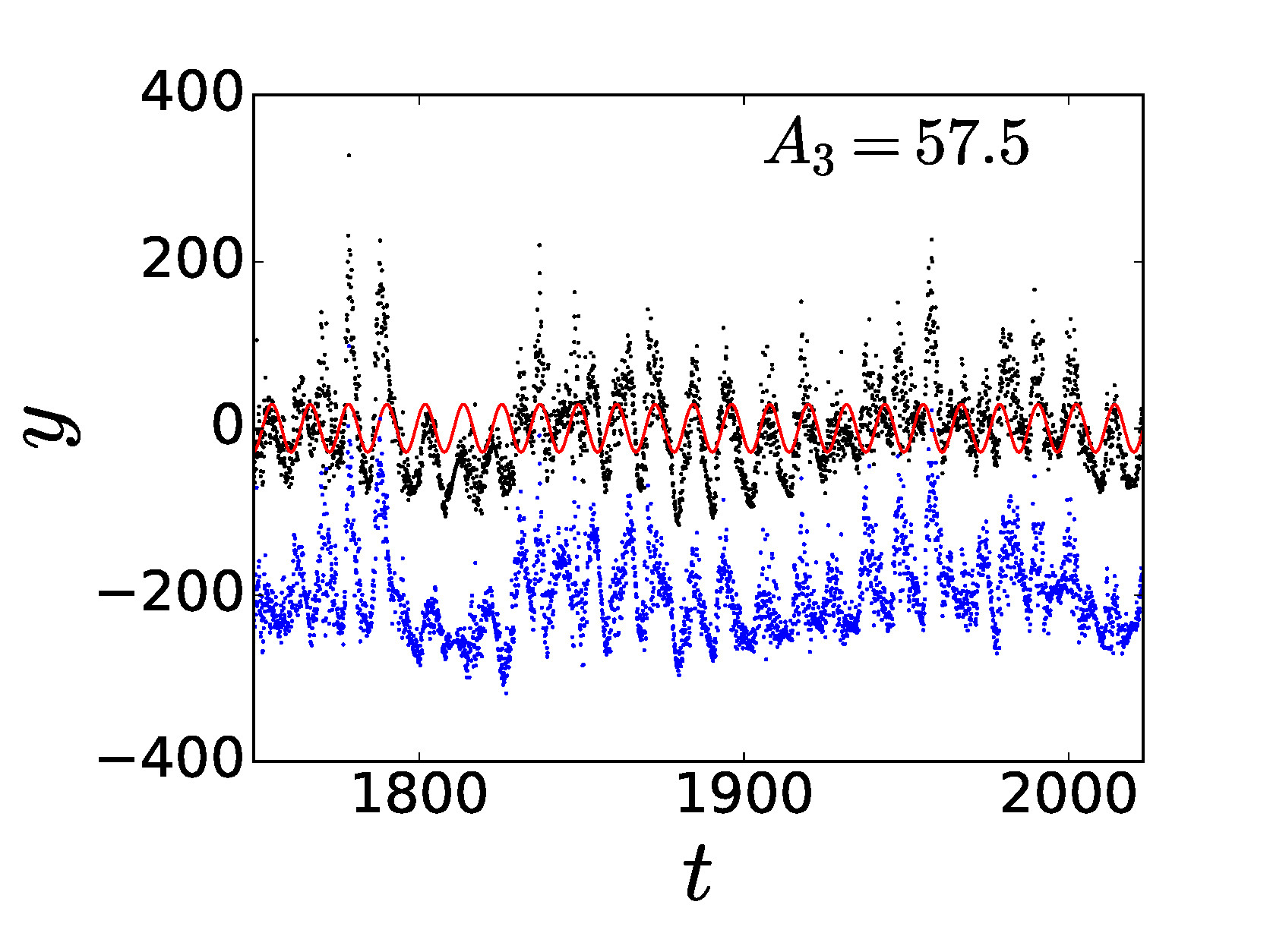} 
 \hspace*{-0.01\textwidth}
 \includegraphics[width=0.25\textwidth,clip=]{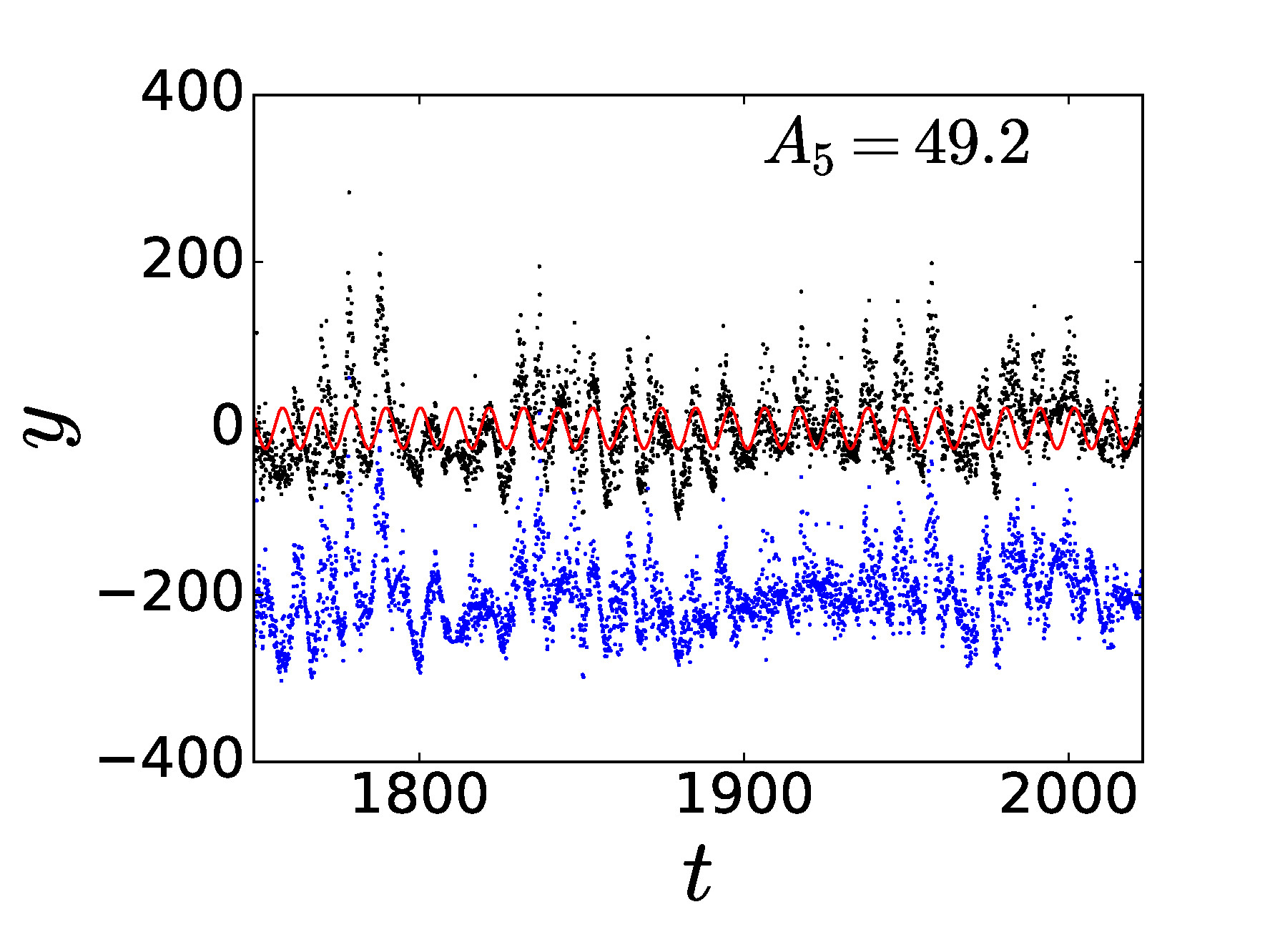} 
 \hspace*{-0.01\textwidth}
 \includegraphics[width=0.25\textwidth,clip=]{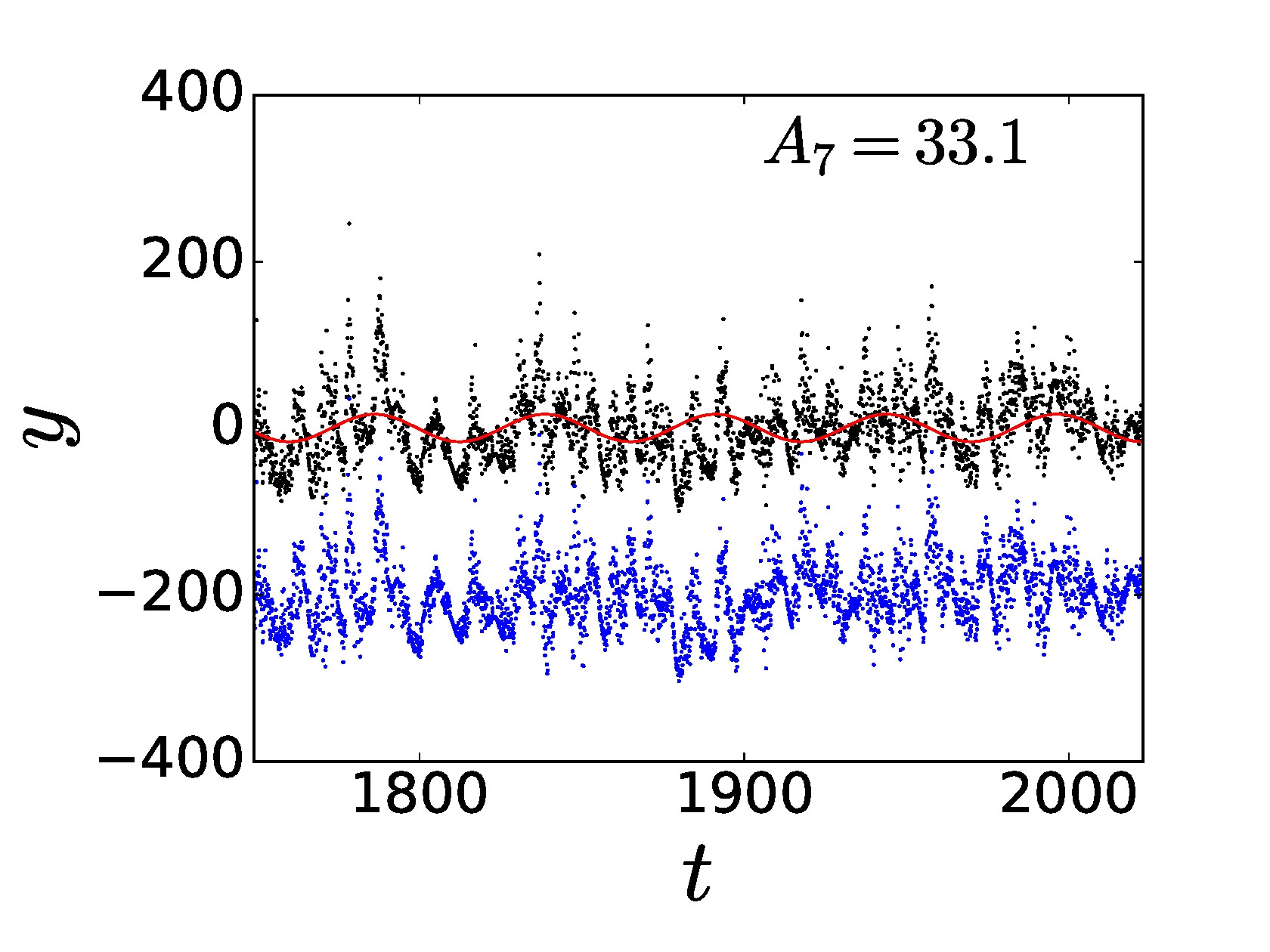} 
  }
\vspace{-0.22\textwidth}
\centerline{\normalsize \bf 
\hspace{0.11\textwidth}   \color{black}{(b)}
\hspace{0.23\textwidth}  \color{black}{(f)}
\hspace{0.22\textwidth}  \color{black}{(j)}
\hspace{0.22\textwidth}  \color{black}{(n)}
  \hfill}
\vspace{0.21\textwidth}
\centerline{\hspace*{0.005\textwidth}
\includegraphics[width=0.25\textwidth,clip=]{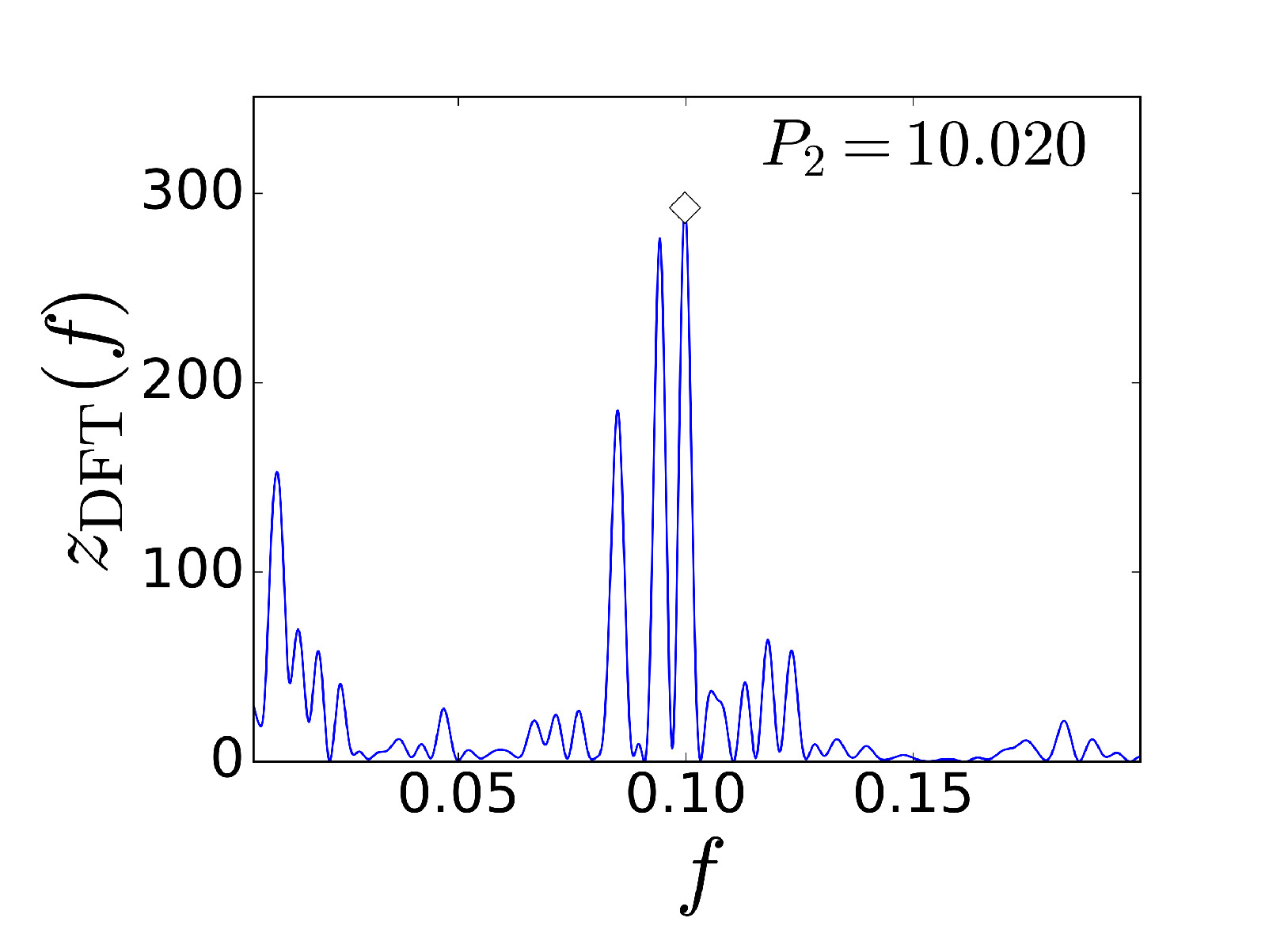} 
 \hspace*{-0.01\textwidth}
 \includegraphics[width=0.25\textwidth,clip=]{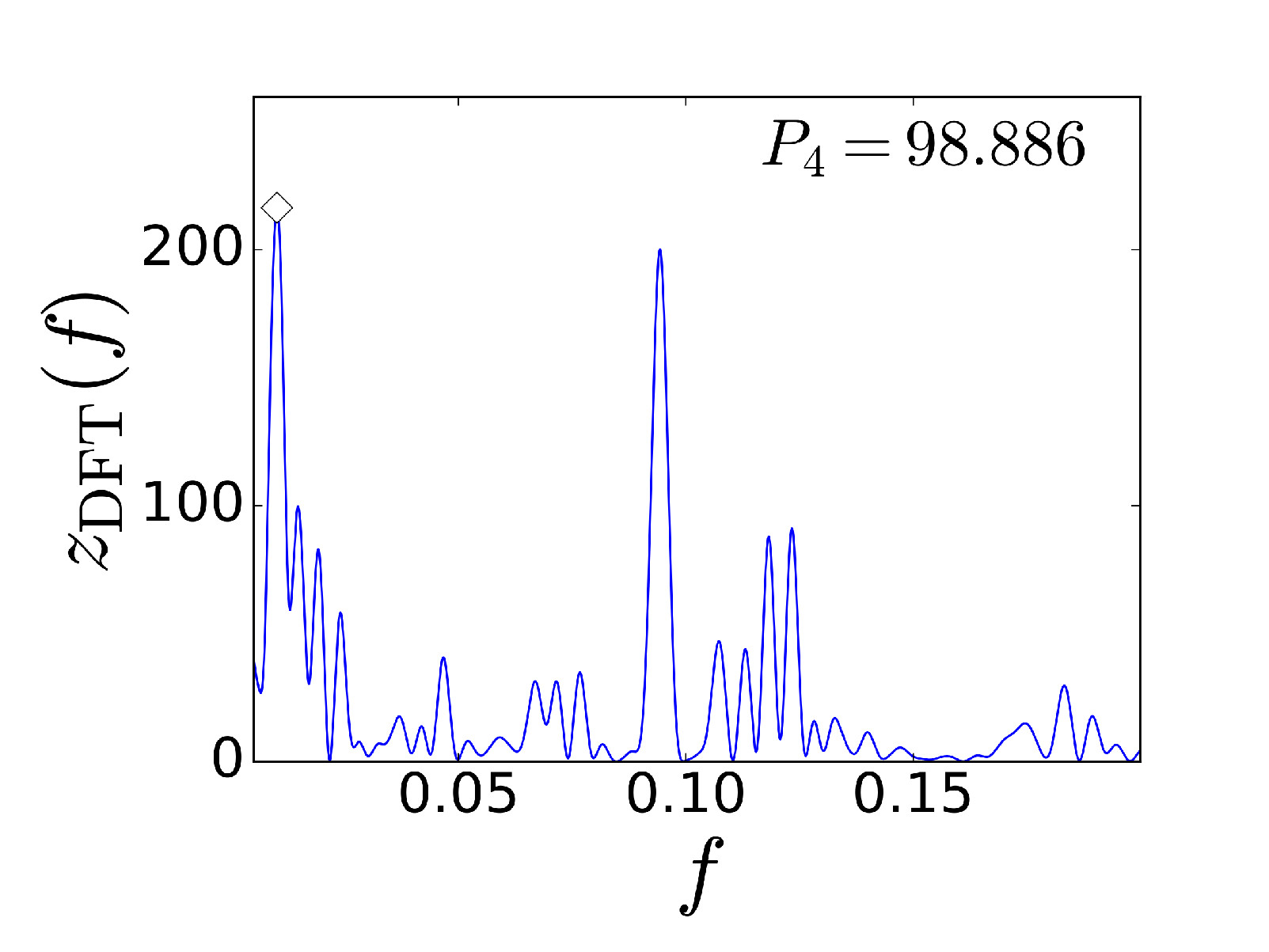} 
 \hspace*{-0.01\textwidth}
 \includegraphics[width=0.25\textwidth,clip=]{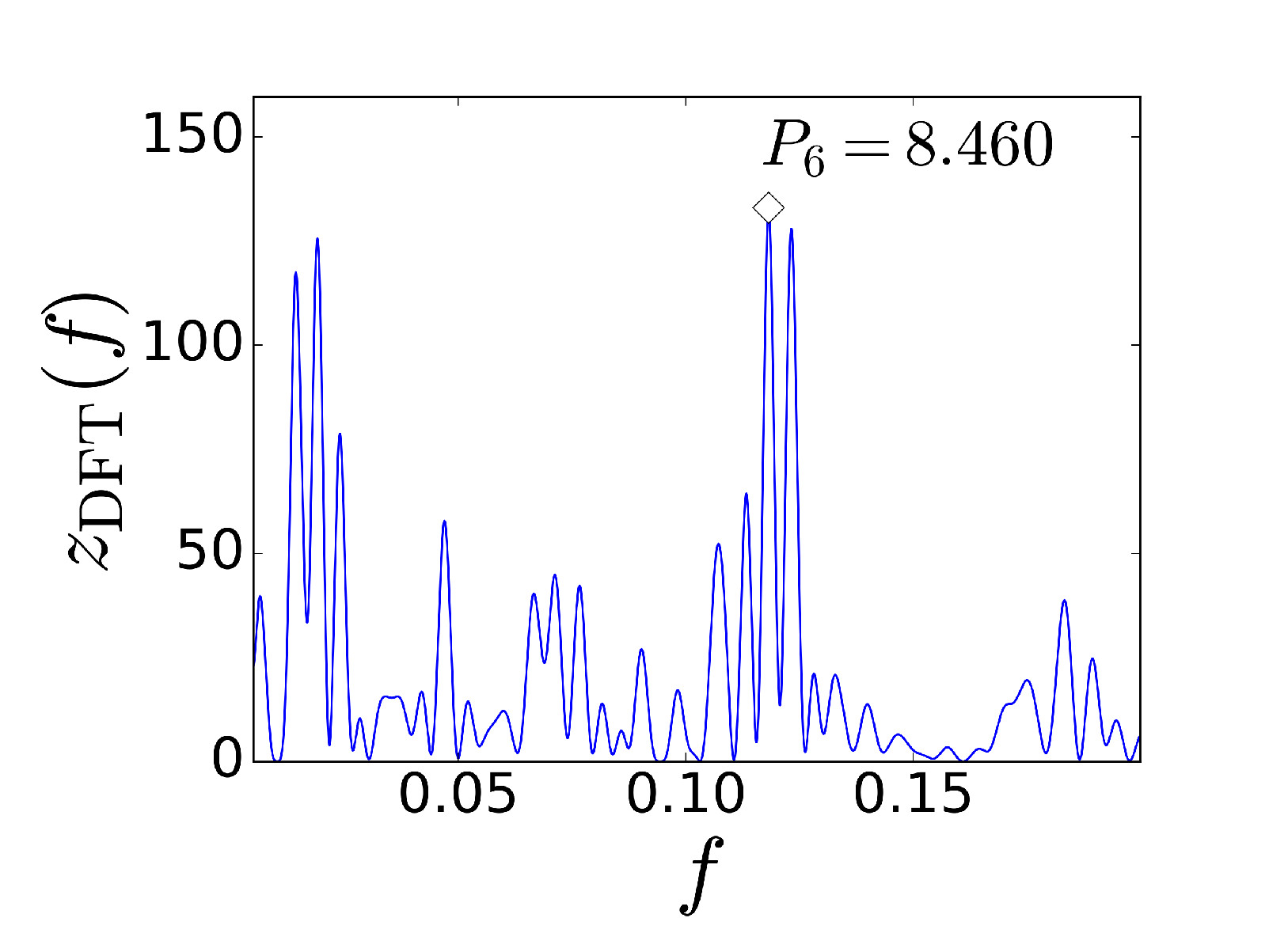} 
 \hspace*{-0.01\textwidth}
 \includegraphics[width=0.25\textwidth,clip=]{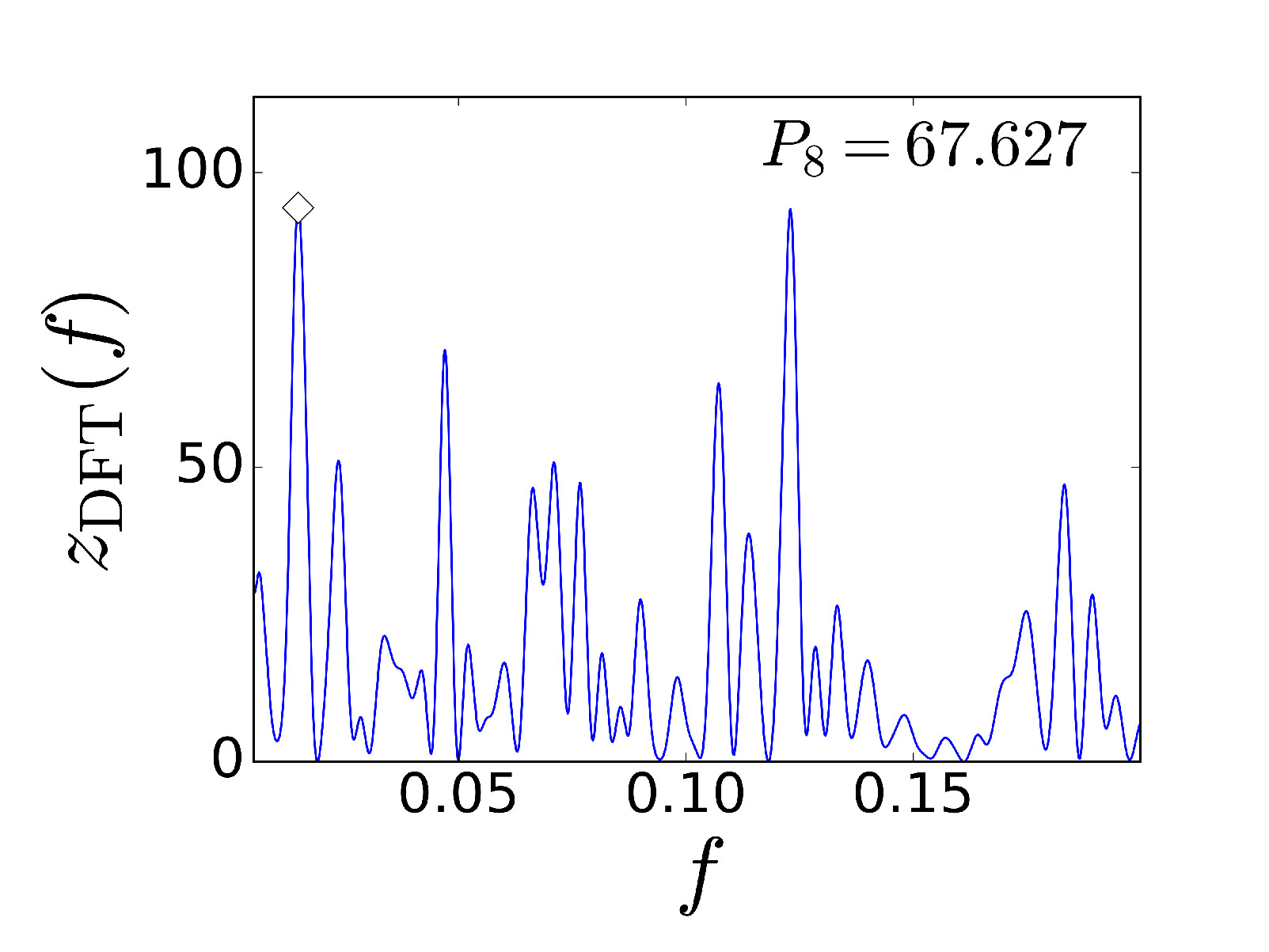} 
}
\vspace{-0.22\textwidth}
\centerline{\normalsize \bf 
\hspace{0.11\textwidth}   \color{black}{(c)}
\hspace{0.23\textwidth}  \color{black}{(g)}
\hspace{0.22\textwidth}  \color{black}{(k)}
\hspace{0.22\textwidth}  \color{black}{(o)}
  \hfill}
\vspace{0.21\textwidth}
\centerline{\hspace*{0.005\textwidth}
 \includegraphics[width=0.25\textwidth,clip=]{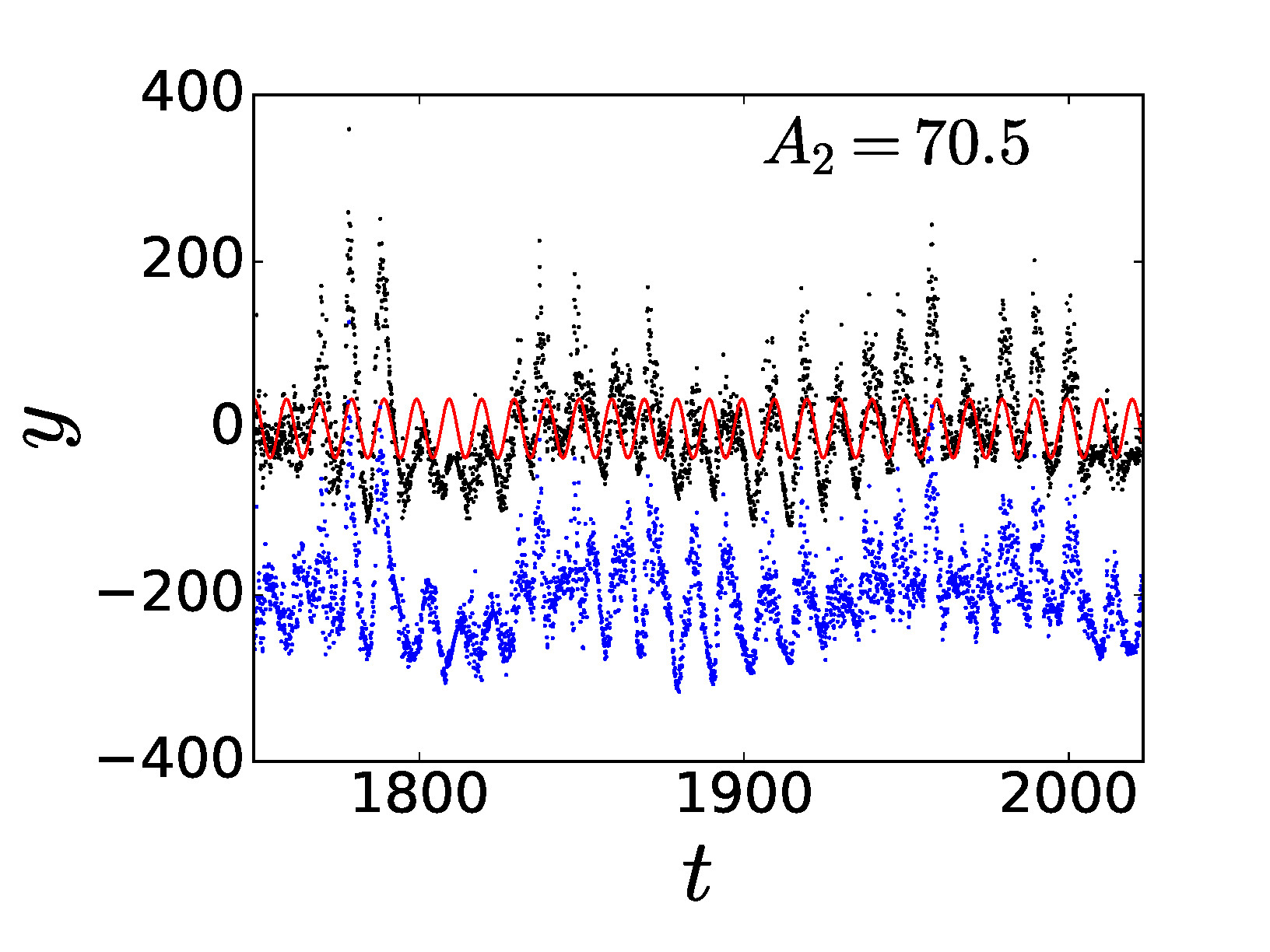}  
 \hspace*{-0.01\textwidth}
 \includegraphics[width=0.25\textwidth,clip=]{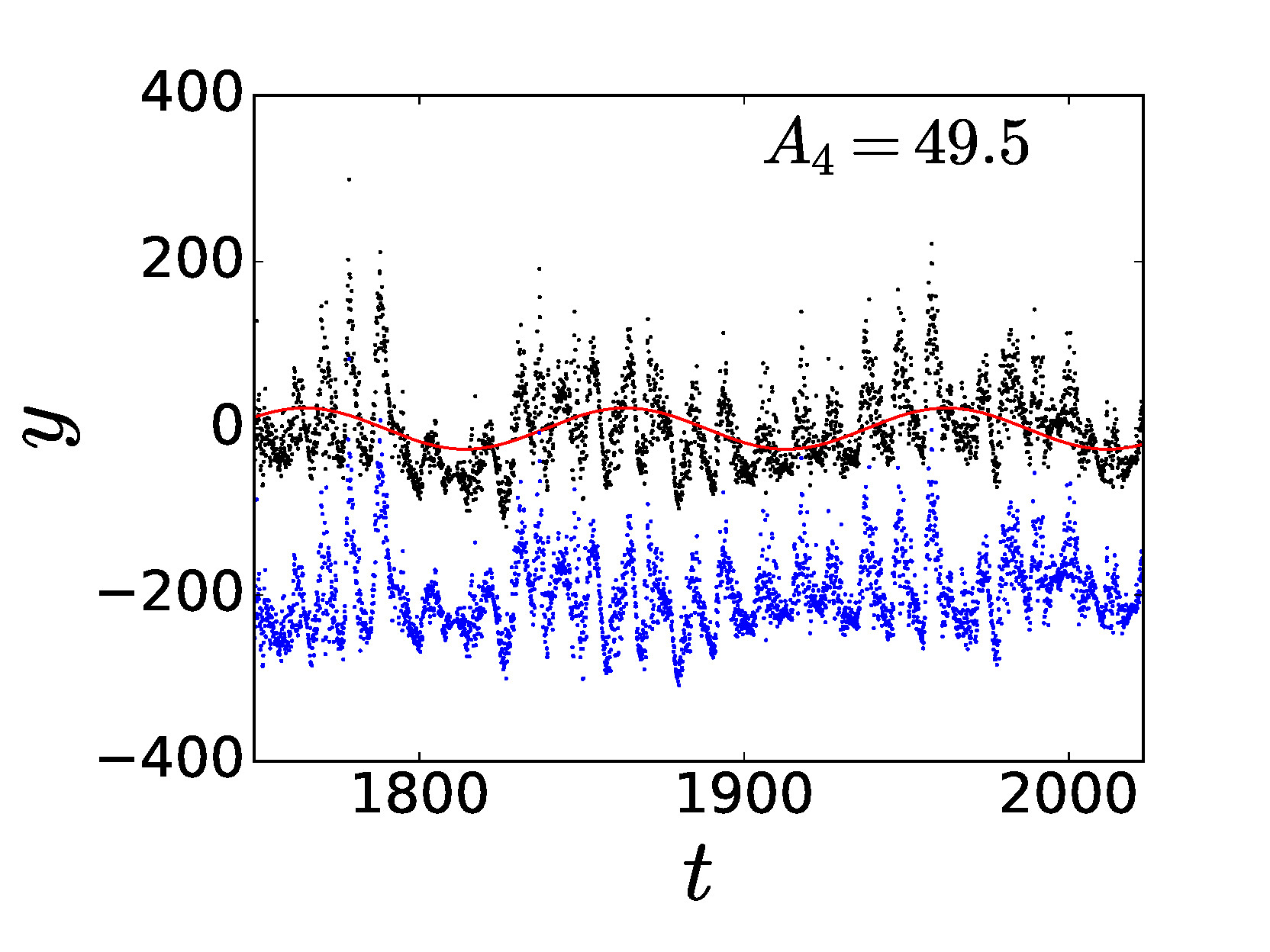} 
 \hspace*{-0.01\textwidth}
 \includegraphics[width=0.25\textwidth,clip=]{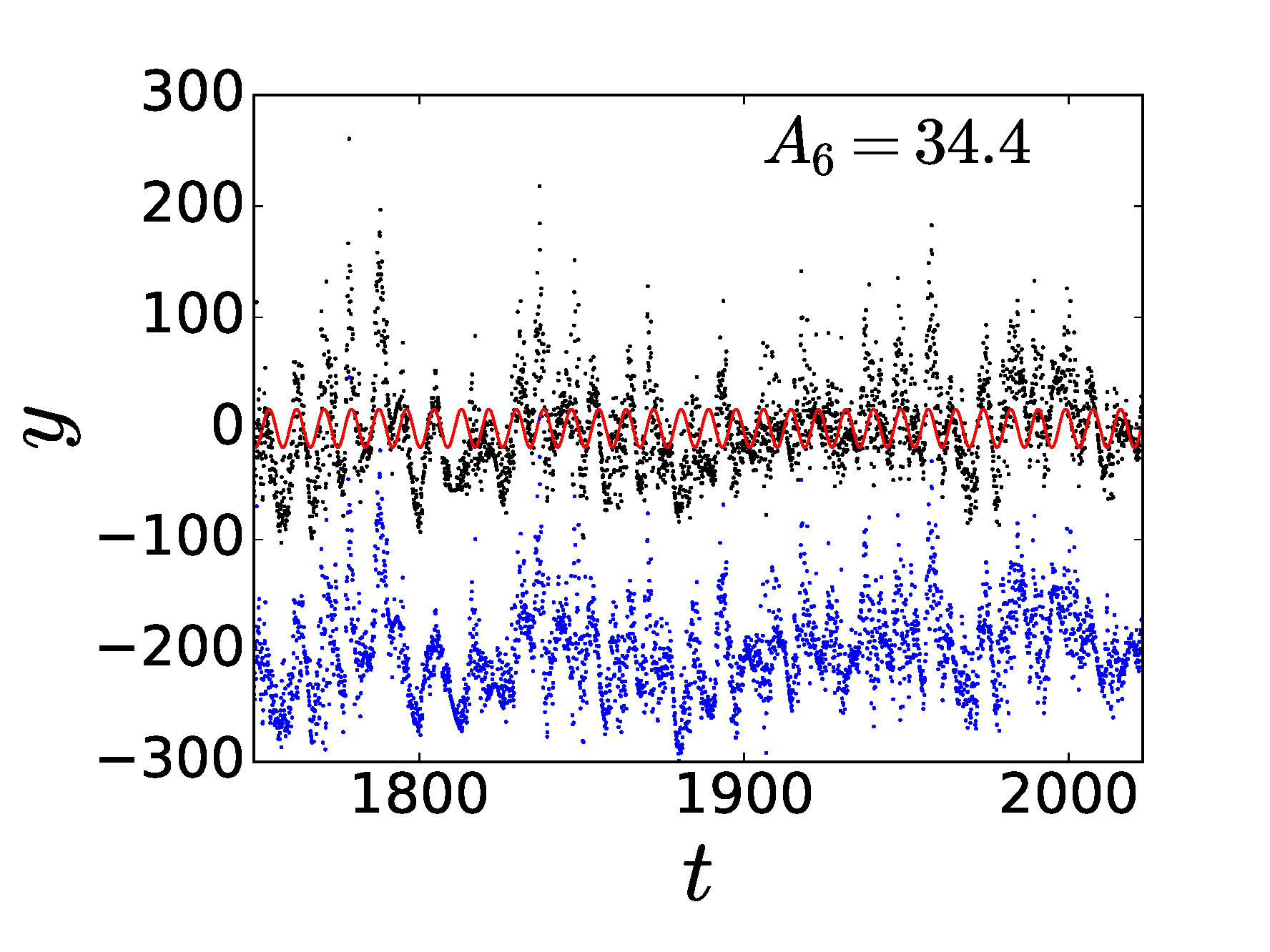} 
 \hspace*{-0.01\textwidth}
 \includegraphics[width=0.25\textwidth,clip=]{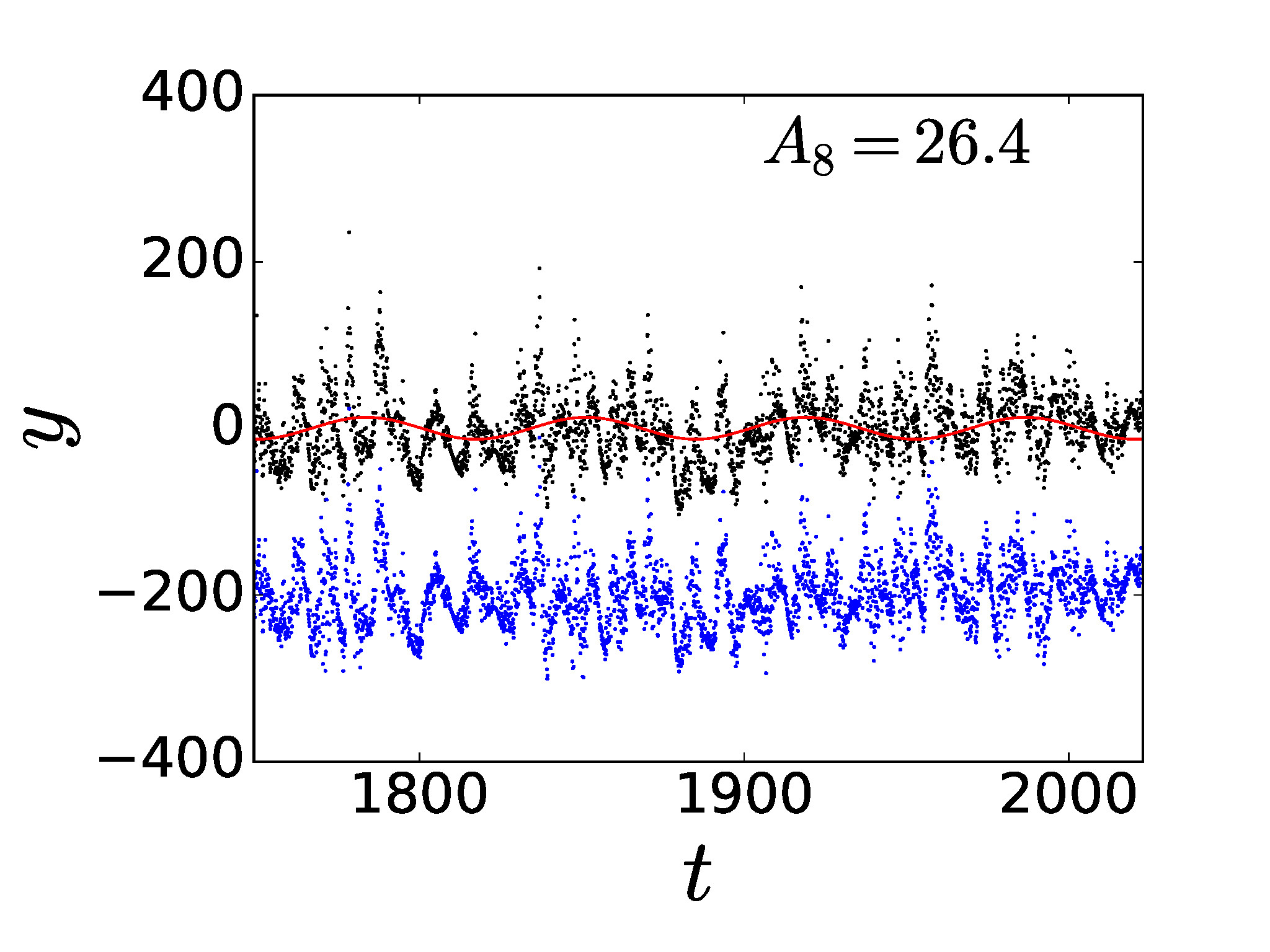} 
}
\vspace{-0.22\textwidth}
\centerline{\normalsize \bf 
\hspace{0.11\textwidth}   \color{black}{(d)}
\hspace{0.23\textwidth}  \color{black}{(h)}
\hspace{0.22\textwidth}  \color{black}{(l)}
\hspace{0.22\textwidth}  \color{black}{(p)}
  \hfill}
\vspace{0.18\textwidth}
\caption{DFT pre-whitening analysis results for \RmonthlyOne ~sample.
  (a) DFT periodogram for original data.
  Diamond denotes peak of periodogram at frequency
  of best period $P_1=11.002$ years.
  Units are $[f]=$ 1/y and $[z_{\mathrm{DFT}}]=$
  dimensionless.
  (b) Pure sine model for original data.
  Black dots denote original $y_i$ data.
  Red curve denotes sine model with period $P_1=11.002$.
  Its peak to peak amplitude is $A_1=93.8$.
  Blue dots denote residuals $\epsilon_i$ offset to
  level -200.
  Units are $[t]=$ years and $[y]=[\epsilon]=[A]=$ dimensionless.
  (c-d) Second signal detected
  from first sample of residuals. 
  (e-f) Third signal detected from second sample of residuals.
  (g-h) Fourth signal detected from next sample of residuals.
  (i-j) Fifth signal. (k-l) Sixth signal.
  (m-n) Seventh signal. (o-p) Eight signal.
  Notations for all panel pairs from ``c-d'' to ``o-p'' are same
  as in panel pair ``a-b''.} 
\label{FigDFTSunSpots}
\end{figure}

The DFT pre-whitening results for the
sample \RmonthlyOne ~are shown
in Fig. \ref{FigDFTSunSpots}.
The first five \SignalOne, \SignalTwo, \SignalThree,
\SignalFour ~and \SignalFive ~signals detected
using DFT pre-whitening
(Fig. \ref{FigDFTSunSpots} a-j) are the
same as those detected using the five pure sine
signal DCM model (Table \ref{TableCompare}: Column 3).
The only minor difference is that DCM detects signal
\SignalFive ~before signal \SignalFour.
Both methods give nearly the same amplitudes for these
five strongest signals.
The sixth-strongest detected signals are different:
the \SignalSeven ~signal for DCM
(Table \ref{TableCompare}: Column 3)
and the \SignalSix ~signal for DFT
(Fig. \ref{FigDFTSunSpots} k-l).
The seventh-strongest detected
signals are also different,
the \SignalEight ~signal for DCM and the \SignalSeven ~signal for DFT.
DCM detects only seven signals from sample
\RmonthlyOne, because the unstable
\M=8 model is 
rejected (Table \ref{TableRmonthly2000K410R14}:``$\UM$'').
DCM analysis of predictive data sample
\RmonthlyTwo, which is a subsample of all data sample
\RmonthlyOne, indicates that the correct
number of real signals is six
(Fig. \ref{FigRmonthly2000Sine}).
We conclude that  DCM and DFT detect
the same five strongest
signals from sample \RmonthlyOne.

        \begin{table}
      \caption{
        Significance estimates for DCM and DFT period detections
        from \RmonthlyOne ~sample.
        (1) Detection order. (2) DCM period $P$.
        (3) DCM critical level $Q_{\mathrm{F}}$ (Equation \ref{EqFisher}).
        (4) DFT period $P$. (5) DFT term
        $E_{\mathrm{DFT}}$ (Equation \ref{EqDFTSigni}).
        (6) DFT false alarm probability $Q_{\mathrm{DFT}}$
         in Ref\cite{Hor86} (Their equation 22)
        Note that DCM model
        for period 8.456 is unstable (`$`\UM$'').
      }\label{TableSigni}
      \begin{center}
      \begin{tabular}{ccccccc}
        \hline
        & \multicolumn{2}{c}{DCM} & & \multicolumn{2}{c}{DFT}  \\
        \cline{2-3} \cline{5-7}
        & $P$     & $Q_F$& & $P$    & $E_{\mathrm{DFT}}$ & $Q_{\mathrm{DFT}}$ \\
        (1) & (2) & (3) & & (4) & (5) & (6)\\
        (-) & (y) & (-) & & (y) & (-) & (-) \\
        \hline
      1 & 11.0324 & --   & & 11.002 & $1.8\times10^{-162}$ & 0 \\
      2 & 10.0001 & \REJ & & 10.020 & $5.5\times10^{-128}$ & 0 \\
      3 & 11.807  & \REJ & & 11.798 & $1.7\times10^{-103}$ & 0 \\
      4 & 99.92   & \REJ & & 98.886 & $1.7\times10^{-90}$  & 0 \\
      5 & 10.569  & \REJ & & 10.602 & $4.0\times10^{-102}$ & 0 \\
      6 & 52.66   & \REJ & & 8.460  & $2.2\times10^{-58}$  & 0\\
      7 & 8.1087  & \REJ & & 52.567 & $8.9\times10^{-59}$  & 0\\
      8 & 8.456 $\UM$&\REJ& & 67.627 & $4.0\times10^{-41}$  & 0\\
        \hline
      \end{tabular}
      \end{center}
    \end{table}

It is essential to understand,
why the DFT pre-whitening analysis of 
\RmonthlyOne ~sample does succeed,
regardless of all five DFT drawbacks discussed above?
The limitation of Equation \ref{EqPrevent} is encountered,
because the difference between the
\SignalFive ~and \SignalOne ~signal
frequencies is smaller than $f_0$
 (Table \ref{TableIndependent}).
All four 
 \SignalTwo, \SignalFive, \SignalOne ~and
 \SignalThree ~signal 
 frequencies are closely packed
 within an interval of $(27.38-23.09)f_0=4.29 f_0$
 independent frequencies.
  The interference of these four signals enhances the
 misleading effects of
 Equations \ref{EqMisleadOne} and \ref{EqMisleadTwo}.
 The amplitudes of
 \SignalTwo, \SignalFive ~and
 \SignalThree ~signals are nearly equal. 
 This causes many
 abrupt large $\Delta \phi \approx 0.5$
 phase shifts during $\Delta T=274$ 
 (Equation \ref{EqAbrupt}).
Nevertheless,
the DFT pre-whitening analysis succeeds, because
signal \SignalOne ~amplitude $A_1=93.8$
is crucially larger
than signal
\SignalTwo, \SignalFive ~and
\SignalThree ~amplitudes. 
As predicted by Equation \ref{EqMisleadTwo}, this

\SignalOne ~signal is detected first.
After the pre-whitening subtraction
of this strongest signal,
the remaining 
\SignalTwo, \SignalFive ~and
\SignalThree ~signals  no longer suffer from
the $f_0$ limitation of Equation \ref{EqPrevent}. 
The two highest amplitude $A_2=70.5$ and $A_3=57.4$
signals \SignalTwo ~and \SignalThree ~are now
``comfortably separated'' by $(27.38-23.09)f_0=4.29f_0$
independent frequencies.
As predicted by the
relation of Equation \ref{EqMisleadTwo}, 
these two signals
are detected next (Fig. \ref{FigDFTSunSpots}c-f).
The frequency difference between the  
remaining \SignalFive ~and \SignalFour ~signals
is so large that DFT can easily detect them.
Our DFT pre-whitening analysis of
  sample \RmonthlyOne ~succeeds,
  because one strong \SignalOne ~signal dominates,
  and the remaining signals are
  detected in a suitable order.
  This successful DFT analysis also confirms
    that the data does not contain any misleading
    long-term trends. 

    Our Table \ref{TableSigni}
    gives the significance estimates for the DCM and DFT
    period detections from the \RmonthlyOne ~sample.
    The DCM gives no direct significance estimate
    for the first detected strongest 11.032 year period.
    After {Wolf} discovered this \SignalOne
    ~signal\cite{Wol52},
    it has been re-detected perhaps more than a million times.
    If this 11.032 year signal is not an artefact,
    the DCM detections for the next six periods
    are extremely significant
    because the Fisher-test always
    rejects the $H_0$ hypothesis 
    at critical levels
    $Q_{\mathrm{F}}$ \REJ ~(Equation \ref{EqFisher}).
    The DCM model for the eighth strongest 8.456 year period is
    unstable (Table \ref{TableRmonthlyK410R14}, \M=8). 

    For the DFT analysis,
    our Table \ref{TableSigni} 
    gives the values for the term
    \begin{equation}
      E_{\mathrm{DFT}}={\mathrm{e}}^{-(n/2)(1+\xi^{-1})^{-1}},
      \label{EqDFTSigni}
    \end{equation}

    \noindent
    where $\xi=(A/2)^2/(2 s_{\epsilon}^2)$,
    $A$ is the sine peak to peak amplitude and
    $s_{\epsilon}$ is the standard deviation residuals
    (variance of noise).
%
    This $ E_{\mathrm{DFT}}$ parameter 
    is used to 
    compute the false alarm probability $Q_{\mathrm{DFT}}$
    in Ref\cite{Hor86} (Their Equation 22).
    All $E_{\mathrm{DFT}}$ values
    are so close to zero that all false alarm probabilities
    $Q_{\mathrm{DFT}}$ are practically equal to zero.
    This means that all DFT period detections are
    extremely significant.

    In short, 
    all DCM and DFT period detections
    in Table \ref{TableSigni} 
    are absolutely certain,
    if the data contain only white noise.
    One could argue that our significance
    estimates can be
    unreliable because we do not consider the
    possibility that the data contains red
    noise \cite{Vau05}. 
    We test the $H_0$ hypothesis in our DCM analysis
    with the Fisher-method (Equation \ref{EqFisher}).
    If the data contained red noise,
    it would influence
    both the simple model and the complex model
    of the Fisher-test.
    Hence, our the DCM analysis results are
    correct with or without the red noise.
    We admit that the red noise could contaminate
    our DFT analysis significance estimates,
    but then again we detect
    the same periods with the DCM and the DFT.
    For some reason,
    the extreme significance estimates
    for the DFT
    period detections from
    the sunspot data have not caught
    the attention that they would have deserved 
    in the earlier studies, like Ref\cite{Zhu18}.
    
Finally, we repeat
      the daily sunspot data DFT pre-whitening
      analysis in Ref\cite{Zhu18}.
      The authors did not detect
the \SignalFour ~signal because 
they searched for periods between 2 and 18 years.
For our tested period interval between 5 and 200 years,
DFT detects the five strongest
10.87, 10.14, 110.6, 12.00, and 7.99 year signals
in the daily sunspot data.
These five periods are closest to the five best
periods that DCM detects in the weighted data
(Table \ref{TableCompare}: Columns 7 and 13).
The probable cause for the minor numerical value
differences is 
that DFT tends to detect less accurate real period
values than DCM (Sections \ref{SectDFTEqual}
and \ref{SectDFTUnequal}). 
Our DFT detection of the 110.6 signal from these daily sunspot
data removes the two
weak 8.92 and 5.44 year signals detected in
Ref\cite{Zhu18}.
We do not apply DCM to the daily sunspot data
for a simple reason.
    For a cluster of parallel computation processors,
    it takes about two days
    to compute the four signal DCM model
    for one monthly sunspot data sample.
    For one daily sunspot data sample,
    this four signal DCM model computation
    would take months, and the computations of many
    samples would take years.

    We conclude that the DCM and DFT detect the same
      strictly
      periodic signals from the sunspot data.
      Both methods give extreme significance estimates
      for these signals.
      These period detections are
      absolutely certain. 
      In earlier studies,
        the DFT, like all other methods
        that search for one period at the time,
        has failed to detect the exact correct numerical period values.
        The DCM does not fail, as illustrated
      in Figs. \ref{FigDFTDCMOne} and  \ref{FigDFTDCMTwo}, where
      the vertical dashed lines (simulated periods)
      intersect the diamonds (detected periods) much more often
      for DCM than for DFT.

\setcounter{section}{6} 
\setcounter{subsection}{6}
\subsection{Prediction for current sunspot cycle 25}
\label{SectDCMthree}
DCM analysis results
for the \RmonthlyNew ~and \CmonthlyNew ~samples
are given in Tables
\ref{TableRmonthly2019K410R14} and
\ref{TableCmonthly2019K410R14}.
The respective electronic datafiles are
\PRtext{Rmonthly2019.dat} and \PRtext{Cmonthly2019.dat}.

\begin{center}
  {\bf How to repeat the DCM analysis of this
    Section \ref{SectDCMthree}}
\end{center}
The model \M=1
analysis in Table 
\ref{TableRmonthly2019K410R14} can be repeated
with two python commands. \\
\PRtext{cp Rmonthly2019K110R14.dat dcm.dat} \\
\PRtext{python dcm.py} \\
The model \M=2 analysis requires the commands \\
\PRtext{cp Rmonthly2019K210R14.dat dcm.dat} \\
\PRtext{python dcm.py} \\
All analysis in Tables
\ref{TableRmonthly2019K410R14}-\ref{TableCmonthly2019K410R14}
can be repeated in a similar fashion.

\clearpage

\begin{table}
  \caption{\RmonthlyNew ~analysis for pure sinusoids $(K_2=1)$. 
    Otherwise as in Table \ref{TableRmonthly2000K410R14}.}
    \label{TableRmonthly2019K410R14} 
\begin{scriptsize}
     \begin{center}
    \begin{adjustbox}{angle=90}
      \begin{tabular}{lcccccccccc}
      \hline
     &  & \multicolumn{4}{c}{Period analysis} & & \multicolumn{3}{c}{Fisher-test} &  \\
\cline{3-6} \cline{8-10}
\multicolumn{11}{c}{Data: Non-weighted original data ($n=3249, \Delta T=271$: \PR{Rmonthly2019.dat})} \\
\M                    &
\RModel{K_1,K_2,K_3}  &
$P_1$ (y)             &
$P_2$ (y)             &
$P_3$ (y)             &
$P_4$ (y)             &
                      &
\M=2                  &
\M=3                  &
\M=4                  &
                     \\ 
             &
$\eta$ (-)   &
$A_1$ (-)    &
$A_2$ (-)    &
$A_3$ (-)    &
$A_4$ (-)    &
             &
$F_R$ (-)    &
$F_R$ (-)    &
$F_R$ (-)    &
Control file \\ 
                       &
$R$ (-)                &
$t_{\mathrm{min,1}}$ (y) &
$t_{\mathrm{min,1}}$ (y) &
$t_{\mathrm{min,1}}$ (y) &
$t_{\mathrm{min,1}}$ (y) &
                       &
$Q_F$ (-)              &
$Q_F$ (-)              &
$Q_F$ (-)              &
                       \\ 
(1)    & 
(2)    & 
(3)    & 
(4)    & 
(5)    & 
(6)    & 
       & 
(7)    & 
(8)    & 
(9)    & 
(10) \\
      \hline
\M=1 &\RModel{1,1,0}      &$11.0018\PM0.0068$&     -           & -                & -              &&$\uparrow    $&$\uparrow    $&$\uparrow     $&                             \\
        &       4         &$93.0\PM3.5      $&     -           & -                & -              &&     x        &   x          &   x           &  \PR{Rmonthly2019K110R14.dat}    \\
        &$   1.14\TM10^7 $&$1755.62\PM0.12  $&     -           & -                & -              &&$1.5\TM10^{-10}$&\REJ         &\REJ           & \\ 
\hline
\M=2 &\RModel{2,1,0}     &$10.614\PM0.012  $&$11.004 \PM0.010$& -                & -              &&-             &$\uparrow    $&$\uparrow     $&                             \\
        &      7         &$70.5\PM2.4       $&$93.1\PM3.4     $& -                & -              &&-             &    x          &    x          &  \PR{Rmonthly2019K210R14.dat}    \\
        &$   9.43\TM10^6 $&$1752.17\PM0.21  $&$1755.45\PM0.15 $& -                & -              &&-             &$   x         $&\REJ          & \\ 
\hline
\M=3 &\RModel{3,1,0}     &$10.0016\PM0.0057$&$10.9962\PM0.0049$&$11.782\PM0.011  $& -              &&-             & -            &$\uparrow     $&                             \\
        &      10        &$75.2\PM2.6      $&$101.4\PM2.2     $&$61.1\PM3.2      $& -              &&-             & -            &  x            &  \PR{Rmonthly2019K310R14.dat}    \\
        &$   7.87\TM10^6$&$1754.293\PM0.0.080$&$1756.036\PM0.078$&$1760.65\PM0.15$& -              &&-             & -            &\REJ           & \\ 
\hline
\M=4 &\RModel{4,1,0}     &$10.013\PM0.011  $&$10.548\PM0.019 $&$11.0059\PM0.0070$&$11.812\PM0.015  $&&-             & -            & -             &                               \\
        &        13      &$66.1 \PM2.3     $&$51.6 \PM3.1   $&$100.5\PM2.4       $&$56.0\PM2.5     $&&-             & -            & -             &  \PR{Rmonthly2019K410R14.dat}    \\
        &$6.88\TM10^6$   &$1754.03\PM0.14  $&$1753.32\PM0.22 $&$1755.899\PM0.096$&$1760.29\PM0.16  $&&-             & -            & -             & \\
\hline
\multicolumn{11}{c}{Data: Non-weighted residuals of \M=4 ($n=3249, \Delta T=271$: \PR{Ryearly2019K410R14Residuals.dat})} \\
\hline
\M=5 &\RModel{1,1,-1}     &$98.68\PM0.0.59  $&     -          & -              & -              && -            & -            &     -         &                             \\
        &       3         &$49.5\PM2.2      $&     -          & -              & -              && -            & -            &     -         &  \PR{Ryearly2019yK11-1R58.dat}    \\
        &$    5.83\TM10^6 $&$1814.17\PM0.67 $&     -          & -              & -              && -            & -            &     -         &  \\ 
\hline
\end{tabular}
\end{adjustbox}
\end{center}
\end{scriptsize}
\addtolength{\tabcolsep}{+0.05cm}
\end{table}    

  \begin{table}
  \caption{\CmonthlyNew ~analysis for pure sinusoids $(K_2=1)$. 
    Otherwise as in Table \ref{TableRmonthly2000K410R14}.}
    \label{TableCmonthly2019K410R14} 
\begin{scriptsize}
     \begin{center}
    \begin{adjustbox}{angle=90}
      \begin{tabular}{lcccccccccc}
      \hline
     &  & \multicolumn{4}{c}{Period analysis} & & \multicolumn{3}{c}{Fisher-test} &  \\
\cline{3-6} \cline{8-10}
\multicolumn{11}{c}{Data: Weighted original data ($n=2420, \Delta T=201.7$: \PR{Cmonthly2019.dat})} \\
\M                    &
\CModel{K_1,K_2,K_3}  &
$P_1$ (y)             &
$P_2$ (y)             &
$P_3$ (y)             &
$P_4$ (y)             &
                      &
\M=2                  &
\M=3                  &
\M=4                  &
                     \\ 
             &
$\eta$ (-)   &
$A_1$ (-)    &
$A_2$ (-)    &
$A_3$ (-)    &
$A_4$ (-)    &
             &
$F_{\chi}$ (-)    &
$F_{\chi}$ (-)    &
$F_{\chi}$ (-)    &
Control file \\ 
                       &
$\chi^2$ (-)           &
$t_{\mathrm{min,1}}$ (y) &
$t_{\mathrm{min,1}}$ (y) &
$t_{\mathrm{min,1}}$ (y) &
$t_{\mathrm{min,1}}$ (y) &
                       &
$Q_F$ (-)              &
$Q_F$ (-)              &
$Q_F$ (-)              &
                       \\ 
(1)    & 
(2)    & 
(3)    & 
(4)    & 
(5)    & 
(6)    & 
       & 
(7)    & 
(8)    & 
(9)    & 
(10) \\
      \hline
\M=1 &\CModel{1,1,0}      &$10.8248\PM0.0090$&     -           & -                & -              &&$\uparrow    $&$\uparrow    $&$\uparrow     $&                             \\
        &       4         &$121.8\PM2.9     $&     -           & -                & -              &&     x        &   x          &   x           &  \PR{Cmonthly2019K110R14.dat}    \\
        &$   3.65\TM10^6 $&$1823.812\PM0.093$&     -           & -                & -              &&$1.5\TM10^{-10}$&\REJ         &\REJ           & \\
\hline
\M=2 &\CModel{2,1,0}     &$10.170\PM0.012   $&$10.8402\PM0.0074$& -                & -              &&-             &$\uparrow    $&$\uparrow     $&                             \\
        &      7         &$56.0\PM2.2       $&$118.5\PM3.0     $& -                & -              &&-             &    x          &    x          &  \PR{Cmonthly2019K210R14.dat}    \\
        &$   2.96\TM10^6 $&$1822.25\PM0.16  $&$1823.722\PM0.080$& -                & -              &&-             &$   x         $&\REJ          & \\
\hline
\M=3 &\CModel{3,1,0}     &$10.126\PM0.013  $&$10.8542\PM0.0081$&$121.9\PM1.4     $& -              &&-             & -            &$\uparrow     $&                             \\
        &      10        &$59.3\PM2.7      $&$121.2\PM2.2     $&$48.9\PM3.0      $& -              &&-             & -            &  x            &  \PR{Cmonthly2019K310R14.dat}    \\
        &$   2.46\TM10^6$&$1822.79\PM0.0.12$&$1823.556\PM0.082$&$1907.8\PM1.0    $& -              &&-             & -            &\REJ           & \\ 
\hline
\M=4 &\CModel{4,1,0}     &$10.066\PM0.013  $&$10.8577\PM0.0070$&$11.870\PM0.026$&$116.7\PM1.6     $&&-             & -            & -             &                               \\
        &        13      &$59.9 \PM2.2     $&$117.0 \PM1.8   $&$43.1\PM1.6     $&$52.9\PM2.0     $&&-             & -            & -             &  \PR{Cmonthly2019K410R14.dat}    \\
        &$2.08\TM10^6$   &$1823.52\PM0.12  $&$1823.590\PM0.079$&$1818.87\PM0.21$&$1911.07\PM0.91  $&&-             & -            & -             & \\
\hline
\multicolumn{11}{c}{Data: Weighted residuals of \M=4 ($n=2420, \Delta T=201.7$: \PR{Cyearly2019K410R14Residuals.dat})} \\
\hline
\M=5 &\CModel{1,1,-1}     &$8.002\PM0.0.017  $&     -          & -              & -              && -            & -            &     -         &                             \\
        &       3         &$25.9\PM2.0      $&     -          & -              & -              && -            & -            &     -         &  \PR{Cyearly2019yK11-1R58.dat}    \\
        &$   1.92\TM10^6 $&$1818.14\PM0.25 $&     -          & -              & -              && -            & -            &     -         &  \\ 
\hline
\end{tabular}
\end{adjustbox}
\end{center}
\end{scriptsize}
\addtolength{\tabcolsep}{+0.05cm}
\end{table}    
  
\clearpage

\section{Statistics for planetary
  signal identification
  \label{SectIdentification}}

Here, we describe the statistics
of integer multiple orbital period identification.
We will show that it is
difficult to identify the integer multiple orbital
periods of Mercury and Venus from the sunspot data.
These identifications are easier
for the Earth and Jupiter.
The connections between these integer multiple orbital
periods and the planetary motions are also discussed.

Before we begin the search for integer multiple orbital periods,
the following possible misunderstandings must be corrected.
  Firstly, all detected signals
  are already given in Table \ref{TableCompare}.
    We are not fitting multiples of planetary signals into the data.
  Secondly, the identification of these multiples is not
    numerology.
    We are only trying to identify the most obvious multiples
    that could be connected to the planets.

\begin{table}
  \caption{Signals detected in all data samples.
    (1) Signal parameters.
    (2-6) Samples:
    pure sine model $(K_2=1)$
    and
    double wave models  $(K_2=2)$.
    Three double sinusoid
    signals are denoted with ``\DW''.
       For example, seventh strongest ``(7)'' signal
        \SignalEight ~for \RmonthlyOne ~$(K_2=1)$
        has period
        $P\pm\sigma_P=8.1087\pm0.0076$
        in Earth years.
        Primary minima and maxima of this signal
        are at years $t_{\mathrm{min,1}}=1751.29\pm0.14$ and
        $t_{\mathrm{max,1}}=1755.34\pm0.14$.
        Rounds $\mathcal{P}\pm\sigma_{\mathcal{P}}$ values are
        $(P\pm \sigma_P)/\Pmer=33.674\pm0.032$,
        $(P \pm \sigma_P)/\Pven=13.180\pm0.012$ and
       $(P\pm \sigma_P)/\Pjup=0.68359\pm0.00064$.
    }

  \label{TableMultiples} 
    \begin{scriptsize}
      \addtolength{\tabcolsep}{-0.05cm}
      \renewcommand{\arraystretch}{0.80}
      \begin{center}
      \begin{tabular}{ccccccc}
       \hline
    (1)      & (2)                   & (3)                  & (4)               & (5)             & (6)            \\          
             & \RmonthlyOne: $K_2=1$ & \RmonthlyOne: $K_2=2$& \CmonthlyOne: $K_2=1$ & \RyearlyOne : $K_2=1$    & \CyearlyOne: $K_2=1$  \\
      \hline   
             & \multicolumn{5}{c}{Signal \SignalEight ~$\cong 13 \times \Pven \cong 8 \times \Pear$} \\
Rank             & (7)               & (8)                  & (5)               &                 & (5)               \\
$P              $&$8.1087\pm0.0076  $&$8.169\pm0.014       $&$8.009\pm0.017    $&                 &$8.005\pm0.058    $\\
$t_{\mathrm{min,1}}$&$1751.29\pm0.14  $&$1750.90\pm0.51      $&$1818.07\pm0.24   $&                 &$1826.09\pm0.68   $\\
$t_{\mathrm{max,1}}$&$1755.34\pm0.14  $&$1754.37\pm0.26      $&$1822.08\pm0.24   $&                 &$1822.09\pm0.70   $\\
$P/\Pmer         $&$33.674\pm0.032   $&$33.924\pm0.058     $&$33.260\pm0.070   $&                 &$33.24\pm0.24     $\\
$P/\Pven         $&$13.180\pm0.012   $&$13.279\pm0.023     $&$13.018\pm0.028   $&                 &$13.012\pm0.094   $\\
$P/\Pjup         $&$0.68359\pm0.00064$&$0.6887\pm0.0012    $&$0.6752\pm0.0014  $&                 &$0.6748\pm0.0049  $\\
      \hline
             & \multicolumn{5}{c}{Signal \SignalSix ~$\cong 35 \times \Pmer$ } \\
 Rank            &                   & (6)                 &                   & (7)              &                  \\
$P              $&                   &$16.753 \pm0.027$  \DW&                   &$8.466\pm0.016  $&                   \\
$t_{\mathrm{min,1}}$&                  &$1759.19\pm0.23     $&                   &$1707.10\pm0.35 $&                   \\
$t_{\mathrm{max,1}}$&                  &$1754.70\pm0.26     $&                   &$1702.86\pm0.35 $&                   \\       
$P/\Pmer         $&                  &$69.57 \pm0.11  $\DW &                   &$35.158\pm0.066 $&                   \\
$P/\Pven         $&                  &$27.232\pm0.044 $\DW &                   &$13.761\pm0.026 $&                   \\
$P/\Pjup         $&                  &$1.4123\pm0.0023$\DW &                   &$0.7137\pm0.0013$&                   \\
       \hline
             &  \multicolumn{5}{c}{Signal \SignalTwo ~$\cong 41.5 \times \Pmer \cong 10 \times \Pear$} \\
Rank             & (2)               & (2)                  & (2)               & (2)             & (2)             \\
$P              $&$10.0001\pm0.0081$ &$20.0062\pm0.0088$\DW &$10.0659\pm0.0077 $&$9.975\pm0.017  $&$10.058\pm0.026 $\\
$t_{\mathrm{min,1}}$&$1754.16\pm0111  $&$1754.284\pm0.063     $&$1823.52\pm0.12  $&$1704.63\pm0.35 $&$1823.55\pm0.25 $\\
$t_{\mathrm{max,1}}$&$1749.16\pm0.11  $&$1749.282\pm0.067     $&$1818.48\pm0.12  $&$1709.62\pm0.34 $&$1818.52\pm0.26 $\\                             
$P/\Pmer        $&$41.529\pm0.034   $&$83.082\pm0.036   $\DW&$41.802\pm0.032   $&$41.424\pm0.070 $&$41.77\pm0.11   $\\
$P/\Pven        $&$16.255\pm0.013   $&$32.520\pm0.014   $\DW &$16.362\pm0.012   $&$16.214\pm0.028 $&$16.349\pm0.042 $\\
$P/\Pjup        $&$0.84304\pm0.00068$&$1.68658\pm0.00074$\DW &$0.84858\pm0.00065$&$0.8409\pm0.0014$&$0.8479\pm0.0022$\\
       \hline
                 & \multicolumn{5}{c}{Signal \SignalFive ~$\cong 44 \times \Pmer$}    \\
Rank             & (5)               & (5)                  &                   & (3)             &                 \\
$P              $&$10.569\pm0.020   $&$10.5407\pm0.0060    $&                   &$10.659\pm0.031 $&                 \\
$t_{\mathrm{min,1}}$&$1753.03\pm0.27  $&$1754.40\pm0.13      $&                   &$1709.27\pm0.44 $&                 \\
$t_{\mathrm{max,1}}$&$1758.31\pm0.26  $&$1758.307\pm0.091    $&                   &$1703.94\pm0.46 $&                 \\
$P/\Pmer        $&$43.891\pm0.083   $&$43.774\pm0.025      $&                   &$44.26\pm0.13   $&                 \\
$P/\Pven        $&$17.180\pm0.032   $&$17.1338\pm0.0098    $&                   &$17.326\pm0.050 $&                 \\
$P/\Pjup        $&$0.89100\pm0.0017$&$0.88861\pm0.00050    $&                   &$0.8986\pm0.0026$&                 \\
       \hline
                 & \multicolumn{5}{c}{Signal \SignalOne ~$\cong 11\times \Pear$} \\
Rank             & (1)                & (1)                  & (1)               & (1)              & (1)             \\
$P               $&$11.0033\pm0.0064 $&$10.9878\pm0.0051    $&$10.8585\pm0.0048 $&$10.981\pm0.020  $&$10.863\pm0.022 $\\
$t_{\mathrm{min,1}}$&$1755.897\pm0.076 $&$1756.678\pm0.077    $&$1823.580\pm0.053 $&$1700.86\pm0.24  $&$1823.56\pm0.20 $\\
$t_{\mathrm{max,1}}$&$1750.395\pm0.078 $&$1750.262\pm0.087    $&$1818.150\pm0.055 $&$1706.35\pm0.23  $&$1828.99\pm0.19 $\\
$P/\Pmer         $&$45.695\pm0.026   $&$45.630\pm0.021      $&$45.093\pm0.020   $&$45.602\pm0.083 $&$45.112\pm0.091 $\\
$P/\Pven         $&$17.886\pm0.010   $&$17.8605\pm0.0083    $&$17.6504\pm0.0078 $&$17.849\pm0.032 $&$17.658\pm0.036 $\\
$P/\Pjup         $&$0.92761\pm0.00054$&$0.92630\pm0.00043   $&$0.91540\pm0.00040$&$0.9257\pm0.0017$&$0.9158\pm0.0018$\\
       \hline
             & \multicolumn{5}{c}{Signal \SignalThree ~$\cong 49 \times \Pmer \cong 1\times \Pjup$} \\
                 & (3)               & (4)                  & (4)               & (5)             & (4)             \\
$P              $&$11.806\pm0.012   $&$11.770\pm0.011      $&$11.863\pm0.021   $&$11.820\pm0.027 $&$11.856\pm0.068 $\\
$t_{\mathrm{min,1}}$&$1760.33\pm0.13  $&$1749.51\pm0.15      $&$1818.92\pm0.20   $&$1701.01\pm0.41 $&$1818.95\pm0.69 $\\
$t_{\mathrm{max,1}}$&$1754.42\pm0.14  $&$1754.40\pm0.21      $&$1824.85\pm0.19   $&$1706.92\pm0.41 $&$1824.88\pm0.66 $\\       
$P/\Pmer        $&$49.028\pm0.050   $&$48.879\pm0.046      $&$49.265\pm0.087   $&$49.09\pm0.11   $&$49.24\pm0.28   $\\
$P/\Pven        $&$19.190\pm0.020   $&$19.132\pm0.018      $&$19.283\pm0.034   $&$19.213\pm0.044 $&$19.27\pm0.11   $\\
$P/\Pjup$        &$0.9953\pm0.0010  $&$0.99224\pm0.00093   $&$1.0001\pm0.0018  $&$0.9964\pm0.0023$&$0.9995\pm0.0057$\\
      \hline
                  & \multicolumn{5}{c}{Signal \SignalSeven $\cong 4.5 \Pjup$} \\
Rank              & (6)               &                    &                   & (6)             &                   \\
$P               $&$52.66\pm0.27     $&                    &                   &$53.83\pm0.76   $&                   \\
$t_{\mathrm{min,1}}$&$1759.63\pm0.70   $&                    &                   &$1704.5\pm2.7   $&                   \\
$t_{\mathrm{max,1}}$&$1785.94\pm0.60   $&                    &                   &$1731.4\pm2.3   $&                   \\
$P/\Pmer         $&$218.7\pm1.1      $&                    &                   &$223.5\pm3.2    $&                   \\
$P/\Pven         $&$85.60\pm0.44     $&                    &                   &$87.5\pm1.2     $&                   \\
$P/\Pjup         $&$4.439\pm0.023    $&                    &                   &$4.538\pm0.064  $&                   \\
      \hline
             & \multicolumn{5}{c}{Signal \SignalNine $\cong 5.5 \times \Pjup$ or $6.0\times \Pjup$} \\
Rank              &                   & (7)                &                   & (8)              &                  \\
$P               $&                   &$143.39\pm0.99   $\DW&                   &$66.7\pm1.4      $&                  \\
$t_{\mathrm{min,1}}$&                   &$1887.84\pm0.80    $&                   &$1753.0\pm3.8    $&                  \\
$t_{\mathrm{max,1}}$&                   &$1784.2\pm1.1      $&                   &$1719.6\pm4.4    $&                  \\      
$P/\Pmer         $&                   &$595.5\pm4.1   $\DW &                   &$277.0\pm5.8     $&                  \\
$P/\Pven         $&                   &$233.1\pm1.6   $\DW &                   &$108.4\pm2.3     $&                  \\
$P/\Pjup         $&                   &$12.088\pm0.083$\DW &                   &$5.62\pm0.12     $&                  \\
       \hline
             & \multicolumn{5}{c}{Signal \SignalFour $\cong$ Synodic periods of Equation \ref{EqConnections} (Table \ref{TableConnections})} \\
                 & (4)               & (3)                  & (3)               & (4)             & (3)             \\
$P              $&$99.92\pm0.57     $&$104.23\pm0.68       $&$116.7\pm1.3      $&$101.4\pm2.4    $&$115.6\pm3.6    $\\
$t_{\mathrm{min,1}}$&$1811.99 \pm0.84 $&$1811.37\pm0.68      $&$1911.02\pm0.74   $&$1710.5\pm3.4   $&$1910.4\pm2.1   $\\
$t_{\mathrm{max1,}}$&$1762.0\pm1.1    $&$1844.00\pm0.48      $&$1852.7\pm1.1     $&$1761.2\pm2.5   $&$1852.6\pm2.6   $\\
$P/\Pmer        $&$415.0\pm2.4      $&$432.8\pm2.8         $&$484.6\pm5.4      $&$421\pm10       $&$480\pm15       $\\
$P/\Pven        $&$162.42\pm0.93    $&$169.4\pm1.1         $&$189.7\pm2.1      $&$164.8\pm3.9    $&$187.9\pm5.8    $\\
$P\Pjup         $&$8.424\pm0.048    $&$8.787\pm0.057       $&$9.84\pm0.11      $&$8.55\pm0.20    $&$9.74\pm0.30    $\\
       \hline
     \end{tabular}
     \end{center}
      \renewcommand{\arraystretch}{1.00}
      \addtolength{\tabcolsep}{+0.05cm}
    \end{scriptsize}
            \end{table}
%
   
\begin{table}
  \caption{Detected planetary signal candidates. 
    (1) Sample ($K_2=1\equiv$ pure sine model,
    $K_2=2\equiv$ double wave model), (2) Time span.
    (3) Period. (4) Rounds (Equations \ref{EqPP} and \ref{EqPPsigma}).
    (5) Rounds deviation (Equation \ref{EqPPzero}).
    (6) Relative rounds deviation (Equation \ref{EqRelativeError}).
    (7) Mean anomaly
    migration $\Delta M/360^{\mathrm{o}}$ (Equation \ref{EqMigration}).
    (8) Planet revolutions during $\Delta T$.
    (9) Relative mean anomaly
    migration $\Delta M/360^{\mathrm{o}}$ (Equation \ref{EqRelMigr}).
    (10) Displaying figure and symbol colour.
  }
\label{TablePromising}
\begin{scriptsize}
\addtolength{\tabcolsep}{-0.09cm}
\renewcommand{\arraystretch}{0.90}
\begin{center}
\begin{tabular}{lccccccccl}
  & & & & & & & \\
  \hline
  (1) & (2) & (3) & (4) & (5) & (6) & (7) & (8) & (9) & (10) \\
       
  Sample ($K_2$) 
  & $\Delta T$
  & $P\pm\sigma_P$
      & $\mathcal{P}\pm\sigma_{\mathcal{P}}$
        & $\Delta \mathcal{P}$
        & $\Delta \mathcal{P}_{\mathrm{rel}}$
          & ${{\Delta M}\over{360^{\mathrm{o}}}}$
            & ${{\Delta T}\over{P_{\mathrm{Planet}}}}$
      & ${{\Delta M_{\mathrm{rel}}}\over {360^{\mathrm{o}}}}$
      & Fig. \\
  & (y) & (y) & (-) & (-) & (-) & (-) & (-) & (-) & \\ 
  \hline
  \multicolumn{10}{c}{Signal \SignalEight: ~$\cong 13 \times \Pven$}          \\
\hline
\RmonthlyOne $(K_2=1)$ & 273.8 & $8.1087\pm0.0076$& $13.180\pm0.012$& 0.180 & 15.0 & 6.1  & 445 & 0.014  & \ref{FigAnomaliesOne}a: red \\
\RmonthlyOne $(K_2=2)$ & 273.8 &  $8.169\pm0.014 $& $13.279\pm0.023$& 0.279 & 12.1 & 9.4  & 445 & 0.021  & \ref{FigAnomaliesOne}a: blue \\
\CmonthlyOne $(K_2=1)$ & 204.8 & $8.009\pm0.017  $& $13.018\pm0.028$& 0.018 & 0.90 & 0.46 & 333 & 0.0014 & \ref{FigAnomaliesOne}a: green \\
\CyearlyOne  $(K_2=1)$ & 203.0 & $8.005\pm0.058  $& $13.012\pm0.094$& 0.012 & 0.13 & 0.30 & 330 & 0.00092& \ref{FigAnomaliesOne}a: cyan  \\  
\hline
  \multicolumn{10}{c}{Signal \SignalEight: ~$\cong 8 \times \Pear$}          \\
\hline
\RmonthlyOne $(K_2=1)$ & 273.8 & $8.1087\pm0.0076$& $8.1087\pm0.0076$&0.1087 & 14.3  & 3.7  & 273.8 & 0.013  & \ref{FigAnomaliesOne}b: red \\
\RmonthlyOne $(K_2=2)$ & 273.8 & $8.169\pm0.014  $& $8.169\pm0.014  $&0.169  & 12.1  & 5.7  & 273.8 & 0.021  & \ref{FigAnomaliesOne}b: blue \\
\CmonthlyOne $(K_2=1)$ & 204.8 & $8.009\pm0.017  $& $8.009\pm0.017  $&0.009  & 0.53  & 0.23 & 204.8 & 0.0011 & \ref{FigAnomaliesOne}b: green \\b
\CyearlyOne  $(K_2=1)$ & 203.0 & $8.005\pm0.0058 $& $8.005\pm0.058  $&0.005  & 0.086 & 0.13 & 203.0 & 0.00062& \ref{FigAnomaliesOne}b: cyan \\
\hline
  \multicolumn{10}{c}{Signal \SignalSix: ~$\cong 35 \times \Pmer$}          \\
\hline
\RyearlyOne $(K_1=1)$  & 321.0 & $8.466\pm016    $& $35.158\pm0.066 $& 0.158 & 2.4  & 6.0   &1333 & 0.0045   & None                           \\
\hline
  \multicolumn{10}{c}{Signal \SignalTwo: ~$41.5 \times \Pmer$}          \\
\hline
\RmonthlyOne $(K_2=1)$ & 273.8 & $10.0001\pm0.0081  $&$41.529\pm0.034$& 0.029   & 0.85& 0.79 &1137 & 0.00070 & \ref{FigAnomaliesOne}c: red \\
\CmonthlyOne $(K_2=1)$ & 204.8 &$10.0659\pm0.0077   $&$41.802\pm0.032  $& 0.302 & 9.4 & 6.2  & 851 & 0.0072  & \ref{FigAnomaliesOne}c: green \\
\RyearlyOne  $(K_2=1)$ & 321.0 &$9.975\pm0.017      $&$41.424\pm0.070  $&-0.076 & 1.1 & -2.4 &1333 & 0.0018  & \ref{FigAnomaliesOne}c: yellow \\
\CyearlyOne  $(K_2=1)$ & 203.0 &$10.058\pm0.026     $&$41.77\pm0.11    $&0.27   & 2.4 & 5.4  & 843 & 0.0065  & \ref{FigAnomaliesOne}c: cyan  \\
\hline
  \multicolumn{10}{c}{Signal \SignalTwo: ~$\cong 10.0 \times \Pear$}          \\
\hline
\RmonthlyOne $(K_2=1)$ & 273.8 & $10.0001\pm0.0081  $&$10.0001\pm0.0081$&0.0001 &0.012&0.0027&273.8 & 0.000010 & \ref{FigAnomaliesOne}d: red \\
\CmonthlyOne $(K_2=1)$ & 204.8&$10.0659\pm0.0077   $&$10.0659\pm0.0077 $&0.0659 & 8.6 & 1.3  &204.8 &0.0065    & \ref{FigAnomaliesOne}d: green \\
\RyearlyOne  $(K_2=1)$ & 321.0 &$9.975\pm0.017      $&$9.975\pm0.017   $&-0.025 & 1.5 & -0.80 &321 & 0.0025    &  \ref{FigAnomaliesOne}d: yellow \\
\CyearlyOne  $(K_2=1)$ & 203.0 &$10.058\pm0.026     $&$10.058\pm0.026  $&0.058  & 2.3 & 1.2   &203 & 0.0058    &  \ref{FigAnomaliesOne}d: cyan \\
\hline
  \multicolumn{10}{c}{Signal \SignalFive: ~$\cong 44 \times \Pmer$}          \\
\hline
\RmonthlyOne $(K_2=1)$ & 273.8 &$10.569\pm0.020     $&$43.891\pm0.083    $& -0.109 & 1.3 & -2.8 & 1137 & 0.0025 &\ref{FigAnomaliesTwo}a: red \\ 
\RmonthlyOne $(K_2=2)$ & 273.8 &$10.5407\pm0.0060   $&$43.774\pm0.025    $& -0.226 & 9.0 & -5.9 & 1137 & 0.0052 &\ref{FigAnomaliesTwo}a: blue  \\
\RyearlyOne  $(K_2=1)$ & 321.0 &$10.659\pm0.031     $&$44.26\pm0.13      $&  0.26  & 2.0 & 7.8 & 1333  & 0.0059 &\ref{FigAnomaliesTwo}a: yellow \\
\hline
  \multicolumn{10}{c}{Signal \SignalOne: ~$\cong 11.0 \times \Pear$}          \\
\hline
\RmonthlyOne $(K_2=1)$ & 273.8 &$11.0033\pm0.0064  $&$11.0033\pm0.0064   $& 0.0033 & 0.52& 0.082 & 273.8 & 0.00030 &\ref{FigAnomaliesTwo}b: red \\
\RmonthlyOne $(K_2=2)$ & 273.8 &$10.9878\pm0.0051  $&$10.9878\pm0.0051   $& -0.0122&  2.4 & -0.30 & 273.8 & 0.0011 &\ref{FigAnomaliesTwo}b: blue \\
\CmonthlyOne $(K_2=1)$ & 204.8 &$10.8585\pm0.0048  $&$10.8585\pm0.0048   $& -0.1415&  29  & -2.7  & 204.8 & 0.013  &\ref{FigAnomaliesTwo}b: green \\
\RyearlyOne  $(K_2=1)$ & 321.0 &$10.981 \pm0.020   $&$10.981 \pm0.020    $& -0.019 &   0.95& 0.56 & 321.0  & 0.0017&\ref{FigAnomaliesTwo}b: yellow   \\
\CyearlyOne  $(K_2=1)$ & 203.0 &$10.863 \pm0.022   $&$10.863 \pm0.022    $& -0.137 & 6.2   & -2.6 & 203.0  & 0.013 &\ref{FigAnomaliesTwo}b: cyan   \\
  \hline
  \multicolumn{10}{c}{Signal \SignalThree: ~$\cong 49.0 \times \Pmer$}          \\  
\hline
\RmonthlyOne $(K_2=1)$ & 273.8 &$11.806\pm0.012   $&$49.028\pm0.050      $&0.028 & 0.56 & 0.65  & 1137& 0.00057  &\ref{FigAnomaliesTwo}c: red \\
\RmonthlyOne $(K_2=2)$ & 273.8 &$11.770\pm0.0.011 $&$48.879\pm0.046      $&-0.121& 2.6  & -2.8  & 1137& 0.0025   &\ref{FigAnomaliesTwo}c: blue \\
\CmonthlyOne $(K_2=1)$ & 204.8 &$11.863\pm0.021   $&$49.265\pm0.087      $&0.265 & 3.0  & 4.6   & 850 & 0.0054   &\ref{FigAnomaliesTwo}c: green \\
\RyearlyOne  $(K_2=1)$ & 321.0 &$11.820\pm0.027   $&$49.09  \pm0.11      $&0.09  & 0.82 & 2.4   & 1333& 0.0018   &\ref{FigAnomaliesTwo}c: yellow \\
\CyearlyOne  $(K_2=1)$ & 203.0 &$11.856\pm0.068   $&$49.24\pm0.28        $&0.24  & 0.86 & 4.1   & 843 & 0.0049   &\ref{FigAnomaliesTwo}c: cyan \\
\hline
  \multicolumn{10}{c}{Signal \SignalThree: ~$\cong 1.0 \times \Pjup$}          \\  
\hline
\RmonthlyOne $(K_2=1)$ & 273.8 &$11.806\pm0.012   $&$0.9953\pm0.0010     $&-0.0047 & 4.7  & -0.11 & 23 & 0.0047   &\ref{FigAnomaliesTwo}d: red \\
\RmonthlyOne $(K_2=2)$ & 273.8 &$11.770\pm0.011    $&$0.99224\pm0.00093   $&-0.0078 & 8.3  & -0.18 & 23 & 0.0078  &\ref{FigAnomaliesTwo}d: blue \\
\CmonthlyOne $(K_2=1)$ & 204.8 &$11.863\pm0.021    $&$1.0001\pm0.0018     $& 0.00010& 0.056& 0.0017& 17 & 0.00010 &\ref{FigAnomaliesTwo}d: green \\
\RyearlyOne  $(K_2=1)$ & 321.0&$11.820\pm0.027     $&$0.9964\pm0.0023     $&-0.0036 & 1.6  & -0.098& 27 & 0.0036  &\ref{FigAnomaliesTwo}d: yellow \\
\CyearlyOne  $(K_2=1)$ & 203.0&$11.856\pm0.068     $&$0.9995\pm0.0057     $&-0.00050& 0.088& -0.0086&17 & 0.00050 &\ref{FigAnomaliesTwo}d: cyan \\
\hline
  \multicolumn{10}{c}{Signal \SignalSeven: ~$\cong 4.5 \times \Pjup$}          \\  
\hline
\RmonthlyOne $(K_2=1)$ & 273.8 &$52.66\pm0.27    $&$4.439\pm0.023      $&-0.061  & 2.6  & -0.32  & 23 & 0.014     & None  \\
\RyearlyOne  $(K_2=1)$ & 321.0 &$53.83\pm0.76    $&$4.538\pm0.064      $& 0.038  & 0.59 &  0.23  & 27 & 0.0084    & None  \\
\hline
  \multicolumn{10}{c}{Signal \SignalNine: ~$\cong 5.5 \times \Pjup$}            \\  
\hline
  \RyearlyOne  $(K_2=1)$ & 321.0 &$66.7\pm1.4      $&$5.62\pm0.12     $& 0.12  & 1.0  & 0.57    & 27 & 0.021      & None\\
\hline
\end{tabular}
\end{center}
\addtolength{\tabcolsep}{+0.04cm}
\renewcommand{\arraystretch}{1.00}
\end{scriptsize}
\end{table}

We have introduced the following nine signal abbreviations
\SignalOne,
\SignalTwo,
\SignalThree,
\SignalFour,
\SignalFive,
\SignalSix,
\SignalSeven,
\SignalEight ~and
\SignalNine ~in Table \ref{TableCompare},
where these signals
were arranged in
the order of decreasing strength,
or equivalently, decreasing signal amplitude.
The numerical value given
in the signal subscript refers
to the approximate
signal period in Earth years.

These nine signals are now rearranged into
the order of increasing periods
\SignalEight,  \SignalSix, \SignalTwo,
\SignalFive, \SignalOne, \SignalThree,
\SignalSeven, \SignalNine ~and \SignalFour
~in our next Tables \ref{TableMultiples}
and \ref{TablePromising}.
We study only the signals detected
in all data samples,
because we get the same results for 
the predictive data samples
(i.e., the subsets of all data samples).
We select only those all data samples,
where DCM detects more than two signals.
The period $(P)$,
the primary minimum epoch $(t_{\mathrm{min,1}})$
and the primary maximum epoch $(t_{\mathrm{min,1}})$
of all
\SignalEight,  \SignalSix, \SignalTwo,
\SignalFive, \SignalOne, \SignalThree,
\SignalSeven, \SignalNine ~and \SignalFour ~signals
are given in Table \ref{TableMultiples}.
The units of these three parameters are Earth years.

We are accustomed to measuring time and periodicity
in Earth years.
The Earth year signal
periods detected in all data samples
are transformed into Mercury, Venus and Jupiter year signal
periods in Tables \ref{TableMultiples} and \ref{TablePromising}.
It is easier to identify integer multiple  periods
from these values.
The known planet period
$P_{\mathrm{Planet}}$  
in Earth years is used to compute
four dimensionless parameters
$\mathcal{P}$,
$\sigma_{\mathcal{P}}$,  
$\Delta \mathcal{P}$ and 
$\Delta \mathcal{P}_{\mathrm{rel}}$
(Equations \ref{EqPP} -\ref{EqRelativeError}).

For any detected period $P\pm\sigma_P$, the rounds error is
$\sigma_{\mathcal{P}} \propto \sigma_P/P_{\mathrm{Planet}}$.
This rounds error $\sigma_{\mathcal{P}}$
is larger for the inner planets Mercury and Venus
having having $P_{\mathrm{Planet}}< \Pear=1$,
and smaller for Jupiter having $P_{\mathrm{Planet}}> \Pear=1$.
For example, the period $P=8.1087\pm0.0076$ gives the following
$\sigma_{\mathcal{P}}=0.032$ (Mercury),
$\sigma_{\mathcal{P}}=0.012$ (Venus),
$\sigma_{\mathcal{P}}=0.0076$ (Earth)
and
$\sigma_{\mathcal{P}}=0.000064$ (Jupiter)
rounds error values
(Table \ref{TableMultiples}: \RmonthlyOne, $K_2=1$ model, signal \SignalEight).
This means that the period of Jupiter is easiest to
identify from the $\mathcal{P}$ rounds values.
On the other hand, the identifications of Mercury and 
Venus from the $\mathcal{P}$ rounds values
are the most difficult ones, and especially in this order.
Furthermore, the correct $\mathcal{P}_0$ rounds value
becomes uncertain,
if the round error $\sigma_{\mathcal{P}}$ exceeds 1/2,
i.e. half a revolution around the Sun.
This complication effect is strongest for the longest
detected
periods $P$, which have the largest $\sigma_P$ errors
in Tables \ref{TableMultiples} and \ref{TablePromising}.

The larger rounds error $\sigma_{\mathcal{P}}$ values
of Mercury and Venus
cause another effect that makes
the identification of their signals
even more difficult.
If the rounds deviation 
$\Delta \mathcal{P}$ values
are drawn from
a Gaussian distribution
having a mean $\mathcal{P}_0$ and
a standard deviation $\sigma_{\mathcal{P}}$,
the probability
\begin{eqnarray}
  P(|\Delta \mathcal{P}| \ge |\Delta \mathcal{P}'|)
\label{EqProbP}
\end{eqnarray}
for $|\Delta \mathcal{P}|$ exceeding
any fixed
large rounds deviation
$|\Delta \mathcal{P}'|$
value
 increases for
 larger rounds error $\sigma_{\mathcal{P}}$ values.
 In other words, large  $|\Delta \mathcal{P}|$ values
 are more probable for Mercury and Venus
 than for the Earth and Jupiter.
This increases the probability
of stronger mean anomaly migration $\Delta M$
for  Mercury and Venus
(see Equation \ref{EqMigration}).

The primary
minima and the primary maxima of the signals
in Table \ref{TableMultiples}
occur at multiples
\begin{eqnarray}
T_{\mathrm{min,k}} & = & t_{\mathrm{min,1}} + k P   \label{EqTMIN} \\
T_{\mathrm{max,k}} & = & t_{\mathrm{max,1}} + k P, \label{EqTMAX}
\end{eqnarray}
where $k=1, 2, ...$ are integers.
The units of $P$, $t_{\mathrm{min,1}}$ and $t_{\mathrm{max,1}}$
are Earth years. The relation
$P = (\mathcal{P}_0 + \Delta \mathcal{P}) P_{\mathrm{Planet}}$ gives
\begin{eqnarray}
  T_{\mathrm{min,k}} & = &
  t_{\mathrm{min,1}}+
  k P =
  t_{\mathrm{min,1}}+
                           k ~(\mathcal{P}_0
  +\Delta \mathcal{P}) ~P_{\mathrm{Planet}} =
   t_{\mathrm{min,1}}+ 
                           k ~\mathcal{P}_0 ~P_{\mathrm{Planet}}
  + k ~\Delta \mathcal{P} ~ P_{\mathrm{Planet}}
                            \nonumber \\
 T_{\mathrm{max,k}} & = &
  t_{\mathrm{max,1}}+
  k P =
  t_{\mathrm{max,1}}+
  k ~(\mathcal{P}_0 +\Delta \mathcal{P}) ~P_{\mathrm{Planet}} =
   t_{\mathrm{max,1}}+ 
            k ~\mathcal{P}_0 ~P_{\mathrm{Planet}}
            + k ~\Delta \mathcal{P} ~ P_{\mathrm{Planet}}
  \nonumber
\end{eqnarray}
The last term
\begin{eqnarray}
  \Delta t =k ~\Delta \mathcal{P} ~P_{\mathrm{Planet}}
  \label{EqDeltat}
\end{eqnarray}
is the time difference between the multiples
$t_{\mathrm{min,1}}+k~P$ and
$ t_{\mathrm{min,1}}+ k ~\mathcal{P}_0  ~P_{\mathrm{Planet}}$,
and between the multiples
$t_{\mathrm{max,1}}+k~P$ and 
$ t_{\mathrm{max,1}}+ k ~\mathcal{P}_0  ~P_{\mathrm{Planet}}$.

The data time span $\Delta T$ contains
$\Delta T/P$ multiples of $P$.
Using $k= \Delta T/P $ in Equation \ref{EqDeltat} gives
\begin{eqnarray}
  \Delta t =
 {{\Delta \mathcal{P} \Delta T}\over{P}}  ~P_{\mathrm{Planet}}
  \nonumber
\end{eqnarray}
time difference
during $\Delta T$.
We convert time difference $\Delta t$
into planet ``mean anomaly migration''
\begin{eqnarray}
  \Delta M =
  {{\Delta t} \over {P_{\mathrm{Planet}}}} \times 360^{\mathrm{o}}
  = {{\Delta \mathcal{P}  \Delta T} \over{P}} \times 360^{\mathrm{o}},
\label{EqMigration}
\end{eqnarray}
where the units of $\Delta M$ are degrees.
For all $\Delta T$ and $P$ combinations,
this relation shows that larger mean anomaly
$\Delta M$ migration occurs
for larger $\Delta \mathcal{P}$ values,
which are most probable for Mercury and Venus
(Equation \ref{EqProbP}).
The ``relative mean anomaly migration''
during  each $P_{\mathrm{Planet}}$ revolution is
\begin{eqnarray}
  {\Delta M_{\mathrm{rel}}} =
  |\Delta M| / ({ {\Delta T} \over {P_{\mathrm{Planet}}}}) \times 360^{\mathrm{o}}
  =  |\Delta \mathcal{P}| {{~P_{\mathrm{Planet}}} \over {P}} \times 360^{\mathrm{o}}.
\label{EqRelMigr}
\end{eqnarray}

  For any particular signal,
  the period $P$ detected
  in each sample determines the linear mean anomaly migration
  $\Delta M$ during the
  $\Delta T$ time span of the data (Equation \ref{EqMigration}).
  The slope of this linear $M(t)$
  mean anomaly migration is 
  $(\Delta \mathcal{P}/P)\times 360^{\mathrm{o}}$
  (Equation \ref{EqMigration}).
  Let us assume that these mean anomaly migration lines
  \begin{eqnarray}
{{\Delta\mathcal{P}} \over {P}} (t-t_C) \times 360^{\mathrm{o}} 
    = M(t)-M_C
    \label{EqConvergencePoint}
  \end{eqnarray}
  of different samples
  intersect close to a ``convergence point'' $(t_C,M_C)$.
  This means that the mean anomaly
  value of each line is $M(t) \approx M_C$ at epoch $t=t_C$.
  The slope of all these $M(t)$ lines approaches zero,
  if $\Delta \mathcal{P} \rightarrow 0
  \equiv \mathcal{P}     \rightarrow \mathcal{P}_0
  \equiv P \rightarrow P_{\mathrm{planet}}$.
  In this case, all lines fulfil 
\begin{eqnarray}
  M(t) \approx M_C
  \label{EqMC}
\end{eqnarray}
for all $t$ values. 
If the mean anomaly $M(t)$ lines
for the same signal are
detected in different samples,
and these lines show a convergence point $(t_C,M_C)$,
then the periods $P$,
the minimum epochs $t_{\mathrm{min,1}}$
and
the maximum epochs $t_{\mathrm{max,1}}$
of all these signals
can be connected to each other.
The presence of this convergence point 
  confirms that
  the individual signals detected
  in different samples
  can represent the same real signal
  having the same period,
  and the same phase.
The statistical fluctuation of
$|\Delta \mathcal{P}|$ rounds deviations
(Equation \ref{EqProbP}) complicate
the detection of this ``convergence point connection''.
We have already shown that the probability for larger
$|\Delta \mathcal{P}|$ values increases
when $\sigma_{\mathcal{P}}$
increases for smaller
$P_{\mathrm{Planet}}$ periods.
These $|\Delta \mathcal{P}|$
fluctuations are largest for Mercury and Venus.
Therefore, the mean anomaly migration $\Delta M$ lines
of these inner planets are ``messier'',
and their convergence point locations
are more difficult to detect.
Hence, the detection of Jupiter and the Earth
from the {\it same}
sunspot data is easier than
the detection of Venus and Mercury.

The relations of Equations \ref{EqMigration} - \ref{EqMC}
are formulated for the mean anomaly $M$,
because this parameter has a linear connection to time.
We compute the true anomaly
$\nu_{\mathrm{min}}$
and
$\nu_{\mathrm{max}}$
values
from the mean anomaly values $M$ 
for the minimum epochs
$t_{\mathrm{min,1}}$
and the maximum epochs
$t_{\mathrm{max,1}}$
(Equations \ref{EqTMIN} and \ref{EqTMAX})
given in Table \ref{TableMultiples}.
The mean anomaly $M$ and the true anomaly
$\nu$ value differences are small
for Venus, the Earth and Jupiter,
which have nearly circular orbits.
The large Mercury's orbit eccentricity $e=0.20$ can
cause clear $\nu$ deviations from the
linear $M$ trend
(e.g. Fig. \ref{FigAnomaliesTwo}c).

We search for signatures of
Mercury, Venus, the Earth and Jupiter
from the
\SignalEight,  \SignalSix, \SignalTwo,
\SignalFive, \SignalOne, \SignalThree,
\SignalSeven, \SignalNine ~and \SignalFour
~signals.
The results for this ``wide search'' of
36 alternatives are given in
Table \ref{TableMultiples}.
We summarise the most promising
planetary signal candidate detections
in Table \ref{TablePromising}.
The three double sinusoid signals ``\DW''
  are given
  in Table \ref{TableMultiples}.
  We exclude
  these double sinusoids 
  from Table \ref{TablePromising},
  because the minor asymmetries of these
  double sinusoid $2 \times P$
  period curves can mislead the identification
  of convergence points
  (Equation \ref{EqConvergencePoint}).

In our next figures, we plot
the $\nu_{\mathrm{min}}$ and $\nu_{\mathrm{max}}$ values
for numerous sample, signal and planet combinations.
In every figure, we use the same symbol colour for the 
signals detected in the same sample.
All samples end in the same year 2022,
but they begin at different years between
1700 and 1815.
This complicates the detection of convergence points,
if they are located outside some sample(s).
We solve this problem by computing 
the $\nu_{\mathrm{min}}$ and $\nu_{\mathrm{max}}$ values
of all samples for the whole time interval between the
years 1700 and 2022.
The symbols for the
$\nu_{\mathrm{min}}$ and $\nu_{\mathrm{max}}$ values
{\it inside} each sample are highlighted with a black
circle around the coloured symbol.
This black circle is missing from the symbols
representing $\nu_{\mathrm{min}}$ and $\nu_{\mathrm{max}}$
values {\it outside} each sample.
We can use these outside values to check,
if the signals detected in each sample can
predict the presence of a convergence point
before the beginning of this sample.

The detected signals are the
$h_i(t)$ functions (Equation \ref{Eqharmonictwo}) in the 
sum $h(t)$ of all signals (Equation \ref{Eqharmonicone}).
  For the symmetric pure sine model $h_i(t)$ functions,
  the separation between the primary
  minimum epoch $t_{\mathrm{min,1}}$
  and the primary maximum epoch $t_{\mathrm{max,1}}$ 
  is exactly $P/2$.
  If the period of the detected signal $P$ is
  an {\it even} multiple of
  the planet's period $P_{\mathrm{Planet}}$,
  the mean anomaly $M$ values are equal at all minima
  $T_{\mathrm{min,k}}$ and maxima $T_{\mathrm{max,k}}$
  (Equations \ref{EqTMIN} and \ref{EqTMAX}).
  Except for
  the possible deviations caused
  by the planetary orbit eccentricity,
  the $\nu_{\mathrm{min}}$ and $\nu_{\mathrm{max}}$
  values will also be equal. 
  This ``overlapping effect'' occurs in
  Figs.
    \ref{FigAnomaliesOne}b,
    \ref{FigAnomaliesOne}d and
    \ref{FigAnomaliesTwo}a.
  If the detected period $P$ is an {\it uneven}
  $P_{\mathrm{Planet}}$ multiple, the differences between
  the mean anomaly $M$ values of all minima
  $T_{\mathrm{min,k}}$ and maxima $T_{\mathrm{max,k}}$
  are equal to 180 degrees.
  The planetary orbit eccentricity can, again,
  cause deviations from this 180 degrees difference
  between the
  true anomalies
  $\nu_{\mathrm{min}}$ and $\nu_{\mathrm{max}}$.
  This 180 degrees
  ``separation effect'' occurs in Figs.
  \ref{FigAnomaliesOne}a,
    \ref{FigAnomaliesTwo}b,
    \ref{FigAnomaliesTwo}c and
    \ref{FigAnomaliesTwo}d.    
  The asymmetries of double wave models can also cause
  deviations from the overlapping effect and
  180 degrees separation effect,
  like those seen in the blue circles of Figs.
    \ref{FigAnomaliesOne}b,
    \ref{FigAnomaliesTwo}a,
    \ref{FigAnomaliesTwo}b
    and
    \ref{FigAnomaliesTwo}d.

\subsection{Signal $S_{8^{\mathrm{y}}}$} 

The shortest period
\SignalEight ~signal is close to
$13 \times \Pven$
and
 $8 \times \Pear$
(Tables \ref{TableMultiples} and \ref{TablePromising}).
The known period ratio is $8 \times \Pear/\Pven=13.0039$.
Due to statistical $\Delta \mathcal{P}$ fluctuations
(Equations \ref{EqProbP} and \ref{EqMigration}),
the true anomaly migration curves of Venus
in Fig. \ref{FigAnomaliesOne}a should appear less regular
than those of Earth in Fig. \ref{FigAnomaliesOne}b.
This is indeed the case.

\subsubsection{Signal \SignalEight
  ~$\cong 13\times \Pven$} \label{SectVenus}

The two rounds
$\mathcal{P}=13.018\pm0.028$ and $13.012\pm0.094$
values for the \CmonthlyOne ~and \CyearlyOne ~samples
differ
$\Delta \mathcal{P}_{\mathrm{rel}}=0.90$ and 0.13
from $13\times\Pven$ 
(Table \ref{TablePromising}: Signal \SignalEight).
The mean anomaly migration is 
$\Delta M/360^{\mathrm{o}}=0.46$ and 0.30 revolutions
during 205 and 203 years, respectively.
The green and cyan circles highlighted
with black circles show this
weak positive upwards
$\nu_{\mathrm{min}}$ and $\nu_{\mathrm{max}}$
true anomaly migration of Venus inside these
\CmonthlyOne ~and \CyearlyOne ~samples
(Fig. \ref{FigAnomaliesOne}a).

The other two rounds
values for sample
\RmonthlyOne ~(pure sine and double wave models) are
$\mathcal{P}=13.180\pm0.012$ and $13.279\pm0.023$.
They differ $\Delta \mathcal{P}_{\mathrm{rel}}=15.0$ and 12.1
from $13\times\Pven$ (Table \ref{TablePromising}: Signal \SignalEight).
For these two $\mathcal{P}$
rounds values, 
the strong positive upwards
mean anomaly migration $\Delta M/360^{\mathrm{o}}$
  is 6.1 and 9.4 revolutions
  during 274 years $\equiv$ 445 Venus revolutions.
  These values are equal to the relative mean anomaly
  migration
  $\Delta M_{\mathrm{rel}}/360^{\mathrm{o}}=0.014$ and 0.021
  during one revolution of Venus.
  We show the respective true anomaly
  $\nu_{\mathrm{min}}$ and $\nu_{\mathrm{max}}$
  migration in 
  Fig. \ref{FigAnomaliesOne}a (red and blue circles).

  The uneven $13 \times \Pven$ multiple causes the
    180 degree separation effect of $\nu_{\mathrm{min}}$
    and $\nu_{\mathrm{max}}$.
The migration lines of large circles denoting the
$\nu_{\mathrm{max}}$ values
show a convergence point
(Equation \ref{EqConvergencePoint}), which is
highlighted with a large dotted blue circle
at $t_C=1820$ years and $\nu_C=315$ degrees.
The migration lines of smaller circles denoting
the $\nu_{\mathrm{min}}$ values
converge at $t_C=1820$ years and $\nu_C=135$ degrees
in the centre of the red large dotted circle.
These two convergence points are at the beginning
of \CmonthlyOne ~and \CyearlyOne ~samples,
where the black circles begin to highlight
the green and cyan circles.
The detection of the two
convergence points confirms that all four
\SignalEight ~signals in Table \ref{TablePromising}
can  be connected, regardless of the large true
anomaly $\nu$ migration
of the two signals detected in 
\RmonthlyOne ~(pure sine and double wave models).
The apparently messy migration of red and blue circles
in Fig. \ref{FigAnomaliesOne}a can arise,
as expected, from the statistical
$\Delta \mathcal{P}$ fluctuations
caused by the short orbital period
of Venus (Equation \ref{EqProbP}).

\begin{figure}  
\centerline{\hspace*{0.015\textwidth}
          \includegraphics[width=0.515\textwidth,clip=]{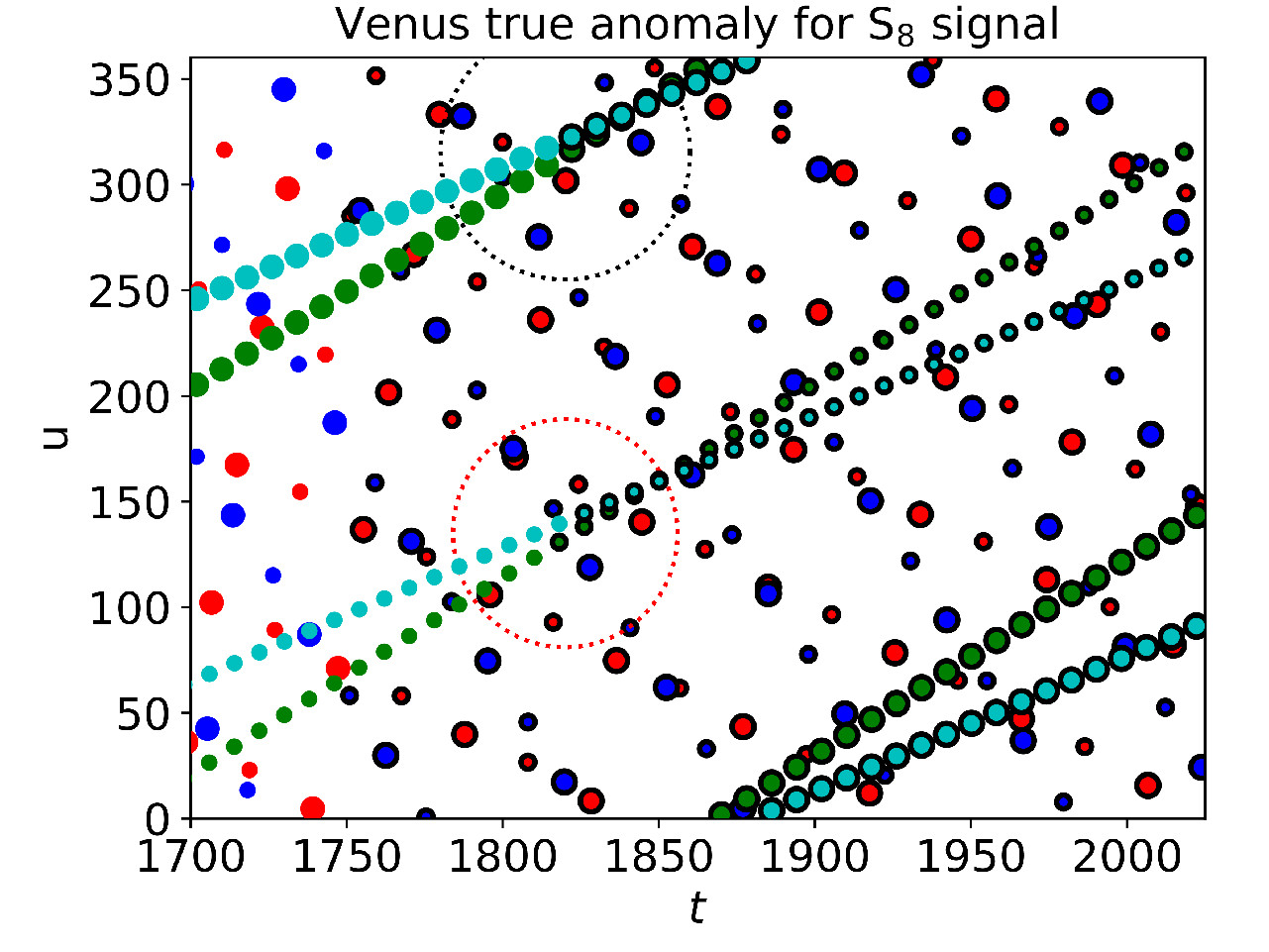} 
         \hspace*{-0.03\textwidth}
         \includegraphics[width=0.515\textwidth,clip=]{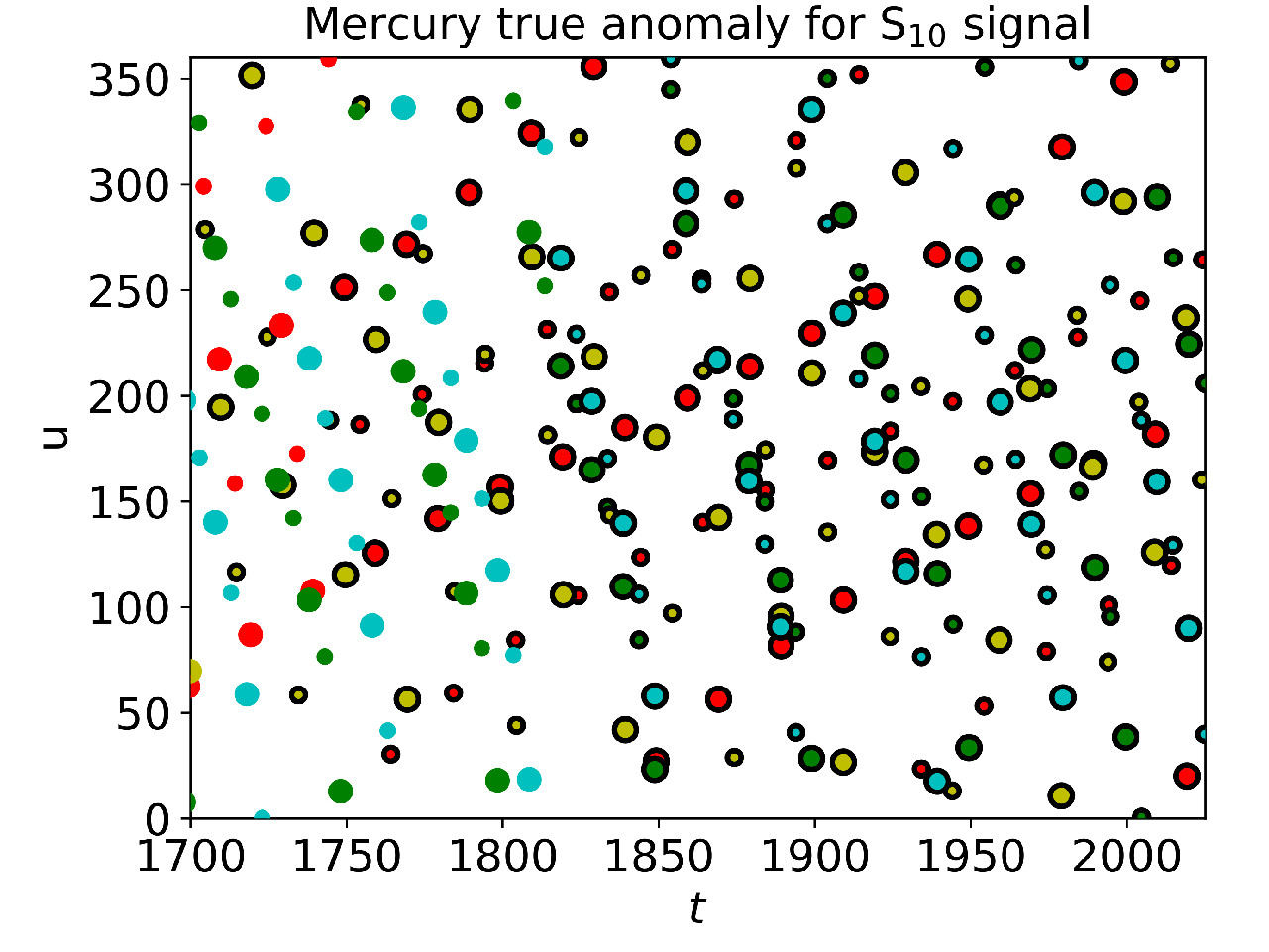} 
        }
\vspace{-0.35\textwidth}
\centerline{\Large \bf 
\hspace{0.42\textwidth}  \color{black}{(a)}
\hspace{0.43\textwidth}  \color{black}{(c)}
\hfill}
\vspace{0.31\textwidth}    
\centerline{\hspace*{0.015\textwidth}
          \includegraphics[width=0.515\textwidth,clip=]{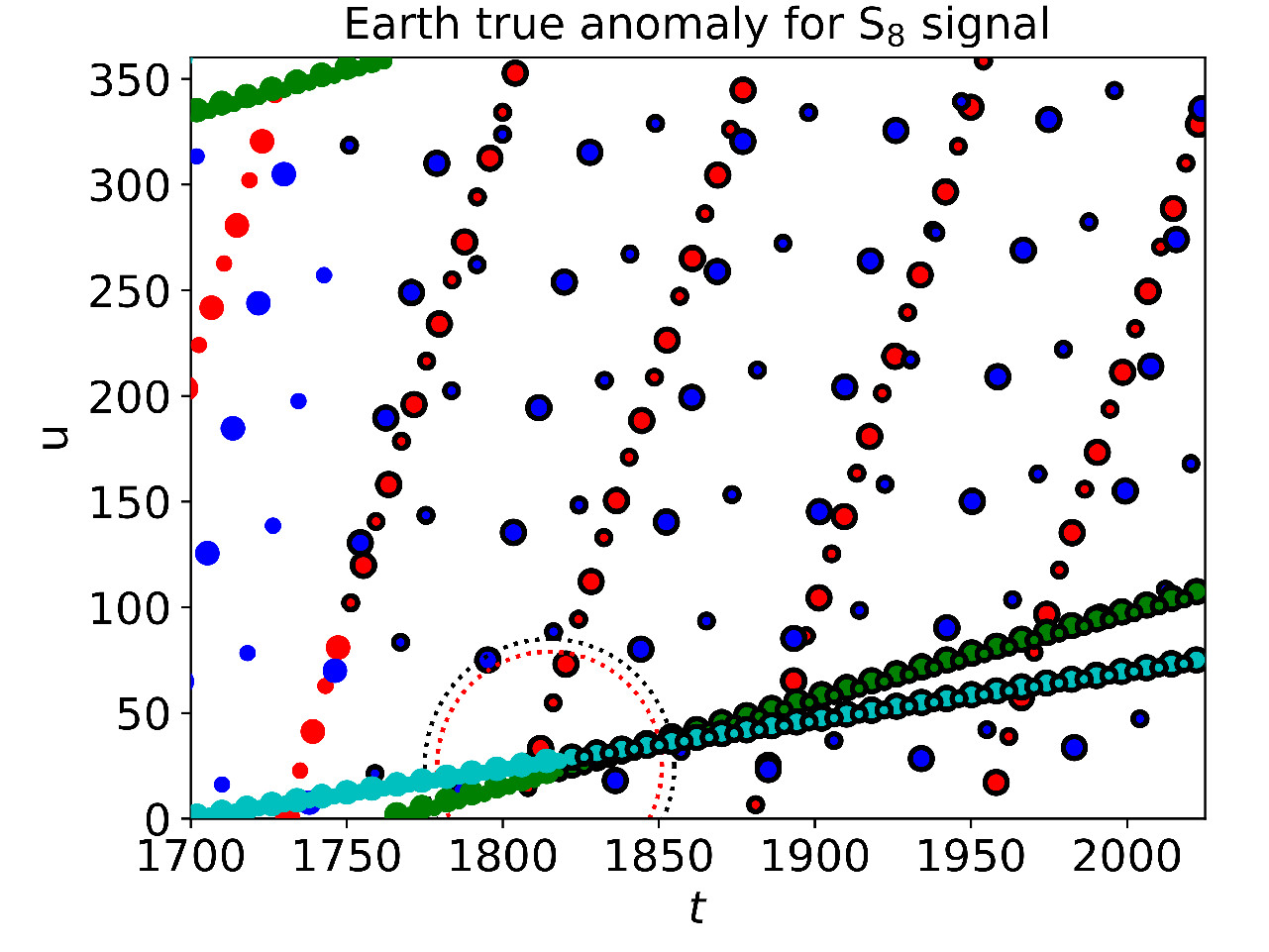} 
         \hspace*{-0.03\textwidth}
          \includegraphics[width=0.515\textwidth,clip=]{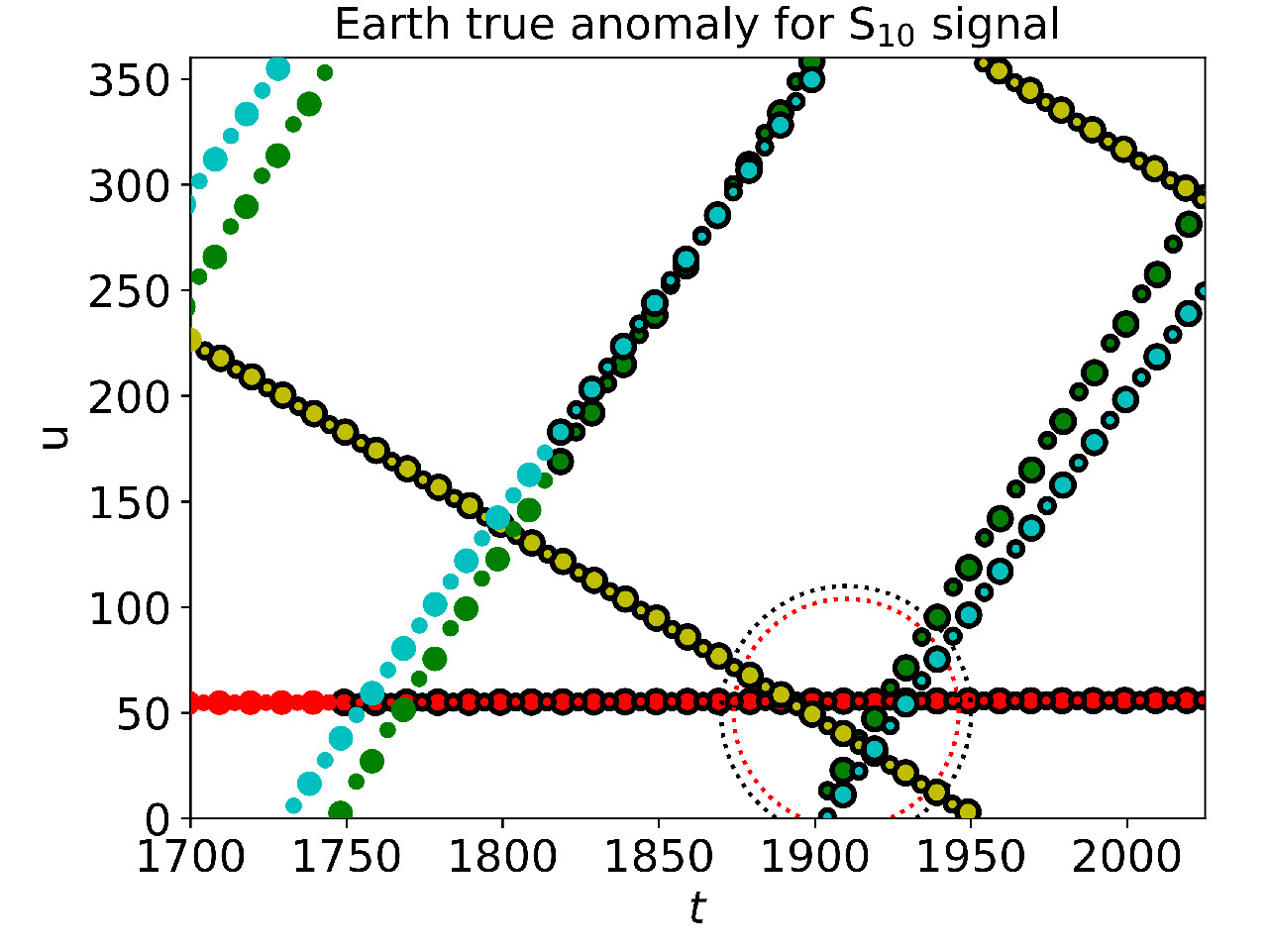}  
        }
\vspace{-0.35\textwidth}   
\centerline{\Large \bf     
\hspace{0.42 \textwidth}   \color{black}{(b)}
\hspace{0.43 \textwidth}   \color{black}{(d)}
   \hfill}
 \vspace{0.30\textwidth}    
 \caption{True anomalies
   $\nu_{\mathrm{min}}$ (small circles)
   and
   $\nu_{\mathrm{max}}$ (large circles)
   for time point multiples
   $T_{\mathrm{min,k}}$ and  $T_{\mathrm{max,k}}$
   (Equations \ref{EqTMIN} and \ref{EqTMAX}).
   We compute these 
   multiples from
   periods $(P)$,
   primary minimum epochs  $(t_{\mathrm{mim,1}})$
   and
   primary maximum epochs $(t_{\mathrm{mim,1}})$
   given in Table \ref{TableMultiples}.
   Symbol colours for different samples are
   red (\RmonthlyOne: pure sines),
   blue (\RmonthlyOne: double waves),
   green (\CmonthlyOne: pure sines),
   yellow (\RyearlyOne: pure sines),
   and
   cyan (\CyearlyOne: pure sines).   
   Symbols for
     $\nu_{\mathrm{min}}$ 
   and
   $\nu_{\mathrm{max}}$
     values inside each sample
     are highlighted with a black circle.
   (a) Venus' true anomalies
   for signal \SignalEight.
   (b) Earth's  true anomalies
   for signal \SignalEight.
   (c) Mercury's true anomalies
   for signal \SignalTwo.
   (d) Earth's true anomalies
   for signal \SignalTwo.
   Units of X-axis and Y-axis are years and degrees,
   respectively. } 
         \label{FigAnomaliesOne}
       \end{figure}

\subsubsection{Signal \SignalEight
  ~$\cong 8 \times \Pear$}

The two $P\pm\sigma_P=8.009\pm0.017$ and $8.005\pm0.058$
Earth year periods 
detected in the \CmonthlyOne ~and \CyearlyOne ~samples
are equal to $8 \times \Pear$ comfortably
within $\pm 1 \sigma_{\mathcal{P}}$
(Table \ref{TablePromising}: Signal \SignalEight).
The mean anomaly migration
$\Delta M/360^{\mathrm{o}}$
is only 0.23 and 0.13 revolutions
in over two centuries
(Fig. \ref{FigAnomaliesOne}b:
green and cyan highlighted circles).

For the other two periods
(\RmonthlyOne: pure sine and double wave models),
the rounds values $\mathcal{P}=8.1087\pm0.0076$ and
$8.169\pm0.014$ differ
$\Delta \mathcal{P}_{\mathrm{rel}}=14.3$
and 12.1 from $8 \times \Pear$
(Table \ref{TablePromising}: Signal \SignalEight).
The respective $\Delta M_{\mathrm{rel}}$
relative mean anomaly migration values are
0.013 and 0.021 during one revolution of Earth.
Hence, these true anomaly 
$\nu_{\mathrm{min}}$ and $\nu_{\mathrm{min}}$ values
of Earth
migrate about 3.7 and 5.7 revolutions
during 274 years
(Fig. \ref{FigAnomaliesOne}b:
red and blue highlighted circles).

 For the three symmetric pure sine models,
 the true anomaly
$\nu_{\mathrm{min}}$ and $\nu_{\mathrm{max}}$ 
lines show the overlapping effect, because 
$8 \times \Pear$ in
an even number multiple
(Fig. \ref{FigAnomaliesOne}b:
red, green and cyan symbols).
Due to the asymmetric signal
\SignalEight ~shape of the double
wave model for sample \RmonthlyOne,
the lines of small and large blue circles
do not overlap.

The true anomaly $\nu_{\mathrm{min}}$
and $\nu_{\mathrm{max}}$
migration curves of Earth  converge
in the centre of the blue and red dotted circles
at $t_C=1815$ years and $\nu=25$ degrees
(Fig. \ref{FigAnomaliesOne}b).
Due to the aforementioned double wave signal asymmetry,
the line of large blue circles deviates from the converging
line of small blue circles.
Although the convergence point is {\it outside} samples
\CmonthlyOne ~and \CyearlyOne,
it is nicely covered by the ``predictive''
green and cyan symbols
not highlighted with black circles.
We can identify this convergence point regardless of
  the large true anomaly migration of the pure sine model and
  the double wave model for sample \RmonthlyOne.
This means that all four  \SignalEight ~signals' period $(P)$,
primary minimum $(t_{\mathrm{min,1}})$
and primary maximum $(t_{\mathrm{max,1}})$ values
for different samples in Table \ref{TableMultiples}
can be connected to the same signal. 

We detect the
convergence points for the signal candidates
of both Venus and the Earth.
This indicates that the repetitive
relative motions of these two planets
may cause this \SignalEight ~signal.

\subsection{Signal S$_{8.^{\mathrm{y}}4}$
  $\cong 35 \times \Pmer$}  

This sixth-strongest signal is detected only in sample
\RyearlyOne ~(Table \ref{TablePromising}: Signal \SignalSix).
The rounds value $\mathcal{P}=35.158\pm0.066$ deviates
$\Delta \mathcal{P}_{\mathrm{rel}}=2.4$ from
$35\times\Pmer$.
We want to draw attention to this \SignalSix ~signal,
because the $\mathcal{P}=69.57\pm0.11$
rounds value of another signal,
the double sinusoid signal for
sample \RmonthlyOne ~$(K_2=2)$,
is quite close to $2 \times 35 \times \Pmer$
(Table \ref{TableMultiples}: Signal \SignalSix).
We show no figure for this \SignalSix ~signal candidate
of Mercury, because one migration line 
can not have a convergence point.

\subsection{Signal \SignalTwo}

The second-strongest signal \SignalTwo ~period is
approximately
$41.5 \times \Pmer$
and
$10 \times \Pear$.
The known ratio is
$10 \times \Pear/\Pmer=45.528$.
Considering the statistical
$|\Delta \mathcal{P}|$ fluctuations
(Equations \ref{EqProbP} and \ref{EqMigration}),
the expected migration curves of Mercury
should be much messier than those of the Earth.
These expectations are ``amply rewarded''
when  Figs. \ref{FigAnomaliesOne}c and
 \ref{FigAnomaliesOne}d are compared.

\subsubsection{Signal \SignalTwo
  ~$\cong 41.5 \times \Pmer$}

For clarity, as well as
to avoid possible misunderstandings,
we define $\mathcal{P}_0$ in Equation \ref{EqPPzero} as
the integer closest to $\mathcal{P}$.
The combined tidal forces of two planets
are strongest when they are aligned to
the same line with respect to the Sun.
This can happen when these planets are at the
same side, or at opposite sides,
of the Sun. Therefore, the connection
$\mathcal{P}_0 \cong 41.5 \times \Pmer$
draws our attention.

The true anomaly $\nu_{\mathrm{min}}$ and $\nu_{\mathrm{min}}$
migration curves for Mercury appear truly messy
(Fig. \ref{FigAnomaliesOne}c).
These true anomaly migration curves are
not fully linear due to
the large eccentricity of Mercury's orbit.
However, the reason for the messy impression is not
eccentricity or strong migration,
because even the largest
relative mean anomaly migration is only
$\Delta M_{\mathrm{rel}}=6.2$ revolutions
in a total of 851 revolutions during
over two centuries
~(Table \ref{TablePromising}: Sample \CmonthlyOne).
The $41.5 \times \Pmer$ value divides
all $\nu_{\mathrm{min}}$ and $\nu_{\mathrm{max}}$ true
anomaly curves into two parts.
First part contains uneven multiples of 
$41.5 \times \Pmer$ revolutions, and the
second part contains even multiples of
$83 = 2 \times 41.5 \times \Pmer$ revolutions.
The true anomaly difference between these two
parts is about half a revolution.
This division effect causes the messy impression
in Fig. \ref{FigAnomaliesOne}c.
It  would make no sense to search for a convergence
point (Equation \ref{EqConvergencePoint})
from this mixture of true anomaly lines.

For one sample \CmonthlyOne,
this planetary signal candidate can be questioned,
because
the rounds $\mathcal{P}=41.802\pm0.032$ value
having $\Delta \mathcal{P}_{\mathrm{rel}}=9.4$
differs strongly from $41.5 \times \Pmer$
(Table \ref{TablePromising}: Signal \SignalTwo).
However, the relative rounds deviation values for
the other three samples,
$0.85 \le \Delta \mathcal{P}_{\mathrm{rel}} \le 2.4$,
support the $41.5 \times \Pmer$ relation.
The signal detection
$\mathcal{P}=41.529\pm0.034$
for sample \RmonthlyOne ~is
extremely accurate, only $\Delta M_{\mathrm{rel}}=0.00070$
during one revolution of Mercury.
The mean anomaly migration of
the respective  highlighted
red circles is only 0.79 revolutions during
1137 revolutions (Fig. \ref{FigAnomaliesOne}c).

\subsubsection{Signal \SignalTwo
  ~$\cong 10 \times \Pear$}

The relative rounds deviation 
$\Delta \mathcal{P}_{\mathrm{rel}}=0.012$,
1.5 and 2.3 values
of three samples
support the $10 \times \Pear$ relation
for \SignalTwo ~signal
(Table \ref{TablePromising}: 
\RmonthlyOne, \RyearlyOne, \CyearlyOne).
For the fourth sample \CmonthlyOne,
the rounds $\mathcal{P}=10.0659\pm0.0077$
value has a large relative rounds deviation
$\Delta \mathcal{P}_{\mathrm{rel}}=8.6$.
However, the mean anomaly migration
$|\Delta M/360^{\mathrm{o}}|$ values
for all four samples are smaller than
1.3 revolutions during two,
or even three, centuries.

The true anomaly
$\nu_{\mathrm{min}}$ and $\nu_{\mathrm{max}}$
migration curves of Earth
are exceptionally clear
(Fig. \ref{FigAnomaliesOne}d).
The lines of small and large circles of every
sample show the overlapping effect caused by the
even number  $10 \times \Pear$ multiple.
This overlapping effect is even more pronounced than in
Fig. \ref{FigAnomaliesOne}b,
because the results for all four samples
are based on pure sine models.
The true anomaly
$\nu_{\mathrm{min}}$ and $\nu_{\mathrm{max}}$
migration convergence point of
all four signals is at $t_C=1910$ years and
$\nu \approx
M_C=$ 50 degrees
(Fig. \ref{FigAnomaliesOne}d:
centre of blue and red dotted circles).
This convergence point is {\it inside} all four samples.
All these four
\SignalTwo ~signals in different samples
can undoubtedly originate from one and the same signal.

\begin{figure}  
\centerline{\hspace*{0.015\textwidth}
         \includegraphics[width=0.515\textwidth,clip=]{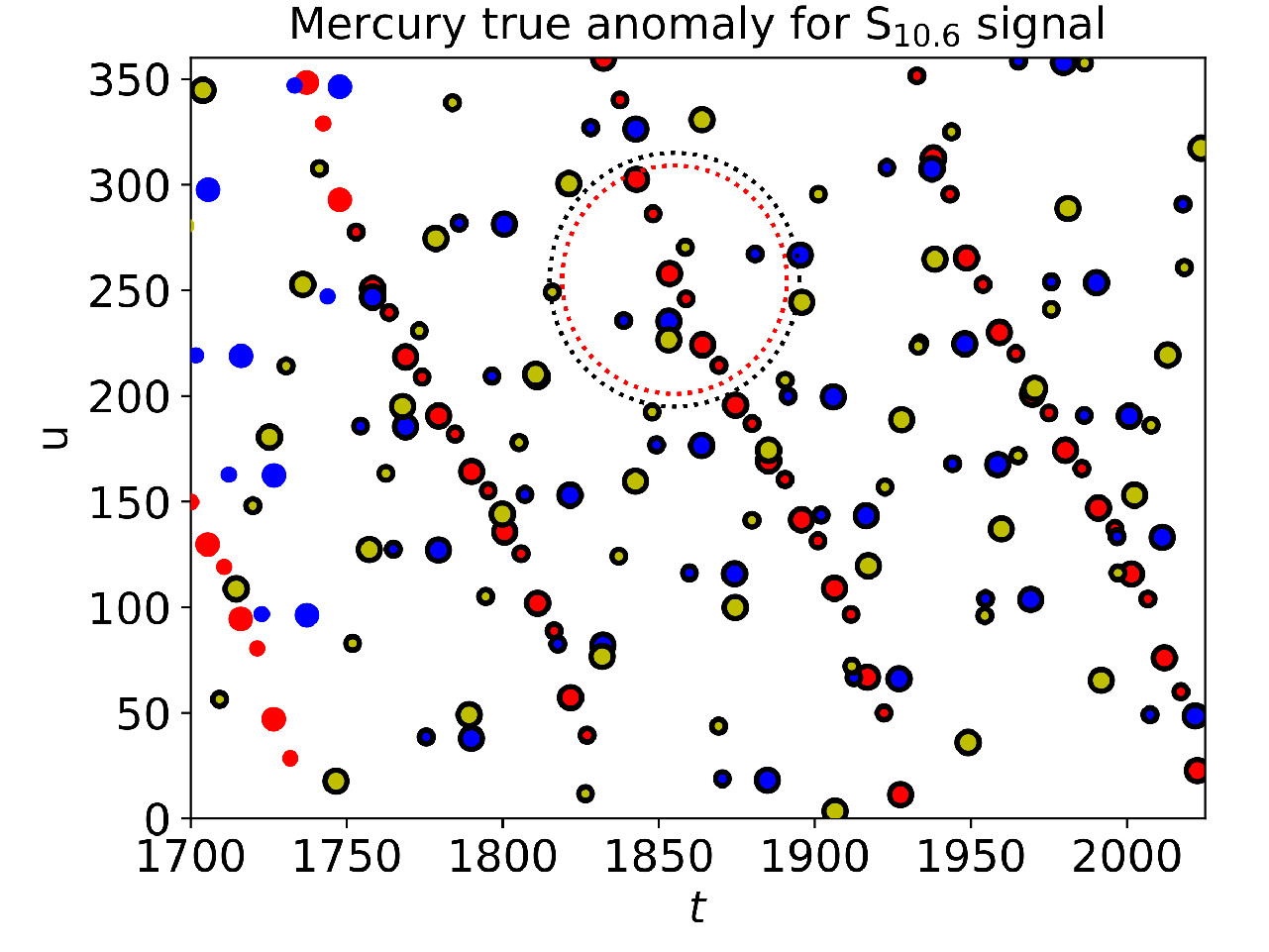}   
         \hspace*{-0.03\textwidth}
         \includegraphics[width=0.515\textwidth,clip=]{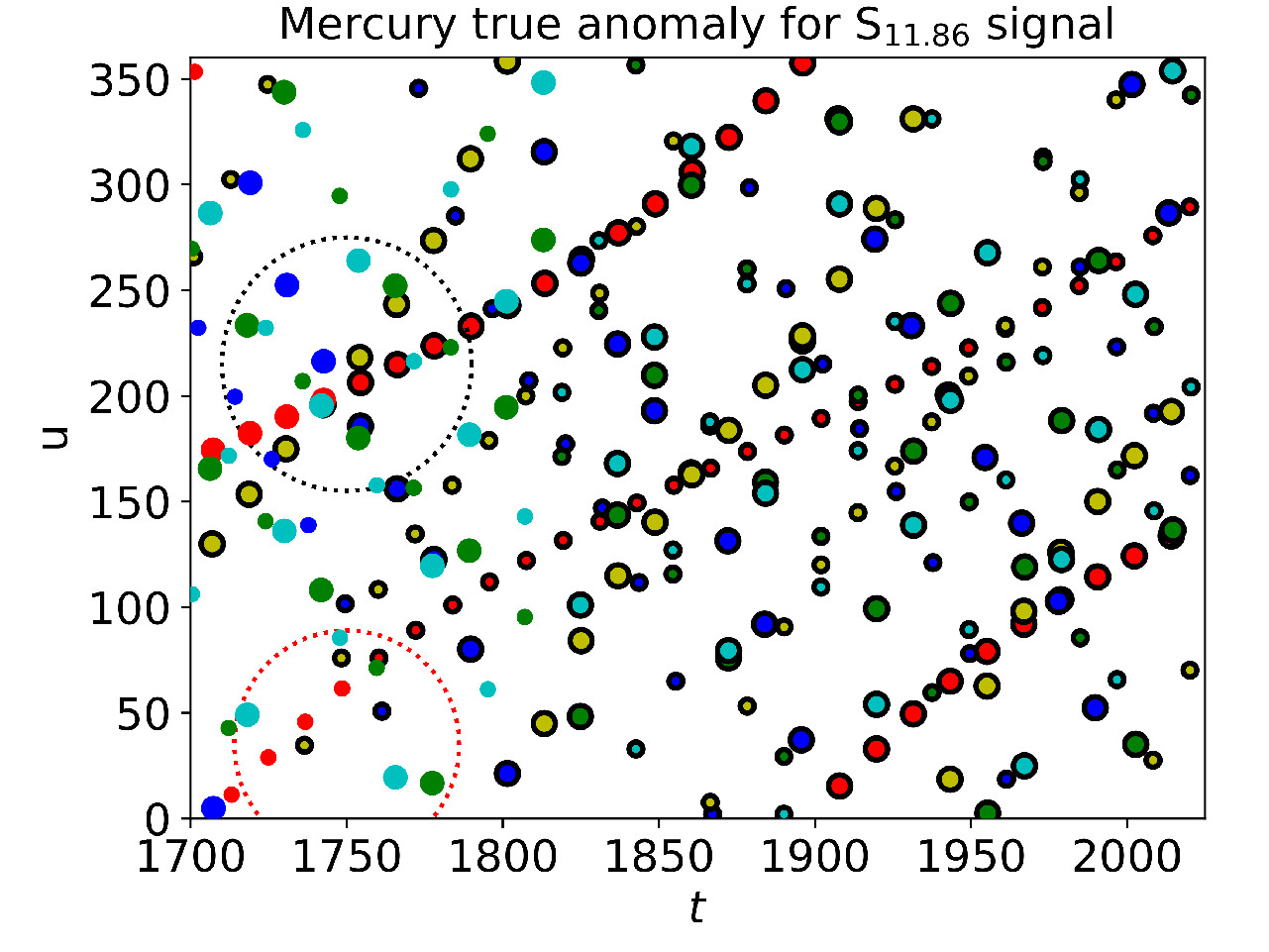} 
        }
\vspace{-0.35\textwidth}
\centerline{\Large \bf 
\hspace{0.42\textwidth}  \color{black}{(a)}
\hspace{0.43\textwidth}  \color{black}{(c)}
\hfill}
\vspace{0.31\textwidth}    
\centerline{\hspace*{0.015\textwidth}
         \includegraphics[width=0.515\textwidth,clip=]{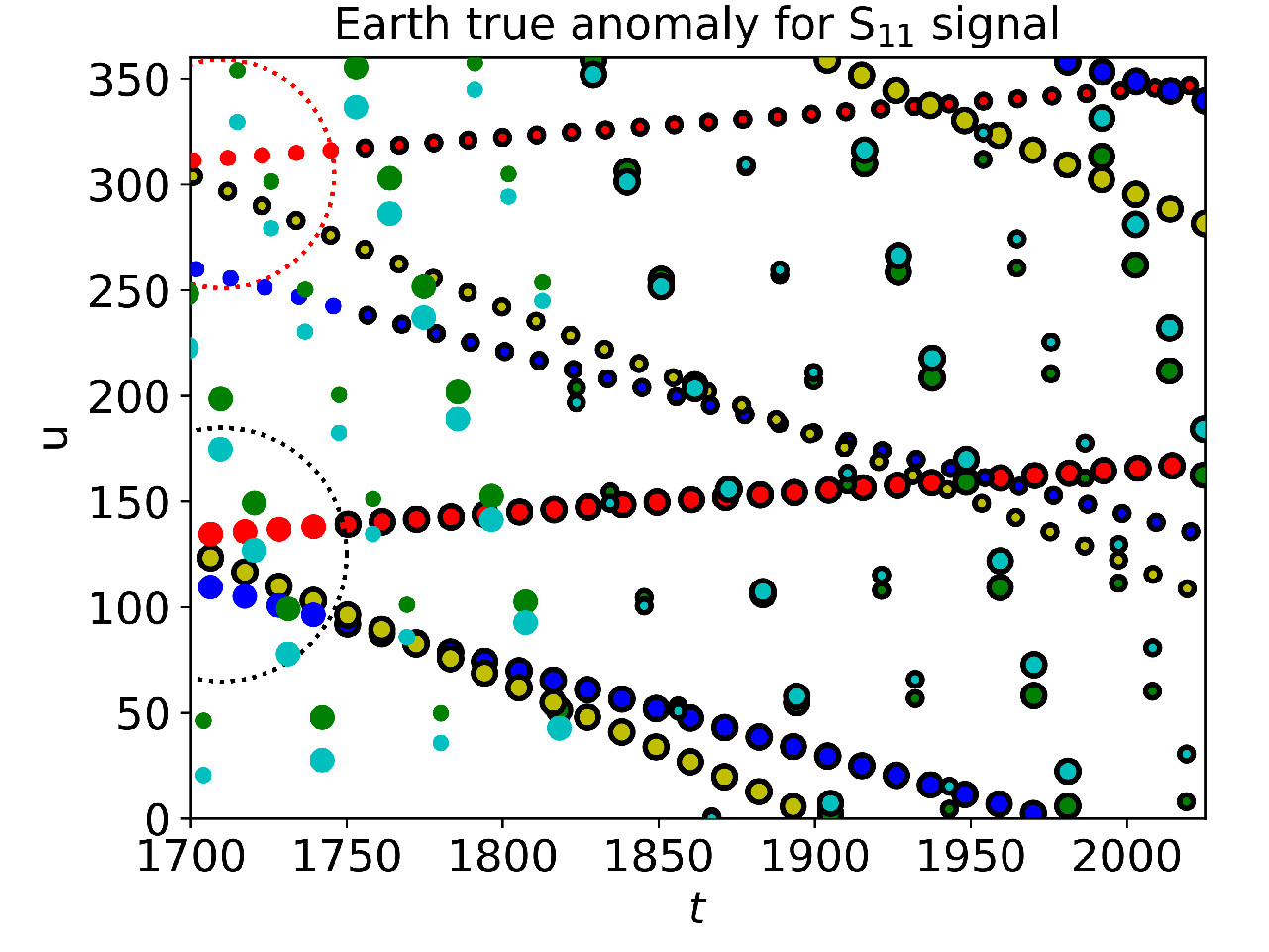}  
         \hspace*{-0.03\textwidth}
         \includegraphics[width=0.515\textwidth,clip=]{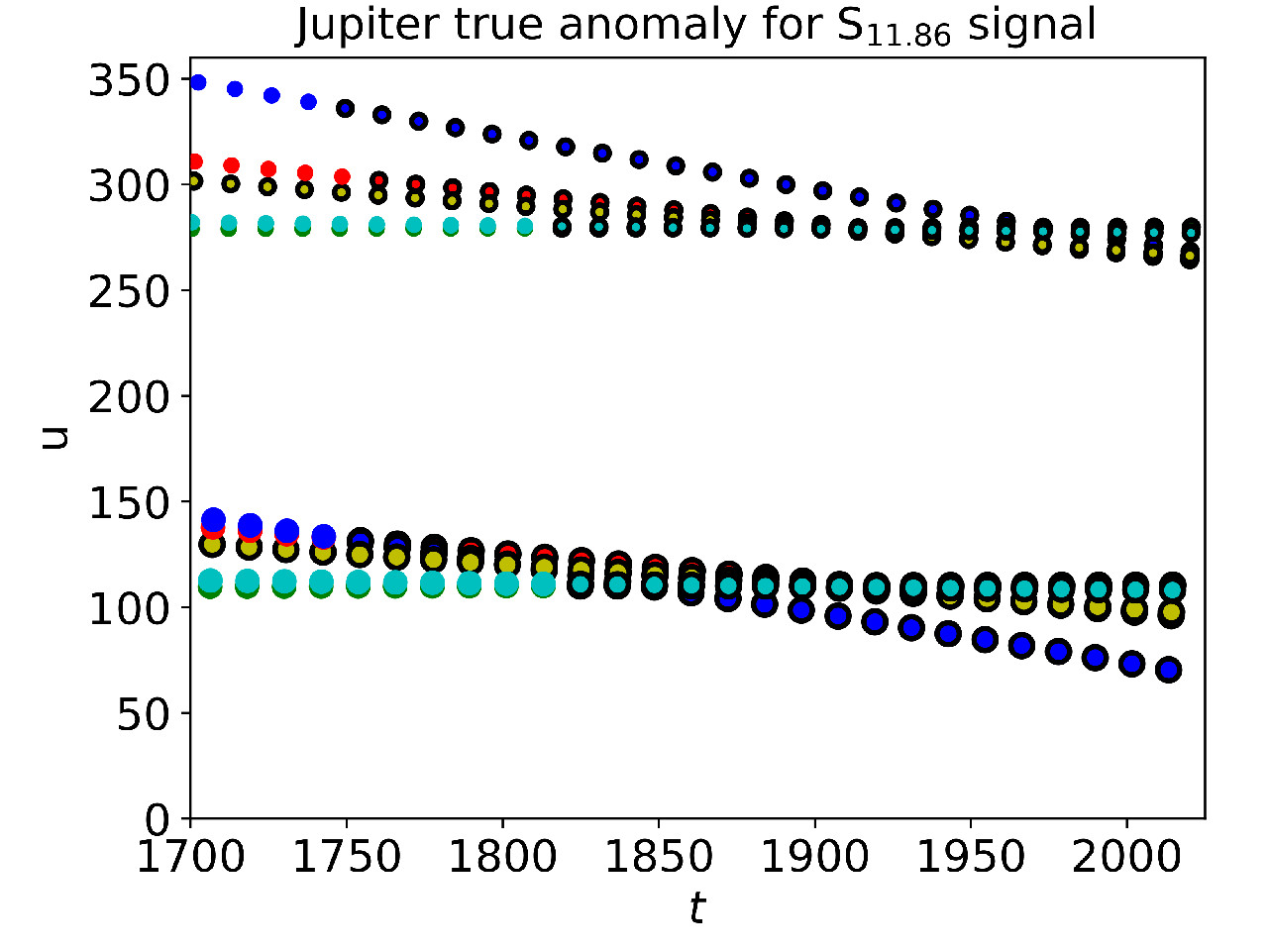} 
        }
\vspace{-0.35\textwidth}   
\centerline{\Large \bf     
\hspace{0.42 \textwidth}   \color{black}{(b)}
\hspace{0.43 \textwidth}   \color{black}{(d)}
   \hfill}
 \vspace{0.30\textwidth}    
 \caption{(a) Mercury's true anomalies
   for signal \SignalFive.
   (b) Earth's true anomalies
   for signal \SignalOne.
   (c) Mercury's true anomalies
   for signal \SignalThree.
   (d) Jupiter's true anomalies
   for signal \SignalThree.
    Otherwise as in Fig. \ref{FigAnomaliesOne}.} 
         \label{FigAnomaliesTwo}
       \end{figure}

\subsection{Signal \SignalFive $\cong 44 \times \Pmer$}

The signal \SignalFive ~period is close to
$44 \times \Pmer=$ 10.595 years.
The identification of the convergence point
of this signal is quite complicated.
It is not easy to perceive the correct 
true anomaly migration directions of the signals
detected in three different samples
(Fig. \ref{FigAnomaliesTwo}a).
The mean anomaly migration values
$\Delta M/360^{\mathrm{o}}=-2.8$, -5.9 and 7.8
reveal that the red and the blue circles migrate
downwards, and the yellow circles migrate upwards
(Table \ref{TablePromising}, Signal \SignalFive).
The large and small circles of each sample 
show the overlapping effect, because this signal 
is an even number multiple $44 \times \Pmer$.
The red and yellow circles of the symmetric pure sine model
for samples \RmonthlyOne ~and \RyearlyOne ~follow
this regularity.
However, the blue small circles for the
asymmetric double wave model of
sample \RmonthlyOne ~break this regularity.
Using all the above-mentioned
information, we identify the convergence
points of both $\nu_{\mathrm{min}}$ and $\nu_{\mathrm{max}}$
true anomalies
at $t_C=1855$ years and $\nu \approx M_C=255$ degrees
(Fig. \ref{FigAnomaliesTwo}a:
centre of blue and red dotted lines).
Due to the strong migration,
the same pattern is repeated at
the approximate 
year and degree coordinates (1760,220) and (1940,280).

The rounds values $\mathcal{P}=43.891\pm0.083$
and $44.26\pm0.13$
differ only $\Delta \mathcal{P}_{\mathrm{rel}}=1.3$
and 2.0 from
$44 \times \Pmer$ (Table \ref{TablePromising}:
Signal \SignalFive).
The  $\mathcal{P}=43.774\pm0.0824$ value for 
sample \RmonthlyOne ~double wave model has a large 
$\Delta \mathcal{P}_{\mathrm{rel}}=9.0$
relative rounds deviation.
However,
all three relative mean anomaly migration
$\Delta M_{\mathrm{rel}}$
values 0.0025, 0.0052 and 0.0059 are nearly the same.

\subsection{Signal \SignalOne ~$\cong 11 \times \Pear$}

This strongest \SignalOne ~signal
is close to $11 \times \Pear$.
The non-weighted DCM analysis gives the
rounds values $\mathcal{P}=11.0033\pm0.0064$,
$10.9878\pm0.0051$ and $10.981 \pm 0.021$
(Table \ref{TablePromising}: \RmonthlyOne, \RyearlyOne).
The respective relative rounds deviation
$\Delta \mathcal{P}_{\mathrm{rel}}=0.52$, 0.24 and 0.95 values 
support the $11 \times \Pear$ relation.
The rounds values $\mathcal{P}=10.8585\pm0.0048$
and $10.863\pm0.022$ obtained from the weighted DCM analysis,
however, show large $\Delta \mathcal{P}_{\mathrm{rel}}=29$
and 6.2 relative rounds deviations
(Table \ref{TablePromising}: \CmonthlyOne, \CyearlyOne).

The uneven $11 \times \Pear$ multiple
  causes the 180 degrees
separation effect of the
$\nu_{\mathrm{min}}$ and
$\nu_{\mathrm{max}}$ true anomalies.
Regardless of the above-mentioned
large $\Delta \mathcal{P}_{\mathrm{rel}}$
values for samples \CmonthlyOne ~and \CyearlyOne,
we obtain very convincing estimates for 
the convergence points of the true anomaly
$\nu_{\mathrm{min}}$ and $\nu_{\mathrm{max}}$
migration curves (Fig. \ref{FigAnomaliesTwo}b).
The maxima converge at $t_C=1710$ years
and $\nu \approx M_C=125$ degrees.
Only the highlighted yellow circles of \RyearlyOne
~reach this convergence point, because it is {\it outside}
samples \RmonthlyOne, \CmonthlyOne ~and \CyearlyOne.
Yet, the large ``predictive''
red, blue, green and cyan circles 
migrate through the large dotted blue circle
surrounding this convergence point
(Fig. \ref{FigAnomaliesTwo}b).
The small circles show that the minima
$\nu_{\mathrm{min}}$  converge
at $t_C=1710$ and $\nu \approx M_C=305$ degrees.
The small blue circles
denoting the asymmetric \RmonthlyOne ~double wave model,
represents the only migration curve that
deviates from this convergence point.
Both convergence points in the year 1710
  are located about 50 years before the beginning of
  \RmonthlyOne ~and \CmonthlyOne ~samples
  (the highlighted red and blue circles)
  and
  about 100 years before the beginning of
  \CmonthlyOne ~and \CyearlyOne ~samples
  (the highlighted green and cyan circles).
  The successful convergence of the
  ``predictive'' red, blue, green and cyan
  circles (not highlighted with black circles)
  indicates that
  all five detected \SignalOne ~signals can represent
  one and the same real signal.
  
\subsection{Signal \SignalThree}

The \SignalThree ~signal period is close
to $49.0 \times \Pmer$ and $1.0 \times \Pjup$.
The known period ratio is $\Pjup/\Pmer=49.2608$.
This pair of \SignalThree ~signal multiples
displays the most dramatic statistical
$|\Delta \mathcal{P}|$ fluctuation effect
(Equations \ref{EqProbP} and \ref{EqMigration}).
The messy Mercury's migration curves
(Fig. \ref{FigAnomaliesTwo}c)
have exceptionally regular
Jupiter's migration curve counterparts
(Fig. \ref{FigAnomaliesTwo}d).
        
\subsubsection{Signal \SignalThree
  ~$\cong 49.0 \times \Pmer$}

None of the five
$\mathcal{P}$ rounds values differs more than
$3 \Delta \mathcal{P}_{\mathrm{rel}}$
from $49.0\times \Pmer$
(Table \ref{TablePromising}: Signal \SignalThree).
No other planetary signal candidate
shows this degree of regularity. 
All relative mean anomaly migration
values fulfil $\Delta M_{\mathrm{rel}}\le0.0054$.
For this particular  $\Delta M_{\mathrm{rel}}$ parameter,
this is also the best result among all
planetary signal candidates.
Even the largest mean anomaly
migration $\Delta M/360^{\mathrm{o}}$
is only 4.6 revolutions during 850
revolutions of Mercury.
The smallest mean anomaly migration
for sample \RmonthlyOne ~$(K_2=1)$
is only 
$\Delta M/360^{\mathrm{o}}=0.65$ during 1137 revolutions.
The twisted large and small red circle curves
of this sample 
nicely illustrate 
Mercury's orbit eccentricity effect
(Fig. \ref{FigAnomaliesTwo}c).

The uneven $49 \times \Pmer$ multiple causes
the 180 degrees separation effect of
$\nu_{\mathrm{min}}$ and  $\nu_{\mathrm{max}}$
values
(Fig. \ref{FigAnomaliesTwo}c).
The red, green, yellow and cyan circles of
four samples show positive upwards migration.
The blue circles denoting the double wave model
results for sample \RmonthlyOne ~show
the only negative downwards migration.
The curves of the large highlighted
red and yellow circles
denoting the $\nu_{\mathrm{max}}$ values
{\it inside} samples
\RmonthlyOne ~$(K_2=1)$ and \RyearlyOne ~intersect
at the convergence point $t_C=1750$ and
$\nu \approx M_C=215$ degrees.
The respective large ``predictive''
blue, green and cyan circle curves,
which are {\it outside} samples
\RmonthlyOne ~$(K_2=2)$,
\CmonthlyOne ~and \CyearlyOne,
also intersect at the centre
of the large blue dotted circle
surrounding this convergence point.
The respective  $\nu_{\mathrm{min}}$
convergence point is at $t_C=1750$ and
$\nu \approx M_C=35$ degrees.
The two detected convergence points indicate
that all five \SignalThree ~signals detected in
different samples can represent one and the
same signal.

\begin{figure}  
\centerline{\hspace*{0.015\textwidth}
         \includegraphics[width=0.515\textwidth,clip=]{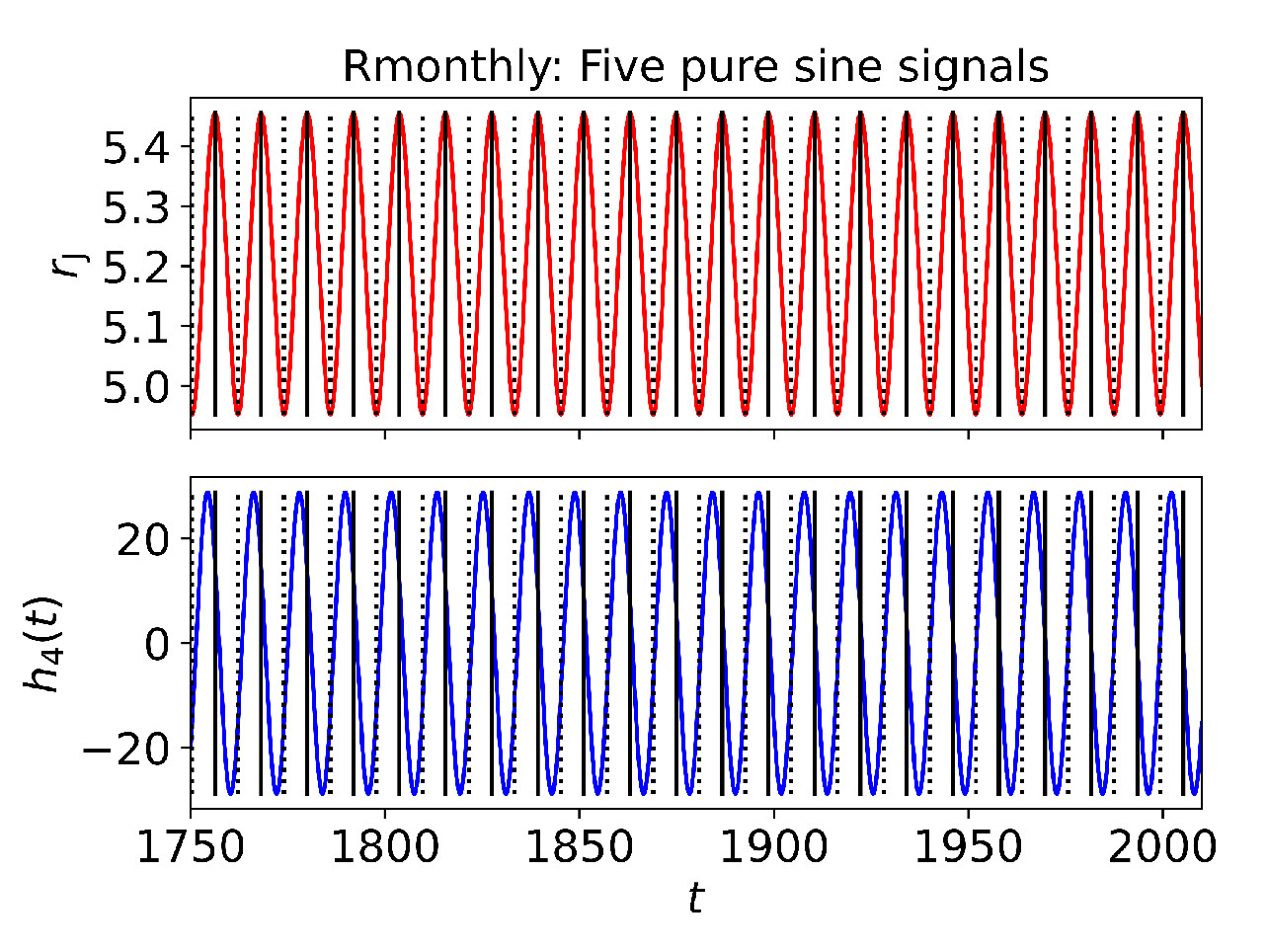} 
         \hspace*{-0.03\textwidth}
         \includegraphics[width=0.515\textwidth,clip=]{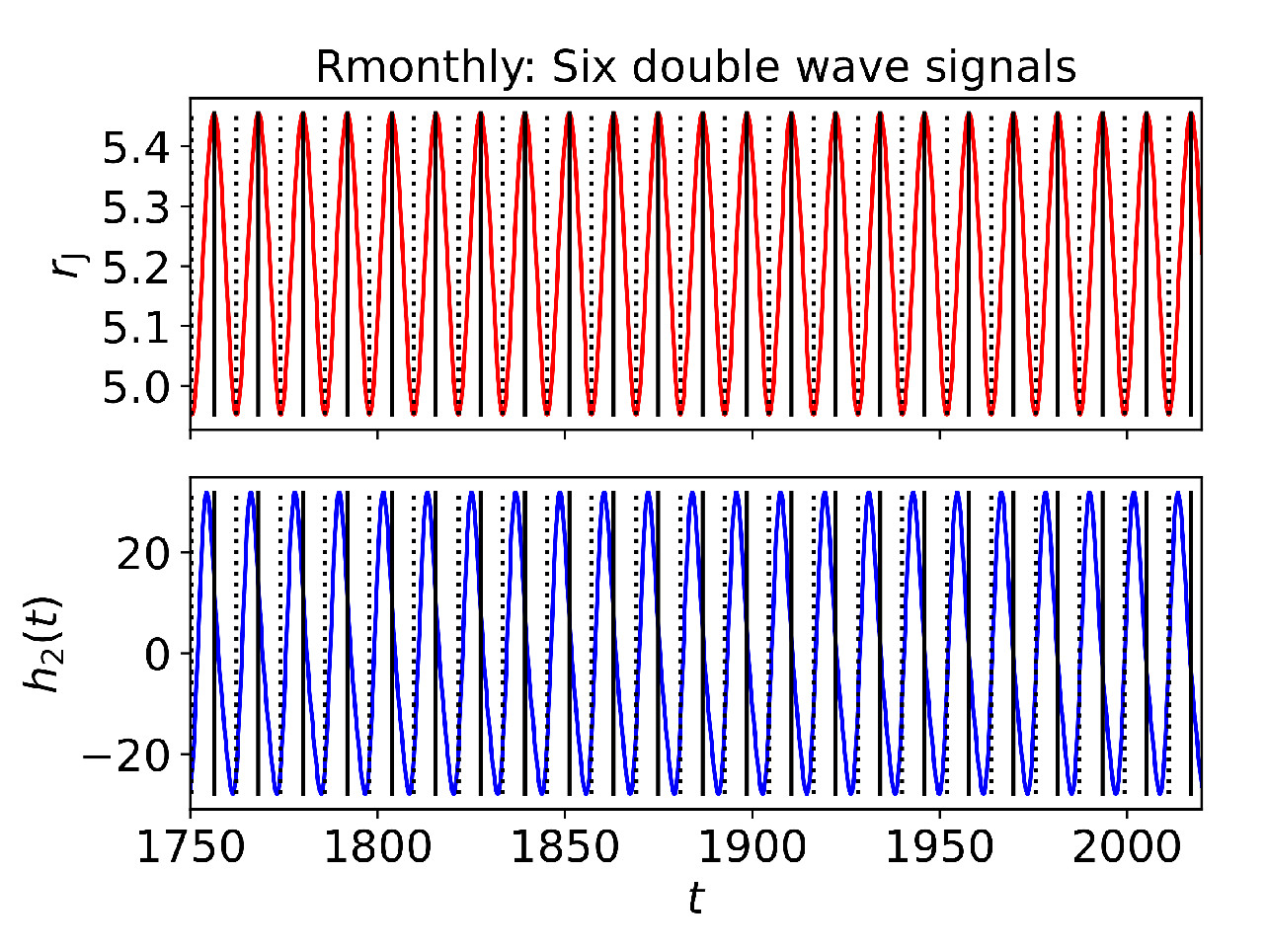} 
        }
\vspace{-0.39\textwidth}
\centerline{\Large \bf 
\hspace{0.43\textwidth}  \color{black}{(a)}
\hspace{0.43\textwidth}  \color{black}{(b)}
\hfill}
\vspace{0.34\textwidth}    
\centerline{\hspace*{0.015\textwidth}
         \includegraphics[width=0.515\textwidth,clip=]{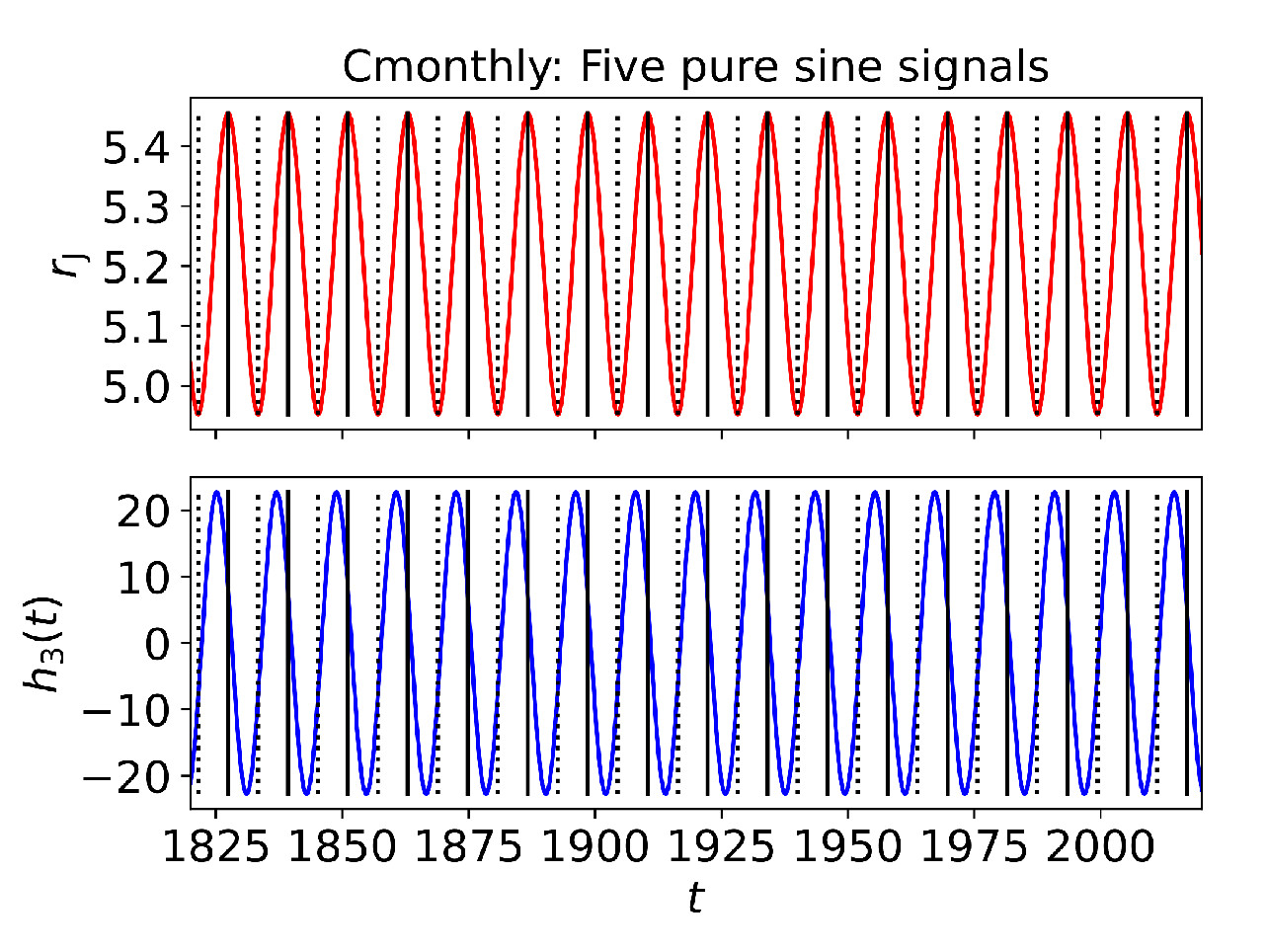} 
         \hspace*{-0.03\textwidth}
         \includegraphics[width=0.515\textwidth,clip=]{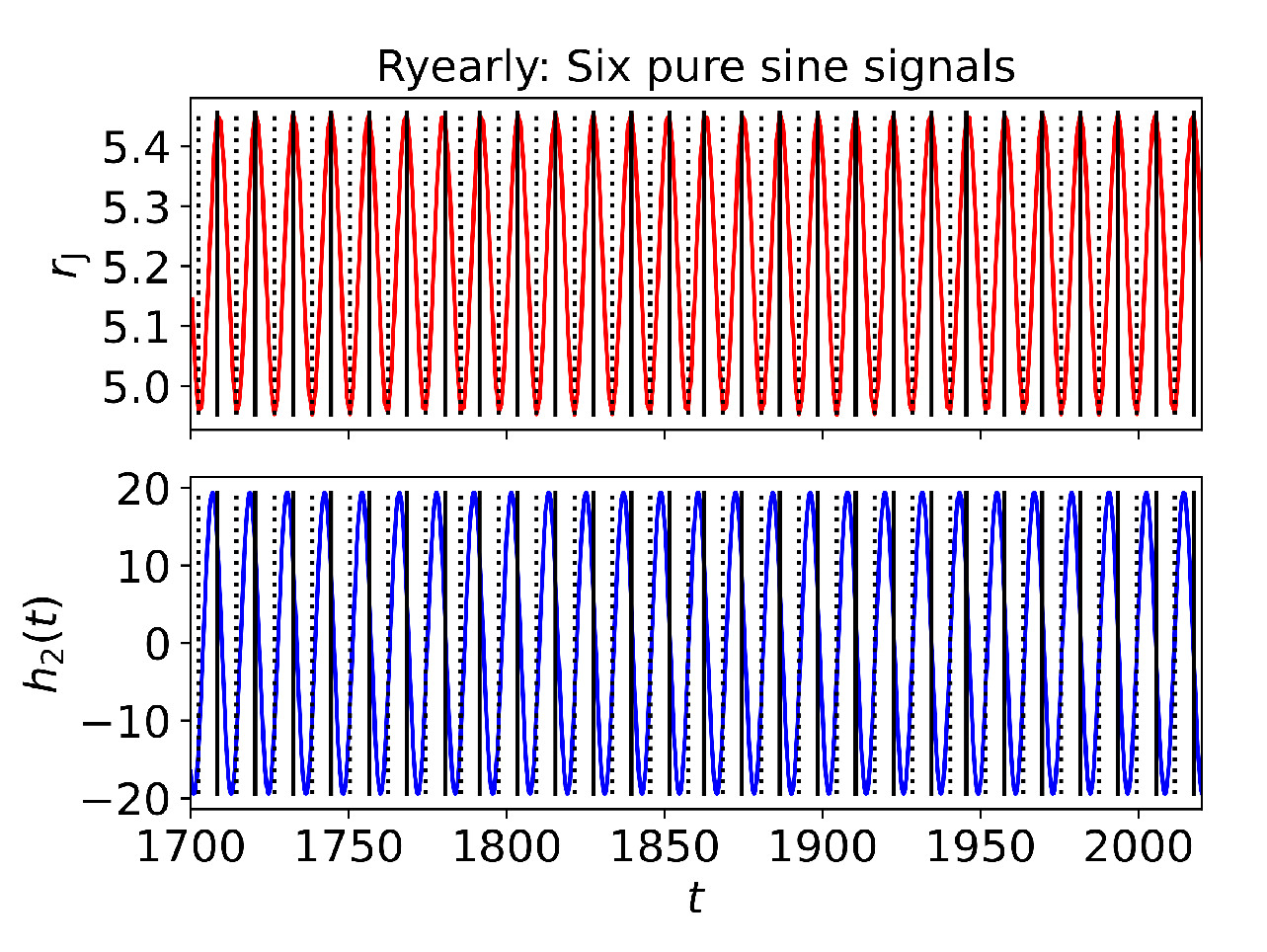} 
        }
\vspace{-0.39\textwidth}   
\centerline{\Large \bf     
\hspace{0.43 \textwidth}   \color{black}{(c)}
\hspace{0.43 \textwidth}   \color{black}{(d)}
   \hfill}
 \vspace{0.20\textwidth}    
\vspace{0.14\textwidth}    
\centerline{\hspace*{0.015\textwidth}
         \includegraphics[width=0.515\textwidth,clip=]{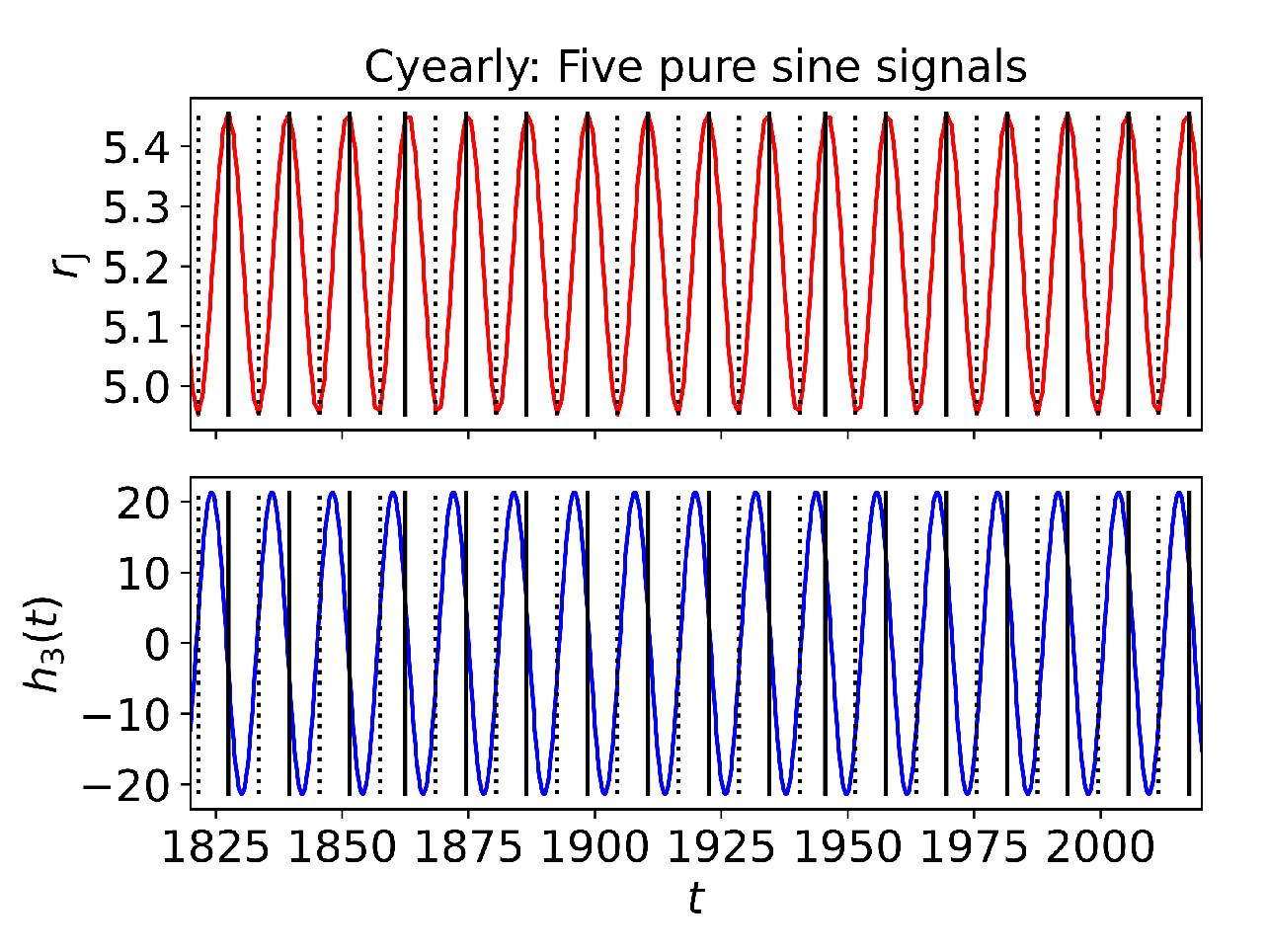}
         \hspace*{-0.03\textwidth}
        }
\vspace{-0.39\textwidth}   
\centerline{\Large \bf     
\hspace{0.68 \textwidth}   \color{black}{(e)}
   \hfill}
 \vspace{0.33\textwidth}    
 \caption{Signal \SignalThree ~connection to
   Jupiter's distance $r_{\mathrm{J}}$ from the Sun.
          (a) 
          Upper panel: Red curve shows distance
          between the Sun and Jupiter.
          Aphelion and perihelion epochs are denoted
          with vertical black
          continuous and dotted lines, respectively.
          Units are x-axis $[t]=$ years and y-axis
          $[r_{\mathrm{J}}]=$ AU.
          Lower panel: Blue curve shows
          simultaneous $h_4(t)$ pure sine
          $11.807$ year signal in \RmonthlyOne
          ~(Table \ref{TableRmonthlyK410R14}).
          Aphelion and perihelion notations are as in upper panel.
          Units are x-axis $[t]=$ years and y-axis
          $[h_4(t)]=$ dimensionless.
          (b) $h_2(t)$ double wave $11.770$ year signal
           ~(Table \ref{TableRmonthlyK420R14}).
           Otherwise as in ``a''.
          (c) $h_3(t)$ pure sine $11.863$ year signal
          in \CmonthlyOne
          ~(Table \ref{TableCmonthlyK410R14}).
           Otherwise as in ``a''.
          (d) $h_2(t)$ pure sine $11.826$ year signal
          in \RyearlyOne
          ~(Table \ref{TableRyearlyK410R14}).
          Otherwise as in ``a''.
          (e) $h_3(t)$ pure sine $11.856$ year signal
          in \CyearlyOne
          ~(Table \ref{TableCyearlyK410R14}).
           Otherwise as in ``a''.  } 
          \label{FigJupiterDistance}
        \end{figure}

\subsubsection{Signal \SignalThree
  ~$\cong 1.0 \times \Pjup$}
\label{SectJupiter}

Three rounds
$\mathcal{P}=1.0001\pm0.0018$,
$0.9964\pm0.0023$ and $0.9995\pm0.0057$ values
support the $1 \times \Pjup$ relation
(Table \ref{TablePromising}: Signal \SignalThree).
Two rounds values for \RmonthlyOne,
$\mathcal{P}=0.9953\pm0.0010$ and
$0.99224\pm0.00093$, differ
$\Delta P_{\mathrm{rel}}=4.7$ and 8.3
from this relation.
However, all mean anomaly migration values
fulfil $|\Delta M|\le0.18$ revolutions.

The uneven $1\times\Pjup$ multiple causes the
180 degree separation effect
between Jupiter's true anomaly
$\nu_{\mathrm{min}}$
and $\nu_{\mathrm{max}}$ values
(Fig. \ref{FigAnomaliesTwo}b).
All true
anomaly migration curves are so regular that
there is no need to search for convergence points.
The small blue circles, which denote
the $\nu_{\mathrm{min}}$ values of the asymmetric
double wave model for sample \RmonthlyOne,
show the largest deviation from the general migration
trends. The regular
convergence of Jupiter's true anomalies
definitely indicates that one and the
same signal is detected in all five different samples.
The strongest amplification of the 
  \SignalThree ~sunspot signal
occurs close to true anomaly $\nu=\nu_{\mathrm{max}}=120^{\circ}$.
The damping of this signal occurs about 6 years (11.86/2) later,
close to true anomaly $\nu=\nu_{\mathrm{min}}=300^{\circ}$.

The detection of the 
convergence points for the signal
candidates of Mercury and Jupiter
  indicates that the
relative motions of these two planets
can cause this \SignalThree ~signal.

The maxima of all five \SignalThree ~signals
in Table \ref{TablePromising} occur between
Jupiter's perihelion and aphelion, which are
denoted with dotted and continuous vertical lines
in Fig. \ref{FigJupiterDistance}.
All \SignalThree ~signal
minima are between Jupiter's
aphelion and perihelion.
The phases of all five \SignalThree ~signals are stable,
and stay in phase with Jupiter's orbital motion.
The signal amplitudes in the monthly and the yearly samples are
nearly the same.
This \SignalThree ~signal is clearly connected to the
orbital motion of Jupiter.
The distance between the Sun and Jupiter
modulates the number of sunspots.

All these results confirm
an irrefutable deterministic
connection between the sunspot cycle and Jupiter's orbital 
motion.
  If these results do not represent the {\it direct}
  detection of Jupiter from the sunspot data,
  then, what have we detected?
  It would be quite a coincidence
  if we had managed to formulate a flawed
  statistical method that would accidentally detect
  Jupiter's period from the sunspot data.
 
\subsection{Signal \SignalSeven ~$\cong 4.5 \Pjup$} 

The two rounds $\mathcal{P}=4.439\pm0.023$
and $4.538\pm0.064$ differ only
$\Delta \mathcal{P}_{\mathrm{rel}}=2.6$ and 0.59
from $4.5 \times \Pjup$
(Table \ref{TablePromising}: \SignalSeven).
We merely mention this regularity, but we do not
search for converge points from the migration
of only two curves.

\subsection{Signal \SignalNine
  ~$\cong 5.5 \times \Pjup$ or  $\cong 6.0\times \Pjup$}

This weakest one of all detected signals
has a rounds value $\mathcal{P}=5.62\pm0.12$
(Table \ref{TablePromising}: Signal \SignalNine).
We mention this \SignalNine
~signal because the double sinusoid
signal period $P=143.30\pm0.99$
(Table \ref{TableMultiples}: Signal \SignalNine) has
a rounds value $\mathcal{P}=12.088\pm0.083$,
which is close to $2 \times 6 \times \Pjup$.

\setcounter{page}{30} 

~
\end{document}